\pdfoutput=1
\documentclass[12pt,oneside,letterpaper,final]{emorythesis}
\usepackage[colorlinks=true,
           linkcolor=blue,
           urlcolor=blue,
           citecolor=blue]{hyperref}
\usepackage{caption}
\usepackage{graphicx,xcolor,color}	
\usepackage{amssymb}
\usepackage[numbers,square,comma,sort&compress]{natbib} 
\usepackage{amsmath}
\usepackage{soul}
\usepackage{mathtools}
\usepackage{multirow}
\usepackage{amssymb}
\usepackage{amsbsy}
\usepackage{dcolumn}
\usepackage{bm}
\usepackage{amsfonts}
\usepackage{tabu}
\usepackage{multirow}
\usepackage{amsthm}
\usepackage{tikz}
\usepackage{bibunits}
\usepackage{pdfpages}

\graphicspath{{Figures/}}
\begin{document}

\dsp
\includepdf[pages={1,2}]{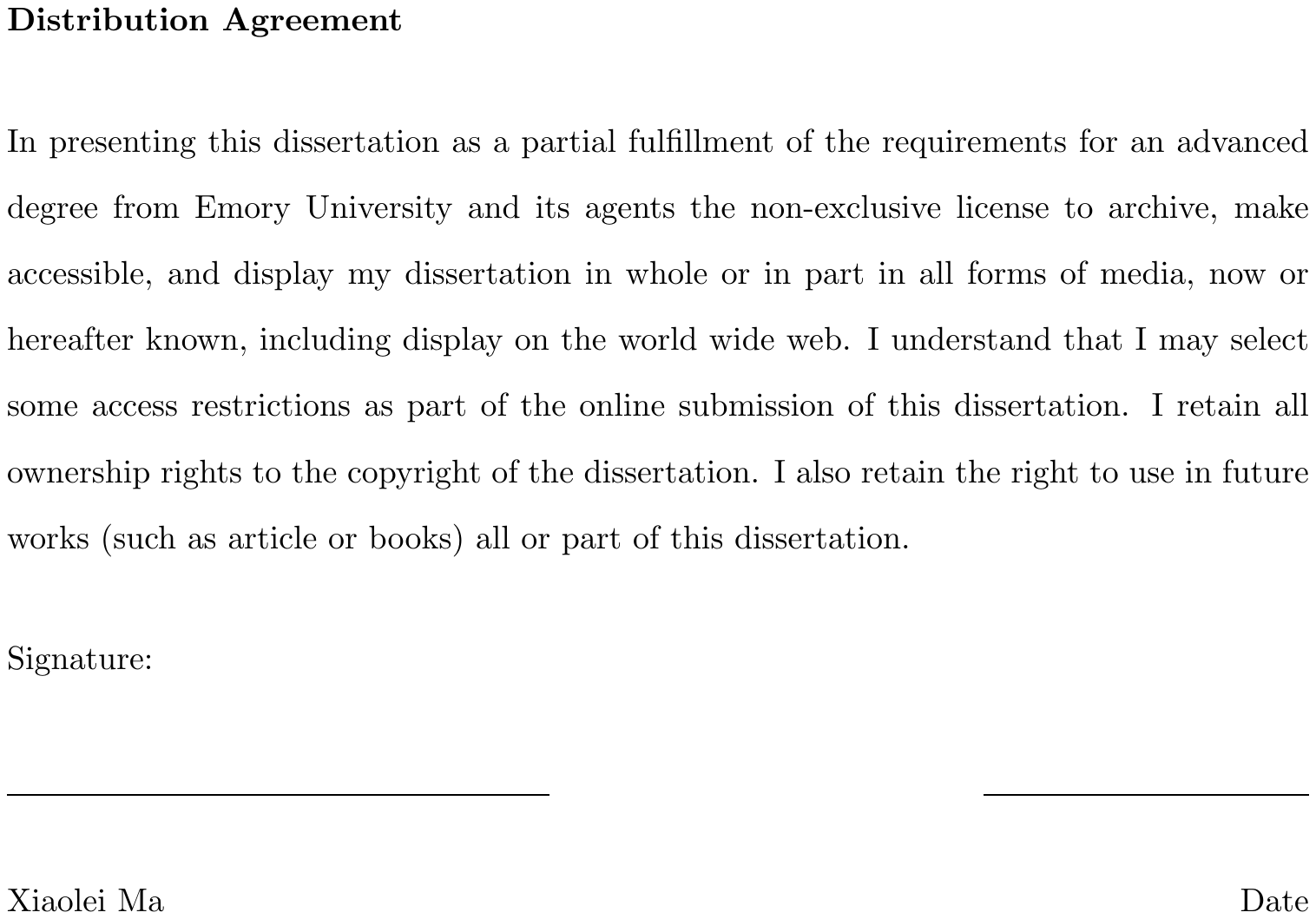}

\title{Experimental investigations on the nonequilibrium dynamics of pattern formation in fluid and granular systems}
\author{Xiaolei Ma}
\degreemonth{} 
\degreeyear{2019}
\degree{Doctor of Philosophy}
\field{Physics}
\department{Physics}
\advisor{Justin Clifford Burton} 

\maketitle

\begin{abstract}

Patterns are quotidian in nature and occur on multiple length scales. Distinct multiscale patterns are generally a consequence of nonequilibrium dynamical processes associated with particular mechanical and hydrodynamic instabilities, which play a vital role in shaping the pattern geometry. In this thesis, I report experimental investigations on the pattern formation in a few examples of fluid and granular systems, and uncover the underlying mechanisms that give rise to those patterns.

Leidenfrost drops are known to experience star-shaped oscillations with little damping. However, the underlying mechanism remains unclear. Here I report that the hydrodynamic coupling between the rapid evaporated vapor flow and vapor-liquid interface excites the star-shaped oscillations, suggesting a purely hydrodynamic origin. In addition, I also give an analytical explanation for an oscillatory ``breathing mode'' found in small Leidenfrost drops.

Polygonal desiccation crack patterns are commonly observed in natural systems. However, it is unclear whether similar crack patterns spanning multiple length scales share the same underlying physics. I also report experimental investigation on polygonal cracks in drying suspensions of micron-sized particles. In cornstarch-water mixtures, multi-scale crack patterns were observed due to two distinct desiccation mechanisms. In addition, we find that the characteristic area of the polygonal cracks ($A_p$), and film thickness ($h$) obey a universal power law, $A_p=\alpha h^{4/3}$. Thus we provide a robust framework for understanding multiscale polygonal crack patterns.

Finally, I report experimental results on sedimentation of non-Brownian particles in viscous fluids, which is crucial in both nature and industrial processes. We observed an effective repulsion between particles with nonuniform density in both two-body and many-body systems, in contrast to particles with uniform density. This trend holds true in two and three dimensions. In addition, we also characterize the statistical properties of the sedimentation patterns of particles in three dimensions. Our results also shed light on the potential for controlling the uniformity of  particle layers after sedimentation.

The patterns I report in this thesis represent typical examples in fluid and granular systems that are driven by nonequilibrium dynamics, and the underlying mechanisms we uncover are expected to enhance our understanding of how these seemingly simple patterns can arise in natural systems.

\end{abstract}

\makecoverpage


\newpage
\clearpage
\phantomsection
\addcontentsline{toc}{section}{Table of Contents}
\tableofcontents

\listoffigures
\listoftables

\begin{citations}

\vspace{0.4in}

Chapter 2 contains research from the following 3 publications:
\begin{quote}
$\bullet$\textit{``The many faces of a Leidenfrost drop''}, \textbf{Xiaolei Ma}, Juan-Jos\'{e} Li\'{e}tor-Santos, Justin C. Burton, \textit{Physics of Fluids}, \textbf{27}, 091109, (2015) \\   

$\bullet$\textit{``Star-shaped oscillations of Leidenfrost drops''}, \textbf{Xiaolei Ma}, Juan-Jos\'{e} Li\'{e}tor-Santos, Justin C. Burton, \textit{Physical Review Fluids}, \textbf{2}, 031602(R), (2017) \\
  
$\bullet$\textit{``Self-organized oscillations of Leidenfrost drops''}, \textbf{Xiaolei Ma}, Justin C. Burton, \textit{Journal of Fluid Mechanics} \textbf{846}, 263 (2018)
\end{quote}

Chapter 3 contains research from the following publication:
\begin{quote}
$\bullet$\textit{``Universal scaling of polygonal desiccation crack patterns''}, \textbf{Xiaolei Ma}, Janna Lowensohn, Justin C. Burton, \textit{Physical Review E}, \textbf{99}, 012802 (2019)
\end{quote}

Chapter 4 contains research from the following paper:
\begin{quote}
$\bullet$\textit{``Interactions between non-Brownian particles with nonuniform density during sedimentation in viscous fluids''}, \textbf{Xiaolei Ma}, Justin C. Burton, to be submitted to \textit{Physical Review Fluids}
\end{quote}


\end{citations}

\begin{acknowledgments}

With the submission of this thesis, I would like to acknowledge the people who have played a key role on my way to pursing a Ph.D. degree in physics. 

First of all, I wish to thank Prof. Justin Burton, my thesis advisor and role model, for the wonderful research training under his supervision in the last five years. Discussing science with Justin is a really enjoyable experience during which I not only learned specific technical tools for research but more importantly got the spirit of thinking about problems as a physicist and to be an independent researcher. Besides the role as an academic advisor, he is also a great friend in my ordinary life, the activities he organized outside Emory campus are so much fun and made my life here really wonderful. It is his patience, bright personality, research passion, immense knowledge and financial support that lead to this thesis, and the times with him at Emory will always be appreciated in the rest of my life. 

I am especially grateful to Prof. Eric Weeks, one of my thesis committee board, both for the constructive suggestions he made for my research activities, and the permission and convenience he provided to do experiment in his lab. In addition, I also want to thank the rest of my thesis committee board: Prof. Alberto Fern\'{a}ndez-Nieves, Prof. Hayk Harutyunyan and Prof. Connie Roth for their insightful suggestions on my research. Besides, I wish to acknowledge my colleague Dr. Juan-Jos\'{e} Li\'{e}tor-Santos for his mentoring and extensive discussions with me when I initially joined the Burton Lab, and my colleagues Cong Cao and Carlos Orellana for many discussions about rheology. I also want to thank the current and past Burton Lab members: Nicholas Cuccia, Stephen Frazier, Yannic Gagnon, Guram Gogia, Dana Harvey, Janna Lowensohn, Joshua M\'{e}ndez, Asher Mouat, Justin Pye, Dominic Robe, Toyin Thompson, Clay Wood, Rui Wu, and Jiaqi Zheng for the awesome lab environment they created and useful discussions. I really thank you all for the laughters you brought to the lab, which made my times in the Burton Lab unforgettable. 

Special thanks go to my friends outside Emory, especially Prof. Guangyin Jing who has brought me to the field of soft matter. I really appreciate the inspiring and enlightening discussions with him, which also shed light for my future career. I also want to thank Hao Luo and Mo Zhou for the helpful discussions about drying crack patterns. And also my friends Xin Tian, Mingyuan He, Bo Jiang, Jiang Zhang, Long Yue, Zhan Zhao, Di Dong, Weixiao Hou, Fuyou Liao, Xiaohe Yan, Zhihui Yang, thanks for the joyful times with all of you, and the funny stories you shared with me.

Finally and most importantly, I wish to express my heartfelt gratitude to my family, this thesis would certainly not have been possible without their unrequited love, unwavering support, and faithful encouragement. \\


\noindent Xiaolei Ma\\
Emory University, Atlanta, USA \\
Spring 2019

\end{acknowledgments}

\dedication
\begin{quote}

\hsp
\em
\raggedleft

Dedicated to my parents\\

\end{quote}

\copyrightpage

\newpage

\startarabicpagination


\chapter{Introduction}
\label{introduction}

\section{Pattern formation in nature}

\begin{figure}
\begin{center}
\includegraphics[width=6 in]{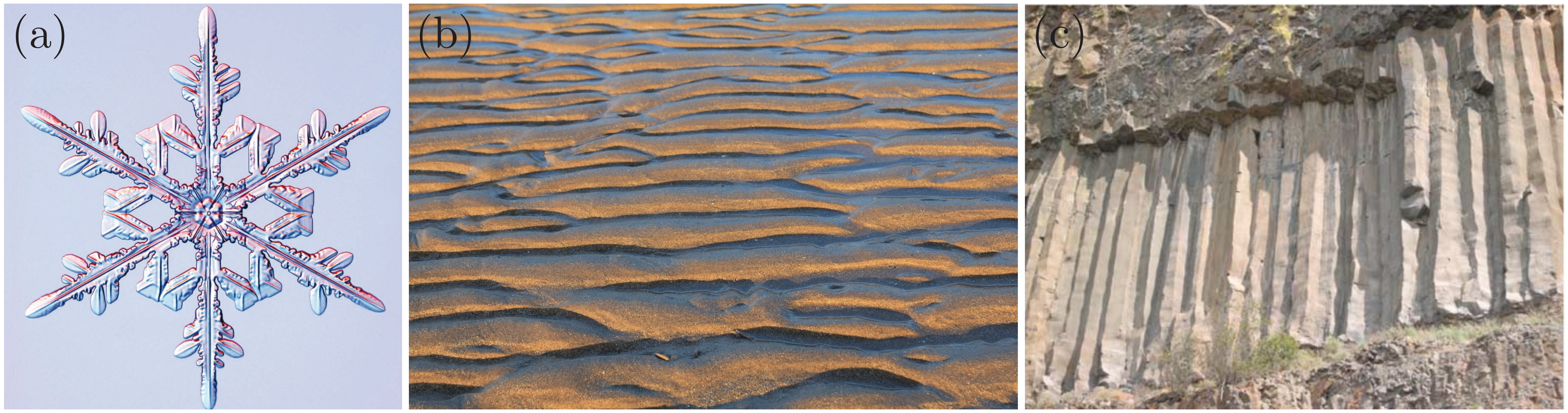}
\caption[Patterns in nature]{Patterns in nature. (a) (b) and (c) show the patterns of a crystalline structure in a snowflake, sand ripples, and a basalt columnar jointing, respectively. Image (a) is reproduced from Ref. \cite{libbrecht2008snowflakes} with permission from Kenneth Libbrecht, (b) is licensed under ``CC0 Public Domain'', and (c) is reproduced with permission from Goehring \textit{et al.} \cite{goehring2008scaling}, copyright (2008) by the American Geophysical Union.}
\label{nature_pattern}
\end{center} 
\end{figure}

The nature is composed of diverse patterns existing from nanoscopic to geophysical scales, such as the examples shown in Fig.\ \ref{nature_pattern}. The cause of patterns can be very complicated depending on particular physical mechanisms. For example, the striking six-fold symmetry of a snowflake, as shown in Fig.\ \ref{nature_pattern}a, is basically a particular signature of crystallization of water molecules. During this process, the water molecules align themselves in order to localize in the lowest energy state. Based on this principle, snowflakes develop hexagonal symmetry, and in reality there are different types of snowflakes depending on the surrounding temperature and humidity \cite{libbrecht2005physics}. Figure\ \ref{nature_pattern}b shows the sand ripples that can be easily observed in a desert or beach. The formation of these nearly periodic patterns originate from the interactions between the fluid (liquid or wind) flow and bed topography. The rough process can be as follows: when the fluid is moving fast, the sand grains will be lifted, gaining translational and rotational energy. In a higher energy state, the grains on the top of a dune are unstable and prone to flow down to dissipate energy when perturbed. As a consequence, the sand grains display such an amazing pattern with the wavelength depending on the fluid flow velocity and depth of the granular material \cite{charru2013sand,cross2009pattern}. A spectacular geologic column joint pattern found in cooled lava flows is shown in Fig.\ \ref{nature_pattern}c. Some well-known examples of those hexagonal fracture patterns can be found for instance in Giant’s Causeway, Northern Ireland, Devils Tower, Wyoming, USA, and Devils Postpile, California, USA. As the hot lava flows are exposed to air, cooling takes place from outside in, generating enormous stress inside the rock. Fractures then form at the surface and grow into the material, which usually takes decades. Depending on the specific environmental settings, the number of sides of the joints can vary and joint structures with the number of sides ranging from 4 to 7 have been observed in reality, nevertheless, hexagonal symmetry is the most common scenario, which is likely due to a particular ordering process \cite{goehring2014cracking,hofmann2015hexagonal}. 

In addition to the amazing patterns observed in non-living systems, biological organisms display a tremendous diversity of patterns, such as the branching pattern of veins in a leaf, the fractal structure in a Romanesco broccoli, the spiral pattern in a nautilus shell, the Turing pattern in a zebra's skin, and the fractures in a crocodile's skin. The origins for the development of the diverse and fascinating patterns observed in biological organisms are quite complicated due to the highly nonequilibrium nature in biological systems which usually involve cell differentiation, cell growth and death, and interaction with environmental conditions, however, many models have to some extent successfully explained the origin of patterns observed in biological systems at the molecular level or based on mechanical principles on the mesoscopic scale, although oversimplified in many cases \cite{koch1994biological,ouyang1991transition,sachs2005pattern,nakamasu2009interactions,boettiger2009neural,fujita2006pattern,milinkovitch2013crocodile}.

Despite the difficulty in fully understanding the formation of patterns in diverse systems in nature, there have been a number of widely applicable theories and models that are now employed as frameworks to describe the generic dynamics of the nonequilibrium growth of patterns in certain systems, for instance, the diffusion-limited aggregation of Brownian particles \cite{witten1981diffusion,garcia2012growth} and the spinodal decomposition of a mixture of liquids \cite{cahn1961spinodal,cahn1965phase}. Although our understanding of the fascinating patterns observed in nature is pretty much in its infancy, there are growing interests in uncovering the underlying mechanisms behind these nonequilibrium patterns physically and mathematically \cite{langer1980instabilities,cross1993pattern,aranson2006patterns,hoyle2006pattern,cross2009pattern}. In Sections\ \ref{patterns_fluids}, \ref{patterns_colloids} and \ref{patterns_sedimentation}, I will present some fundamental background for pattern formation in particular systems of fluids, colloidal suspensions, and sedimentation, respectively.

\section{Patterns in fluids}
\label{patterns_fluids}

The world of fluids is rich in intriguing patterns, which can essentially be explained by hydrodynamic theories \cite{lamb1993hydrodynamics,drazin1982hydrodynamic,charru2011hydrodynamic}. Generally, the origin of patterns observed in incompressible fluid systems can be understood by the Navier-Stokes equation: 
\begin{equation} 
\frac{\partial \textbf{u}}{\partial t}+(\textbf{u}\cdot \nabla)\textbf{u}=\textbf{g}-\frac{1}{\rho}\nabla p+\nu \nabla^2 \textbf{u},
\label{NS_eq}
\end{equation}
In most cases, the fluid of interest is assumed to be incompressible, i.e.,
\begin{gather}
\label{density_time}
\frac{\partial \rho}{\partial t}=0,\\
\label{density_space}
\nabla \rho=0,
\end{gather}  
and given the condition of mass conservation:
\begin{equation} 
\frac{\partial\rho}{\partial t}+\nabla\cdot(\rho\textbf{u})=\frac{\partial\rho}{\partial t}+\nabla\rho\cdot\textbf{u}+\rho(\nabla\cdot\textbf{u})=0,
\label{mass_conservation}
 \end{equation}
we can simply arrive:
\begin{equation} 
\frac{D\rho}{Dt}=\frac{\partial \rho}{\partial t}+\textbf{u}\cdot\nabla\rho=-\rho(\nabla\cdot \textbf{u})=0,
\label{material_derivative} \end{equation}
which yields the continuity equation:
\begin{equation} 
\nabla \cdot \textbf{u}=0,
\label{incompressibility}
 \end{equation}
where \textbf{u} is the flow velocity, $\rho$ is the fluid density, $\nu=\eta/\rho$ is the kinematic viscosity wherein $\eta$ is the dynamic viscosity, $\textbf{g}$ is acceleration field of the volume forces, and $p$ is the pressure. In certain regimes, the viscous force is dominant over inertial forces and gravity, and the nonlinear, advective inertial terms in Eq.\ \ref{NS_eq} can be safely ignored, leading to: 
\begin{equation} 
\nabla p=\eta \nabla^2 \textbf{u},
\label{stokes_eq}
\end{equation}
which is known as the Stokes equation. Equation\ \ref{stokes_eq} is much easier to deal with both mathematically and numerically than Eq.\ \ref{NS_eq}, and most hydrodynamic phenomena in this regime can be analytically understood due to the linearity of Eq.\ \ref{stokes_eq}. 
\begin{figure}
\begin{center}
\includegraphics[width=1\textwidth]{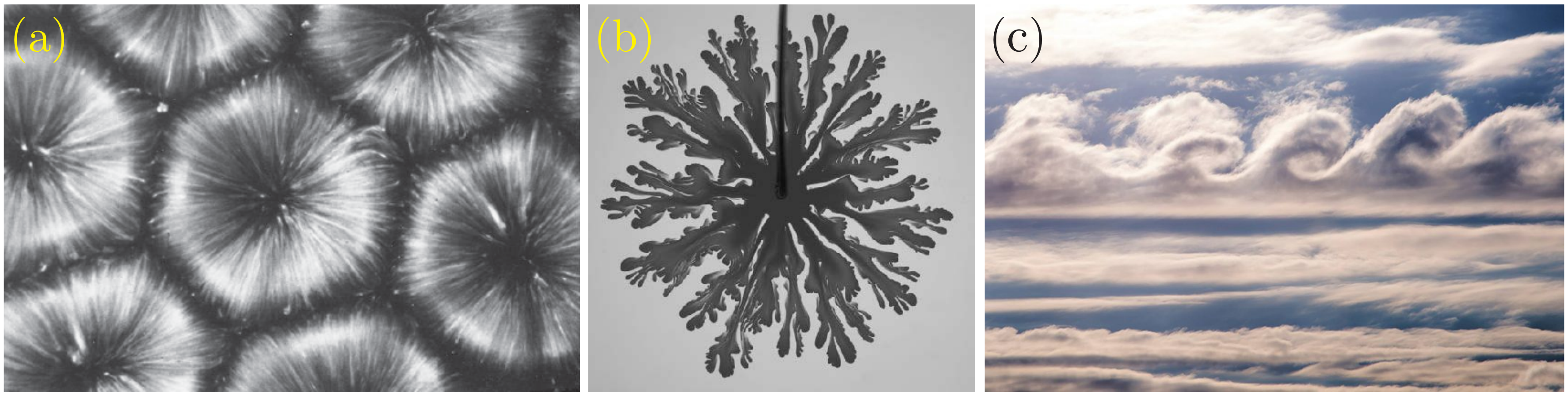}
\caption[Patterns driven by hydrodynamic instabilities]{Patterns driven by hydrodynamic instabilities. (a) shows the hexagonal convection cell pattern when heating a liquid from below with the top surface exposed to air, which is driven by the B\'{e}nard-Marangoni thermocapillary instability, (b) shows the viscous fingering pattern when imposing a less viscous fluid into a more viscous fluid in a Hele-Shaw cell, which is driven by the Saffman-Taylor instability, (c) shows the wave pattern of a cloud layer near New York when the pressure difference across the interface of cloud and air is beyond a critical value, which is driven by the Kelvin-Helmholtz instability. Image (a) is reproduced with permission from Ref.\ \cite{guyon2001physical}, copyright (2015) by the Oxford University Press, (b) is reproduced with permission from Bischofberger \textit{et al.}\ \cite{bischofberger2014fingering}, copyright (2014) by the Springer Nature, and image (c) credit to Paul Chartier.}
\label{hydrodynamic_instability_pattern}
\end{center} 
\end{figure}
\begin{figure}[!tbph]
\begin{center}
\includegraphics[width=6 in]{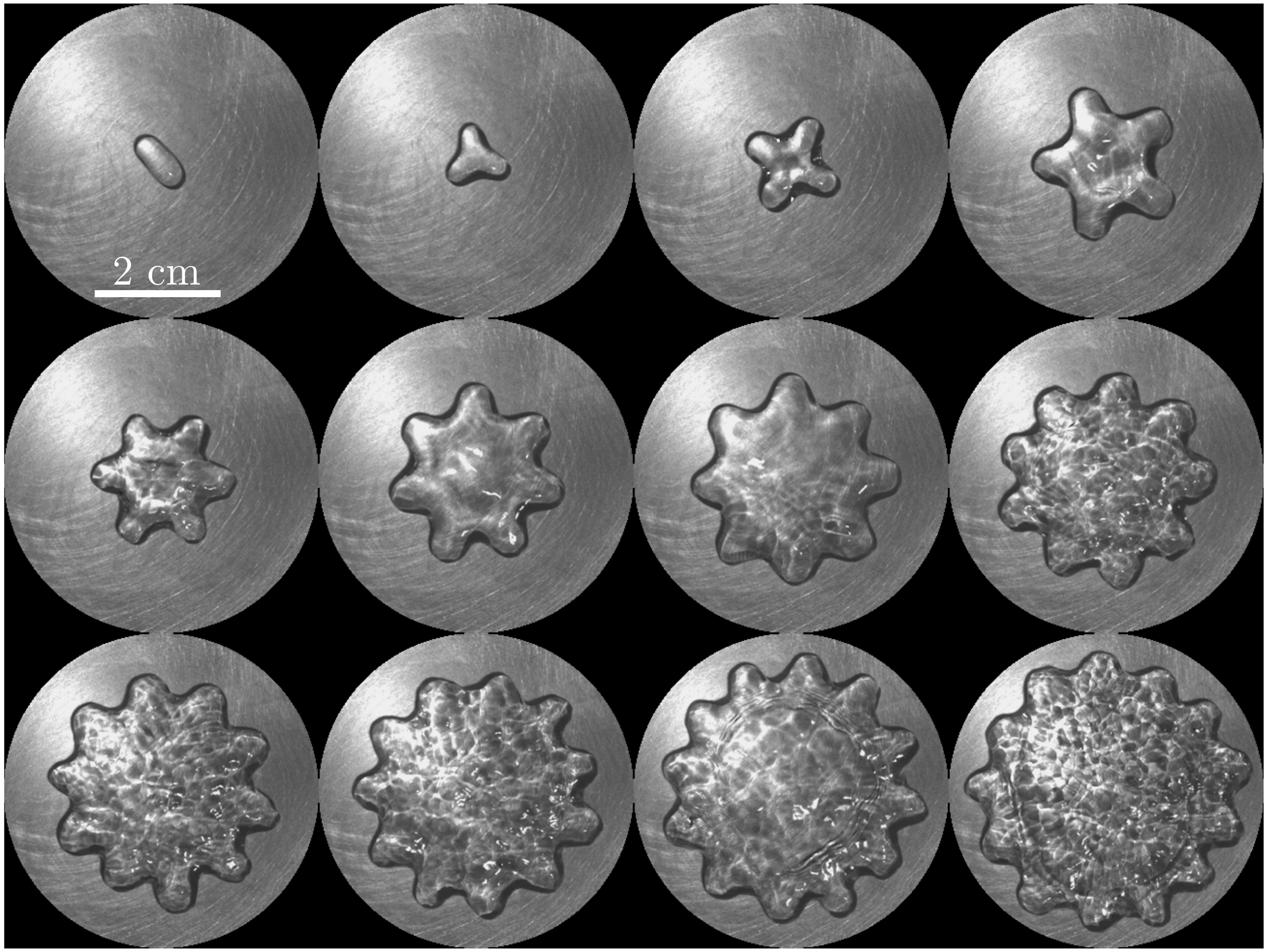}
\caption[Star-shaped oscillation modes of Leidenfrost water drops]{Star-shaped oscillation patterns of Leidenfrost drops on a curved, aluminum surface with a temperature of $\approx$ 350 $^\circ$C. Oscillation modes with $n = 2$ to $n = 13$ are shown when the lobes are at their maximum displacement. The scale bar applies to all images. The images are reproduced with permission from Ma \textit{et al.} \cite{ma2015many}, copyright (2015) by the American Institute of Physics.}
\label{oscillation_mode_water}
\end{center} 
\end{figure}

Figure\ \ref{hydrodynamic_instability_pattern}a shows the hexagonal convection cells which can be observed when heating a liquid from below with the top surface exposed to air. Due to the perturbation, the warm fluid will move upward, thus bringing local perturbation to the temperature of the free surface and leads to a surface tension gradient at the interface resultantly. The surface tension gradient induces a Marangoni stress, which drives the fluid to spread out thus forming the hexagonal pattern. In this systems, the resisting forces are caused by the liquid viscosity and thermal diffusion, which tend to stabilize the temperature variations in the fluid and make the fluid more homogeneous. Usually, a dimensionless Marangoni number defined as the ratio of surface tension effect to viscous and thermal diffusive effects is used to characterize to the relative importance, and when the Marangoni number reaches a critical value, hexagonal cell patterns should occur between the bottom layer and the top free surface with the size of the cell depending on the thickness of the fluid layer, which is the so-called B\'{e}nard-Marangoni thermocapillary instability \cite{benard1901tourbillons,rayleigh1916lix,marangoni1871ausbreitung,schatz2001experiments,maroto2007introductory,guyon2001physical}. Another example is the viscous fingering patterns that usually occur when a less viscous liquid is injected into a more viscous liquid in a rectangular configuration, for instance, a Hele-Shaw cell, as shown in Fig.\ \ref{hydrodynamic_instability_pattern}b. Once the critical injecting velocity of the less viscous liquid is reached, an instability will be excited and lends to the front to be unstable for growth of fingers, which consequently initiates a fingering pattern known as the Saffman-Taylor instability \cite{saffman1958penetration,homsy1987viscous}. Figure\ \ref{hydrodynamic_instability_pattern}c shows the wave pattern in a cloud layer, which is driven by a shear flow at the interface composed of two fluids with different densities (heavy fluid on the bottom and light fluid on the top). Basically, a shear flow at the interface will cause a pressure gradient in the vicinity of the interface, then the interface will be unstable and heavy fluid will be lifted up whereas the light fluid be pushed down, thus forming vortex-like wave pattern in the could layer. This is known as the Kelvin-Helmholtz instability \cite{charru2011hydrodynamic}. Besides the pattern shown in Fig.\ \ref{hydrodynamic_instability_pattern}c, the Kelvin-Helmholtz instability also manifests in other well-known examples, such as Jupiter's Red Spot, the Sun's corona, and in ocean waves \cite{marcus1988numerical,foullon2011magnetic,smyth2012ocean}.

In addition to the gorgeous patterns displayed in Fig.\ \ref{hydrodynamic_instability_pattern}, isolated drops are also able to show fascinating patterns. Leidenfrost drops are known for the star-shaped oscillations that spontaneously form, as shown in Fig.\ \ref{oscillation_mode_water}. The Leidenfrost effect occurs when depositing a liquid drop onto a hot surface well above the boiling point of the liquid. Then the drop will levitate on a vapor cushion fed by the drop evaporation, consequently the Leidenfrost drops will undergo a long lifetime as the vapor insulates the liquid from the hot solid substrate \cite{biance2003leidenfrost,quere2013leidenfrost}.

The capillary length is a length scale that characterizes the relative importance of gravitational force and surface tension. Assuming a spherical liquid drop with a size, $l$, surface tension, $\gamma$, and density, $\rho_l$, then the balance of surface tension force, $\gamma l$, and gravitational force, $\rho_lgl^3$, leads to a characteristic length scale $l_c=\sqrt{\gamma/\rho_lg}$, which is known as the capillary length. For drop with a size larger than $l_c$ the gravitational force will dominate over surface tension forces and the drops will be flatten by gravity. More interestingly, for drops with a radius much larger than $l_c$, the thickness of the drops, $h$, will become constant, i.e., $h\approx 2l_c$, which can be simply derived by balancing the surface tension force, $\gamma/(h/2)$, and hydrostatic force, $\rho_lgh/2$, per unit length near the edge of the drop. Large Leidenfrost drops easily form large-amplitude, star-shaped oscillations as shown in Fig.\ \ref{oscillation_mode_water}, which is generally due to the complex coupling between the solid, liquid and vapor phases. As will be shown in Chapter\ \ref{Leidenfrost_stars} in more detail, the star-shaped oscillations of Leidenfrost drops are parametrically driven by the pressure oscillations in the vapor beneath the drop. The pressure oscillations are initiated by the capillary waves with a characteristic wavelength (frequency) traveling from the drop center to the edge and the capillary waves are excited by a strong shear due to the rapidly flowing evaporated vapor. This suggests that the star-shaped oscillations of Leidenfrost drops are purely hydrodynamic in origin \cite{ma2017star,ma2018self}. 

\newpage
\section{Patterns in colloidal films}
\label{patterns_colloids}

\begin{figure}
\begin{center}
\includegraphics[width=6 in]{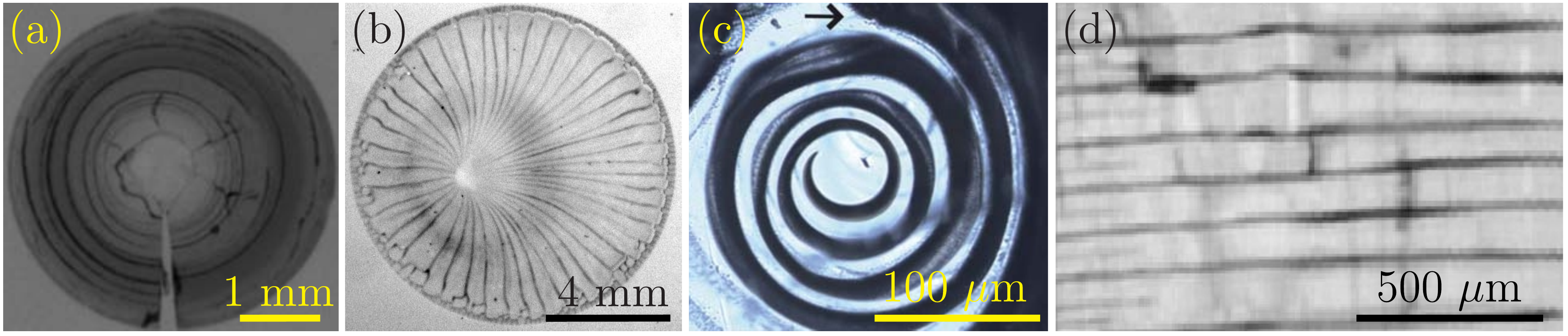}
\caption[Crack patterns by drying colloidal suspensions]{(a) and (b) show the circular and radial cracks by drying nanoscopic silica-water drops on a polystyrene substrate and a glass substrate, respectively. (c) shows spiral cracks by drying a nanoscopic latex-water suspension in a circular container with the top surface exposed to air. (d) shows the parallel cracks by drying nanoscopic silica-water suspension in a quasi-two-dimensional thin chamber. Images (a) and (b) are reproduced with permission from Jing \textit{et al.} \cite{jing2012formation}, copyright (2012) by the American Chemical Society, (c) is reproduced with permission from Lazarus \textit{et al.} \cite{lazarus2011craquelures}, copyright (2011) by the Royal Society of Chemistry, and (d) is reproduced with permission from Dufresne \textit{et al.} \cite{dufresne2003flow}, copyright (2003) by the American Physical Society.}
\label{colloidal_cracks}
\end{center} 
\end{figure}

Crack patterns by drying are ubiquitous in nature, artistic works and industrial processes \cite{routh2013drying}. Typical examples include the complex crack network in a dried blood drop \cite{brutin2011pattern}, craquelures in old paintings \cite{giorgiutti2015striped,goehring2017drying}, T/Y-shaped cracks in dried mud \cite{goehring2010evolution}, and polygonal terrain cracks \cite{goehring2014cracking,Maarry2010mars,el2014potential}. Practically, a broad range of applications, such as thin film coating, forensics, and controllable surface patterning rely on the knowledge of the physical processes that determine crack patterns \cite{prosser2012avoiding,hatton2010assembly,liu2016surface,nam2012patterning,zeid2013influence}. To understand the underlying mechanisms behind those versatile, multiscale cracks, drying colloidal suspensions has been widely employed as a model system in the laboratory. Colloid generally refers to a type of multiphase system wherein particles (solid, liquid or gas) with a size between $10^{-9}$ m to $10^{-6}$ m are dispersed in a liquid. Prototypical colloidal systems include milk, blood and ink in our daily lives. One of the most important properties of colloidal particles is that they are susceptible to thermal fluctuations, which leads to Brownian motion due to the random collision with liquid molecules \cite{russel1991colloidal}. However, many crack patterns in natural systems contain particles with sizes from 10 $\mu$m to 1 mm, so larger particles are also used to investigate crack pattern formation. In contrast to colloidal particles, the major interactions between between these granular particles are from friction and collision, and when the particles size is close to 1 $\mu$m, other factors such as van der Waals forces, humidity are also likely to play a role in the interaction between granular particles \cite{de1999granular,andreotti2013granular,herminghaus2013wet,franklin2016handbook}.

Heretofore, a large number of types of crack patterns have been observed by drying colloidal suspensions under various conditions, and those results indicate that many factors can influence the selection of crack patterns \cite{routh2013drying,goehring2015desiccation}. Figure\ \ref{colloidal_cracks} shows a few typical crack patterns observed in drying colloidal suspensions in recent years. Different substrate wettability can lead to distinct types of crack patterns as indicated by Figs.\ \ref{colloidal_cracks}a and \ref{colloidal_cracks}b which show circular and radial cracks by drying colloidal drops on polystyrene (hydrophobic) and glass (hydrophilic) substrates, respectively \cite{jing2012formation}. The film thickness can also play a role in selecting the crack pattern as shown in Fig.\ \ref{colloidal_cracks}c in which spiral crack patterns are observed in drying a latex-water suspension in a circular container with the top surface exposed to air with a film thickness of $\approx$ 200 $\mu$m, in contrast, only polygonal cracks can be observed in films with thickness of $\approx$ 20 $\mu$m in the same experimental setup \cite{lazarus2011craquelures}. In addition to the open system shown in Figs.\ \ref{colloidal_cracks}a, \ref{colloidal_cracks}b and \ref{colloidal_cracks}c, when the colloidal suspensions are confined in a quasi-two-dimensional space, an array of roughly periodically spaced cracks are usually observed with drying as shown in Fig.\ \ref{colloidal_cracks}d \cite{dufresne2003flow}.

Besides the crack patterns shown in Fig.\ \ref{colloidal_cracks}, a large variety of crack patterns have been reported in dried suspensions \cite{allain1995regular,inasawa2012self,goehring2011wavy,nandakishore2016crack,kiatkirakajorn2015formation,neda2002spiral,vermorel2010radial,darwich2012highly, jing2012formation,xu2010imaging,giorgiutti2015striped,giorgiutti2016painting,bohn2005hierarchicalI,bohn2005hierarchicalII,pauchard2003morphologies, boulogne2013annular,allen1987desiccation,gauthier2010shrinkage}. The diversity in observed patterns depends on numerous factors such as film geometry \cite{lazarus2011craquelures,nandakishore2016crack}, particle mechanics \cite{nawaz2008effects}, liquid additives \cite{pauchard1999influence,liu2014tuning}, preparation history \cite{nakahara2006transition,nakahara2006imprinting}, solvent volatility \cite{boulogne2012effect,giorgiutti2014elapsed}, and external fields \cite{khatun2012electric,pauchard2008crack}. The difficulty in understanding pattern selection and propagation lies in the combinations of elastic instabilities, plastic deformations and multiphase interactions in the such systems \cite{goehring2013plasticity,goehring2015desiccation,goehring2017drying}.

Despite this broad range of crack patterns, we know surprisingly little about what controls the size and hierarchy of commonly observed polygonal cracks, which are visible on both the micro- and macro-scales. Although multiple experimental, numerical and theoretical studies have reported specific scalings between the crack spacing and film thickness considering the mechanical, thermodynamic, hydrodynamic, and statistical properties of cracking, however they often lead to contradictory results. Given the importance of understanding crack patterns, in Chapter\ \ref{drying_cracks} I report experimental results on the mechanisms underlying polygonal crack patterns by drying micro-sized particulate suspensions. We found distinct mechanisms lead to multiscale crack patterns in drying cornstarch-water suspensions, and the crack area and film thickness of the multiscale cracks we observed in our experiment obey a universal power law that can be derived from a balance of stored elastic energy in the film and the surface energy released upon cracking \cite{ma2018universal}.

\newpage
\section{Patterns driven by sedimentation}
\label{patterns_sedimentation}

\begin{figure}
\begin{center}
\includegraphics[width=1\textwidth]{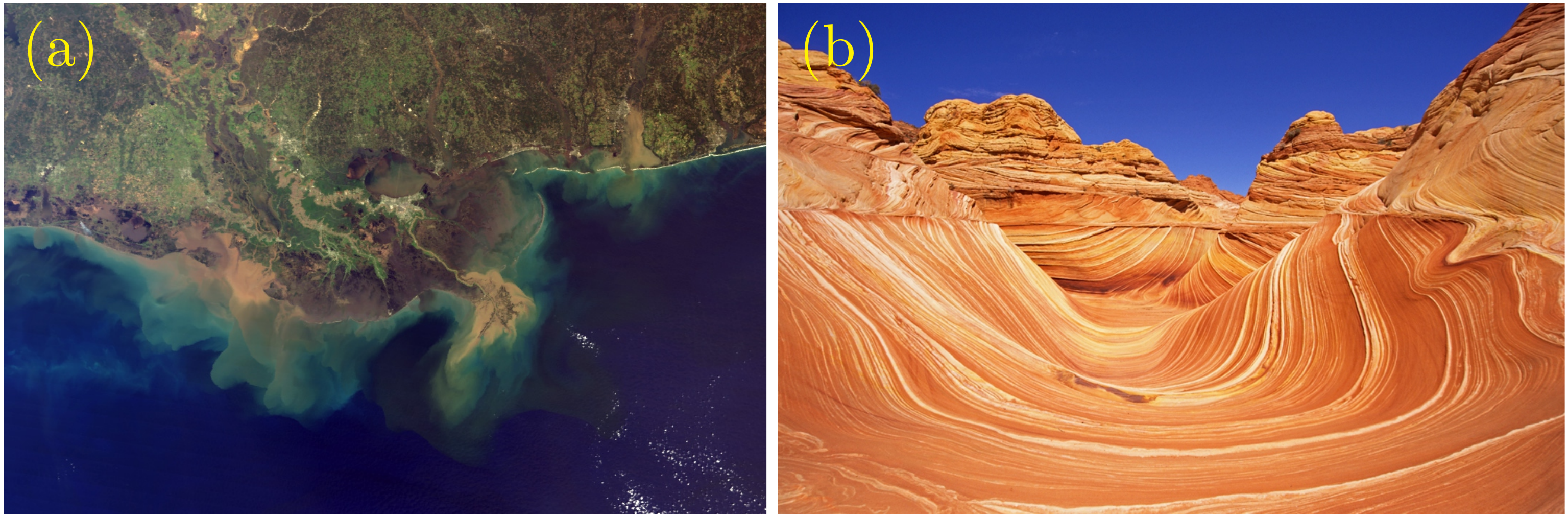}
\caption[Patterns driven by sedimentation]{(a) The Mississippi River Delta, showing the sediment plumes from the Mississippi and Atchafalaya Rivers, USA, (b) shows the sedimentary rocks located in The Wave, Vermillion Cliffs, Arizona, USA. Images (a) and (b) are taken from Wikipedia licensed under CC-BY-2.5.}
\label{sedimentation_in_nature}
\end{center} 
\end{figure}

Sedimentation of particles in fluids plays a key role in shaping natural patterns \cite{allen1970physical,vanoni2006sedimentation}, for example the formation of river deltas (Fig.\ \ref{sedimentation_in_nature}a) and geological deposits of sandstone and siltstone (Fig.\ \ref{sedimentation_in_nature}b). In industrial applications, sedimentation is also widely used in the processes of decontamination, neutralization, particle separation and biological purification. Given its crucial role, sedimentation has been extensively studied with a particular focus on the settling velocity fluctuations of mono/poly-disperse, Brownian/non-Brownian particles, density fluctuations, the effects of particle geometry, system size, and many-body hydrodynamics \cite{batchelor1972sedimentation,davis1985sedimentation,guazzelli2011fluctuations,piazza2014settled,mucha2004model}, however, the sedimentation is still far from being fully understood due to the complex long-range hydrodynamic, many-body interactions between particles. 

Practically, one of the important issues is to precisely control particle clustering and the uniformity of particulate films after sedimentation, which relies on one's ability to control particle interactions. Recently, it has been theoretically suggested that controlling these interactions can be done by tuning the particle geometry and mass distribution of non-Brownian particles in Stokes flows \cite{goldfriend2017screening}. The key idea is that as the particle geometry center is different from the center of mass, the particles will response anistropically to the fluid disturbance generated by the density fluctuations. This will eventually generate an effective repulsion between particles, suppressing the density fluctuations. In contrast, for symmetric particles (e.g., sphere) with uniform density distributions there is no such repulsion between particles, which is due to the isotropic response to the fluid disturbance endowed by the highly symmetrical geometry. In Chapter\ \ref{sedimentation_pattern}, I will show some experimental evidence related to those theoretical predictions.

\begin{figure}
\begin{center}
\includegraphics[width=3in]{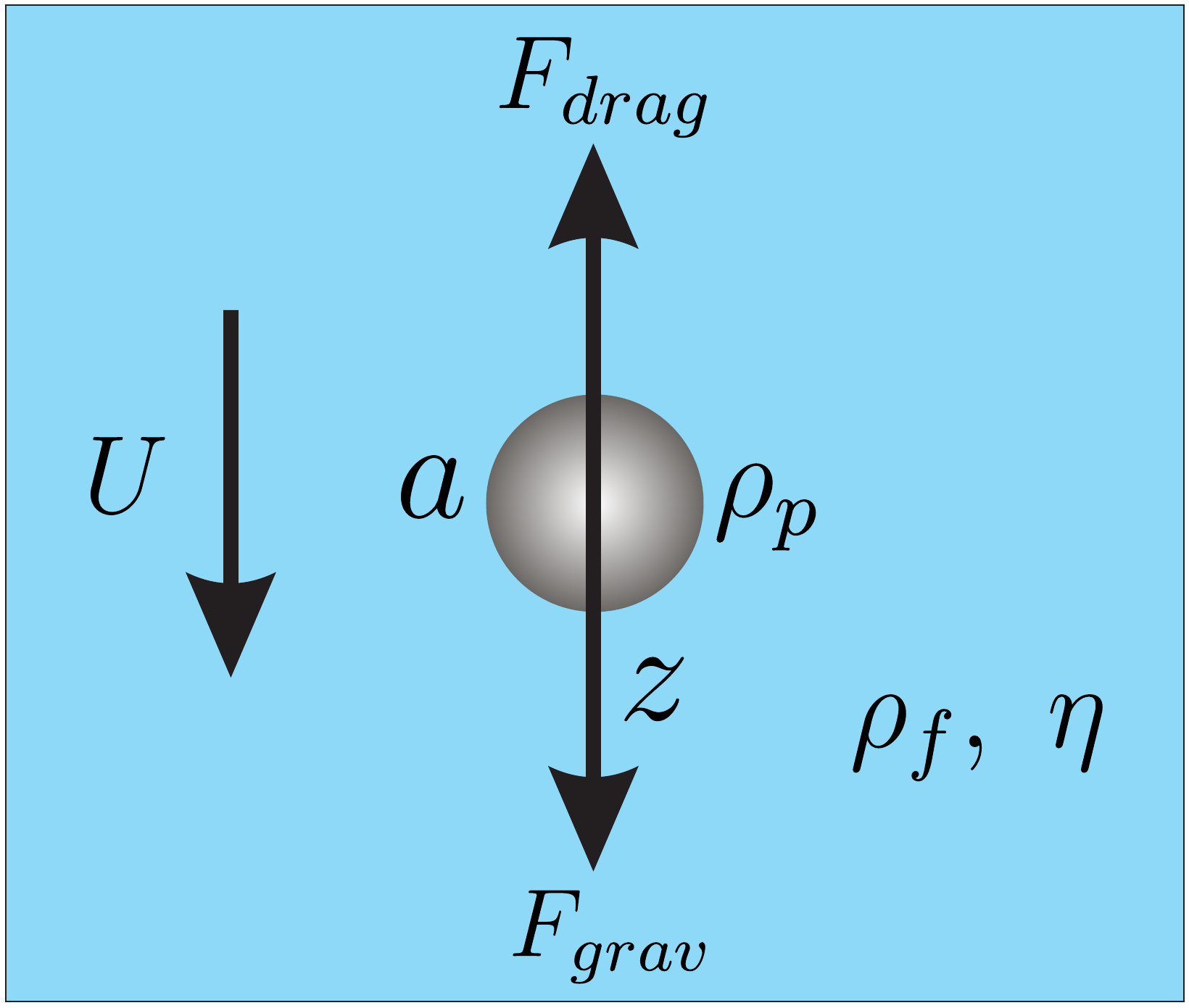}
\caption[Sketch of the sedimentation of a spherical particle in a viscous fluid]{Sketch of the sedimentation of a spherical particle with density $\rho_p$, radius $a$ and velocity $U$ in the vertical direction $z$ in a viscous fluid with density $\rho_f$ and dynamic viscosity $\eta$.}
\label{sketch_sedimentation}
\end{center} 
\end{figure}

In typical experimental, theoretical and computational studies, viscous fluids are used for sedimentation in order to simplify the analysis, in which the inertial forces can be ignored, and the sedimentation dynamics of particles are merely determined by Stokes flow (i.e., Eq.\ \ref{stokes_eq}). For a spherical particle settling down in a very viscous fluid as illustrated in Fig.\ \ref{sketch_sedimentation}, the particle moves very slowly and the motions can be considered steady state. Then balancing the drag, gravitational, and buoyancy forces leads to: 
\begin{equation} 
6\pi\eta a U=\frac{4}{3}(\rho_p-\rho_f)\pi a^3g,
\label{force_balance}
\end{equation}
and the steady state sedimentation velocity (Stokes velocity) is: 
\begin{equation} 
U_{stokes}=\frac{2}{9}\frac{a^2}{\eta}(\rho_p-\rho_f)g.
\label{sedimentation_velocity}
\end{equation}

One of the most interesting properties of sedimentation in viscous fluids is the long-range hydrodynamic nature of the interactions between particles. This can be easily derived by considering Eqs.\ \ref{incompressibility} and \ref{stokes_eq} with appropriate boundary conditions, and the disturbance velocity and pressure fields around a sphere in a spherical coordinate system ($r$, $\theta$, $\phi$), the solution is \cite{lautrup2011,guazzelli2011physical}: 
\begin{gather}
\label{disturbance_velocity_r}
u_r=\left( 1-\frac{3}{2}\frac{a}{r}+\frac{1}{2}\frac{a^3}{r^3} \right) U \cos \theta,\\
\label{disturbance_velocity_theta}
u_\theta=-\left( 1-\frac{3}{4}\frac{a}{r}-\frac{1}{4}\frac{a^3}{r^3} \right) U \sin \theta,\\
\label{disturbance_velocity_phi}
u_\phi=0,\\
\label{disturbance_pressure}
p=-\frac{3}{2}\eta\frac{a}{r^2}U\cos \theta,
\end{gather} 
where $U$ is the uniform stream velocity at $r\rightarrow\infty$.

\begin{figure}
\begin{center}
\includegraphics[width=3in]{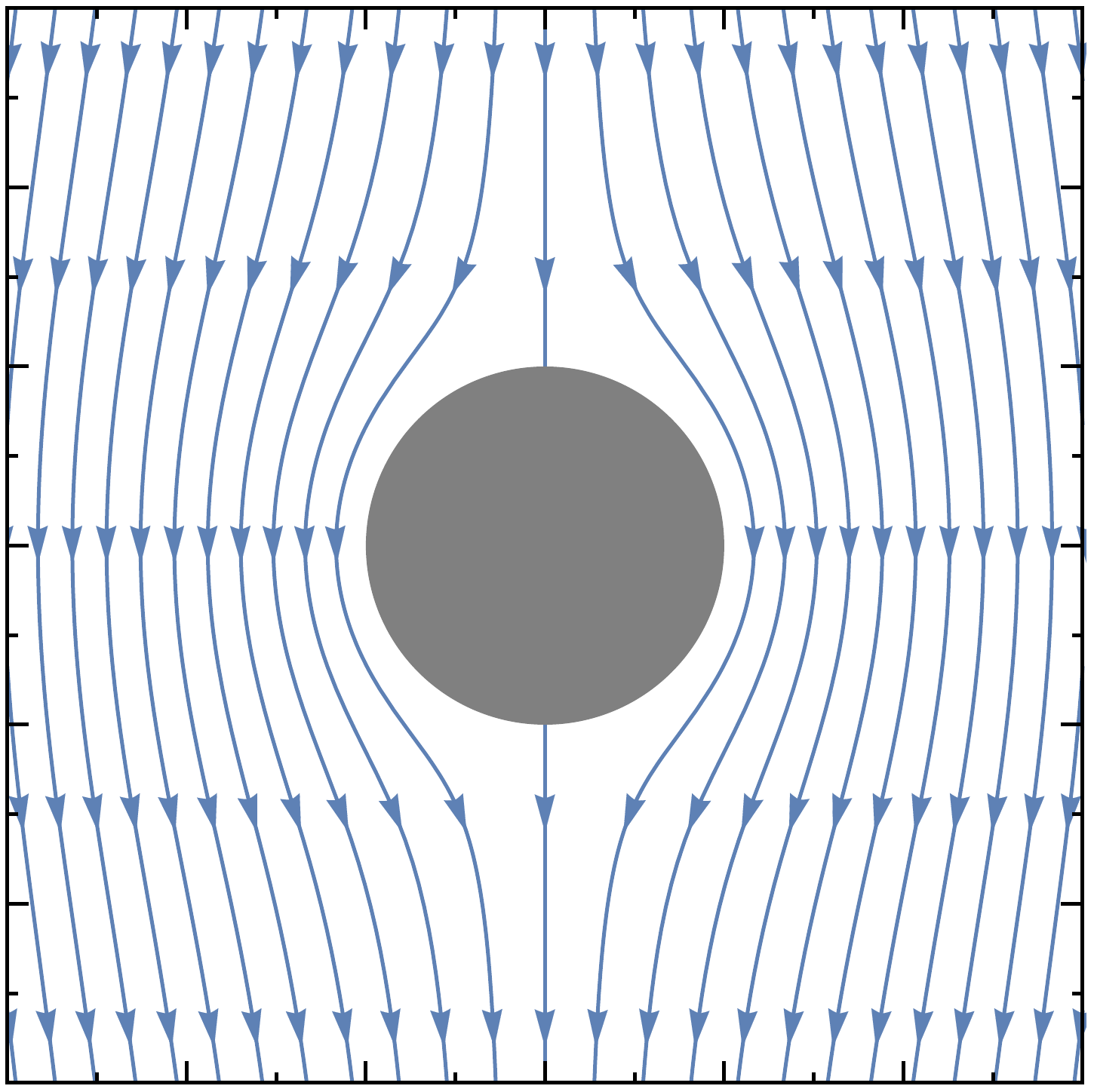}
\caption[Sketch of Stokes flow around a sphere in a Cartesian coordinate system]{Sketch of Stokes flow around a sphere based on Eqs.\ \ref{disturbance_velocity_r} to \ref{disturbance_pressure} in a Cartesian coordinate system. Note the horizontal axis represents the $x$ direction, whereas the vertical axis represents the $z$ direction.}
\label{stokes_flow_sphere}
\end{center} 
\end{figure}

Figure\ \ref{stokes_flow_sphere} shows the two dimensional view of the Stokes flow around a solid sphere in a Cartesian coordinate system based on Eqs.\ \ref{disturbance_velocity_r} to \ref{disturbance_pressure}. What is different here from the pattern of potential flow around a sphere is that the perturbation created by a translating sphere decays slower and extends further due to the term $O(1/r)$, which is what we mean by the long-range nature of the interactions.   

To leading orders, Eqs.\ \ref{disturbance_velocity_r} and \ref{disturbance_velocity_theta} are reduced to: 
\begin{equation}
\label{disturbance_velocity_1}
u\sim1-\frac{1}{r}\simeq \frac{1}{r},\\
\end{equation} 
whereas Eq.\ \ref{disturbance_pressure} is simplified to: 
\begin{equation}
\label{disturbance_pressure_1}
p\sim\frac{1}{r^2}.
\end{equation}

This long-range hydrodynamic characteristic, as indicated by Eq.\ \ref{disturbance_velocity_1}, can result in interesting and complex phenomena in many-body systems, such as collective motions, large-scale velocity and structure fluctuations in the sense of the dynamic coupling between hydrodynamics and the positions and orientations of particles, and considerable progress has been made regarding those problems  \cite{jeffery1922motion,xue1992diffusion,segre1997long,ramaswamy2001issues,guazzelli2011fluctuations,piazza2014settled,padding2004hydrodynamic,jung2006periodic,batchelor1972sedimentation,
hinch1977averaged,koch1991screening,caflisch1985variance,brenner1999screening,hunter2012physics,chajwa2018kepler}. More recent attention has been paid to controlling the particle interactions during sedimentation so as to control the uniformity of particle cluster patterns after sedimentation.

One of the promising pathways for realizing controllable sedimentation is to tune the density distributions as theoretically suggested by Goldfriend \textit{et al.} \cite{goldfriend2017screening}. Using linear stability analysis, the authors predict that the particles (non-Brownian) with different density distributions respond very differently to the fluid disturbance, and there should be an effective repulsion between elongated particles with nonuniform density distributions, in contrast to particles with uniform density distributions, for example, a sphere or a rod will align with the fluid flow (see Fig.\ \ref{response_to_flow} for comparison). This will create a lateral motion of the particles because of the greater downward gravitational torque of the heavier part compared to the lighter adjacent part. This torque pulls the particle from the preferred alignment direction, consequently pulling the particles to move towards the regions of lower settling velocities and lower densities. This lateral motion is theoretically promising in suppressing density fluctuations so as to realize controllable uniformity of particles after sedimentation.

In Chapter\ \ref{sedimentation_pattern}, I report our we experimental investigation on the effect of particle density distributions on the sedimentation dynamics. In the experiment, we used equal-sized aluminum and steel balls to construct particles with different density distributions. We observed an effective repulsion between the particles with nonuniform density distributions in both two-body and many-body systems, in contrast to the counterpart of particles with uniform density distributions. Our results suggest that the theoretical predictions by Goldfriend \textit{et al.} \cite{goldfriend2017screening} are basically correct despite that the density fluctuations are only restricted to the $x$-$y$ plane (see Fig.\ \ref{sketch_sedimentation}). It is expected that our results could open a novel pathway for tunable route towards the uniformity of particulate films via sedimentation.

The examples of pattern formation I report in this thesis represent the typical scenarios in fluid and granular systems driven by nonequilibrium dynamics, though in the real world the systems are quite complex and diverse. The underlying mechanisms giving rise to the pattern formation in fluid and granular systems we uncover in laboratory experiments (Chapters\ \ref{Leidenfrost_stars}, \ref{drying_cracks} and \ref{sedimentation_pattern}) can enhance our understanding of how nonequilibrium dynamics shape the nature, and are also expected to serve as directive tools for manually controlling pattern formation for multiple purposes.

\chapter{Oscillations of Leidenfrost drops}
\label{Leidenfrost_stars}

\section{Introduction}

\begin{figure}
\begin{center}
\includegraphics[width=4.6 in]{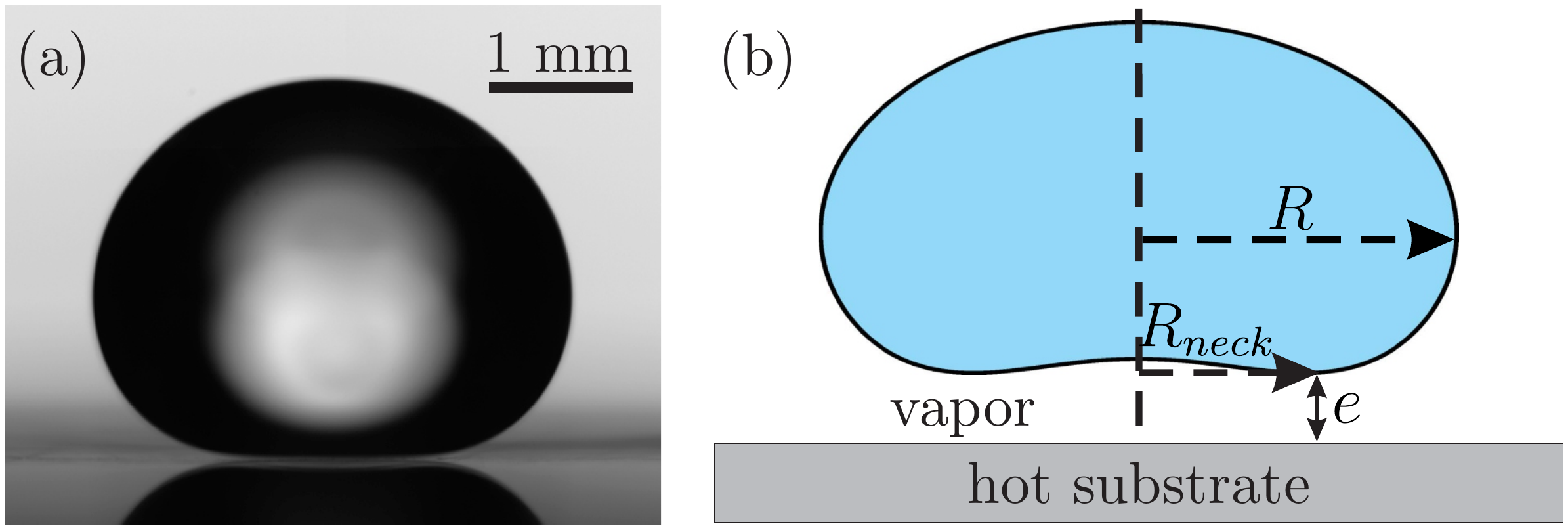}
\caption[Leidenfrost effect and the geometry of Leidenfrost drops]{(a) shows an image of a Leidenfrost water drop on a solid substrate with temperature of 320 $^\circ$C, (b) illustrates the geometry of the Leidenfrost drop, where $R$ represents the maximum radius of the drop, $R_{neck}$ represents the radius of the neck region which is closest to the hot solid substrate, and $e$ represents the gap height between the bottom surface of the drop and the hot solid substrate, which is also the average thickness of the vapor layer fed by drop evaporation. Image (a) is reproduced with permission from Burton \textit{et al.} \cite{burton2012geometry}, copyright (2012) by the American Physical Society.
} 
\label{leidenfrost_effect}
\end{center} 
\end{figure}

When a millimeter-scale volatile liquid drop (e.g., water) is deposited on a sufficiently hot solid, it can survive for minutes due to the presence of a thermally-insulating layer of evaporated vapor beneath the drop. This phenomenon is known as the Leidenfrost effect for example the Leidenfrost drop shown in Fig.\ \ref{leidenfrost_effect}a, and the geometry of the Leidenfrost drop is illustrated in Fig.\ \ref{leidenfrost_effect}b. In this levitated state, commonly known as the Leidenfrost regime \cite{leidenfrost1756aquae}, the supporting vapor layer is maintained by the sustained evaporation of the liquid, and individual drops are free to undergo frictionless motion due to the absence of liquid-solid contact.

The Leidenfrost effect can be easily observed by placing a water drop onto a hot pan over a cook stove in the kitchen, and has been the subject of numerous fundamental and applied studies due to the complex and rich interactions between the solid, liquid, and vapor phases \cite{quere2013leidenfrost}. Examples include the evaporation dynamics and geometry of the drop \cite{burton2012geometry,biance2003leidenfrost,myers2009mathematical,pomeau2012leidenfrost,xu2013hydrodynamics,sobac2014leidenfrost,hidalgo2016leidenfrost,maquet2016leidenfrost,wong2017non}, the stability of the vapor-liquid interface \cite{duchemin2005static,lister2008shape,snoeijer2009maximum,bouwhuis2013oscillating,trinh2014curvature,raux2015successive,maquet2015leidenfrost,van2019asymptotic}, hydrodynamic drag-reduction \cite{vakarelski2011drag,vakarelski2012stabilization,vakarelski2014leidenfrost}, self-propulsion of droplets \cite{linke2006self,dupeux2011viscous,dupeux2011trapping,lagubeau2011leidenfrost,cousins2012ratchet,li2016directional,soto2016surfing,sobac2017self,bouillant2018leidenfrost}, impact dynamics \cite{biance2006elasticity,tran2012drop,castanet2015drop,shirota2016dynamic}, green nanofabrication \cite{abdelaziz2013green}, nanoparticle painting \cite{elbahri2007anti}, particle self-organization \cite{maquet2014organization}, explosion of Leidenfrost liquid mixtures \cite{moreau2019explosive}, chemical reactions \cite{bain2016accelerated}, fuel combustion \cite{kadota2007microexplosion}, quenching process in metallurgy \cite{bernardin1999leidenfrost}, heat transfer \cite{talari2018leidenfrost,shahriari2014heat}, directional transport \cite{agapov2014length,li2016directional}, soft heat engines \cite{waitukaitis2017coupling,wells2015sublimation} and thermal control of nuclear reactors \cite{van1992physics}. The rich phenomena behind Leidenfrost effect essentially stem from the complex coupling between the liquid-vapor interface and the rapid vapor flow beneath the drop, which is however far from being well understood \cite{quere2013leidenfrost}. Here I report our experimental investigation on the coupling dynamics through the study of self-sustained oscillations of Leidenfrost drops.

In many of these examples, transient and self-sustained capillary oscillations play an important role in the dynamics. The interplay between gravity, the flow of vapor beneath the drop, and the liquid surface tension can lead to both small- and large-amplitude oscillations with very little damping. An understanding of these detailed interactions is crucial for the stability of the vapor layer, the failure of which can lead to explosive boiling upon contact with the hot surface. However, the excitation mechanism of these oscillations is complicated by the presence of both hydrodynamic and thermal effects, for example, rapidly-flowing vapor can cause a strong shear stress at the liquid-vapor interface, and temperature gradients in the liquid can lead to convective and Marangoni forces. Here we focus on capillary oscillations in individual Leidenfrost drops, where the shape is mostly determined by the competition between the gravity and surface tension, as measured by the relative size of the drop with respect to the capillary length, $l_c\equiv \sqrt{\gamma/\rho_l g}$, where $\gamma$ and $\rho_l$ denote the surface tension and density of the liquid, and $g$ is the acceleration due to gravity. For drops with radius $R<l_c$, surface tension forces are dominant, and the drop shape is essentially spherical except for a vanishingly small flat region near the solid surface \cite{burton2012geometry}. Caswell \cite{caswell2014dynamics} identified a planar vibrational mode (``breathing" mode) in the neck region of the drop closest to the solid substrate. The oscillation frequencies were found to obey a power law that is not consistent with a general three-dimensional dispersion relation for capillary waves \cite{rayleigh1879capillary}. Here we provide an analytical expression for the breathing mode using a simple model based on gravity and surface tension which shows excellent agreement with the experimental data.

Large Leidenfrost drops form liquid puddles whose thickness is approximately 2$l_c$ (see Section\ \ref{patterns_fluids}). These puddles are known to spontaneously form large-amplitude, star-shaped oscillations \cite{holter1952vibrations,adachi1984vibration,strier2000nitrogen,snezhko2008pulsating,strier2000nitrogen,ma2015many,Maleidenfrost2016}. Similar oscillations have been observed in large drops on periodically-shaken, hydrophobic surfaces \cite{noblin2005triplon,noblin2009vibrations}, drops levitated by an underlying airflow \cite{bouwhuis2013oscillating}, and drops excited by an external acoustic or electric field \cite{shen2010parametric,shen2010parametrically,mampallil2013electrowetting}. In studies where the frequency of external forcing is prescribed, the oscillations are excited by a parametric mechanism \cite{brunet2011star}. The external forcing leads to variations in the drop radius with time. Since the drop radius appears in the dispersion relation for azimuthal, star-shaped oscillations, the evolution of the oscillation amplitude obeys an equation similar to the Mathieu equation. For Leidenfrost drops, however, the mechanism is less clear since there is no prescribed frequency, and the star oscillations are excited and sustained through the heat input and resulting evaporation of the liquid. It has been suggested that the star oscillations may result from modulations of the surface tension of the liquid due to temperature variations \cite{adachi1984vibration,takaki1985vibration,tokugawa1994mechanism}, or perhaps due to convective patterns \cite{snezhko2008pulsating,strier2000nitrogen}. However, \cite{bouwhuis2013oscillating} observed star-shaped oscillations in drops which are supported by an external, steady air flow, suggest that a hydrodynamic coupling between the gas flow and liquid interface initiates the oscillations.

Given the importance of the Leidenfrost effect in basic fluid and thermal transport, or the numerous practical applications (e.g., self-propulsion and drag-reduction), we know surprisingly little about the coupling between the evaporated vapor flow and vapor-liquid interface that lead to rich dynamical phenomena. Here we explore this coupling by investigating the self-organized, star-shaped oscillations of Leidenfrost drops using six different liquids: water, liquid N$_2$, ethanol, methanol, acetone and isopropanol. The liquid drops were levitated on curved surfaces in order to suppress the Rayleigh-Taylor instability, and star-shaped oscillation modes with $n$ = 2 to 13 lobes along the drop periphery were observed. The number of observed modes depended sensitively on the liquid viscosity, whereas the oscillation frequency (wavelength) depended only on the capillary length but not the mode number, substrate temperature, or drop size. Accompanying pressure measurements in the center of the vapor layer indicate that the pressure variation frequency was approximately twice that the drop oscillation frequency for all of the observed modes, consistent with a parametric forcing mechanism. We show that the pressure oscillations are driven by capillary waves of a characteristic wavelength beneath the drop traveling from the drop center to the edge, and such capillary waves can be generated by a strong shear stress at the liquid-vapor interface. Additionally, we find that although thermal convection is expected to be quite strong, the robust frequency (wavelength) of star oscillations is only weakly affected by varying either the substrate or environmental temperature, suggesting that star-shaped oscillations of Leidenfrost drops are hydrodynamic in origin.

\section{Experimental setup}
\label{experiment}
In the experiment, blocks of engineering 6061 aluminum alloy with dimensions of 7.6 cm $\times$ 7.6 cm $\times$ 2.5 cm were used as substrates. Resistive heaters were embedded with high-temperature cement into the aluminum to control the substrate temperature, $T_s$. Six different liquids were used as Leidenfrost drops: deionized water, liquid nitrogen (liquid N$_2$), ethanol, methanol, acetone, and isopropyl alcohol (isopropanol). The relevant liquid properties at the boiling point, $T_ {b}$, such as the surface tension $\gamma$, density $\rho_l$, dynamic viscosity $\eta_l$, and capillary length $l_c$, are listed in Table\ \ref{Leidenfrost_tab:1}. The substrates were heated to different temperatures based on $T_ {b}$ for each liquid. For water, the temperature of the substrate was set from 523 K to 773 K, for ethanol, methanol, acetone, and isopropanol the temperature of the substrate was set to 523 K, whereas the substrate for liquid N$_2$ was not heated due to its extremely low $T_ {b}$. 

\begin{figure}
\begin{center}
\includegraphics[width=4 in]{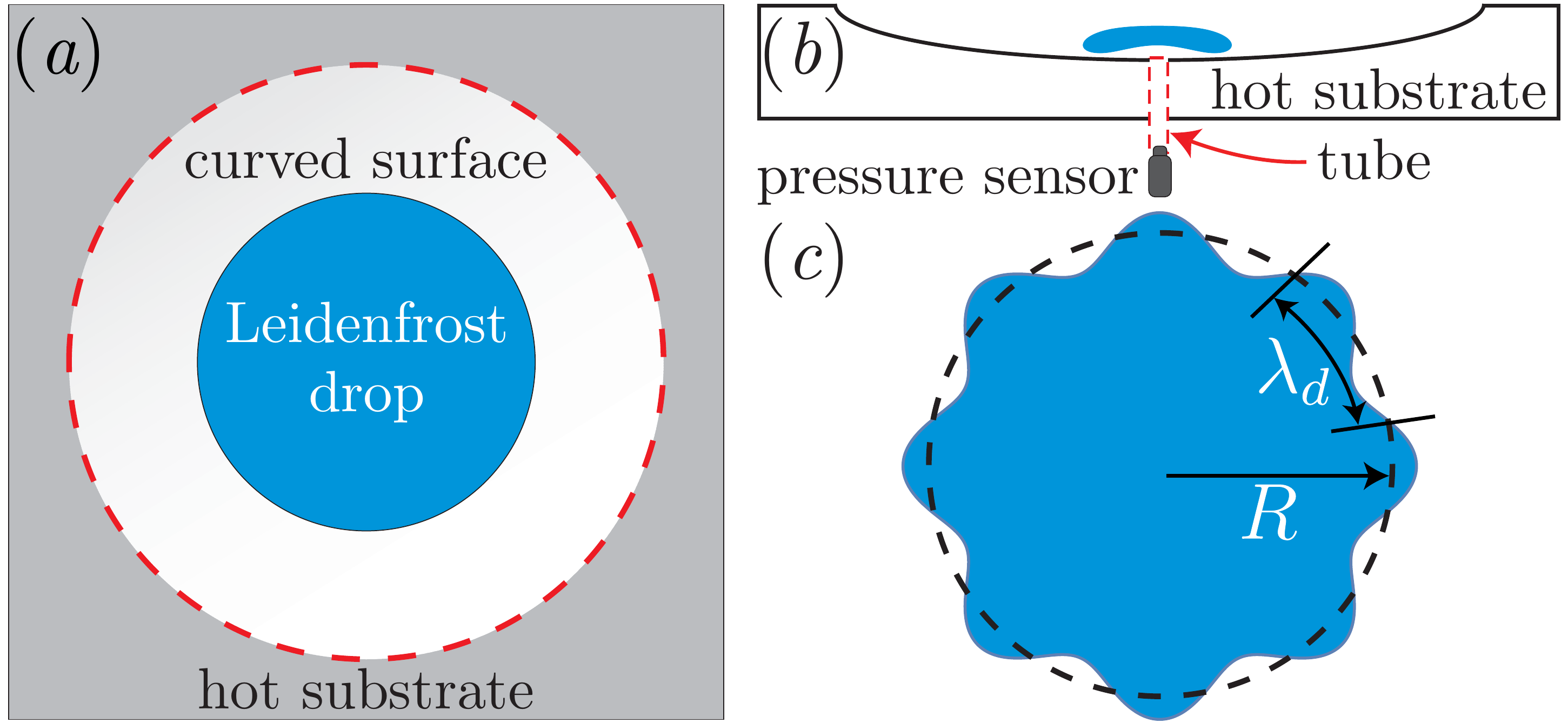}
\caption[Experimental setup for star-shaped oscillations of Leidenfrost drops]{Schematics of the experimental setup and star-shaped oscillation pattern. (a) Top-down view of the substrate whose upper surface is milled into a bowl shape enclosed by the dashed red circle. (b) Cross-sectional view of the substrate. (c) Star-shaped oscillation pattern of a Leidenfrost drop showing the standing wave along the drop periphery, which obeys $2\pi R=n\lambda_d$, where $R$ is the drop radius, $n$ is the number of lobes, and $\lambda_d$ is the wavelength of the standing wave.
} 
\label{setup}
\end{center} 
\end{figure}

The upper surfaces of the substrates were machined into a concave, spherical shape in order to suppress the buoyancy-driven Rayleigh-Taylor instability at the vapor-liquid interface and keep the drops stationary \cite{snoeijer2009maximum,quere2013leidenfrost,trinh2014curvature}. After machining, the roughness of the surface was inspected using optical microscopy with 50$\times$ magnification. By changing the focus, we determined that over a 250 $\mu$m $\times$ 250 $\mu$m surface area, the peak-to-peak roughness was less than 10 $\mu$m, and often much smaller than this value. The curved surfaces for different liquids were designed to satisfy $l_{c}/R_s$ = 0.03, where $R_s$ is the radius of curvature of the surface whose top and cross-sectional views are schematically shown in Figs.\ \ref{setup}a and \ref{setup}b, respectively. Following this principle, we fabricated three types of curved substrates, i.e. one for water, one for ethanol, methanol, acetone, and isopropanol, and one for liquid N$_2$, considering their respective capillary lengths ($l_c$) listed in Table\ \ref{Leidenfrost_tab:1}. For some experiments, a plano-concave, fused silica lens (focal length = 250 mm) was used as the heated substrate in order to allow for optical imaging of the capillary waves underneath the Leidenfrost drop. A T-type thermocouple (maximum measurable temperature: 473 K, tip diameter $\approx$ 0.08 cm, HYP-2, Omega Engineering) was used to measure the internal temperature profile of Leidenfrost drops.

For most of the substrates, a pressure sensor (GEMS Sensors, response time: 5 ms, sensitivity: 2 mV/Pa) was connected to a hole (diameter = 1 mm) at the center of the curved substrate in order to measure the pressure variations in the vapor layer at sample rates of 500-1000 Hz during quiescent and oscillatory phases of drops as illustrated in Fig.\ \ref{setup}b. We used a high-speed digital camera (Phantom V7.11, Vision Research) with a resolution of about 132 pixels/cm to image the motions of drops from above at frame rates of 1000 frames per second. Recorded videos were then analyzed with NIH ImageJ software to obtain the frequency and wavelength of the star-shaped oscillations (A typical star-shaped oscillation mode $n$ = 8 is schematically shown in Fig.\ \ref{setup}c).

\begin{table}
\begin{center}
\def~{\hphantom{0}}
\begin{tabular}{lcccccccc}
     \hline 
      \hline 
     liquid                 &$T_{b}$  &$\gamma$      &$\rho_l$   &$\eta_l$   &$\l_{c}$  &$T_s$ &Modes &$Re_l$\\       
     \hline
     water                &373        &59.0	        &958	    &0.282	&2.5       &523-773 &2-13       &1340     \\
     liquid N$_2$	    &77          &8.90     	 &807	    &0.162	&1.1       &298 &3-5,7        &539      \\
     acetone	          &329       &18.2	       &727	    &0.242	&1.6       &523 &5-10       &601    \\
     methanol	          &338        &18.9	       &748	    &0.295	&1.6       &523 &6-10        &511      \\
     ethanol              &352	     &18.6	       &750	    &0.420	&1.6       &523 &7-11      &355        \\
     isopropanol        &356      &15.7	       &723	    &0.460 	&1.5      &523  &9,10	       &283   \\
     \hline 
      \hline 
\end{tabular}
\caption[Physical properties of liquids at the boiling point $T_b$ (K). Units are as follows: $\gamma$ (mN/m), $\rho_l$ (kg/m$^3$), $\eta_l$ (mPa s), $l_c$ (mm). The last three columns indicate the range of substrate temperatures, $T_s$ (K), used in the experiments, the observed mode numbers ($n$), and the Reynolds number computed in Section\ \ref{stars of different liquids}]{Physical properties of different liquids at the boiling point $T_b$ (K). Units are as follows: $\gamma$ (mN/m), $\rho_l$ (kg/m$^3$), $\eta_l$ (mPa s), $l_c$ (mm). Data was taken from Ref. \cite{lemmon2011nist}. The last three columns indicate the range of substrate temperatures, $T_s$ (K), used in the experiments, the observed mode numbers ($n$), and the Reynolds number computed in Section\ \ref{stars of different liquids}.}
\label{Leidenfrost_tab:1}
\end{center}
\end{table}

\section{Results and discussion}
\label{results and discussions}
\subsection{The geometry of Leidenfrost drops on curved surfaces}
\label{dropshape}

The shape of a Leidenfrost drop is determined by a competition between surface tension and gravity. It has been shown that a nonwetting drop with radius $R \ll l_c$ will exhibit a quasi-spherical profile except for the bottom region where the drop is slightly flattened by gravity, which still holds true for small Leidenfrost drops \cite{burton2012geometry}. In comparison, for large Leidenfrost drops, i.e. $R \gg \l_c$, the shape of the drops is dominated by gravity and resembles a circular puddle with a constant thickness of approximately 2$l_c$ \cite{biance2003leidenfrost}. However, these large puddles are susceptible to the Rayleigh-Taylor instability, which is manifested by bubbles rising from the vapor layer beneath the drop \cite{biance2003leidenfrost}. This instability can be easily understood by considering a system composed of two fluids with the heavy fluid on the top of the light fluid. Such a system is gravitationally unstable since the heavy fluid is prone to fall through the light fluid, and any fluctuation at the initial flat interface of the two fluids will cause pressure imbalance, which in turn amplifies the fluctuation and deforms the drop. In our experiment, we used curved surfaces (see Section\ \ref{experiment}), which essentially act as a restoring-force-like role, in order to suppress the gravitationally driven Rayleigh-Taylor instability and obtain large, stable Leidenfrost drops \cite{snoeijer2009maximum,quere2013leidenfrost,trinh2014curvature}. 

\begin{figure}
\begin{center}
\includegraphics[width=4 in]{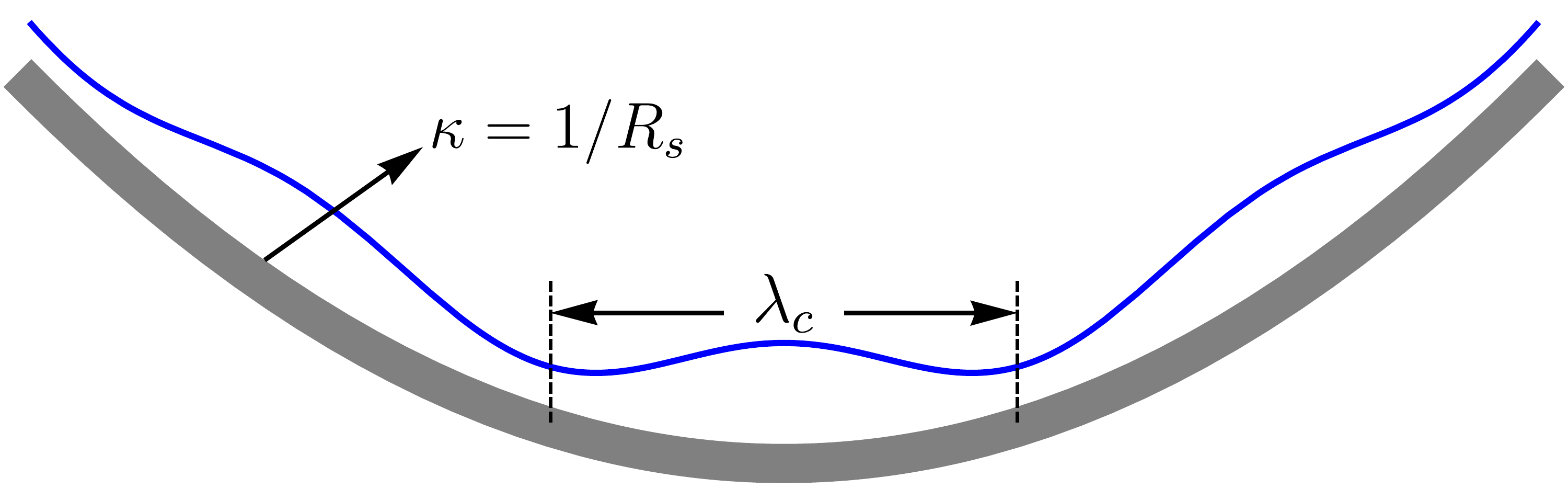}
\caption[Schematic of the suppression of Rayleigh-Taylor instability by surface curvature]{Schematic of the suppression of Rayleigh-Taylor instability at the liquid interface (solid blue curve) by a curved substrate (solid gray curve) with a curvature $\kappa=1/R_s$, where $R_s$ is the radius of curvature defined before, $\lambda_c$ is the capillary wavelength. 
} 
\label{RT_instability}
\end{center} 
\end{figure}
\begin{figure}
\begin{center}
\includegraphics[width=4 in]{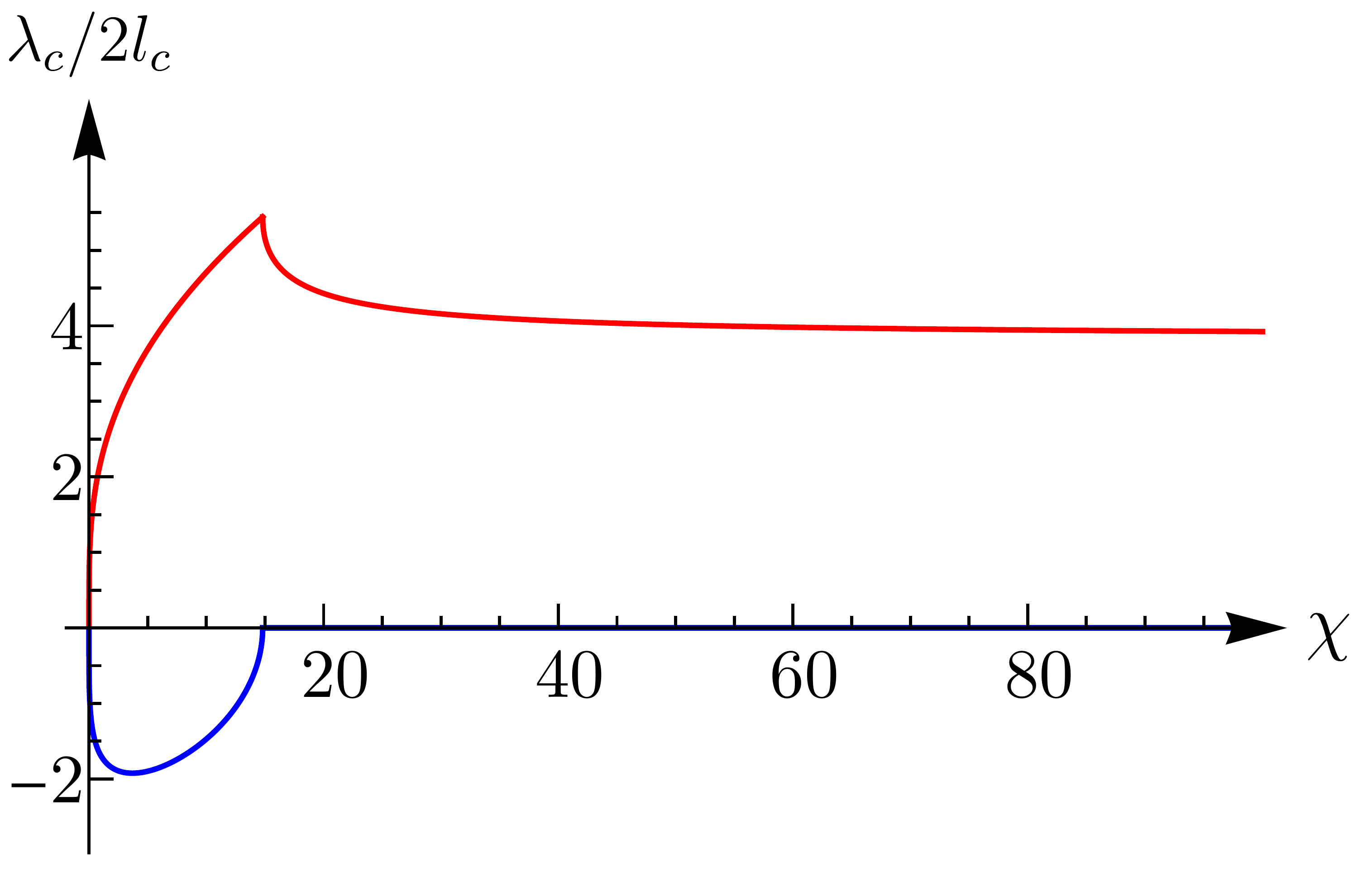}
\caption[Threshold of the perturbation amplitude for the suppression of Rayleigh-Tayor instability]{Plot of Eq.\ \ref{Rneck_scale}. The red and blue curves plot the real and imaginary parts of Eq.\ \ref{Rneck_scale}, respectively.}
\label{pertubation_amplitude}
\end{center} 
\end{figure}

The effects of surface curvature on the Rayleigh-Taylor instability can be seen through the following simple model, similar to the original model used by Biance \textit{et al.} \cite{biance2003leidenfrost}. Assuming the liquid-vapor interface beneath a Leidenfrost drop is perturbed with axisymmetric, sinusoidal variations, as illustrated in Fig.\ \ref{RT_instability}, then the shape function of the new drop interface, $S(r)$, can be expressed in cylindrical coordinate as:
\begin{equation}
S(r)=e+\frac{r^2}{2R_s}+\epsilon \cos \left( \frac{2\pi r}{\lambda_c} \right), 
\end{equation}
where $e$ is the mean vapor layer thickness, $\epsilon$ is the perturbation amplitude, and $\lambda_c$ is the capillary wavelength generated at the bottom of the drop (see Fig.\ \ref{sketch}). We also assume that $R_s\gg e\gg \epsilon$, implying a small perturbation to the equilibrium shape of the drop \cite{duchemin2005static,lister2008shape}. For a stable interface, the pressure at the drop center should be smaller or equal to the pressure at $r\approx\lambda_c/2$ in order to drive liquid back to the center. If the pressure in the vapor layer is constant, then to leading order, the pressures at $r$ = 0, and $r$ = $\lambda_c/2$ are:
\begin{align}
\label{P0}
P_0\approx\gamma \kappa|_{r\rightarrow0}&=2\gamma \left( \frac{1}{R_s} -\frac{4\epsilon \pi^2}{\lambda_c^2}\right),\\
\label{Pneck}
P_1\approx\gamma \kappa|_{r\rightarrow \lambda_c/2}-\rho_l g\Delta z&=\frac{4\epsilon\pi^2 \gamma}{\lambda_c^2}+\frac{2\gamma+\rho_l g \lambda_c}{2R_s}-2\epsilon \rho_l g,
\end{align}  
where the height difference between the two points is $\Delta z=S(r)_{r\rightarrow0}-S(r)_{r\rightarrow \lambda_c/2}=2\epsilon-\lambda_c^2/8R_s$. The condition for stability can be found by equating Eqs.\ \ref{P0} and \ref{Pneck}, leading to an expression for the wavelength:
\begin{equation}
\lambda_c=2\left( 2\epsilon R_s-\frac{\sqrt{2\rho_l g\epsilon R_s(-3\pi^2\gamma +2\rho_l g\epsilon R_s)}}{\rho_l g} \right)^{1/2}.
\label{Rneck}
\end{equation}
Since the perturbation amplitude and radius of curvature always appear as a product, Eq.\ \ref{Rneck} can also be written as:
\begin{equation}
\dfrac{\lambda_c}{l_c}=2\left( 2\chi-\sqrt{2\chi(2\chi-3\pi^2 )} \right)^{1/2},
\label{Rneck_scale}
\end{equation}
where $\chi=\epsilon R_s/l_c^2$. For finite $\epsilon$, in the limit $R_s\rightarrow\infty$, Eq.\ \ref{Rneck} simplifies to $\lambda_c/2\approx 3.85l_c$, in agreement with the prediction in Biance \textit{et al.} \cite{biance2003leidenfrost} for a flat surface. However, the addition of a curved surface couples the perturbation amplitude to the radius of curvature. The quantity $2\chi-3\pi^2$ must be positive in order for $\lambda_c$ to be real, and thus represent the condition for instability. Figure\ \ref{pertubation_amplitude} shows the plots of Eq.\ \ref{Rneck_scale}, in which the red curve shows the real part, whereas the blue curve shows the imaginary part. As we can see from Fig.\ \ref{pertubation_amplitude}, on a curved surface, the perturbation amplitude must satisfy $\chi\gtrsim14.8$, i.e., $\epsilon\gtrsim14.8 l_c^2/R_s$ in order to lead to the Rayleigh-Taylor instability. In this sense, we can conclude that the surface curvature suppresses the Rayleigh-Taylor instability. 

\begin{figure}
\begin{center}
\includegraphics[width=4.2 in]{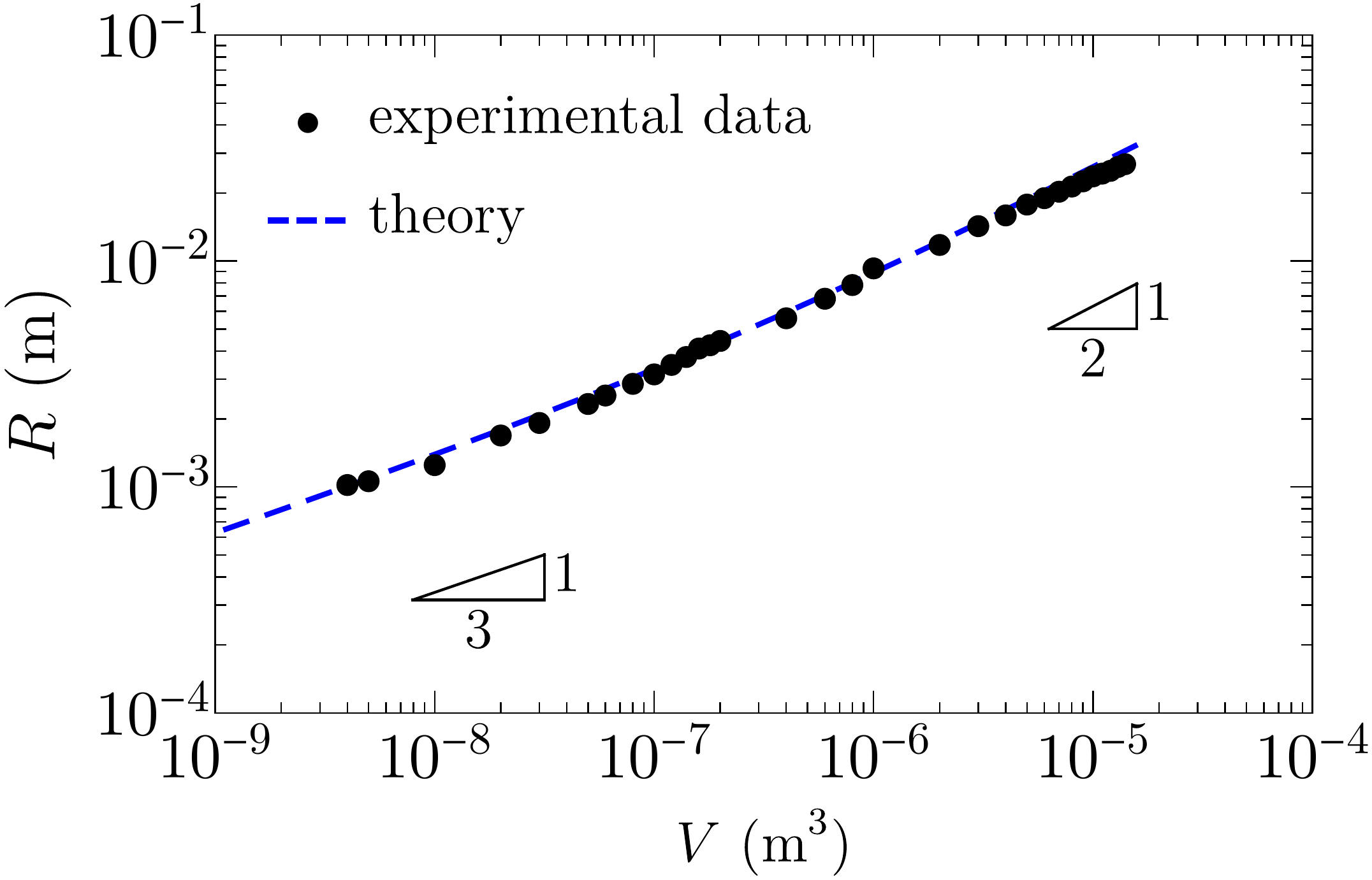}
\caption[Drop radius $R$ versus volume $V$ for star-shaped Leidenfrost water drops]{The variations of drop radius $R$ with respect to volume $V$ for Leidenfrost water drops. The dashed blue line shows the theoretical prediction for a drop on a flat surface assuming that the contact angle of the drop is $180^{\circ}$ as indicated in the text.}
\label{RV}
\end{center} 
\end{figure}

Although the curved surface suppresses the Rayleigh-Taylor instability, it still has an affect on the overall drop shape. We investigated the influence of surface curvature on the drop shape by depositing water drops of given volumes $V$ on a curved surface with $T_s$ = 623 K. We measured the radius $R$ of the drops from recorded images taken immediately after deposition prior to the onset of oscillations. As can be seen in Fig.\ \ref{RV}, stable Leidenfrost water drops with $R\approx 11l_c\approx$ 28 mm are obtained on the curved surface, while on a flat substrate, the maximum radius of a stable Leidenfrost water drop is $R\approx3.95l_c\approx$ 10 mm ($l_c$= 2.5 mm for Leidenfrost water drops, see Table\ \ref{Leidenfrost_tab:1}) \cite{snoeijer2009maximum,biance2003leidenfrost,burton2012geometry}, suggesting that surface curvature plays a crucial role in suppressing the Rayleigh-Taylor instability. The dashed blue line in Fig.\ \ref{RV} shows the theoretical prediction for a drop on a flat surface by solving the Young-Laplace differential equation numerically and assuming that the contact angle of the drop is $180^{\circ}$ \cite{burton2010experimental,burton2012geometry}. The experimental results show excellent agreement with the theoretical prediction, indicating that the shape of the Leidenfrost drops is not strongly affected by the surface curvature. More specifically, when $R<l_c$ (2.5 mm for water), the drops are quasi-spherical and thus $R\propto V^{1/3}$, whereas $R\propto V^{1/2}$ for $R>l_c$, which is expected for puddle-like drops with constant thickness.

However, one can also notice that most of the experimental data is slightly less than the corresponding theoretical prediction, which we attribute to two possible reasons. First, for small drops, the evaporation begins from the moment of deposition on the hot surface, which removes a small amount of water prior to imaging. Second, large drops are thicker in the center due to the underlying curved substrate, leading to a smaller apparent radius for a given volume. Thus, although surface curvature and evaporation may somewhat reduce the drop size, the effects seem to be a minor influence on the overall drop shape.

\subsection{Breathing mode of small Leidenfrost drops}
\label{The breathing mode}

Caswell \cite{caswell2014dynamics} experimentally characterized axisymmetric oscillations in the radius of small Leidenfrost drops using interference imaging. The drops displayed small-amplitude changes in the radius of the flat region near the surface. Due to volume conservation, an increase (decrease) in the drop radius leads to a decrease (increase) in the thickness of the drop. Caswell \cite{caswell2014dynamics} found that the oscillation frequency of the breathing mode, $f_b$, obeyed a distinct power law, $f_b \propto R_0^{-0.68\pm0.01}$, where $R_0$ is the average drop radius during the oscillation. This dependence is distinctly different than the expected three-dimensional dispersion relation for inviscid spherical drops, $f\propto R_0^{-3/2}$ \cite{rayleigh1879capillary}. Here we provide an analytical model which explains this contrast and fits the experimental data with no adjustable parameters. 

\begin{figure}[!tbph]
\begin{center} 
\includegraphics[width=3.5 in]{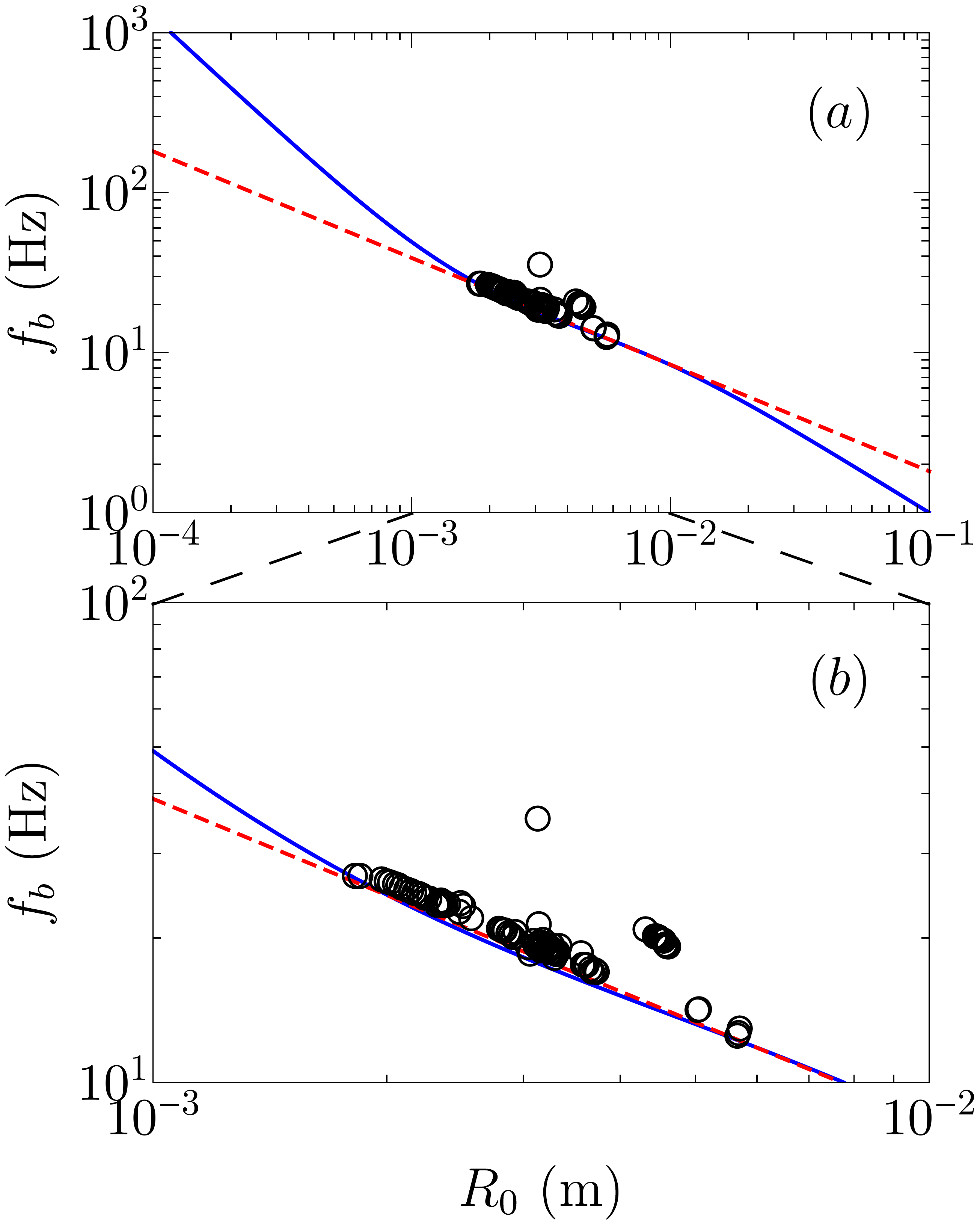}
\caption[Breathing oscillation mode of small Leidenfrost drops]{(a) Breathing mode oscillation frequency $f_b$ of small Leidenfrost drops as a function of the average drop radius $R_0$. The dashed red curve represents the power law $f_b \propto R_0^{-0.68}$, and the solid blue curve represents the dispersion relation from Eq.\ \ref{breathing_dispersion_relation}. (b) A zoomed-in view of the data in (a). With permission, the data shown here are taken from Caswell \cite{caswell2014dynamics}.}        
\label{breathing_mode}
\end{center} 
\end{figure}

To leading order, and due to the axisymmetry of the breathing mode, we model a Leidenfrost drop as an incompressible liquid cylinder of volume = $V$ and a time dependent radius, $R(t)$. We will assume that the bottom of the cylinder is fixed at $z=0$, which is reasonable if the thickness of the vapor layer varies much less than the radius. In cylindrical coordinates, $(r, \phi, z)$, the simplest form for the velocity which satisfies $\nabla\cdot\vec{\bf v}=0$ and the boundary conditions $\vec{\bf v}\cdot\hat{\bf r}|_{r=R(t)}=R'(t)$ and $\vec{\bf v}\cdot\hat{\bf z}|_{z=0}=0$ is:
\begin{equation}
\vec{\bf v}=\dfrac{R'(t)}{R(t)}\left(r,0,-2z\right).
\end{equation}
For simplicity, we will write $R(t)=R$, and the time derivative of $R$ as $R'(t)=R'$. The total kinetic energy of the drop is then:
\begin{align}
T=&\int\frac{1}{2}\rho_l\left| \vec{\bf v} \right|^2\mathrm{d}V=\dfrac{\pi \rho_l R^2}{R'^2}\int_{0}^{R}\int_{0}^{h}r(r^2+4z^2)\mathrm{d}z\mathrm{d}r.
\end{align}
Evaluating the integrals, and using the fact that $V=\pi R^2h$, where $h$ is the time-dependent height of the cylinder, we obtain:
\begin{align}
T=&\dfrac{1}{12}\rho_l V\left( 3+\frac{8V^2}{\pi ^2 R^6} \right)R'^2.
\end{align}
The total potential energy is the sum of gravitational potential energy and surface energy:
\begin{equation} 
U=\dfrac{1}{2}\rho_l Vgh+2\pi\gamma R(h+R)=\dfrac{gV^2\rho_l+4\pi\gamma R\left( V+\pi R^3 \right)}{2\pi R^2}.
\end{equation}  

The equilibrium drop radius, $R_0$, can be found by minimizing the potential energy with respect to $R$:
\begin{equation} 
\dfrac{\mathrm{d}U}{\mathrm{d}R}=0=4\pi\gamma R_0-\dfrac{2 V \gamma}{R_0^2}-\dfrac{g \rho_l V^2}{\pi R_0^3}.
\label{eqR}
\end{equation}  
For large drops where $R\gg h$, the first and third terms must balance. Equating these two terms and using the fact that $V=\pi R^2 h$, we see that:
\begin{equation} 
h\approx2\sqrt{\dfrac{\gamma}{\rho_l g}},\hspace{12 pt}R_0\rightarrow\infty,
\label{heq}
\end{equation}  
which agrees with the expected asymptotic thickness of large Leidenfrost drops \cite{biance2003leidenfrost}. We may now define the Lagrangian of the system as $L=T-U$, and apply the Euler-Lagrange equation to obtain a differential equation for $R$:
\begin{equation} 
R''=\frac{6\rho_l V^2g\pi R^4+12\gamma V \pi^2R^5-24\gamma \pi^3 R^8+24\rho_l V^3R'^2}{8\rho_l V^3R+3\rho_l V\pi^2R^7}.
\label{R''} 
\end{equation}

To proceed further, we will linearize the equation by considering only small oscillations of the radius, $R=R_0(1+\epsilon e^{i \omega t})$, where $R_0$ is equilibrium drop radius found by solving Eq.\ \ref{eqR}, $\epsilon$ is the perturbation amplitude, and $\omega$ is the angular frequency of the oscillation. Assuming that $\epsilon\ll 1$, to leading order Eq.\ \ref{R''} reduces to an expression for the angular frequency:
\begin{equation} 
\omega^2=\dfrac{576\gamma^2(\gamma +\Gamma)+36 \rho_l^2g^2  R_0^4 (\gamma +2 \Gamma )+384 \gamma \rho_l g  R_0^2 (5 \gamma +2 \Gamma )}{\rho_l  R_0^3 \left(1120 \gamma^2+9 \rho_l^2 g^2  R_0^4+192 \gamma  \rho_l g  R_0^2\right)},
\label{breathing_dispersion_relation}
\end{equation}
where $\Gamma\equiv\sqrt{\gamma^2+4\gamma \rho_l g R_0^2}$ and we have substituted $V$ for the equilibrium radius $R_0$ using Eq.\ \ref{eqR}.

Taking typical values of the parameters in Eq.\ \ref{breathing_dispersion_relation} for water, we can plot the oscillation frequency of the breathing mode, $f_b=\omega/2\pi$, as a function of $R_0$, which is shown by the solid blue curve in Fig.\ \ref{breathing_mode}a. The theory shows excellent agreement with the data. For comparison, the power law $f_b \propto R_0^{-0.68}$ found by Caswell \cite{caswell2014dynamics} is plotted as the dashed red line in Fig.\ \ref{breathing_mode}a. A closer view of the comparison between the prediction by Eq.\ \ref{breathing_dispersion_relation} and $f_b \propto R_0^{-0.68}$ can be seen in Fig.\ \ref{breathing_mode}b. The data lie in the transition regime between two asymptotic limits, $R_0\ll l_c$ and $R_0\gg l_c$, which may explain the anomalous power law reported by Caswell \cite{caswell2014dynamics}. In these two limits, Eq.\ \ref{breathing_dispersion_relation} reduces to:
\begin{align} 
&\omega\approx\left(\dfrac{36\gamma}{35\rho_l R_0^3}\right)^{1/2},\hspace{12 pt}R_0\rightarrow 0\\
&\omega\approx\dfrac{4}{R_0}\left(\dfrac{g\gamma}{\rho_l}\right)^{1/4},\hspace{12 pt}R_0\rightarrow\infty\nonumber.
\end{align}  
The scaling for small drops is independent of $g$, which is expected based on their nearly-spherical shape. For large drops where $h\approx2l_c$ (Eq.\ \ref{heq}), the scaling is determined essentially by gravity: $\omega\sim\sqrt{2gh}/R_0$.

\begin{figure}[!tbph]
\begin{center}
\includegraphics[width=5.5 in]{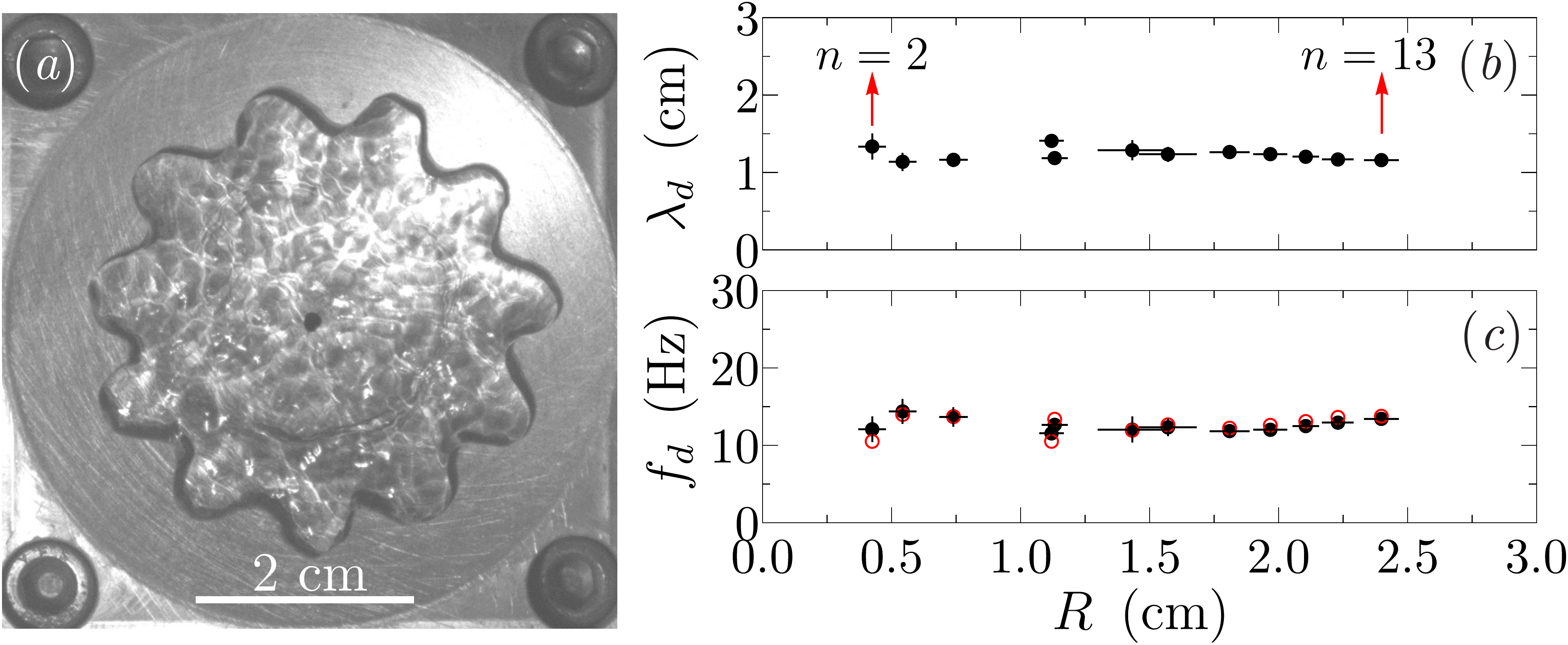}
\caption[Drop oscillation wavelength $\lambda_d$ and frequency $f_d$ versus drop radius $R$ for star-shaped Leidenfrost water drops]{Star-shaped oscillations of Leidenfrost water drops at $T_s$ = 623 K. (a) A snapshot of the oscillation mode with $n = 11$. (b) and (c) show the star-shaped oscillation wavelength $\lambda_d$, and frequency $f_d$, with respect to drop radius $R$. The data points from left to right  in (b) and (c) represent the increasing mode number $n$ as indicated by the red arrows, and the error bars in (b) and (c) are the standard deviations of multiple drops. The data for $\lambda_d$ in (b) are indirectly measured using the relation $2 \pi R = n\lambda_d$, and the red circles in (c) represent the theoretical prediction by Eq.\ \ref{dispersion2D} for drops with different radii.} 
\label{wavelength_frequency}
\end{center} 
\end{figure}

\subsection{Star-shaped oscillations of Leidenfrost water drops}
\label{stars of water drop}

In addition to the breathing mode, large Leidenfrost drops may also develop azimuthal, star-shaped oscillations. Similar oscillations have been observed in a variety of systems involving liquid drops \cite{brunet2011star}. In our experiments with Leidenfrost water drops using a curved substrate, we observe star-shaped oscillation modes with $n$ = 2 to 13 lobes along the drop periphery at $T_s=623$ K. A typical star-shaped oscillation mode ($n=11$) is shown in Fig.\ \ref{wavelength_frequency}a. Figures\ \ref{wavelength_frequency}b and \ref{wavelength_frequency}c show the drop oscillation wavelength $\lambda_d$ and frequency $f_d$ for different $R$, respectively, which are measured by analyzing the high-speed videos of the oscillating drops. Surprisingly, both $\lambda_d$ and $f_d$ remained nearly constant as we varied $R$. Increasing $R$ thus led to an increase in the allowable number of lobes $n$, as indicated by the red arrows. This similar trend also applies to other liquids used in our experiments, which will be discussed in Section\ \ref{stars of different liquids}.

For the free oscillations of an incompressible, axisymmetric spherical drop with infinitesimal deformations, the natural resonance frequency, $f_n$, of the $n$th-mode is given by \cite{rayleigh1879capillary}:
\begin{equation} 
f_n=\frac{1}{2\pi}\sqrt{\frac{\gamma n(n-1)(n+2)}{\rho_l R^{3}}}.
\label{dispersion3D} 
\end{equation}  
For a liquid puddle with $R\gg h\approx$ 2$l_c$, where $h$ is the thickness of the liquid puddle \cite{biance2003leidenfrost}, the resonance frequency $f_n$ takes the form \cite{yoshiyasu1996self}:
\begin{equation} 
f_n=\frac{1}{2\pi}\sqrt{\frac{\gamma n(n^{2}-1)}{\rho_l R^{3}}}\sqrt{{\frac{1}{1+(2-\frac{\pi}{2}+\frac{n-3}{4})\frac{l_c}{R}}}}.
\label{dispersion2D} 
\end{equation}  
The first term under the square root is for a strictly two-dimensional drop, whereas the correction factor, $1/\sqrt{1+(2-\frac{\pi}{2}+\frac{n-3}{4})\frac{l_c}{R}}$, is due to the quasi-two-dimensional nature of the puddle. 

As shown in Fig.\ \ref{wavelength_frequency}, for water drops where we observed $n$ = 2 modes, the average radius was $R\approx$ 4 mm, which is greater than $l_c$. Hence, it is reasonable to use Eq.\ \ref{dispersion2D} to predict the oscillation frequency for all star-shaped modes in our experiments. The comparison between theory and experiment is shown in Fig.\ \ref{wavelength_frequency}c, which indicates an excellent agreement. For smaller values of $n$, both $\lambda_d$ and $f_d$ vary non-monotonically with $R$. There are a few potential reasons for this. One possibility is that nonlinear effects play a more significant role for smaller values of $n$ because the ratio of the mode amplitude to the drop radius is larger \cite{becker1991experimental,smith2010modulation}. However, the excellent agreement with the linear theory, i.e. Eq.\ \ref{dispersion2D}, suggests that nonlinear effects may not be important. 

Instead, we suggest that the behavior may be related to the mode selection mechanism. For a given radius, either the frequency or the wavelength is preferentially selected by the excitation mechanism. Once $f_d$ or $\lambda_d$ is selected, Eq.\ \ref{dispersion2D} will determine the other. As will be shown in Section\ \ref{capillary_waves_sec}, there are strong capillary waves excited at the liquid-vapor interface beneath the drop. These waves produce pressure oscillations which parametrically couple to the star-shaped modes. A determination of the expected behavior of $f_d$ on $R$ is thus complicated by the physics of the vapor flow beneath the drop, nevertheless, we would expect the effects to be less variable for $R\gg\lambda_d$, which agrees with the data shown in Fig.\ \ref{wavelength_frequency}. 

\subsection{Pressure oscillations in the vapor layer}
\label{pressure oscillation}

It has been well-documented that star oscillations of large liquid drops can be initiated by a parametric forcing mechanism \cite{brunet2011star}. For instance, when a liquid puddle is deposited onto a superhydrophobic, vertically-vibrating substrate with a prescribed frequency, then the equation of motion of the drop can be described by an equation similar to Mathieu's equation, and star oscillations will be excited when the drop oscillation frequency is approximately half that the excitation frequency \cite{yoshiyasu1996self,brunet2011star}. In the experiment, we observed star oscillations of Leidenfrost water drops (e.g. see Fig.\ \ref{wavelength_frequency}a) in the absence of an external excitation. Therefore, we hypothesize that the star oscillations are driven by fluctuations in the vapor layer pressure that supports the drop.

In order to explore the dynamics of pressure fluctuations in the vapor layer, we utilized a pressure sensor attached to the center of the curved substrate, schematically shown in Fig.\ \ref{setup}b. We measured the pressure variations in the vapor layer of water drops at $T_s$ = 623 K during the initiation of the star-shaped oscillations. We started recording the pressure immediately after the drops were placed on the substrate, and stopped after the well-defined star-shaped oscillations had been initiated for several seconds. The whole process lasted about 18.5 s. 

\begin{figure}[!tbph]
\begin{center}
\includegraphics[width=1\textwidth]{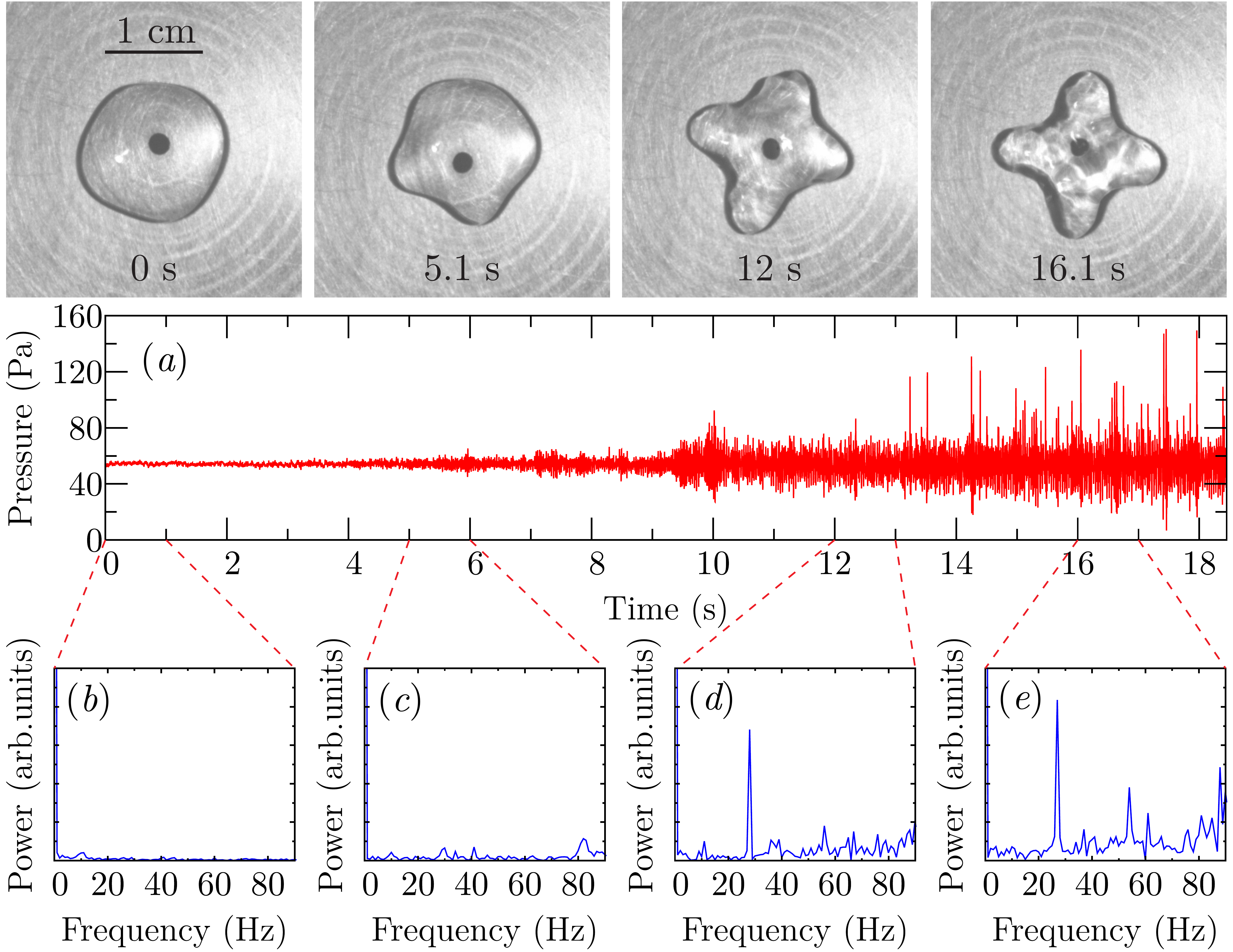}
\caption[Initiation of star-shaped oscillation of a Leidenfrost water drop]{(a) The pressure variations in the vapor layer during the initiation of a 4-mode star-shaped oscillation of a Leidenfrost water drop within $\approx$ 18.5 s.  (b), (c), (d), and (e) represent the Fourier power spectra of the pressure in the vapor layer during the time intervals of 0-1 s, 5-6 s, 12-13 s, and 16-17 s, respectively. The snapshots of the top panel represent the drop profile at 0 s, 5.1 s, 12 s, and 16.1 s, respectively, during the initiation process.} 
\label{parametric_oscillation}
\end{center} 
\end{figure}

The snapshots on the top panel of Fig.\ \ref{parametric_oscillation}a show images of an $n$ = 4 drop at different stages during the oscillation. More specifically, the drop shape was nearly circular immediately upon placement on the substrate ($t$ = 0 s), then the pressure fluctuations in the vapor layer increased as the drop shape developed well-defined lobes. The amplitude continued to grow with time ($t$ =12 s) until the formation of a steady-state, star-shaped oscillation with large amplitude ($t$ = 16.1 s).  

Figure\ \ref{parametric_oscillation}a shows the pressure variations in the vapor layer during the whole process. The mean pressure required to support the drop is $\rho_l gh\approx2\rho_l g l_c= 47$ Pa. However, the mean pressure we measure is slightly larger than this value because the pressure is measured at the center of the substrate where the pressure is larger in order to drive the viscous vapor to flow out to the drop edge. The pressure fluctuations around the mean increase with time until the formation of a steady star-shaped oscillation with a large amplitude (about $t$ = 14 s).

In order to gain insight into the underlying relationship between the oscillations of the drop and pressure in the vapor layer, we performed a Fast Fourier Transform on the pressure data in four different time intervals: 0-1 s, 5-6 s, 12-13 s, and 16-17 s. The results are shown in Figs.\ \ref{parametric_oscillation}b-\ref{parametric_oscillation}e.  Initially, the pressure remains nearly constant, and there are no sharp peaks in the power spectrum (Fig.\ \ref{parametric_oscillation}b). Between $t$ = 5-6 s, the pressure fluctuations become stronger and more periodic, and several small peaks are visible (Fig.\ \ref{parametric_oscillation}c). Then, as the star oscillation is further developed, a sharp peak which is located at $\approx$ 28 Hz shows up in the power spectrum (Fig.\ \ref{parametric_oscillation}d). Finally, when the star oscillation is fully developed, the pressure fluctuations stop growing, and a sharp peak located at $\approx$ 28 Hz dominates the spectrum (Fig.\ \ref{parametric_oscillation}e). 

It is interesting to note that the location of the sharp peak is approximately twice the oscillation frequency of a fully-developed, $n$ = 4 mode Leidenfrost water drop (see Fig.\ \ref{wavelength_frequency}c). This is consistent with a parametric forcing mechanism for the excitation of the star oscillations. Generally, for an oscillator with time-dependent frequency, the oscillation amplitude exponentially increases with time when the driving frequency is approximately twice that the natural frequency, which is known as parametric forcing. A well-known example is the parametric pendulum with a natural frequency of $\sqrt{g/l}/(2\pi)$, and parametric oscillation occurs when the vertical frequency $\sqrt{k/m}/(2\pi)$ of the center of mass of the pendulum is twice that natural frequency, i.e., $\sqrt{k/m}/(2\pi)=\sqrt{g/l}/\pi$ \cite{landau1976course}, where $l$ and $m$ are the length and mass of the pendulum, respectively, whereas $k$ is the spring constant of the pendulum. Moreover, harmonics at higher frequencies are also visible as indicated by the secondary peaks in Fig.\ \ref{parametric_oscillation}e. This is likely due to nonlinear effects involved in the star-shaped oscillations. In addition, the location of the peak is robust. As will be shown in Section\ \ref{thermal effect}, the location of the dominant peak in the power spectrum is mostly independent of the substrate temperature and the environmental temperature. 

\begin{figure}
\begin{center}
\includegraphics[width=4.2 in]{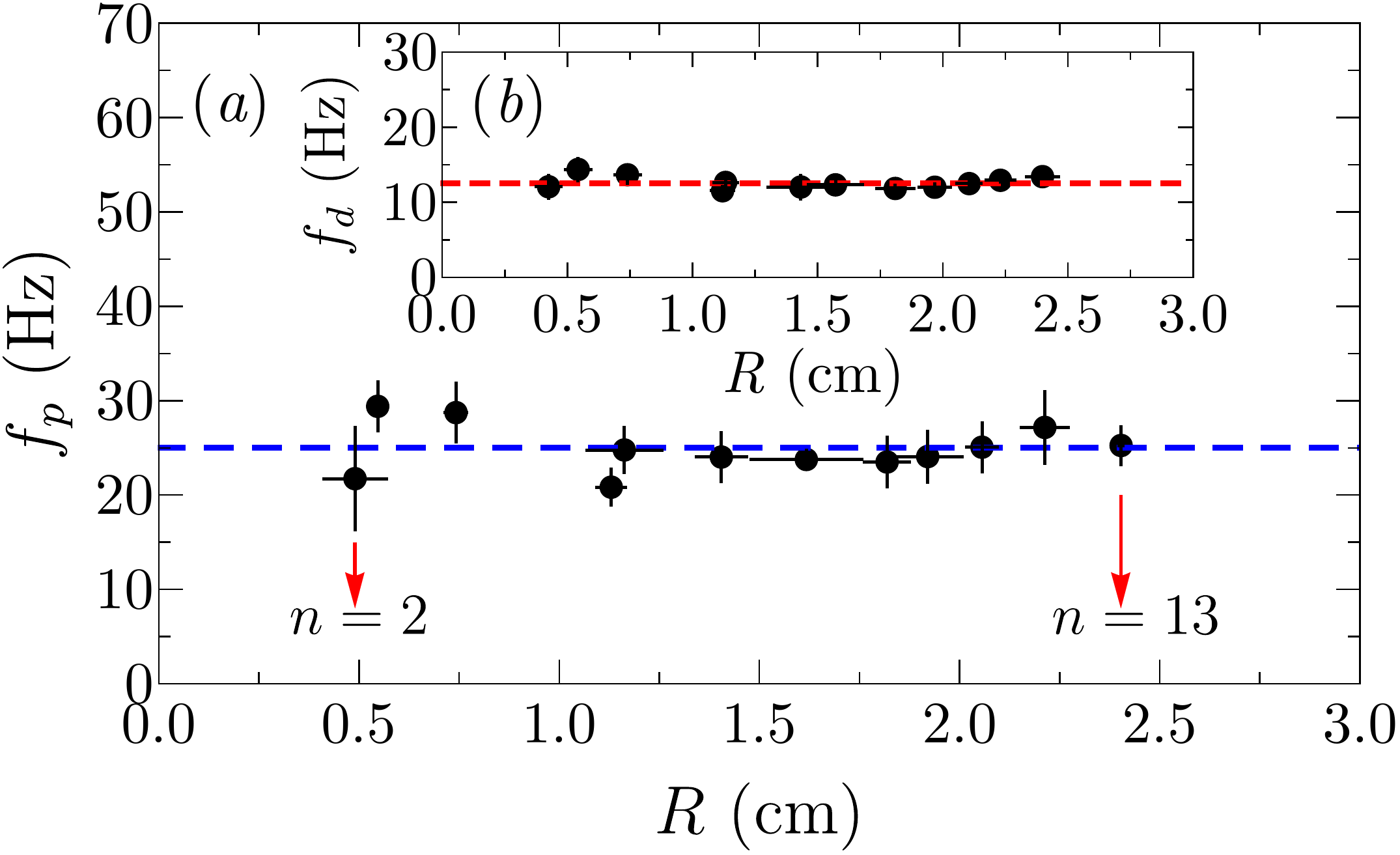}
\caption[Pressure oscillation frequency $f_p$ and drop oscillation frequency $f_d$ versus drop radius $R$]{(a) Summary of the pressure oscillation frequencies $f_p$ in the vapor layer of star-shaped Leidenfrost water drops with different radii at $T_s=623$ K. The data points from left to right represent the oscillation modes from $n=2$ to 13, respectively, as indicated by the red arrows. (b) The drop oscillation frequency $f_d$ with respect to drop radius $R$ (Fig.\ \ref{wavelength_frequency}c). The error bars of $f_p$ are taken from the full-width at half-max of the highest peak in the Fourier power spectra computed over a time interval of 10 s, and the error bars of $R$ are defined as the standard deviation of multiple measurements of different drops after the star oscillations are fully developed for each mode. The dashed lines in (a) and (b) are visual guides to indicate that $f_p\approx2f_d$.
} 
\label{f_vs_R}
\end{center} 
\end{figure}

Figure \ref{f_vs_R}a shows the pressure oscillation frequency $f_p$ in the vapor layer of Leidenfrost water drops for all of the observed modes. The data points from left to right correspond the oscillation modes from $n$ = 2 to 13, as indicated by the red arrows. By comparing this with the drop oscillation frequency $f_d$ (Fig.\ \ref{f_vs_R}b), we find $f_p\approx 2f_d$ as indicated by the dashed blue and red lines in Figs.\ \ref{f_vs_R}a and \ref{f_vs_R}b, respectively. This robust relationship suggests that the star-shaped oscillations are parametrically driven by the pressure \cite{miles1990parametrically,kumar1994parametric,yoshiyasu1996self,brunet2011star,terwagne2011tibetan}. Consider the radial position of a point on the perimeter of the drop during a star-shaped oscillation, $r(t)$. This point will oscillate in time due to the azimuthal standing wave, and obey the following equation:
\begin{equation} 
\frac{\mathrm{d}^2r}{\mathrm{d}t^2}+\omega^2r=0.
\label{dropmotion}
\end{equation}  
To leading order, $\omega$ is the resonant frequency of the mode, Eq.\ \ref{dispersion2D} can be written as:
\begin{equation}
\omega_{0}^{2}=\dfrac{n\left( n^2-1\right) \gamma}{\rho_l R^{3}}.
\label{dispersionrelationatR0}
\end{equation}
For simplicity, we have ignored the correction due to the quasi-two-dimensional nature of the drop. The pressure variations will induce vertical oscillations of the drop, and in turn oscillations of the drop radius:
\begin{equation} 
R=R_0(1+\epsilon \cos\omega_p t), 
\label{Rvariation}
\end{equation}  
where $R_0$ is the average radius of the drop, and $\epsilon$ is the amplitude of the small perturbation. Plugging Eq.\ \ref{Rvariation} into Eq.\ \ref{dispersionrelationatR0}, to leading order in $\epsilon$ we obtain $\omega^2=\omega_0^2(1-3\epsilon\cos\omega_p t )$. Therefore, Eq.\ \ref{dropmotion} now becomes:
\begin{equation} 
\frac{\mathrm{d}^2r}{\mathrm{d}t^2}+\omega_0^2\left( 1-3\epsilon \cos \omega_p t \right)r=0.
\label{dropmotion1}
\end{equation}  

When $\omega_p \approx 2\omega_0$, the parametric resonance is the strongest, i.e. the oscillation amplitude will exponentially increase with time until the star-shaped oscillations are fully developed. This is consistent with our measurements of the pressure oscillation frequencies in the vapor layer (Fig.\ \ref{f_vs_R}), which are approximately twice that the drop oscillation frequencies (Fig.\ \ref{wavelength_frequency}c) that we observed in our experiment. We note that it is possible that the pressure oscillations are a \emph{consequence} of the star oscillations. If the star oscillations were excited by some other mechanism, then the symmetry of the star shape would necessitate a pressure extrema when the star shape reaches its maximum amplitude. However, the only energy which is available to drive the oscillation comes from the evaporation and gas flow in the vapor layer. Thus we expect that the pressure oscillations are the source of the star oscillations. Section\ \ref{capillary_waves_sec} will explore the potential source of the pressure oscillations.

\begin{figure}
\begin{center}
\includegraphics[width=1\textwidth]{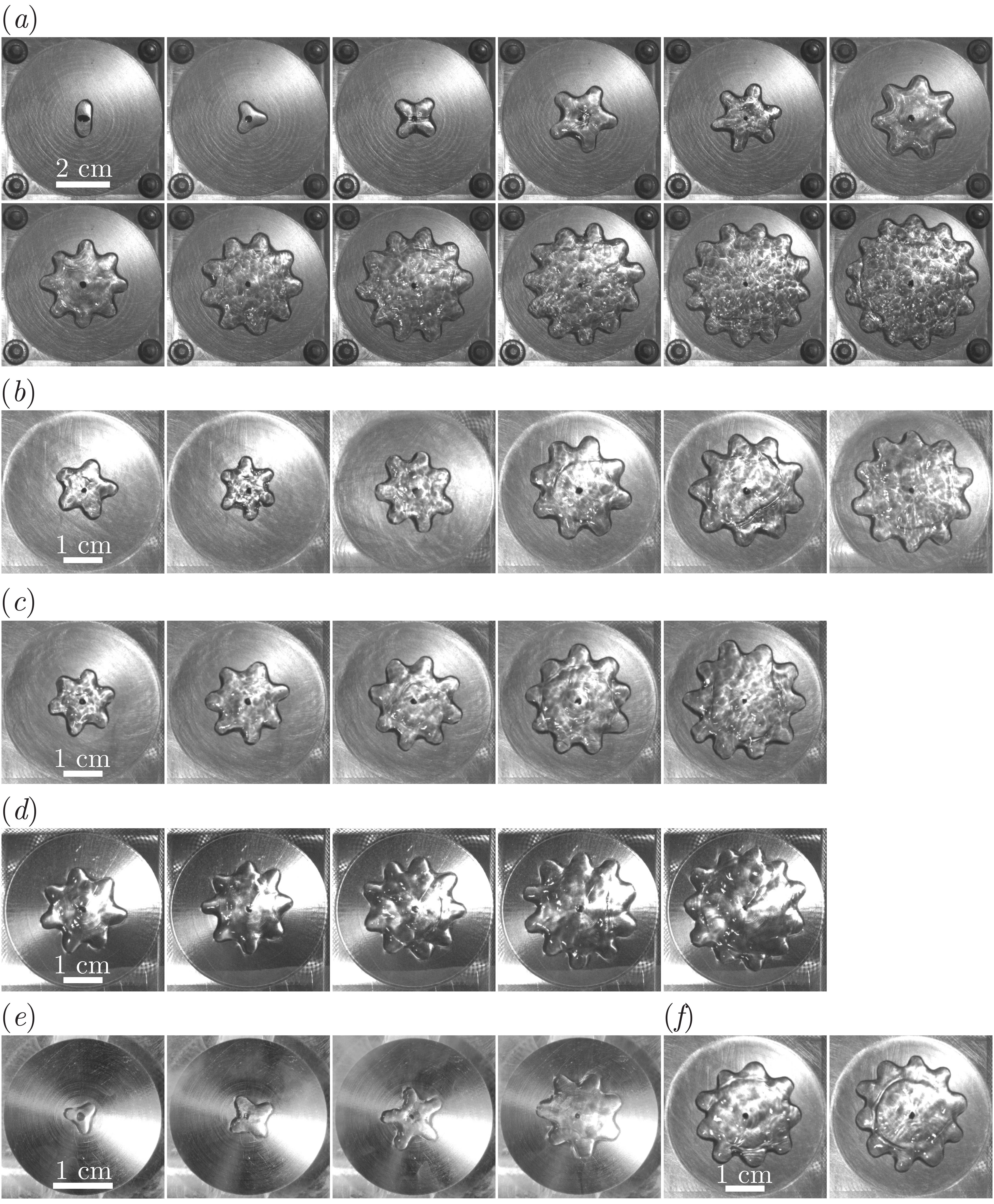}
\caption[Images of star-shaped oscillation modes with $n = 2$ to 13 of six different liquids when the lobes are at their maximum displacement.]{Snapshots of star-shaped oscillation modes with $n = 2$ to 13 of six different liquids when the lobes are at their maximum displacement. Panels (a), (b), (c), (d), (e), and (f) represent water, acetone, methanol, ethanol, liquid N$_2$, and isopropanol, respectively. For water: $T_s = 623$ K, for acetone, methanol, ethanol and isopropanol: $T_s = 523$ K. The substrate for liquid N$_2$ was not heated.
} 
\label{modegallery}
\end{center} 
\end{figure}

\subsection{Star-shaped oscillations of different liquids}
\label{stars of different liquids}

In addition to water, we performed similar experiments with five other liquids: liquid N$_2$, acetone, methanol, ethanol, and isopropanol. For these liquids, the curved surfaces were also machined to satisfy $l_{c}/R_s \approx 0.03$ (see Section\ \ref{experiment}), and the relevant physical properties and substrate temperatures are listed in Table\ \ref{Leidenfrost_tab:1}. Figure\ \ref{modegallery} shows snapshots of star-shaped oscillations for the six different liquids we used in the experiments, where the other five liquids show similar star-shaped patterns to water drops but with different observable oscillation modes. By analyzing the images of the star-shaped oscillations, we find that star-shaped oscillations of acetone, methanol, ethanol, and isopropanol share a similar wavelength $\lambda_d \approx 0.9$ cm and frequency $f_d \approx 14$ Hz, whereas liquid N$_ 2$ shows a wavelength of $\lambda_d \approx 0.6$ cm and frequency $f_d \approx 17$ Hz. This suggests that the star oscillations may depend on the capillary length of the liquid.

\begin{figure}[!tbph]
\begin{center}
\includegraphics[width=4.3 in]{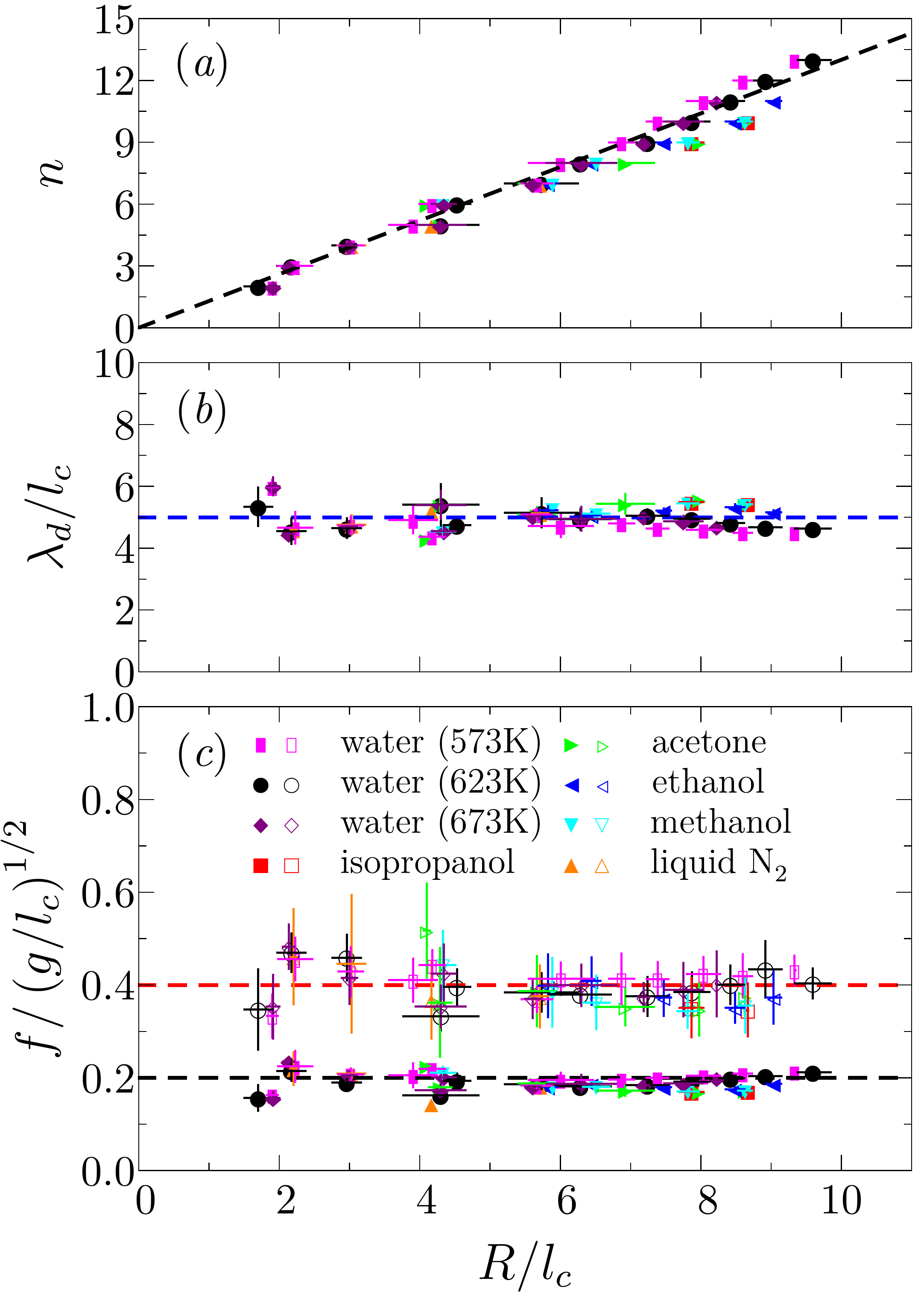}
\caption[Parametric oscillations of star-shaped oscillations of Leidenfrost drops]{Dependence of star-shaped oscillation mode $n$ (a), and dimensionless oscillation wavelength $\lambda_d/l_c$ (b) and frequency $f/\sqrt{g/l_c}$ (c) with respect to the dimensionless drop radius $R/l_c$ for six different liquid. In (c), the solid symbols represent the drop oscillation frequency $f_d$, whereas the open symbols denote the corresponding pressure oscillation frequency $f_p$. The error bars of $R$, $\lambda_d$ and $f_d$ come from the standard deviations of multiple measurements of different drops, and the error bars of $f_p$ are defined as the full width of the highest peak in the power spectrum at half the maximum value. The dashed lines in all panels are visual guides.}
\label{scaling}
\end{center} 
\end{figure}

Figure\ \ref{scaling} shows the dependence of star-shaped oscillation mode $n$, rescaled wavelength $\lambda_d/l_c$, and rescaled frequency $f/\sqrt{g/l_c}$ on the rescaled drop radius $R/l_c$ of the six liquids used in the experiment. The mode number $n$ increases linearly with $R/l_c$ as indicated by the dashed black line in Fig.\ \ref{scaling}a, showing that large modes can only be observed in larger drops, as supported by Fig.\ \ref{modegallery}. The rescaled wavelength $\lambda_d/l_c$ for the six liquids collapses onto a straight line as shown in Fig.\ \ref{scaling}b, which indicates that the wavelength of the star oscillations only depends on $l_c$ of the liquid. Figure\ \ref{scaling}c shows the oscillation frequency of both the azimuthal star oscillations, $f_d$, and the corresponding pressure, $f_p$, in the vapor layer. The solid symbols denote the drop oscillations, whereas open symbols with the same color represent the corresponding pressure oscillations in the vapor layer for a specific liquid. Both $f_d$ and $f_p$ collapse fairly well, and the relationship $f_p \approx 2f_d$ is consistent with the data, as indicated by the dashed black and red lines. Additionally, by comparing different values of $T_s$ for water, Fig.\ \ref{scaling} shows that $\lambda_d$, $f_d$, and $f_p$ are mostly independent of substrate temperature. 

The main influence of $T_s$ is the through the number of observable modes. For instance, at $T_s$ = 673 K, we did not observe oscillation modes $n$ = 12 and $n=13$ in water, which we attribute to the extremely fast evaporation rate of the liquid, as will be discussed in Section\ \ref{capillary_waves_sec}. Although the robust relationship between $f_d$ and $f_p$ suggests that the star oscillations of the drops are parametrically driven by the pressure in the vapor layer, there is an interesting variation in both $f_d$ and $f_p$ at small mode numbers. The data for smaller modes (e.g. $n= 2, 3, 4$) deviates further from the average compared to larger modes. This behavior may be nonlinear in nature since the amplitude of the mode relative to the drop radius is much larger for small $n$ \cite{becker1991experimental,smith2010modulation}. In addition, the quasi-two-dimensional nature of the drop is more important at small $n$, which may also contribute to the variations in this regime. A more quantitative explanation for this dependence is left for future studies.

\begin{figure}
\begin{center}
\includegraphics[width=4.3 in]{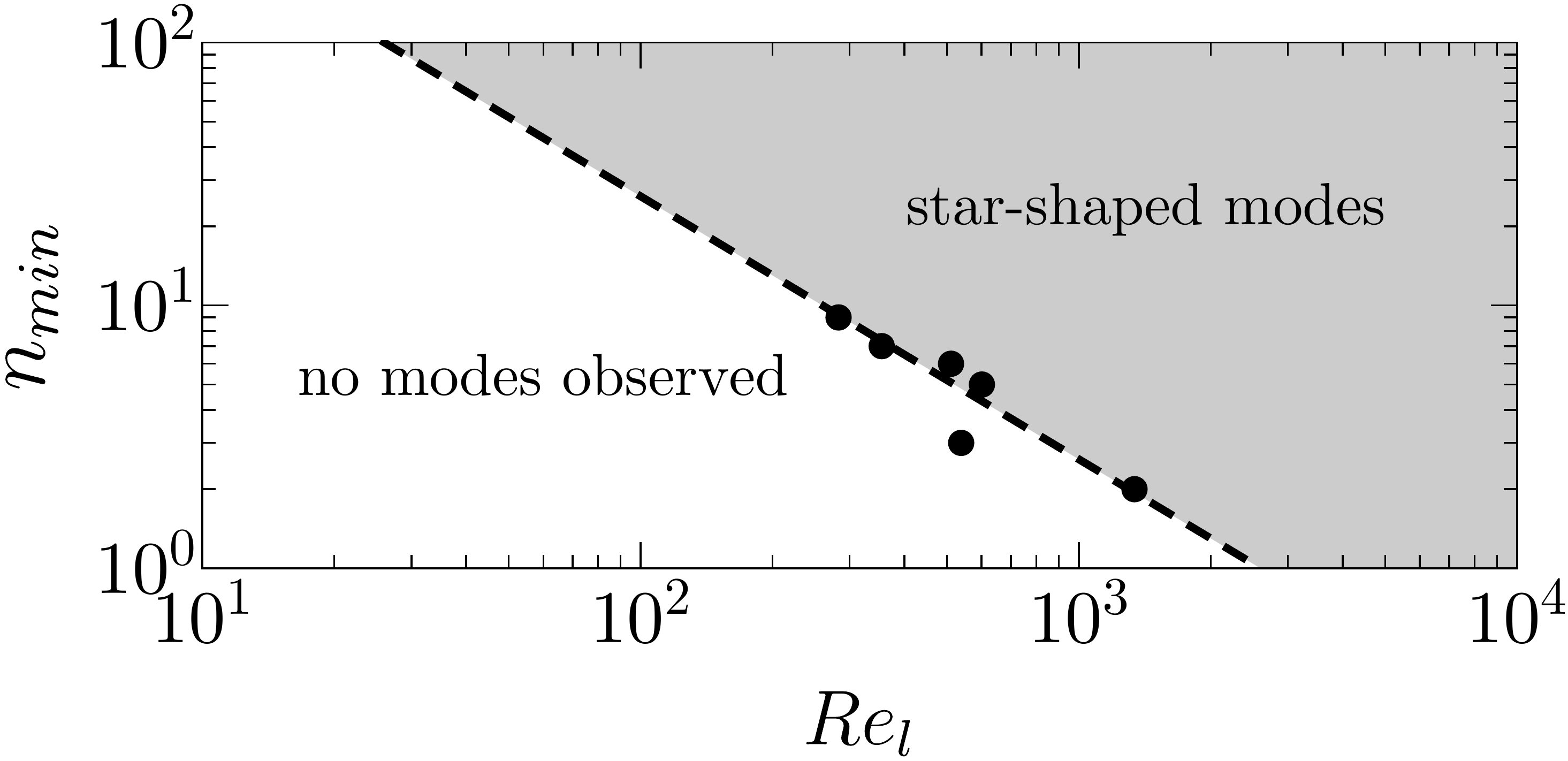}
\caption[Threshold for star-shaped oscillations of Leidenfrost drops]{Scaling behavior for the minimum observed star mode $n_{min}$ with respect to $Re_l$ of different liquids (Table\ \ref{Leidenfrost_tab:1}). The dashed line represents the fit to the data, 2600 $Re^{-1}_l$.
} 
\label{threshold}
\end{center} 
\end{figure}

As shown in Table\ \ref{Leidenfrost_tab:1}, different oscillation modes are observed in different liquids. In particular, some liquids failed to display smaller mode numbers. This may be due to inherent viscous damping that prevents the sustained excitation of large-amplitude oscillations. Following the analysis in Ma \textit{et al.} \cite{Maleidenfrost2016}, the role of damping can be characterized by the Reynolds number associated with the liquid oscillation. For a star oscillation, the characteristic length scale and time scale are $l_c$ and $\sqrt{l_c^3\rho_l/\gamma}$, respectively. Thus, the inertial term in the Navier-Stokes equation is estimated as $|\rho_l(\vec{\bf v}\cdot\nabla)\vec{\bf v}|=\gamma/l_c^2$, and the viscous term is $|\eta_l\nabla^2\vec{\bf v}|=\eta_l/\sqrt{l_c^5\rho_l/\gamma}$. The Reynolds number of the oscillating liquid is defined as the ratio of inertial to viscous terms: $Re_l=\sqrt{l_c\rho_l\gamma}/\eta_l$, and the values of $Re_l$ for different liquids are listed in Table\ \ref{Leidenfrost_tab:1}. 

Figure\ \ref{threshold} shows the dependence of the minimum mode number $n_{min}$ on $Re_l$ for each liquid. The dashed line represents a suggested scaling $Re^{-1}$, indicating that the liquid viscosity damps the oscillations of smaller drops and thus sets the minimum mode number of stable star oscillations. However, the substrate temperature and thus evaporation rate will likely affect the number of observed modes as well. A higher evaporation rate will induce a stronger driving of the oscillation modes, which would reduce $n_{min}$. This can be seen in Fig.\ \ref{threshold} for liquid N$_{2}$, which undergoes much more rapid evaporation due to the large difference between the boiling and substrate temperature, $T_s-T_b$. In addition, as mentioned previously, the stronger evaporation may also inhibit larger modes if the pressure oscillations are less coherent. In the next section we show how the flow in the vapor layer is linked to the pressure oscillations.

\subsection{Origin of pressure oscillations in the vapor layer}
\label{capillary_waves_sec}

As evidenced by Fig.\ \ref{scaling}c, the star-shaped oscillations of Leidenfrost drops are parametrically driven by the pressure variations in the vapor layer. Star oscillations induced by a parametric coupling have been observed in a variety of systems where variations of the drop radius are induced by external, periodically-modulated fields \cite{brunet2011star}. However, in our experiment, there are no obvious external fields. It is then crucial to understand the source of the pressure oscillations. In this section we show how capillary waves of a characteristic wavelength at the liquid-vapor interface induce the pressure oscillations in the vapor layer.

\begin{figure}
\begin{center}
\includegraphics[width=4.2 in]{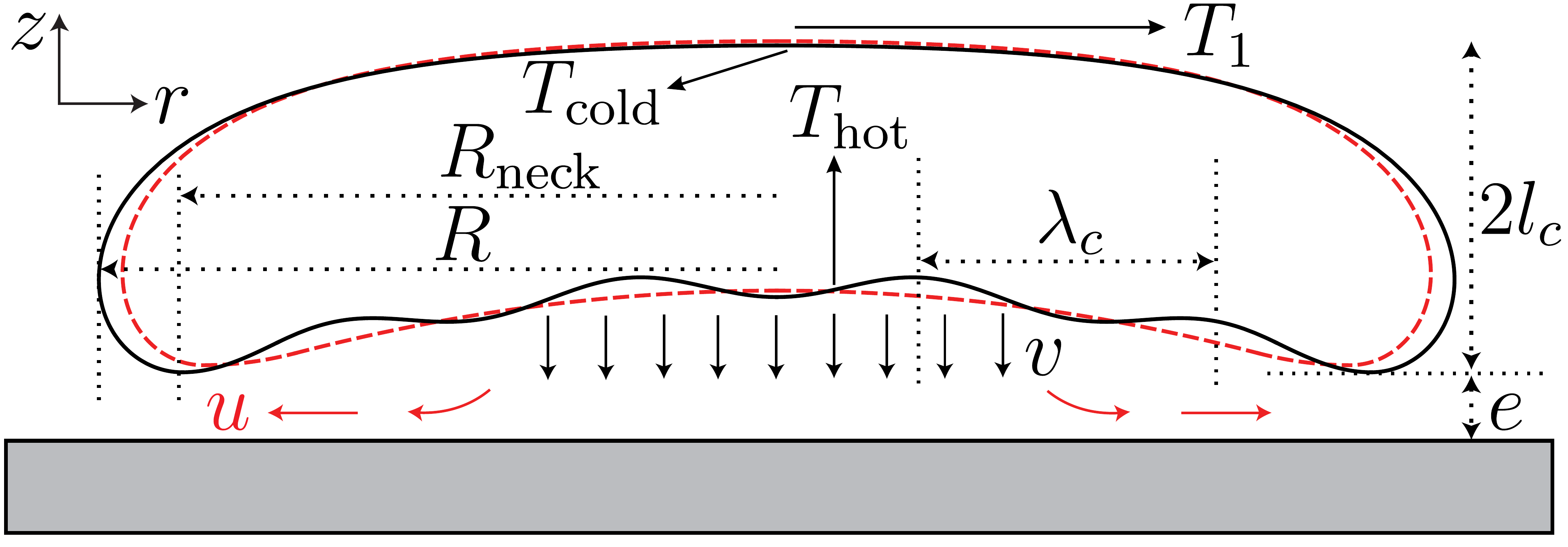}
\caption[Sketch of the geometry of a large Leidenfrost drop]{Sketch of the cross-sectional view of a large Leidenfrost drop levitating on a flat hot surface. The red dashed line and black solid line represent the drop profiles before and after the initiation of capillary waves at the bottom surface of the drop. The symbols are defined in the main text.
} 
\label{sketch}
\end{center} 
\end{figure}

Figure\ \ref{sketch} shows s sketch of a large, axisymmetric Leidenfrost drop with a maximum radius = $R$ and thickness $\approx2l_c$. The mean vertical velocity of the gas at the liquid surface is $v$, $e$ is the mean thickness of the vapor layer, and $u$ is the outward radial velocity of the gas near $r = R$. For such a large Leidenfrost drop, both the bottom surface of the drop and the substrate surface are assumed to be approximately flat to simplify the analysis. Following the model of Biance \textit{et al.} \cite{biance2003leidenfrost}, the mass loss rate of the drop due to evaporation can be expressed as:
\begin{equation} 
\frac{\mathrm{d}m}{\mathrm{d}t}=\frac{\kappa_v}{L}\frac{\Delta T}{e}\pi R^2=\rho_v\pi R^2 v,
\label{evaeq1}
\end{equation}  
where $\kappa_v$ is the thermal conductivity of the vapor, $L$ is the latent heat of the evaporation, $\Delta T=T_s-T_b$, and $\rho_v$ is the density of the vapor at the liquid interface.  Since $e\approx$ 100 $\mu$m, and thus $R\gg e$ \cite{burton2012geometry}, we employ lubrication theory, leading to an expression relating the flow rate to the mean pressure, $P$, in the vapor layer:
\begin{equation} 
\frac{\mathrm{d}m}{\mathrm{d}t}=\rho_v\frac{2\pi e^3}{3\eta_v}P=\rho_v\frac{2\pi e^3}{3\eta_v}2l_c\rho_l g=\rho_v\pi R^2v,
\label{evaeq2}
\end{equation}  
where $\eta_v$ is the dynamic viscosity of the vapor. We note here that parameters such as $\kappa_v$, $\rho_v$, and $\eta_v$ will vary in the vapor layer due to the temperature gradient between the liquid and solid surfaces. For simplicity, we will assume that these represent average values of the vapor layer properties. A more detailed treatment, possibly including full numerical simulations, would be necessary to extend this simplified model.

In addition, due to mass conservation, the volumetric flow rate of gas from the bottom of the drop must be equal to the flow rate exiting the perimeter of drop:
\begin{equation} 
\pi R^2v=2\pi eRu.
\label{massconservation}
\end{equation}  

Solving Eqs.\ \ref{evaeq1}, \ref{evaeq2}, and \ref{massconservation} for $u$, $v$, and $e$ yields:
\begin{eqnarray}
\label{u}
u=\left( \frac{\rho_l g\kappa_v l_c}{3\rho_v\eta_vL} \right)^{1/2}\Delta T^{1/2},\\
\label{v}
v=\left[ \frac{4l_c\rho_l g}{3\eta_v}\left( \frac{\kappa_v \Delta T}{L \rho_v} \right)^{3} \right]^{1/4}R^{-1/2},\\
\label{e}
e=\left( \frac{3\kappa_v \Delta T \eta_v}{4Ll_c\rho_l \rho_vg} \right)^{1/4}R^{1/2}.
\end{eqnarray}  

In this model, the steady-state, linear temperature profile in the vapor layer is valid since the characteristic time scale associated with thermal diffusion, $e^2/D_v\approx$ 1 ms, is smaller than the typical residence time of the gas in the vapor layer, $R/u\approx$ 10 ms, where $D_v$ is the thermal diffusivity of the vapor. Let us assume a water Leidenfrost drop with $R$ = 0.01 m, and for the vapor, we use the properties of steam at 373 K and 1 atm of pressure: $\rho_l$ = 959 kg/m$^3$, $\rho_v$ = 0.45 kg/m$^3$, $\gamma$ = 0.059 N/m, $g$ = 9.8 m/s$^2$, $L$ = 2.26 $\times 10^6$ J/kg, $\eta_v$ = 1.82 $\times$ 10$^{-5}$ Pa s, and $\kappa_v$ = 0.04 W/m/K. Although the properties of the vapor may vary somewhat in the vapor layer due to the temperature gradient, we have verified that this does not significantly affect our analysis. 

\begin{figure}
\begin{center}
\includegraphics[width=5.2 in]{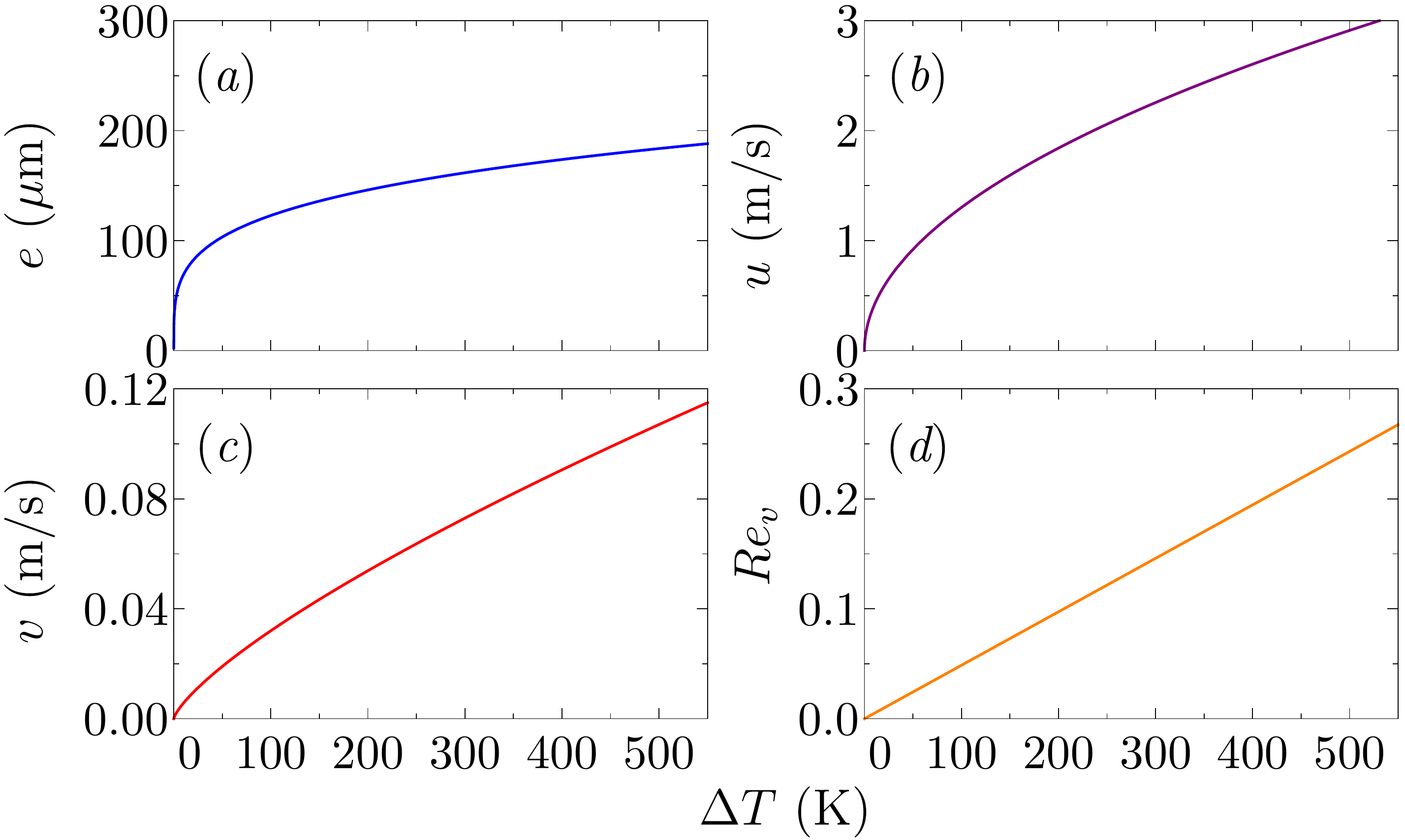}
\caption[Variations of vapor film thickness $e$, radial flow velocity $u$, vertical flow velocity $v$, and the Reynolds number of the vapor $Re_v$ with respect to temperature difference $\Delta T$]{Dependence of vapor film thickness (a), radial velocity (b), vertical velocity (c), and the Reynolds number in the vapor (d) on temperature difference $\Delta T$, respectively.
} 
\label{vaporflow}
\end{center} 
\end{figure}

We can now plot $u$, $v$, and $e$ with respect to $\Delta T$ based on Eqs.\ \ref{u}, \ref{v} and \ref{e}. The results are shown in Figs.\ \ref{vaporflow}a, \ref{vaporflow}b, and \ref{vaporflow}c, respectively. The inertial term in the Navier-Stokes equation is estimated as $|\rho_v(\vec{\bf v}\cdot\nabla)\vec{\bf v}|=\rho_vv^2/e$, and the viscous term as $|\eta_v\nabla^2\vec{\bf v}|=\eta_vv/e^2$. Thus the Reynolds number in the vapor layer is $Re_v=\rho_vve/\eta_v$. Plugging in Eqs.\ \ref{v} and \ref{e}, we arrive at a surprisingly simple expression for the Reynolds number in the vapor: $Re_v=\Delta T\kappa_v/L\eta_v$, which is independent of the drop size. This expression is plotted in Fig.\ \ref{vaporflow}d. Under typical experimental conditions, $Re_v\approx 0.2$, suggesting that the original lubrication flow assumption is valid, although inertial forces are non-negligible.

Based on Fig.\ \ref{vaporflow}, we can estimate the Bernoulli pressure in the vapor layer of Leidenfrost water drops as $\rho_vu^2/2\approx 1$ Pa, which is much smaller than the pressure variations $\approx$ 10 Pa (see Fig.\ \ref{parametric_oscillation}a), suggesting that the inertial force does not account for the pressure variations. The viscous pressure is $P_{\mathrm{vis}}\sim \eta_vvR^2/e^3$, thus the pressure variation due to the local variations of the film thickness is $\Delta P_{\mathrm{vis}}=\frac{\mathrm{d}P_{\mathrm{vis}}}{\mathrm{d}e} \Delta e$. Then $\Delta P_{\mathrm{vis}}=$ 10 Pa corresponds to $\Delta e \approx$ 15 $\mu$m considering the typical values of $v$ (Fig.\ \ref{vaporflow}c) and $R$. Although this local film thickness could also be possibly induced by the vertical motion of the center of mass of the drops, the fact that small and large drops (see Fig.\ \ref{scaling}c) share a nearly constant oscillation frequency suggests this is not the case. Therefore, the pressure variations in the vapor layer are likely to be induced by the local variations of the vapor film thickness.

\begin{figure}
\begin{center}
\includegraphics[width=5 in]{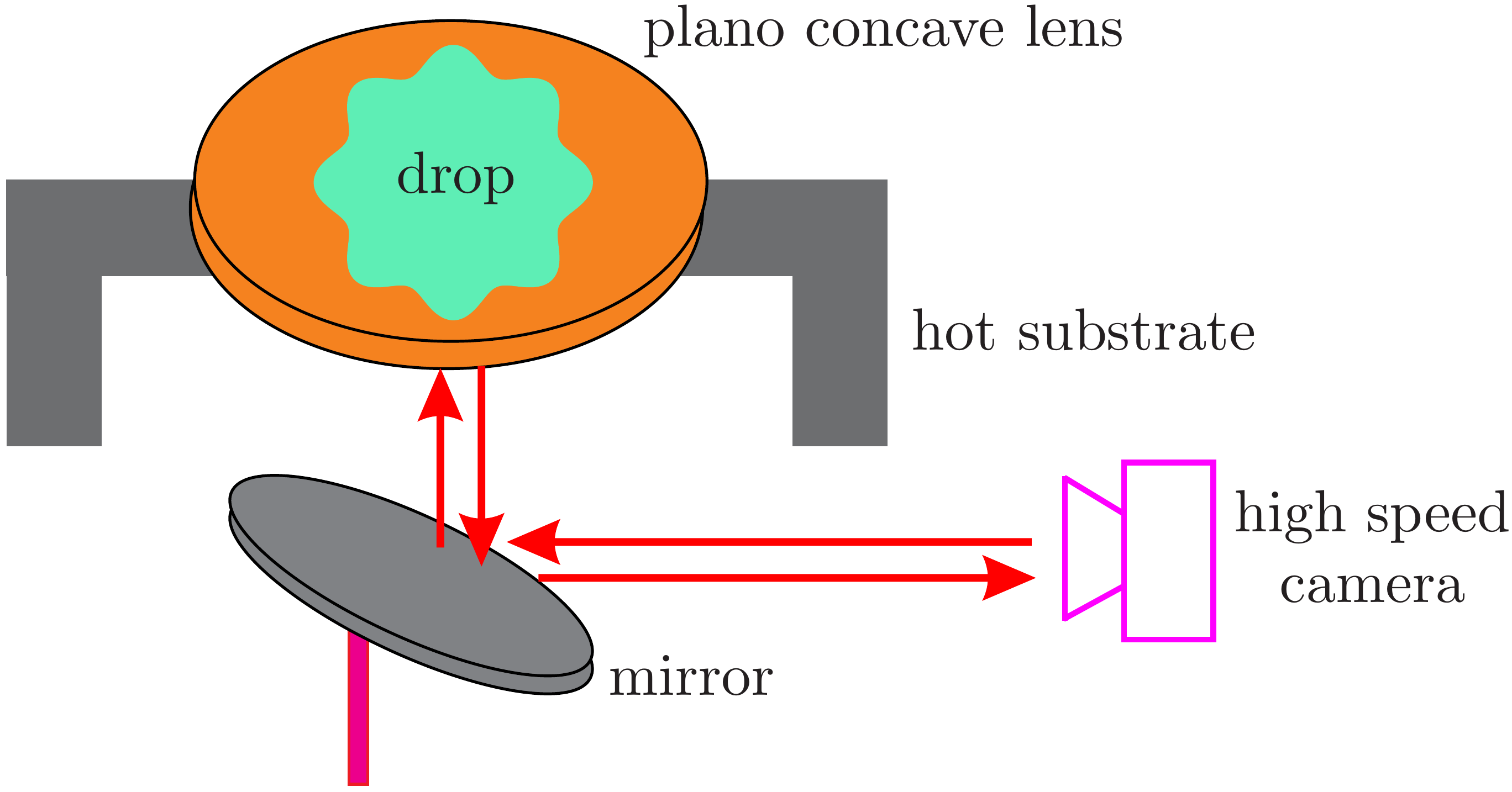}
\caption[Sketch of the experimental setup for imaging capillary waves underneath Leidenfrost drops]{Sketch of the experimental setup for imaging capillary waves underneath Leidenfrost drops. 
\label{imageing_capillary_waves}
} 
\end{center} 
\end{figure}

\begin{figure}
\begin{center}
\includegraphics[width=5 in]{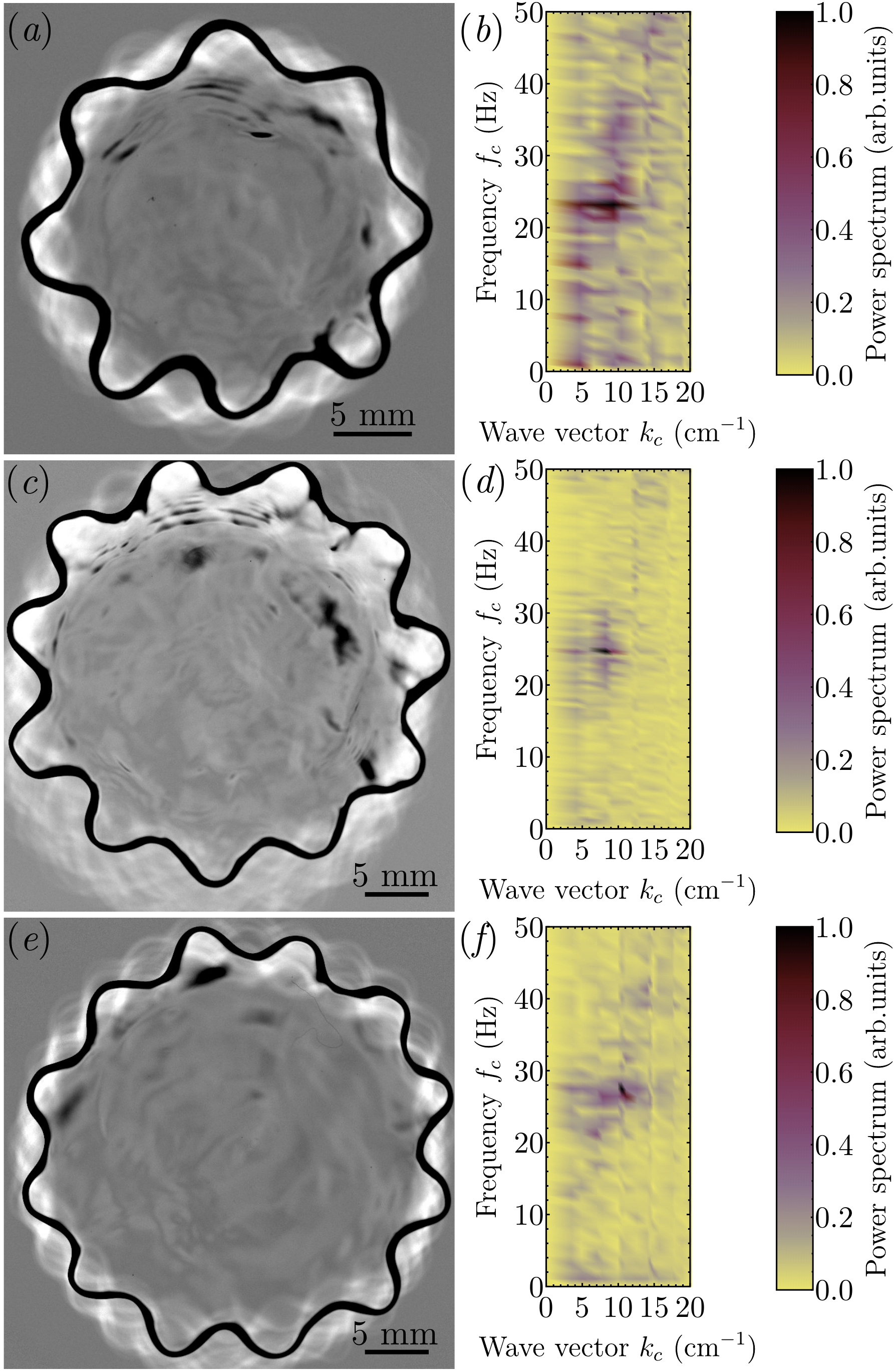}
\caption[Fourier power spectra of the capillary waves underneath star-shaped Leidenfrost drops]{(a) (c) and (e) show the capillary waves imaged beneath 8-mode acetone drop, 11-mode ethanol drop and 12-mode ethanol drop, respectively. (b), (d) and (f) represent the corresponding Fourier power spectra of the capillary waves shown in (a), (c), and (e), respectively. The images of capillary waves were enhanced for visibility, as described in the text. 
\label{capillary_waves}
} 
\end{center} 
\end{figure}

We propose that capillary waves with a characteristic wavelength, $\lambda_c$, traveling from the center to the edge of the drop lead to local variations in vapor film thickness, and thus pressure variations in the vapor layer, as schematically shown in Fig.\ \ref{sketch}. To confirm this possibility, we used a heated, plano-concave, fused silica lens as a substrate to image the capillary waves from below, as schematically shown in Fig.\ \ref{imageing_capillary_waves}. The results for acetone and ethanol Leidenfrost drops are shown in Fig.\ \ref{capillary_waves}. The images were produced by averaging all frames in a given video sequence, then subtracting this background image in order to enhance contrast in the center of the drop. The white ``halo" surrounding the drop is a consequence of this subtraction process. The Fourier spectra for the capillary waves in the central region of the drops were computed in both time and space using the pixel intensity as the signal. Using the Nyquist sampling theorem, the maximum frequency was limited by half the video frame rate (500-1000 Hz), and the maximum spatial frequency was set by half the camera magnification (132 px/cm).

We observed a sharp peak in the measured Fourier spectrum for all the analyzed video sequences. In Figs.\ \ref{capillary_waves}a and \ref{capillary_waves}b (8-mode, acetone), this peak is located at a capillary wave frequency $f_c\approx$ 24 Hz and within a range of wave numbers, $k_c=2\pi/\lambda_c$, from 7 cm$^{-1}$ to 12 cm$^{-1}$, where $\lambda_c$ is the wavelength of the capillary waves. Figures\ \ref{capillary_waves}c and \ref{capillary_waves}d show the capillary waves beneath a 11-mode Leidenfrost ethanol drop and a sharp peak at a capillary wave frequency $f_c\approx$ 25 Hz with $k_c\approx$ 7-10 cm$^{-1}$. Similarly, for the capillary waves beneath a 12-mode Leidenfrost ethanol drop (Fig.\ \ref{capillary_waves}e), a sharp peak exists at $f_c\approx$ 27 Hz  and $k_c\approx$ 10-11 cm$^{-1}$ as shown in Fig.\ \ref{capillary_waves}f. These frequencies show excellent agreement with the corresponding typical pressure oscillation frequencies measured in the vapor layer of star-shaped Leidenfrost ethanol and acetone drops (see Fig.\ \ref{scaling}c). 

It is well-known that capillary waves can be generated at a liquid-vapor interface due to a strong shear stress in the vapor \cite{Miles1957,Zhang1995,Paquier2015,Zeisel2008,chang2002complex}. A similar mechanism underlies the Kelvin-Helmholtz instability, however, the intermediate Reynolds number in the Leidenfrost vapor layer complicates the analysis. Nevertheless, we can estimate the strength of this shear stress. Typically, the ``friction velocity" is generally used to measure the strength of shear, which is defined as $u_*=\sqrt{\tau/\rho_v}$, where $\tau$ is the shear stress at the liquid-vapor interface. The maximum shear stress at the interface is:
\begin{equation} 
\tau=\frac{6\eta_vu}{e},
\label{shear_stress}
\end{equation}
assuming a parabolic-flow profile in the vapor layer with mean velocity $u$ near the edge of the drop. Using water as an example, and plugging Eqs.\ \ref{u} and \ref{e} and the value of $\eta_v$ of water vapor into Eq.\ \ref{shear_stress} yields $u_*\approx 2$ m/s. This friction velocity is quite strong and sufficient to lead to the growth of unstable modes with wavelengths of millimeter scale \cite{Zhang1995,Zeisel2008}.

The general dispersion relation of gravity-capillary waves with a dense upper-layer is: 
\begin{equation} 
f_c=\dfrac{1}{2\pi}\sqrt{\left(-gk_c+\frac{\gamma k_c^3}{\rho} \right)\tanh\left( k_ch \right)},
\label{capillary_wave_eq}
\end{equation}
where $k_c$ = 2$\pi$/$\lambda_c$, and $h\approx2 l_c$ is the thickness of the drop. For simplicity, we have assumed that the normal velocity is zero at the upper surface of the drop.  For a large Leidenfrost water drop whose characteristic pressure oscillation frequency in the vapor layer is $f_p\approx$ 26 Hz (see Fig.\ \ref{f_vs_R}), the corresponding capillary wavelength is calculated to be $\lambda_c\approx 3.03l_c$, so that $k_c\approx$ 8.3 cm$^{-1}$ \cite{Maleidenfrost2016}.  Similarly, for the Leidenfrost acetone and ethanol drops shown in Figs.\ \ref{capillary_waves}a, \ref{capillary_waves}c, and \ref{capillary_waves}e, the corresponding $k_c\approx 13 $ cm$^{-1}$, which is slightly larger than the positions of the peaks indicated in Figs.\ \ref{capillary_waves}b, \ref{capillary_waves}d, and \ref{capillary_waves}f, where $\lambda_c\approx 4l_c$. The agreement between the estimate for $k_c$ and the measurements shown in Fig.\ \ref{capillary_waves} is good considering the simplicity of Eq.\ \ref{capillary_wave_eq}, which is derived using an inviscid, semi-infinite flow in both phases.

The capillary-wave origin for the star-shaped oscillations also agrees with the minimum size of the $n$ = 2 mode. More specifically, the radius we measured in our experiments is $R$ rather than $R_{\mathrm{neck}}$ as illustrated in Fig.\ \ref{sketch}. The relationship between these two lengths is $R=R_{\mathrm{neck}}+0.53l_c$ \cite{snoeijer2009maximum,burton2012geometry}. Thus, the minimum drop size required to fit one capillary wavelength beneath the drop is $2R_{\mathrm{neck}}\approx \lambda_c$. Using $\lambda_c\approx 3.03l_c$ from above, we find that the radius of an $n$ = 2 mode drop should be $R\approx2.05 l_c$. This is in good agreement with Fig.\ \ref{scaling}a, which shows $R/l_c$ is slightly less than 2 for the smallest drops. Taken together, this analysis suggests a purely hydrodynamic origin for the star oscillations based on capillary waves generated by a strong shear stress in the rapidly-flowing vapor beneath the drop.

\begin{figure}
\begin{center}
\includegraphics[width=4 in]{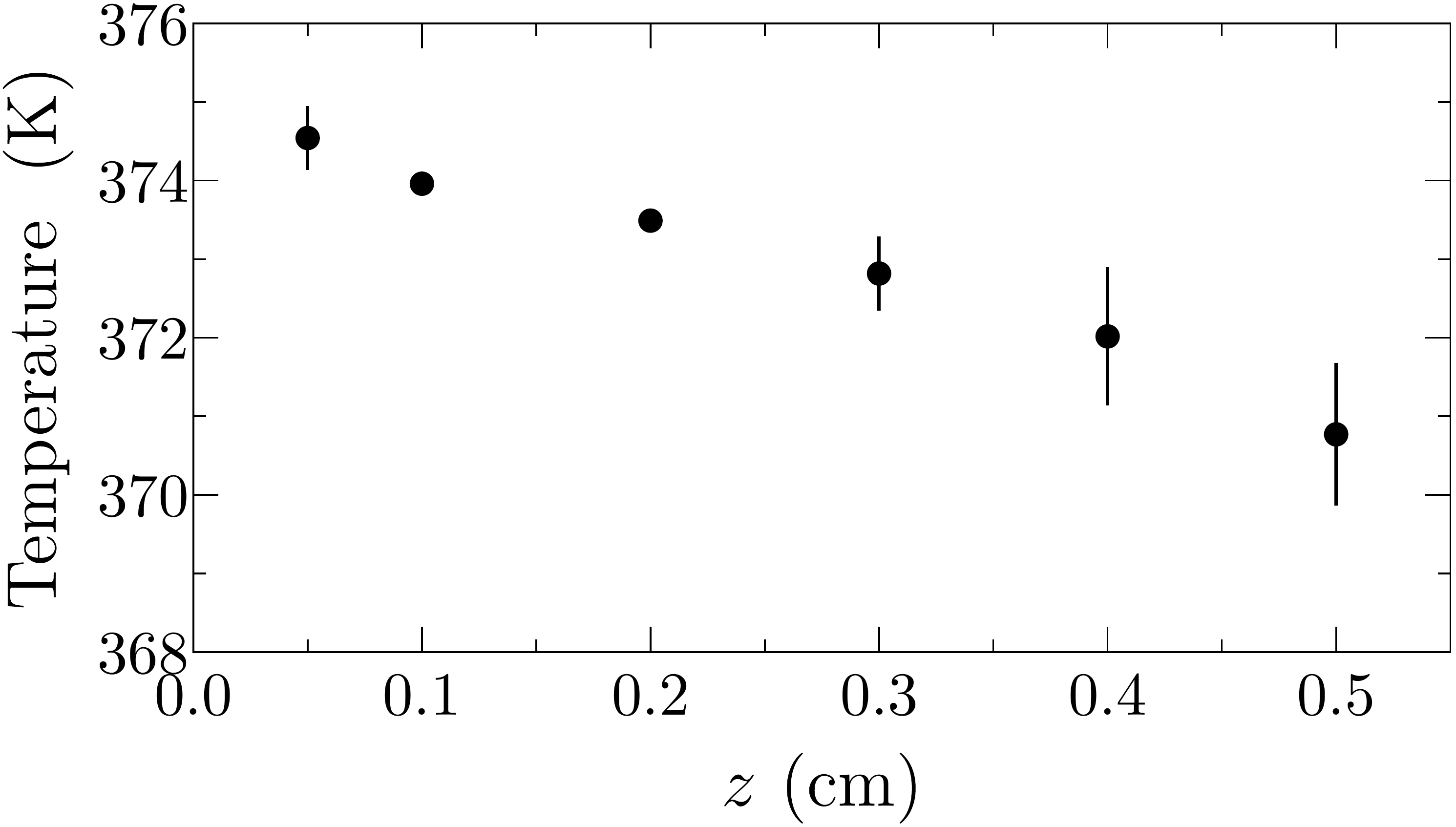}
\caption[Temperature profile of a large Leidenfrost water drop]{The temperature profile of a large Leidenfrost water drop with the volume of 10$^{-6}$ m$^{3}$. The error bars come from the standard deviations of multiple measurements. 
} 
\label{temperature_profile}
\end{center} 
\end{figure}

\subsection{Thermal effects}
\label{thermal effect}

In the Leidenfrost effect, thermal energy is transferred from the substrate to the drop, inducing evaporation and enabling the sustained star-shaped oscillations. Here we consider the role of thermal effects, such as convection in the liquid, which may play a role in the star oscillations. At sufficiently large temperature gradients in liquid layers, convective structures can develop with well-defined length scales. A well-known example is B\'{e}nard-Marangoni convection, where hexagonal patterns are generated in a thin layer of liquid when heated from the bottom \cite{benard1901tourbillons,rayleigh1916lix}. The convection pattern is strongly affected by the variation of surface tension with temperature \cite{marangoni1871ausbreitung,schatz2001experiments,maroto2007introductory}. Given that our data collapses when scaled by the capillary length of the liquid (Fig.\ \ref{scaling}), this particular type of convection may be important for Leidenfrost drops since the capillary length also appears in the length scale which characterizes the size of the convective patterns. 

\begin{figure}[!tbph]
\begin{center}
\includegraphics[width=5.7 in]{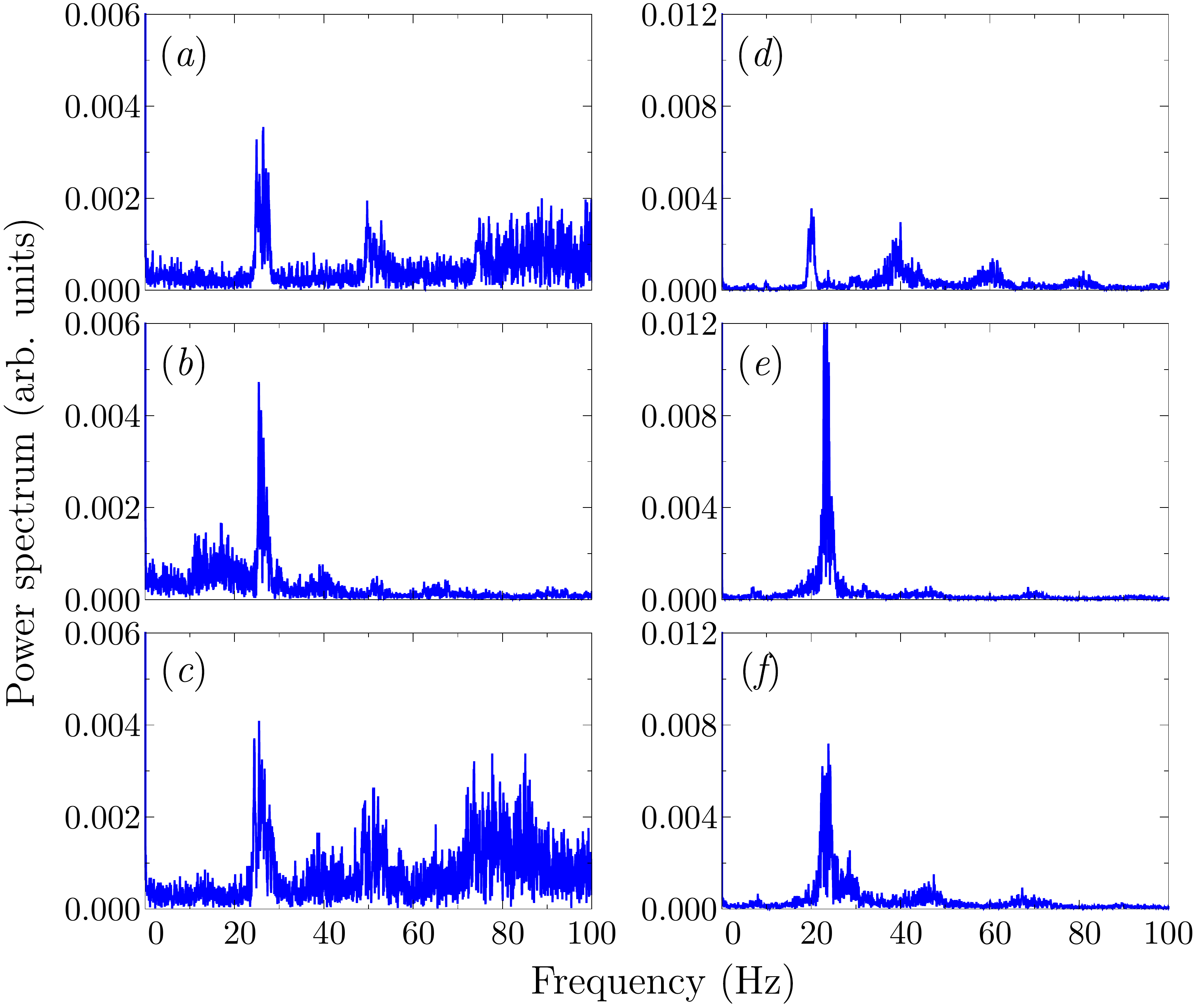}
\caption[Fourier power spectra of the pressure variations in the vapor layer at different environment temperature $T_1$]{Fourier power spectra of the pressure variations in the vapor layer at different environment temperature $T_1$. (a), (b) and (c) represent pressure variations of a 4-mode Leidenfrost water drop at $T_1 \approx$ 367 K, 483 K, and 623 K, respectively. (d), (e) and (f) represent pressure variations of a 5-mode Leidenfrost water drop at $T_1 \approx$ 367 K, 483 K, and 623 K, respectively. All of the data shown here was obtained at the same scanning rate within the same time interval, 20 s.  
} 
\label{thermal_convection}
\end{center} 
\end{figure}

Generally, the Marangoni number, $Ma$, and Rayleigh number, $Ra$, are used to characterize B\'{e}nard-Marangoni convection. These dimensionless numbers are defined as: 
\begin{gather}
\label{ma}
Ma=\frac{\left(\partial \gamma /\partial T\right)\delta Th}{\nu_l \rho D_l},\\
\label{ra}
Ra=\frac{g\alpha \delta T h^3}{\nu_l D_l}.
\end{gather}  
These numbers characterize the strengths of thermocapillary and buoyancy effects, respectively. Additionally, the ratio of thermal conduction inside the liquid to the conduction at the liquid-vapor interface is characterized by the Biot number: 
\begin{equation}
Bi=\frac{\beta h}{\kappa_l}=\frac{T_{\mathrm{hot}}-T_{\mathrm{cold}}}{T_{\mathrm{cold}}-T_1},
\label{bi}
\end{equation}
The definitions of the symbols in $Ma$, $Ra$, and $Bi$ are listed in Table\ \ref{Leidenfrost_tab:2}, as well as their values for water at the boiling point.

\begin{table}
\begin{center}
\def~{\hphantom{0}}
\begin{tabular}{llr}
 \hline
      \hline
     Symbol                                          &Quantity                                       &Value \\ 
     \hline
     $\nu_l$                                         &liquid kinematic viscosity	                      &2.94$\times$10$^{-7}$ m$^2$/s \\
     $D_l$	                                     &liquid thermal diffusivity 	                      &1.743$\times$10$^{-7}$ m$^2$/s   \\
     $\alpha$	                                     &thermal expansion coefficient          &7.52$\times$10$^{-4}$ /K   \\
     $\kappa_l$	                               &liquid thermal conductivity                       &3.2$\times$10$^{-2}$ W/m/K \\
     $h$	                                           &drop thickness	                            &0.005 m  \\
     $\delta T$	                              &temperature difference	                      &$T_{\mathrm{hot}}-T_{\mathrm{cold}}$ \\
     $\beta$	                                    &heat transfer coefficient	               &depends on $\delta T$  \\
     $\partial \gamma/\partial T$	    &surface tension gradient                  &1.46$\times$10$^{-4}$ N/m/K\\
     $T_1$	                                    &surface temperature	               &$>$367 K  \\
     \hline
      \hline
\end{tabular}
\caption[The values of the physical quantities of liquids at the boiling points]{Physical properties of water at the boiling points \cite{lemmon2011nist}.}
\label{Leidenfrost_tab:2}
\end{center}
\end{table}

Figure\ \ref{temperature_profile} shows the temperature profile of a large Leidenfrost water drop levitating on its vapor layer with $T_s= 673$ K, in which the substrate surface is defined as $z=0$. The temperature was measured with a fine-point thermocouple, as described in Section\ \ref{experiment}. We define the temperatures of the bottom and top surfaces of the drop as $T_{\mathrm{hot}}$ and $T_{\mathrm{cold}}$, respectively, and the temperature of the position which is slightly above the top surface of the drop is denoted as the environment temperature, $T_1$, as shown in Fig.\ \ref{sketch}. From Fig.\ \ref{temperature_profile} we can obtain $\delta T$ = $T_{\mathrm{hot}}-T_{\mathrm{cold}}\approx$ 3.8 K, the environment temperature measured to be $T_1=$ 367 K, thus we can calculate $Ma\sim 2\times 10^5$, $Ra\sim 7\times 10^4$, and $Bi\approx$ 1.1. Both the values of $Ma$ and $Ra$ are much larger than their critical values, which are typically of order 100 for the initiation of convective instability \cite{maroto2007introductory,leal2007advanced}. Thus it is possible that thermal convection plays a role in initiating the star-shaped oscillations. In this case, one may expect the star-shaped wavelength, $\lambda_d$, to be related to the critical wavelength of the convective instability. This critical wavelength depends on the Biot number. We implemented a qualitative test for this dependence by wrapping aluminum foil around the substrate and Leidenfrost drop, which dramatically increased the environment temperature, $T_1$, near the top of drop.  

Figures\ \ref{thermal_convection}a and \ref{thermal_convection}d show Fourier power spectra of the pressure oscillations in the vapor layer of Leidenfrost water drops during $n$ = 4 and $n$ = 5 oscillations at $T_1 \approx$ 367 K, respectively. Figures\ \ref{thermal_convection}b and \ref{thermal_convection}e are the spectra of pressure oscillations during $n$ = 4 and $n$ = 5 oscillations at $T_1 \approx $ 483 K, respectively. Finally, Figs.\ \ref{thermal_convection}c and \ref{thermal_convection}f represent the same modes at $T_1$ = 623 K. For $T_1$ = 483 K and 623 K, the direction of heat transfer at the upper surface of the drop has been reversed; energy is added to the drop. Overall, dramatically increasing the surrounding temperature affects the appearance of higher harmonics in the spectra. The behavior is non-monotonic, and is likely due to the highly nonequilibrium conditions (high evaporation rate) induced by the high temperatures of the substrate ($T_s$) and environment ($T_1$). A more detailed understanding of the pressure oscillation spectra is left for future studies. Nevertheless, the position of the main peak in the spectra is independent of $T_1$, indicating that convection and the details of thermal transport in Leidenfrost drops play a secondary role in the star-shaped oscillations.

\section{Summary and outlook}

Both large and small Leidenfrost drops display self-organized oscillations due to the constant input of thermal energy and continuous evaporation and flow beneath the drop. Here we have focused on radial oscillations (i.e. ``breathing" mode) of small Leidenfrost drops, and the large-amplitude, star-shaped oscillations that appear in large Leidenfrost drops. We have characterized the number of observed modes for various volatile liquids, the frequency and wavelength of the oscillations, and the pressure variations in the vapor layer beneath the drops. The number of observed modes is sensitive to the properties of the liquid (see Table\ \ref{Leidenfrost_tab:1}), i.e. the star-shaped oscillations of smaller Leidenfrost drops are dissipated by the liquid viscosity, which sets the minimum oscillation mode number $n_{min}$ that can be observed in experiment. The relationship between the frequency and wavelength agrees very well the quasi-two-dimensional theory proposed by Yoshiyasu \textit{et al.} \cite{yoshiyasu1996self}. The dominant frequency associated with the pressure oscillations is approximately twice the drop oscillation frequency, consistent with a parametric forcing mechanism for the star oscillations. 

One of the main findings of our work is identifying the underlying cause of the pressure oscillations. By imaging the liquid-vapor interface from below the drop, and using a simplified model for the flow in the thin vapor layer, we conclude that capillary waves of a characteristic wavelength, $\lambda_c\approx 4 l_c$, lead to pressure oscillations at the experimentally measured frequency. The flow in the vapor layer is quite rapid. Near the edge of the drop, the mean radial velocity can reach 1-2 m/s or more. In the small gap ($\sim$ 100 $\mu$m) between the liquid and the solid surface, this flow applies a large shear stress to the liquid-vapor interface, and can easily excite capillary waves with millimeter-scale wavelengths. The dispersion relation for the capillary waves then leads to a characteristic frequency for the pressure oscillations, which in turn parametrically drive the star-shaped oscillations. Furthermore, although the vapor flow is inherently driven by evaporation and heat transfer, the substrate and surrounding temperature have little effect on the dominant frequency and wavelength of the oscillations, suggesting they are purely hydrodynamic in origin.

Although the work presented here has focused mostly on the origin of star-shaped oscillations, the coupling of the flow in the vapor layer and the liquid-vapor interface underlies a rich spectrum of dynamical phenomena observed in both Leidenfrost liquid layers and drops. In particular, of key interest is understanding the failure of the Leidenfrost vapor layer which can lead to explosive boiling. If the observed capillary waves beneath the drop act as the precipitant to vapor-layer failure, then it is possible that geometrical tailoring of the surface to be commensurate with $\lambda_c$ may inhibit the generation of capillary waves. In addition, patterned, ratchet-shaped substrates with wavelengths $\approx$ 1-3 mm are known to induce propulsion of small Leidenfrost drops \cite{cousins2012ratchet,linke2006self,lagubeau2011leidenfrost}, however, less is known about transport of large quantities of liquid, and the dependence on the wavelength of the surface patterns. We leave these questions open to future experiments. More generally, our results may offer insight into the direct control of oscillations in levitated drops in many other systems \cite{haumesser2002high,duchemin2005static,paradis2005surface,ishikawa2006noncontact,lister2008shape,langstaff2013aerodynamic}, for example, precise control is crucial when levitating high-temperature or harmful liquids using a gas film. We also expect our results to enhance the understanding of dynamics that couple a thin, supporting gas film, a liquid interface, and a solid surface, a scenario which occurs through forced wetting and gas entrainment in liquid coating \cite{Xu2005,Driscoll2011,marchand2012,Kolinski2012,Liu2013,Liu2015}.

\chapter{Polygonal desiccation crack patterns}
\label{drying_cracks}

\section{Introduction}
In addition to the gorgeous star-shaped oscillatory patterns in Leidenfrost drops, there also exist a large group of amazing patterns driven by drying particulate suspensions. For example, when spill a coffee drop onto a desk, the final residues of the drop after drying form a ring-shaped pattern, which is known as the coffee ring effect. The formation of the coffee ring is driven by the capillary flow drawn from the bulk of the drop to the edge as the drying rate at the drop periphery is greater than of the bulk \cite{deegan1997capillary,deegan2000contact,deegan2000pattern}. For more concentrated particulate suspensions, the particles pack closely into a network during drying, and the particulate films are prone to form cracks if enough capillary stress is stored in the film. This capillary induced phenomena contribute largely to the diverse natural patterns, and one of the well-known category is the desiccation crack patterns \cite{routh2013drying,goehring2015desiccation}.    

\begin{figure}[!]
\begin{center}
\includegraphics[width=4 in]{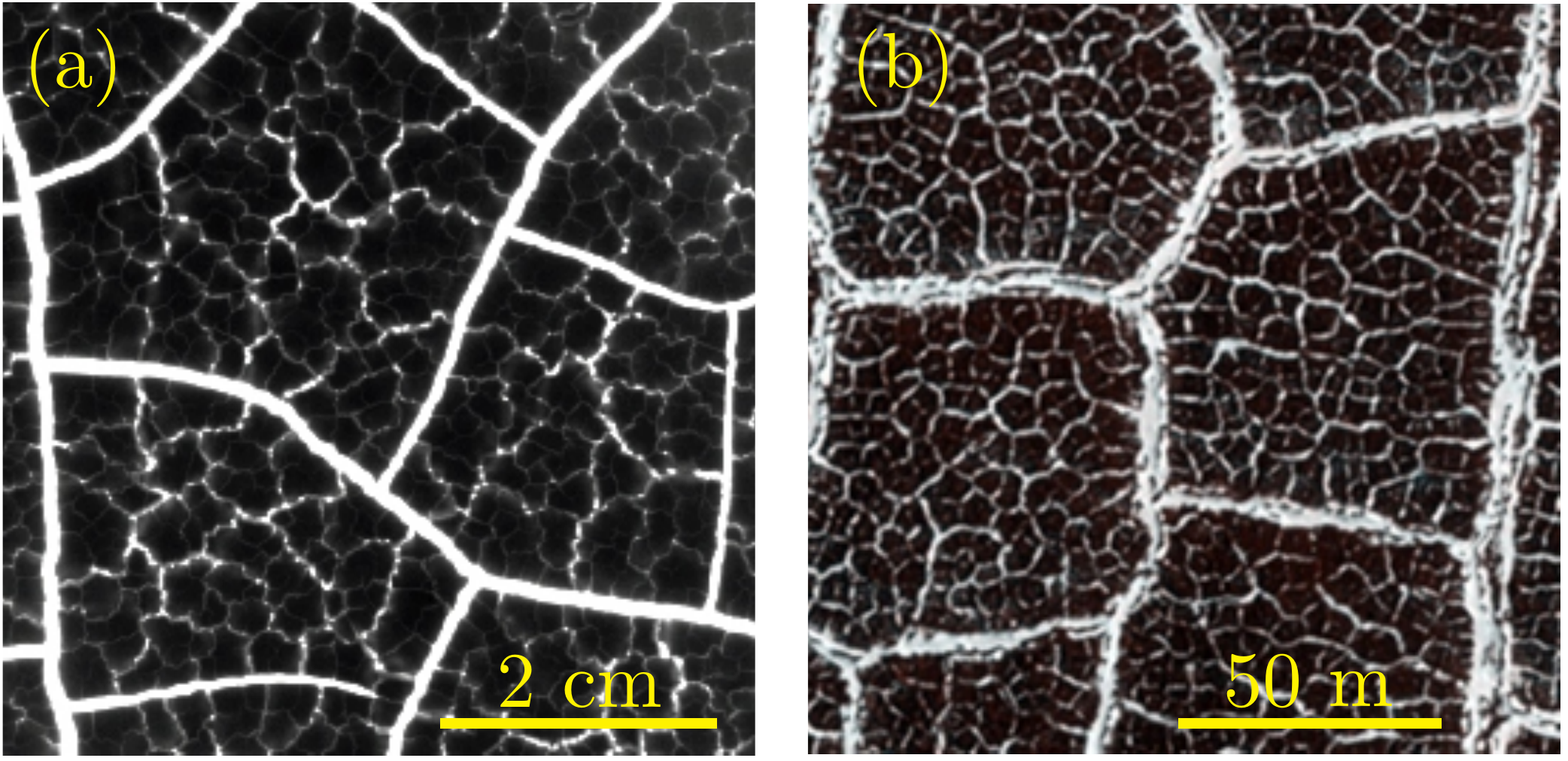}
\caption[Multiscale polygonal crack patterns]{Multiscale polygonal crack patterns. (a) Crack patterns by drying cornstarch-water suspensions in a petri dish. (b) Polygonal terrain in an ancient dried lake be on Mars (HiRISE: PSP\_007372\_2475, image courtesy of NASA/JPL/University of Arizona). 
\label{multiscale_polygonal_cracks}
} 
\end{center} 
\end{figure}

Desiccation crack patterns observed in natural systems span many orders of magnitude in size (Fig.\ \ref{multiscale_polygonal_cracks}) \cite{ball1999self,xu2009drying,goehring2015desiccation,routh2013drying,goehring2013pattern,goehring2013evolving}. Among the large diversity of desiccation crack patterns, polygonal patterns are the most common. Typical examples include the complex crack network in dried blood \cite{brutin2011pattern}, craquelures in old paintings \cite{giorgiutti2015striped,goehring2017drying}, T/Y-shaped cracks in dried mud \cite{goehring2010evolution}, and polygonal terrain cracks \cite{goehring2014cracking,Maarry2010mars,el2014potential}. In particulate suspensions, the formation of desiccation cracks depends on the interplay between order and disorder in granular systems, as well as mechanical instabilities initiated by local, nonequilibrium interactions between the liquid, solid, and vapor phases \cite{routh2013drying,goehring2015desiccation}. Nevertheless, a broad range of practical applications, such as thin film coating, forensics, and controllable surface patterning rely on knowledge of the physical processes that determine crack patterns \cite{prosser2012avoiding,hatton2010assembly,liu2016surface,nam2012patterning,zeid2013influence}. Despite numerous studies which focus on desiccation crack patterns in a diverse range of systems, a fundamental understanding of the characteristic length scales associated with polygonal crack patterns is lacking, and it is not clear if the observed patterns in both microscopic and geologic crack patterns share the same underlying mechanisms.

In the laboratory, drying particulate suspensions, both Brownian and non-Brownian, are model systems for replicating and understanding desiccation cracks in nature. For a crack to form in any material, the mechanical potential energy released during fracture must exceed the energetic cost of creating new surfaces \cite{griffith1921vi}. In homogeneous, isotropic, elastic solids, the dynamics of fracturing have been recently characterized with exquisite detail   \cite{hutchinson1991mixed,fineberg1999instability,bouchbinder2010dynamics,bouchbinder2014dynamics,creton2016fracture}. However, the dynamics of drying-induced cracks are complicated by the lack of material homogeneity and a nonlinear relationship between stress and strain \cite{kitsunezaki2009crack,goehring2013plasticity,goehring2015desiccation,dufresne2006dynamics,lidon2014dynamics,man2008direct,xu2010imaging,xu2013imaging}. This complexity is enhanced by the multi-phase nature of the material: liquid menisci between particles generate heterogeneous shrinkage through capillary pressure, and excess liquid-vapor surface area in the bulk of the material \cite{dufresne2003flow,tirumkudulu2005cracking,singh2007cracking,chiu1993dryingI,chiu1993dryingII}. As a consequence, in addition to polygonal cracks, a large variety of cracks patterns have been reported in dried suspensions \cite{allain1995regular,inasawa2012self,goehring2011wavy,nandakishore2016crack,kiatkirakajorn2015formation,neda2002spiral,vermorel2010radial,darwich2012highly, jing2012formation,xu2010imaging,giorgiutti2015striped,giorgiutti2016painting,bohn2005hierarchicalI,bohn2005hierarchicalII,pauchard2003morphologies, boulogne2013annular,allen1987desiccation,gauthier2010shrinkage}. The variability in observed patterns depends on numerous factors such as film geometry \cite{lazarus2011craquelures,nandakishore2016crack}, particle mechanics \cite{nawaz2008effects}, liquid additives \cite{pauchard1999influence,liu2014tuning}, preparation history \cite{nakahara2006transition,nakahara2006imprinting}, solvent volatility \cite{boulogne2012effect,giorgiutti2014elapsed}, and external fields \cite{khatun2012electric,pauchard2008crack}.

Despite this broad range of crack patterns, we know surprisingly little about what controls the size and hierarchy of commonly observed polygonal cracks, which are visible on both the micro- and macroscales. As shown in Fig.\ \ref{multiscale_polygonal_cracks}, a detailed understanding of desiccation crack patterns can lead to more accurate interpretations of planetary geomorphology, where data are limited to satellite-based imaging \cite{Maarry2010mars,el2014potential}. Thus far, laboratory experiments and numerical models have produced contradictory results as to the mechanism and dependence of the crack spacing on desiccation conditions. For example, for regularly-spaced cracks produced by directional drying, the relationship between the crack spacing, $\lambda$, and material thickness, $h$, is a power law, $\lambda\propto h^\beta$. Experimentally, numerous groups have reported $\beta\leq1$  \cite{allain1995regular,lee2004drying,smith2011effects}, and various theoretical predictions give $2/3\leq\beta\leq1$ \cite{ma2012possible,komatsu1997pattern}. For polygonal patterns, one may expect $A_p\propto \lambda^2\propto h^{2\beta}$. Groisman \textit{et al.} \cite{groisman1994experimental} reported $A_p\propto h^2$ in desiccated suspensions of coffee grinds, although only over a four-fold increase in $h$. Other experimental \cite{shorlin2000alumina} and numerical \cite{leung2000pattern,leung2010criticality} studies have reported similar scalings. Most recently, Flores \cite{flores2017mean} showed that $A_p\propto h^{4/3}$ using a model based solely on a balance of the average stress and surface energy released during cracking. Yet despite this history of investigation, there has been no systematic experimental study of the film's thickness and mechanics on the size of polygonal crack patterns.

The pattern morphology of polygonal cracks is also of interest since it reveals information about the formation and history of the cracking process \cite{goehring2015desiccation}. For example, the distribution of angles at crack junctions \cite{akiba2017morphometric,shorlin2000alumina,groisman1994experimental} and statistical analysis of the correlation length in crack patterns \cite{colina2000experimental} are commonly quantified from images of the surface. Repeated wetting and drying of the material can lead to more ``Y''-shaped junctions rather than ``T''-shaped junctions \cite{goehring2010evolution,goehring2014cracking}. For some commonly used desiccation suspensions such as cornstarch-water  mixtures, crack patterns with two distinct length scales can be identified, as shown in Fig.\ \ref{multiscale_polygonal_cracks}a. For thick samples, the smaller polygons grow into the material, resembling columnar jointing patterns often found in nature \cite{muller1998starch,muller1998experimental,toramaru2004columnar,goehring2009drying,goehring2009nonequilibrium,akiba2017morphometric}. These smaller polygons are also known to coarsen with depth in the material \cite{goehring2005order}. Hierarchical patterns have also been observed in the cracking glaze of ceramics \cite{bohn2005hierarchicalI,bohn2005hierarchicalII}. It is not yet apparent why some materials display hierarchical crack patterns, and some do not. 

Here we present experimental evidence which resolves many of these important, outstanding questions. Our experiments involve analysis of desiccated crack patterns in various granular materials, such as cornstarch and CaCO$_3$, suspended in different volatile liquids: water, isopropanol (IPA), and silicone oil. We use both very thin, quasi-two-dimensional chambers, as well as open petri dishes to dry the samples. For all observed crack patterns, we find that the characteristic polygonal area is consistent with $A_p=\alpha h^{4/3}$ over more than three orders of magnitude in $h$, in agreement with a recent theoretical prediction \cite{flores2017mean} based on a balance of stress and surface energy for crack formation.  This scaling is independent of the shape of the polygons, which varies considerably depending on the material-liquid combination. By characterizing the modulus of the desiccated suspension, we are also able to quantitatively predict the prefactor $\alpha$. For all material-liquid combinations, we only observe multiscale crack patterns in cornstarch-water mixtures. We show that these cracks are due to two distinct desiccation mechanisms. Primary cracks form first due to capillary-induced shrinkage of the material. Secondary cracks form much later, and are due to deswelling of the hygroscopic cornstarch particles. Taken together, these results provide a quantitative pathway for interpreting multiscale polygonal desiccation crack patterns observed in diverse systems, from microscale colloidal films to terrestrial and extra-terrestrial planetary surfaces.

\section{Experimental setup}
We used commercial, polydisperse cornstarch particles from ARGO. The average radius, $R$, of the particles is $\approx$ 5 $\mu$m. We also used CaCO$_3$ particles from OnlineScienceMall with $R\approx$ 1 $\mu$m, and for some experiments, glass beads with $R \approx$ 5 $\mu$m from Miscrospheres-Nanospheres. Particle sizes were measured using optical microscopy. Different fluids such as deionized water, low-viscosity silicone oil (0.65 cSt, ClearCo), and 99.9\% isopropanol (IPA) were used as solvents to prepare particulate suspensions. For samples thicker than $h\approx$ 1 mm, we dried suspensions in polystyrene petri dishes of diameters 14 cm and 8.5 cm. Glass microscope slides were used to build thin, quasi-two-dimensional chambers as discussed in Section\ \ref{sec:secondary_cracks_thin_chamber}. For the thin chambers, the tunable thickness, $h$, was set by vinyl spacers cut from a sheet. Once the vinyl spacers were placed on the edges, a sample of suspension was placed on the microscope slide, and the sample was compressed by a second glass slide and secured mechanically before gluing with optical epoxy. Similar setups have been used by previous authors \cite{allain1995regular,dufresne2003flow,goehring2010solidification,inasawa2012self}. However, one important distinction in our experiments is that due to the relatively slow evaporation in this system, evaporation occurs nearly isotropically around the perimeter of the sample (see Figs.\ \ref{sketch_thin_chamber}a and \ref{sketch_thin_chamber}b), so drying is not uni-directional. 

We used a conventional bright-field microscope to image crack patterns in the thin chambers, and a USB 3.0 digital video camera (Point Grey) connected to a macro lens to image the crack formation during drying of particulate suspensions in petri dishes from above. Recorded images were analyzed using NIH ImageJ software to obtain the area of polygonal cracks and thickness of the dried films. An electronic balance (Omega) was used to monitor the instantaneous mass of suspensions during drying. All experiments were performed at room temperature (20 $^\circ$C) with uncontrolled relative humidity of $\approx$ 60\%. Modulus measurements were obtained by slowly pressing a stainless steel ball of diameter 1.9 cm into the material using a rheometer (TA Instruments AR2000), and recording the applied normal force. 
\begin{figure}[!tbph]
\begin{center}
\includegraphics[width=1\textwidth]{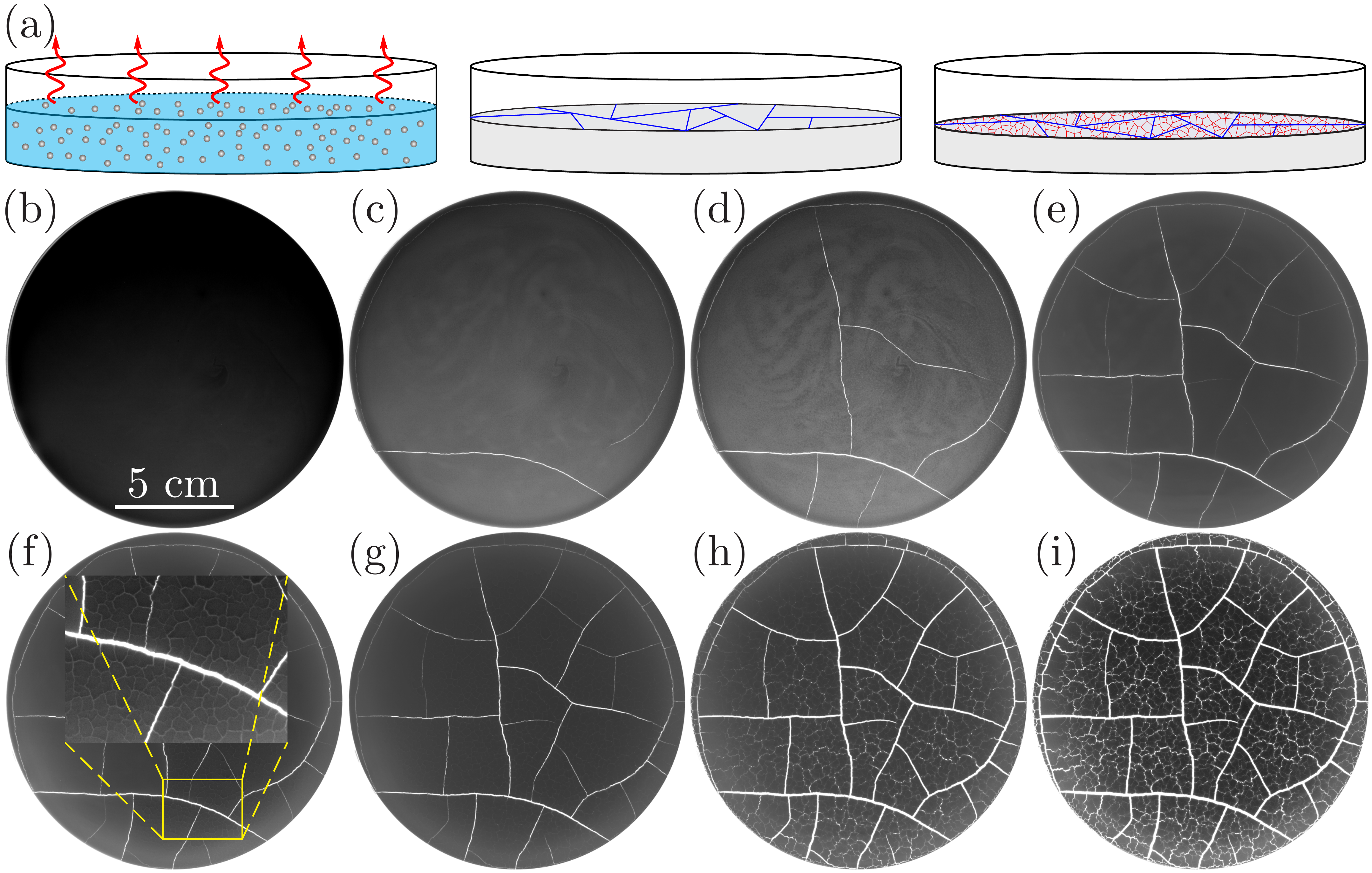}
\caption[Experimental setup for drying cornstarch-water suspensions in a petri dish]{(a) Experimental setup for drying cornstarch-water suspensions in petri dishes where drying occurs at the top interface as indicated by the red arrows. Formation of multiscale crack patterns during drying of a cornstarch suspension ($\phi_i$ = 40\%) in a petri dish: (b) Image of the initial suspension ($t=0$), (c) the primary cracks appear ($t\approx$ 10.5 h), (d) the number of primary cracks increases ($t\approx$ 11 h), (e) the number of primary cracks stops growing ($t\approx$ 22 h), (f) secondary cracks appear ($t\approx$ 24 h), (g) the number of secondary cracks increases ($t\approx$ 29 h), (h) the number of secondary cracks stops increasing though drying still proceeds ($t\approx$ 42.5 h), (i) the final crack pattern ($t\approx$ 72 h). The inset in (f) is a zoomed-in view of the section enclosed by the yellow box. Some of the images here have been enhanced for the best visualization of crack patterns, and the scale bar applies to all images.The final dried film has a thickness of $h\approx$ 0.7 cm.
\label{hierarchical_cracks}
}
\end{center} 
\end{figure}

\section{Results and discussion}  
\subsection{Multiscale cracks in cornstarch-water suspensions}

We prepared cornstarch-water suspensions and dried the samples in petri dishes. We varied the thickness of the films by controlling the initial volume of the suspensions. Above a critical thickness, $h_c$, as will discussed in Section\ \ref{sec:crack_condition}, we observed two distinct crack patterns that appeared at different stages of desiccation. Figure\ \ref{hierarchical_cracks} shows the formation of these multiscale cracks during drying of a cornstarch-water suspension with $\phi_i=40\%$. As shown in Fig.\ \ref{hierarchical_cracks}a, drying occurs at the air-water interface. The primary cracks (blue polygons) first appear, then as drying proceeds, secondary cracks (red polygons) appear within the larger polygons. After the suspension has dried for $t\approx$ 10.5 h, primary cracks first appear (Fig.\ \ref{hierarchical_cracks}c), and the number of primary cracks increases with time (Fig.\ \ref{hierarchical_cracks}d). At $t\approx$ 24 h, secondary cracks are visible (Fig.\ \ref{hierarchical_cracks}f). When $t\approx$ 42.5 h, the number of secondary cracks stops increasing though drying still proceeds (Fig.\ \ref{hierarchical_cracks}h). Finally, the drying contributes to the widening of the existing cracks, as shown in Fig.\ \ref{hierarchical_cracks}h. We note that the primary cracks penetrate completely through the sample when they form, whereas the secondary cracks grow more slowly, and their visibility increases with time. 

As reported by previous authors \cite{leung2000pattern,mizuguchi2005directional,akiba2017morphometric}, the primary cracks are a result of film shrinkage induced by the Laplace pressure on the scale of the particle size. As the water evaporates, menisci form between individual particles of radius $R\approx5$ $\mu$m. Thus the average pressure is reduced in the suspension by $\approx\gamma/R\approx$ 15 kPa, where $\gamma = 72$ mN/m is the surface tension of water.  Since the suspension is partially adhered to the bottom surface of the petri dish, the cracks form almost uniformly over the sample. Without this adhesion, the suspension undergoes isotropic shrinkage, and the number of primary cracks is reduced \cite{groisman1994experimental}. 

As the stress boundary condition at substrate can potentially play a role in the formation of desiccation crack patterns. We examined the effect of the boundary condition on the formation of crack patterns by drying cornstarch-water suspensions in petri dishes. We modified the bottom surfaces of the petri dishes: sandpaper sheets with grit size 120 were used to roughen the petri dish surface so as to make the surface less adhesive. This is a little bit counterintuitive since increasing surface roughness usually increases contact area and makes the surface more adhesive. However, here the boundary stress is mainly contributed by the tangential component of the adhesion, and roughening the substrate surface leads to the structures of troughs and peaks, which decrease tangential component of the adhesion such that the boundary stress decreases. A commercial coating (Rain-X) was applied in order to make the surface more hydrophobic or less adhesive. In addition, a 5-min epoxy layer was applied to the petri dish surface to enhance the adhesive properties.

\begin{figure}[!]
\begin{center}
\includegraphics[width=1\textwidth]{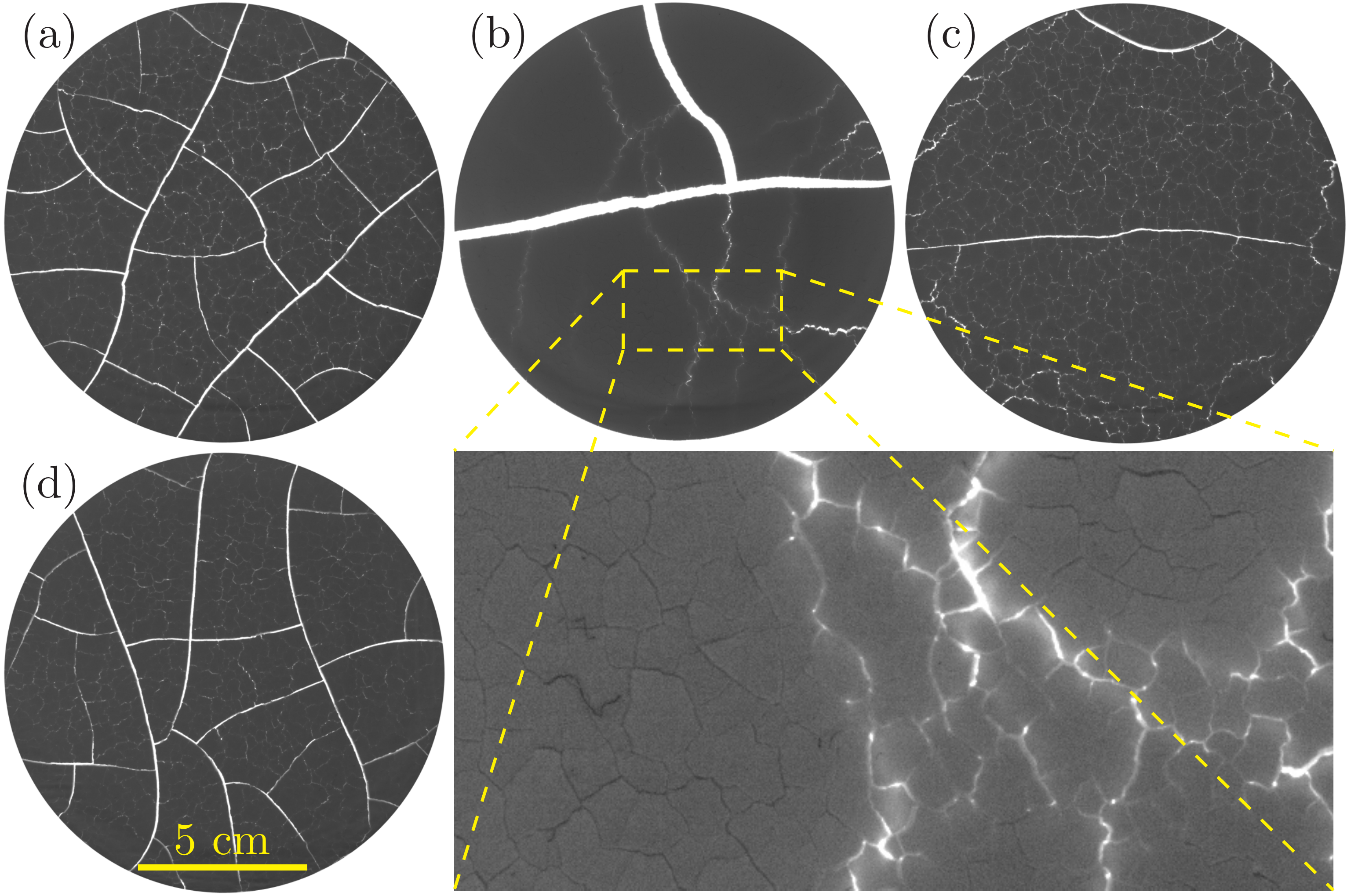}
\caption[Effect of stress boundary condition on crack formation]{Desiccation crack patterns of cornstarch-water suspensions with the same initial volume ($V_i=$ 120 ml) and the same initial volume fraction ($\phi_i=40\%$) dried in petri dishes with different boundary conditions. (a) Image of the dried polygonal cracks without modifying the surfaces of the petri dish. (b) The surface of the petri dish was roughened by sandpaper sheets with grit size 120 in order to decrease the adhesion between the suspension and the substrate since the contact area is reduced. (c) The surface of the petri dish was made hydrophobic by coating with a layer of Rain-X. (d) The surface of the petri dish was coated with a 5-min epoxy layer. The scale bar applies to all images. The zoomed-in image of (b) was enhanced for better visualization of secondary, small-scale cracks. 
\label{boundary_stress_effect}
 } 
\end{center} 
\end{figure}

We prepared cornstarch-water suspensions with the same initial volume ($V_i=$ 120 ml) and the same initial volume fractions ($\phi_i=40\%$), then deposited the suspensions into the prepared petri dishes with different boundary conditions, and the suspensions dried at room temperature. Figure \ref{boundary_stress_effect}a shows the crack patterns of the dried cornstarch-water film without any modification to the petri dish surface as a control experiment. It is evident that the increase of the surface roughness will decrease the number of large-scale cracks as indicated by Fig.\ \ref{boundary_stress_effect}b since increasing the surface roughness decreases the contact area between the suspension and the substrate, and the boundary stress decreases consequently, whereas the small-scale cracks are still observable as shown in the zoomed-in image of a particular region enclosed by a dashed yellow box in Fig.\ \ref{boundary_stress_effect}b. Similarly, as the hydrophobicity was increased (Fig.\ \ref{boundary_stress_effect}c), the number of large-scale cracks dramatically decreased, whereas the small-scale cracks are insensitive to the coating layer. The results shown in Figs.\ \ref{boundary_stress_effect}b and \ref{boundary_stress_effect}c indicate that the decrease in adhesion of the suspension to the substrate leads to dramatic decrease in the number of large-scale, primary cracks (or the increase of polygon area), which is likely due to the more isotropic shrinkage of the film during drying. 

Figure\ \ref{boundary_stress_effect}d shows the polygonal cracks in a petri dish whose surface was coated with an epoxy resin layer. In this case, although the adhesion of the suspension to the substrate has increased, the number of large-scale cracks does not dramatically change compared to Fig.\ \ref{boundary_stress_effect}a. This is not surprising since shrinkage in the film will continue until the yield stress of the granular material is reached so that a crack is formed, and then this process continues. Therefore, although the boundary condition plays a key role in determining the area of the polygonal cracks, weakening the adhesion has the strongest effect, especially if the adhesive stress is smaller than the yield stress of the material. The secondary cracks are generally unaffected by the choice of boundary condition, and the origin of secondary cracks will be discussed in Section\ \ref{sec:particle_deswelling}.

\begin{table}[!]\caption[Physical properties of the parameters in Eq.\ \ref{critical_stress}]{Physical properties of the parameters in Eq.\ \ref{critical_stress}.}
\centering
\begin{tabular}{ccc}
     \hline\hline
     Symbol                                     &Meaning                                  &Value \\ 
     \hline
     \multirow{2}{1em}{$G$}                                        & \multirow{2}{6.5em}{particle shear modulus}                  &4 GPa (cornstarch)\\
                                               &                    &                32 GPa (CaCO$_3$)\\
    
     \multirow{2}{1em}{$R$}                                        &\multirow{2}{6.5em}{particle radius}                     &5 $\mu$m (cornstarch)\\
                                               &                      &              1 $\mu$m (CaCO$_3$)\\
      
    \multirow{3}{1em}{$\gamma$}                                  &\multirow{3}{6.5em}{surface tension}                 &72 mN/m (water)\\ 
								&			 & 				23 mN/m (IPA)\\
								&			 & 				16 mN/m (silicone oil)\\
   $M$                                      &coordination number           &5 \\
     $\phi$               &random close packing           &0.67\\
$h_c$                                   &critical thickness                  &\\
      $\sigma_ c$                                &critical stress                &\\

     \hline\hline
\label{Drying_tab:1}
\end{tabular}
\end{table}

\subsection{Critical condition for primary cracks}
\label{sec:crack_condition}
The appearance of primary cracks for thicker samples of cornstarch and water suspensions, as shown in Fig.\ \ref{hierarchical_cracks}, can be understood in terms of a well-known theory for the initiation of cracks in colloidal thin films \cite{tirumkudulu2005cracking,singh2007cracking}. The theory assumes a no-slip boundary condition between the bottom boundary of the film and the substrate, and predicts a relationship between the critical film thickness and stress when crack should appear:
\begin{equation}
\frac{\sigma_c R}{2 \gamma}=0.1877\left( \frac{2R}{h_c} \right)^{2/3}\left( \frac{G M \phi R}{2\gamma} \right)^{1/3},
\label{critical_stress}
\end{equation}
where the definitions of the parameters in Eq.\ \ref{critical_stress} are listed in Table\ \ref{Drying_tab:1}. Although the particle radius ultimately cancels from Eq.\ \ref{critical_stress}, it is included here so that each term is dimensionless, as in Ref. \cite{singh2007cracking}. We have included typical values for the shear modulus of both cornstarch and CaCO$_3$ taken from the literature \cite{katz2013}, assuming a crystalline form of CaCO$_3$ and a Poisson's ratio of 0.5 for cornstarch. The same values of $\phi$ and $M$ were used for all calculations.

Taking the typical values of the suspensions of cornstarch-water and CaCO$_3$-water used in the experiment (see Table\ \ref{Drying_tab:1}), we can calculate the critical thickness for cracking. Since the stress driving the primary cracks is due to capillary pressure \cite{lee2004drying}, we can assume that $\sigma_c\approx\gamma/R$. Solving for $h_c$, we obtain $h_c\approx$ 1500 $\mu$m for cornstarch, and $h_c\approx$ 400 $\mu$m for CaCO$_3$. 

\begin{figure}
\begin{center}
\includegraphics[width=1\textwidth]{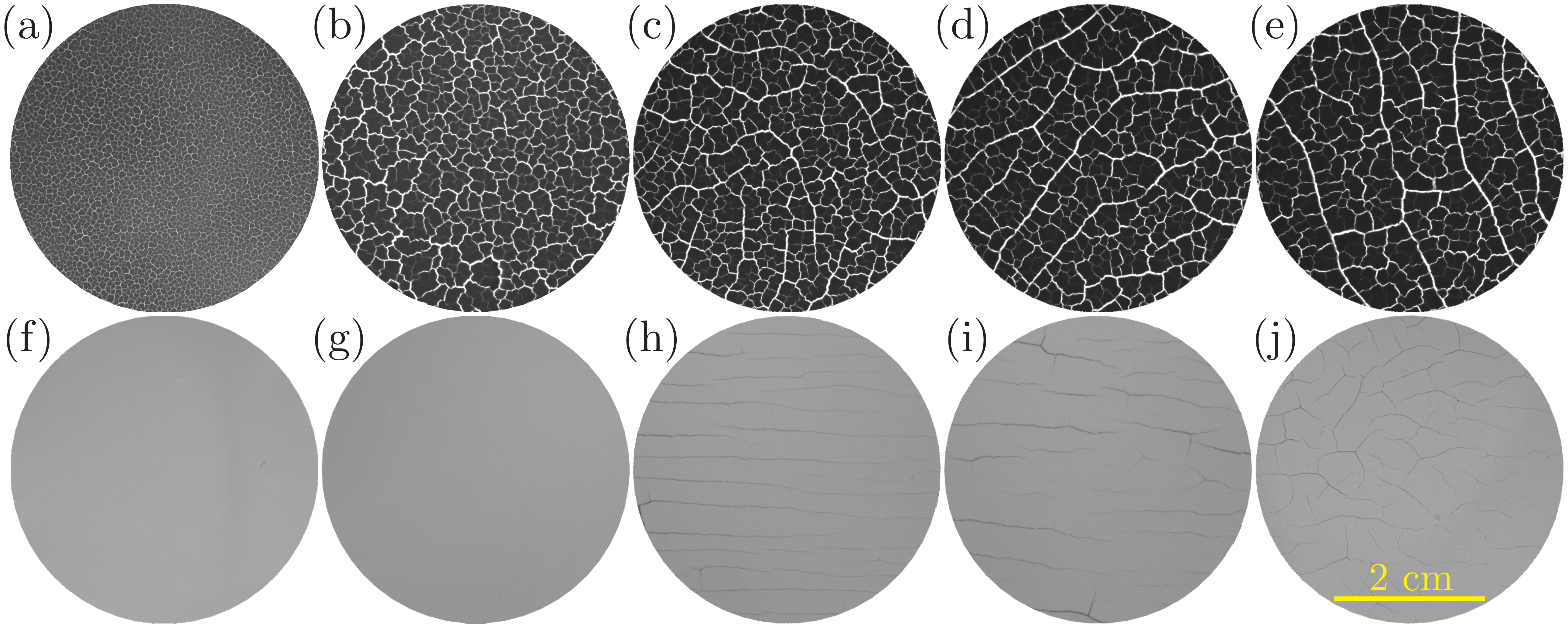}
\caption[Measurement of critical film thickness for crack formation]{Measurement of critical film thickness for cracking of cornstarch-water and CaCO$_3$-water suspensions. Images (a) to (e) show dried cornstarch-water suspensions with thicknesses of 281 $\mu$m, 600 $\mu$m, 864 $\mu$m, 1181 $\mu$m, and 1236 $\mu$m, respectively. Images (f) to (j) show dried CaCO$_3$-water suspensions with thicknesses of 267 $\mu$m, 500 $\mu$m, 553 $\mu$m, 635 $\mu$m, and 701 $\mu$m, respectively. The scale bar applies to all images.
\label{critical_film_thickness}
} 
\end{center} 
\end{figure}

In order to examine whether Eq.\ \ref{critical_stress} accurately predicts the critical film thickness for cracking, we dried both thin films of cornstarch-water and CaCO$_3$-water suspensions in petri dishes, and the results are shown in Fig.\ \ref{critical_film_thickness}. In Figs.\ \ref{critical_film_thickness}a-\ref{critical_film_thickness}b, only small-scale, secondary cracks are observed. In Fig.\ \ref{critical_film_thickness}c, the small-scale, secondary cracks are obvious and the large-scale, primary cracks are about the appear. In Fig.\ \ref{critical_film_thickness}d, both large-scale, primary cracks and small-scale, secondary cracks are obvious. This indicates that the critical film thickness of cornstarch-water suspensions lies in between $h=864$ $\mu$m (Fig.\ \ref{critical_film_thickness}c) and $h=1181$ $\mu$m (Fig.\ \ref{critical_film_thickness}d). In Fig.\ \ref{critical_film_thickness}g, no cracks are visible, whereas in Fig.\ \ref{critical_film_thickness}h cracks appear, suggesting that the critical film thickness for CaCO$_3$-water suspensions lies between 500 $\mu$m and 553 $\mu$m. Both the critical film thicknesses we observed show good agreement with the predictions by Eq.\ \ref{critical_stress}, though slightly different from the predicted values. However, Eq.\ \ref{critical_stress} fails to explain the origin of secondary cracks in dried cornstarch-water suspensions. For the secondary cracks, there is no critical film thickness, and the cracks are visible in samples that are only a few particles thick, as will be discussed in Section\ \ref{sec:secondary_cracks_thin_chamber}, suggesting that the stress is not solely due to capillary pressure, as described in Ref. \cite{lee2004drying}. 

\subsection{Cornstarch particle deswelling drives secondary cracks}
\label{sec:particle_deswelling}
\begin{figure}[!]
\begin{center}
\includegraphics[width=4.5 in]{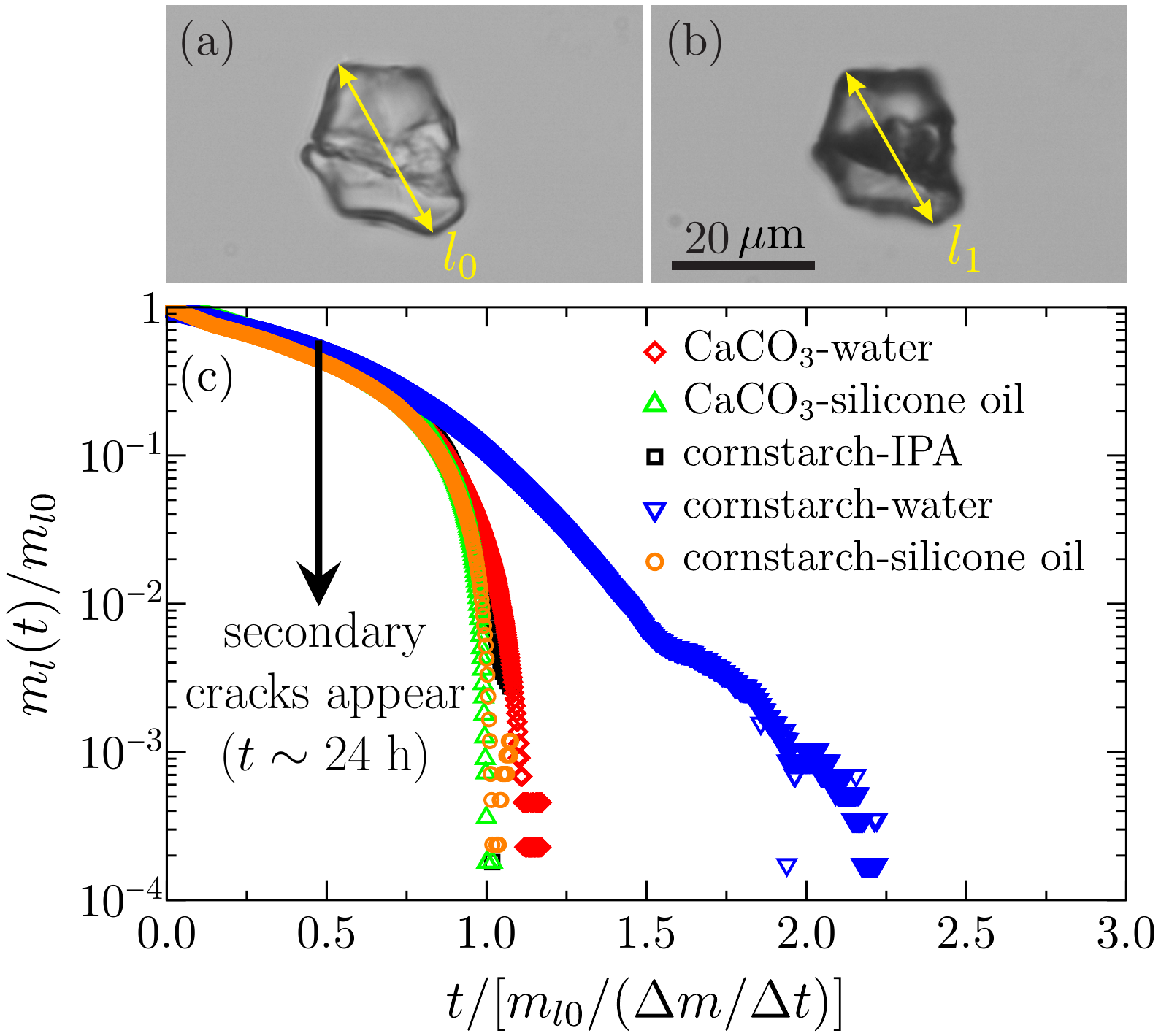}
\caption[Deswelling of cornstarch particles during drying]{Deswelling of a large, individual cornstarch particle from the wet state (a) to the fully dry state (b). The lengths indicated in the images are $l_0$ = 27 $\mu$m and $l_1$ = 24 $\mu$m. (c) Normalized evaporation rate of suspensions of cornstarch-water, -IPA and -silicone oil, and CaCO$_3$-water and -silicone oil. The symbols are defined as follows: $m_{l}(t)$ is the instantaneous mass of the liquid in the prepared suspension, $m_{l0}$ is the initial mass of the liquid, and $\Delta m/\Delta t$ is the initial evaporation rate of the liquid at $t=0$. Note that only cornstarch-water suspensions show secondary cracks during drying.
\label{deswelling}
} 
\end{center} 
\end{figure}

Multiscale crack patterns in dried cornstarch-water suspensions have been reported previously \cite{bohn2005hierarchicalI,bohn2005hierarchicalII,akiba2017morphometric}. The secondary cracks in cornstarch-water suspensions have been used as a model system to investigate the formation of geophysical columnar joints \cite{muller1998starch,muller1998experimental,toramaru2004columnar,goehring2008scaling,goehring2009drying,goehring2009nonequilibrium,goehring2013evolving}.
However, to the best of our knowledge, the origin of the different types of cracks is not well-understood. It has been suggested that the small-scale cracks are driven by the spatial nonuniformity of the local shrinkage of the film \cite{mizuguchi2005directional,akiba2017morphometric}, which has never been confirmed. More recently, Goehring \cite{goehring2009drying} showed that the strong separation between two distinct drying regimes dominated by liquid and vapor transport in the particle network could influence the formation of small-scale crack patterns in cornstarch-water suspensions, yet similar physics should apply in other particle networks where small-scale cracks are not observed. Consequently, the underlying mechanism for the formation of multiscale crack patterns remains unclear. 

One of the main results of this work is that distinct polygonal crack patterns are due to distinct shrinkage mechanisms. The initial evaporation of the suspending liquid creates capillary stress at the interface, which shrinks the sample and induces stresses sufficient for cracking (primary cracks). For cornstarch in water, the secondary cracks are driven by a second shrinkage mechanism: the deswelling of the particles. We observed strong deswelling by drying swollen cornstarch particles dispersed in dilute suspensions, and the average deswelling ratio was $\approx$ 5-10\%, as shown in Figs.\ \ref{deswelling}a-\ref{deswelling}b.

In order to examine whether particle deswelling is unique for cornstarch-water suspensions, we prepared suspensions of cornstarch-water, -IPA, and -silicone oil, and suspensions of CaCO$_3$-water and -silicone oil and compared their drying kinetics. We used an electronic balance to record the instantaneous mass of the prepared suspensions during drying, and the results are shown in Fig.\ \ref{deswelling}c. The instantaneous mass of the liquid, $m_l(t)$, is normalized by the initial mass of the liquid, $m_{l0}$ (vertical axis), and the drying time $t$ is normalized by the initial evaporation rate of the liquid, $m_{l0}/(\Delta m/\Delta t)$ (horizontal axis). It can be easily seen in Fig.\ \ref{deswelling}c that the normalized drying dynamics of all of the suspensions follow the same curve, except for cornstarch and water. The drying dynamics are much slower at late times for cornstarch in water, and the sample takes more than twice as long to dry. This discrepancy suggests that the cornstarch particles are deswelling in the later drying stage, so that evaporation depends on diffusion of water out of the individual particles. Finally, the point where the drying rate of the cornstarch-water suspensions start to deviate from the other four suspensions is exactly when small-scale, secondary cracks show up during drying cornstarch-water suspensions in petri dishes ($t\sim 24 $ h, see Fig.\ \ref{hierarchical_cracks}), indicating that the secondary cracks are driven by deswelling-induced shrinkage.

Generally, the drying kinetics of granular suspensions follow a similar pattern, i.e., initially the saturated state, in which the drying rate is only determined by the evaporation at the top free surface, next drying rate almost keeps constant and then decreases followed by a saturation state as the liquid content is nearly completely gone, and these characteristics are essentially determined by the fluid-pore interactions, pore size distributions and the connectivities of pores \cite{silverstein1997studies,lu2004unsaturated,goehring2015desiccation}. Therefore, the normalized drying rate for various particle-liquid suspensions shown in Fig.\ \ref{deswelling} furthermore suggests the unique small-scale, secondary cracks are driven by the deswelling of cornstarch particles in water given the differentiated drying behavior of cornstarch particles.

\begin{figure}[!tbph]
\begin{center}
\includegraphics[width=1\textwidth]{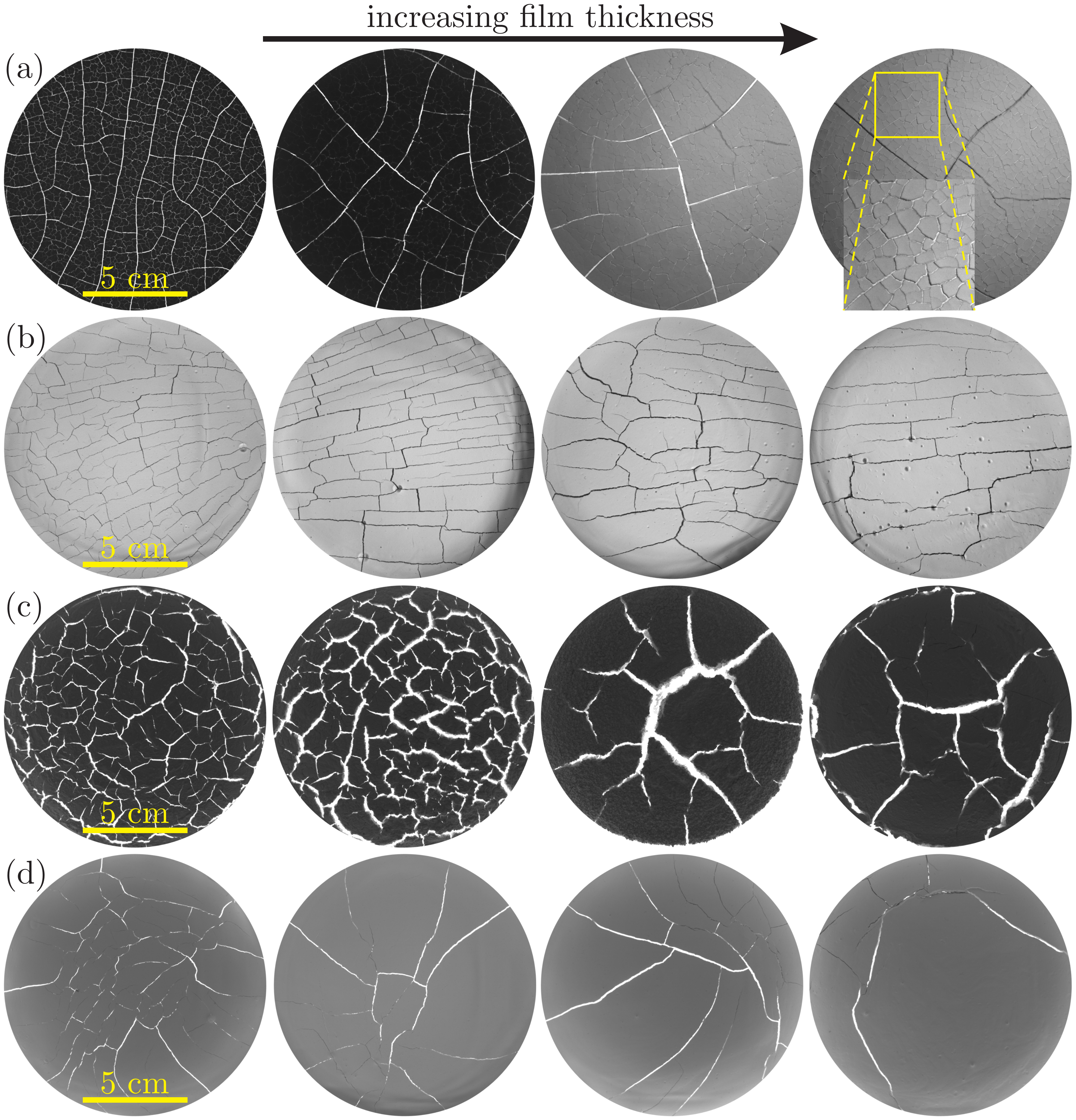}
\caption[Gallery of polygonal desiccation crack patterns of different particulate suspensions]{Multiscale crack patterns observed in dried films of cornstarch and CaCO$_3$ particles suspended in different fluids. Panel (a) shows images of polygonal crack patterns in dried film of cornstarch in water with film thicknesses of 2, 7, 10, and 20 mm, from left to right, respectively. Panels (b), (c) and (d) show dried crack patterns of films of  CaCO$_3$ suspended in water, silicone oil, and IPA, respectively. In (b) the film thicknesses from left to right are 1.5, 2, 3, 5.5 mm; in (c) from left to right are 2.5, 4, 7.5, 9 mm; and in (d) from left to right are  3.1, 5, 5.5, 9.5 mm.
\label{allcracks}
} 
\end{center} 
\end{figure}

\subsection{Primary cracks in different particle suspensions}
\label{sec:multi-scale cracks}

Above the critical thickness, we explored primary crack patterns in various suspensions. We dried suspensions of cornstarch and CaCO$_3$ in water, silicone oil, and IPA in petri dishes. We varied the initial volumes of the suspensions in order to obtain different film thicknesses and polygon areas. Figure\ \ref{allcracks}a shows the multiscale cracks observed in dried cornstarch-water films, for different film thicknesses. For primary polygonal cracks, the average polygon area increases with thickness, whereas for secondary cracks, the average polygon area initially increases with film thickness and then saturates for large $h$. For CaCO$_3$-water suspensions (Fig.\ \ref{allcracks}b), the polygonal pattern is not isotropic, and has a preferred direction. This anisotropy is well-known, and is likely due to particle chain formation induced by drying \cite{akiba2018}. For CaCO$_3$ particles in both silicone oil and IPA (Figs.\ \ref{allcracks}c and \ref{allcracks}d), the cracks are distinctly different than those observed in water, nevertheless, the characteristic size of the polygons increase with thickness. This dependence will be discussed in detail in Section\ \ref{sec:universal_scaling}.

\begin{figure}
\begin{center}
\includegraphics[width=5 in]{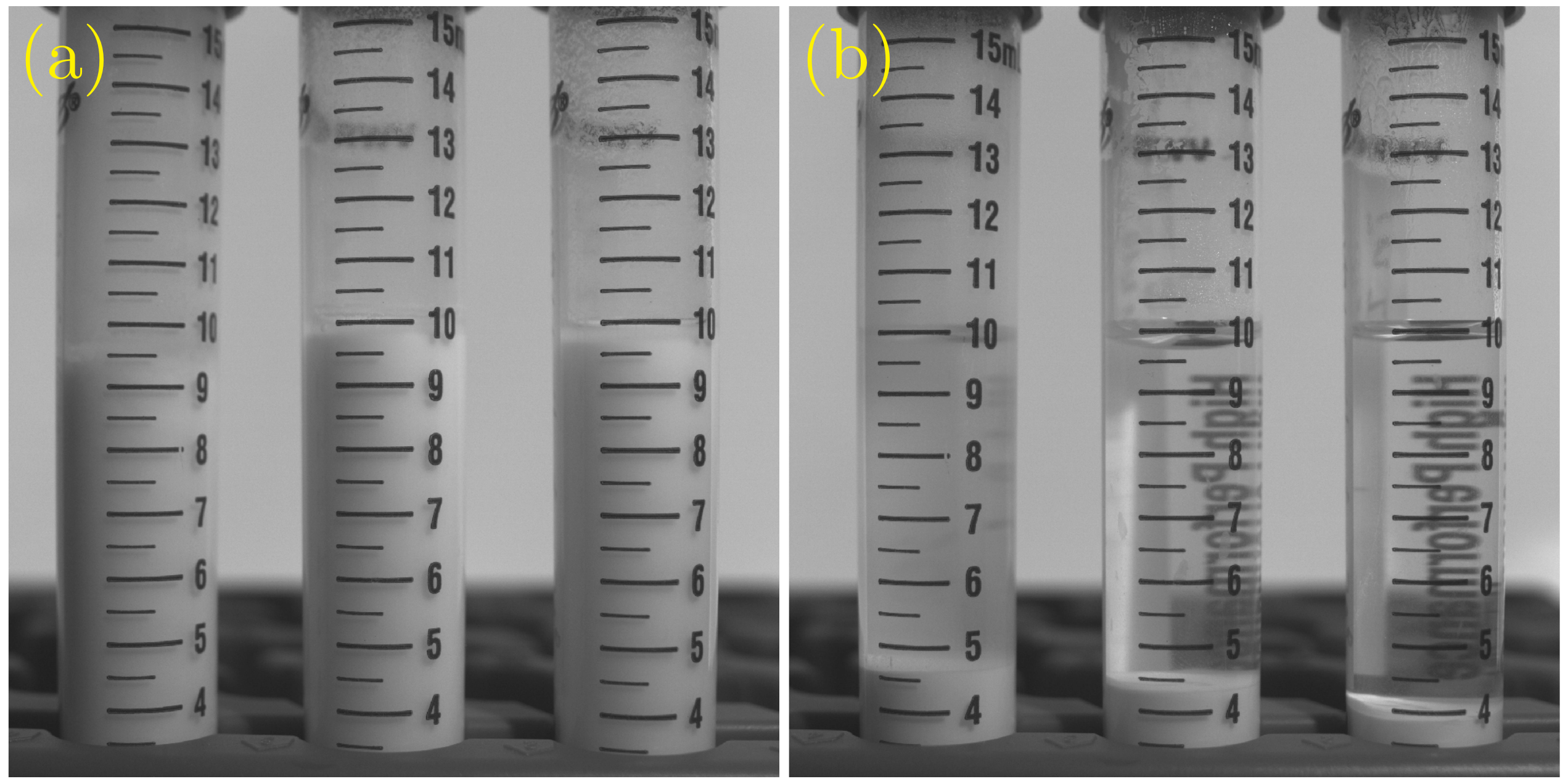}
\caption[Packing behavior of cornstarch particles dispersed in different liquids]{Suspensions of cornstarch-silicone oil, cornstarch-water, and cornstarch-IPA (from left to right) with the same same initial volume ($V_i=$ 10 ml), and the same initial volume fraction ($\phi_i=20\%$). (a) and (b) show the images of the initial suspensions and the suspensions after being centrifuged, respectively.
\label{cornstarch_in_fluids}
} 
\end{center} 
\end{figure}

Although the liquids used in our experiments have different surface tensions and vapor pressures, we suspect that some of the differences in patterns are mostly due to particle-liquid interactions and surface energies. In order to examine these assumptions, we have tested the packing ability of various particle-liquid combinations, and found significant differences between the same particles in different solvents (Fig.\ \ref{cornstarch_in_fluids}). We prepared cornstarch-water, cornstarch-silicone oil, and cornstarch-IPA suspensions with the same initial volume ($V_i=$ 10 ml), and the same initial volume fraction ($\phi_i=20\%$). The samples were then centrifuged at 2000 rpm for 2 mins. Longer centrifuge times did not change the result. According to particle number conservation,
\begin{equation}
\phi_f V_f=\phi_i V_i,
\label{particle_conservation}
\end{equation}
where $\phi_f$ and $V_f$ are the final volume fraction and the total volume of the particle suspension after being centrifuged, respectively. Figure\ \ref{cornstarch_in_fluids}a shows images of the initial cornstarch suspensions, and the images of the cornstarch suspensions after being centrifuged are shown in Fig.\ \ref{cornstarch_in_fluids}b. From Fig.\ \ref{cornstarch_in_fluids}b we find that $V_f\approx$ 4.6 ml, 4.4 ml, and 3.9 ml for cornstarch particles in silicone oil, water, and IPA, respectively. Therefore, we can calculate that $\phi_f\approx 43\%$ for cornstarch-silicone oil, $\phi_f\approx 45\% $ for cornstarch-water, and $\phi_f\approx 51\% $ for cornstarch-IPA, respectively. The results shown in Fig.\ \ref{cornstarch_in_fluids} suggest that particles suspended in different liquids show different packing abilities, depending on the particle-liquid interactions.

\begin{figure}
\begin{center}
\includegraphics[width=5 in]{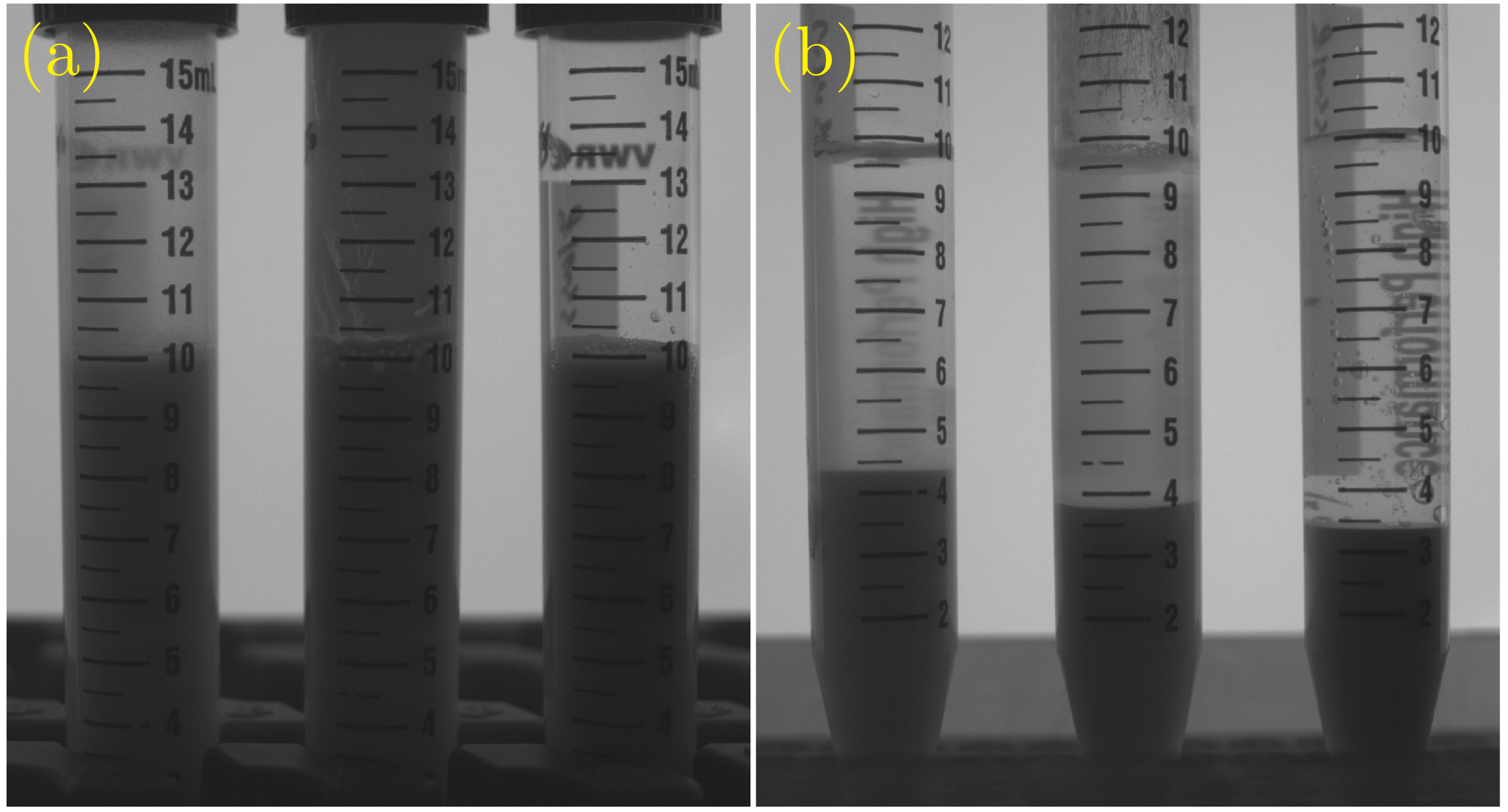}
\caption[Packing behavior of cornstarch, CaCO$_3$ and glass beads particles dispersed in water]{Suspensions of cornstarch-water, CaCO$_3$-water, and glass beads-water (from left to right) with the same initial volume fraction ($\phi_i=20\%$), and the same initial volume ($V_i$= 10 ml). (a) and (b) show the images of the initial suspensions and images of the suspensions after being centrifuged, respectively.
\label{particles_in_water}
} 
\end{center} 
\end{figure}

Additionally, we measured the packing abilities of different particles suspended in water. To do this, we prepared the suspensions of cornstarch-water, CaCO$_3$-water, and glass beads-water with the same initial volume ($V_i=10$ ml), and the same initial volume fraction ($\phi_i=20\%$) as shown in Fig.\ \ref{particles_in_water}a from left to right, respectively. Figure\ \ref{particles_in_water}b shows the images of the three suspensions after being centrifuged. From Fig.\ \ref{particles_in_water}b, we can obtain the the final volume $V_f$ of different particles in water, i.e., $V_f\approx 4.4$ ml for cornstarch, $V_f\approx 3.8$ ml for CaCO$_3$, and $V_f\approx 3.4$ ml for glass beads. Therefore, using Eq.\ \ref{particle_conservation} we can calculate the final volume fraction $\phi_f$ for different particles in water, i.e., $\phi_f\approx 45\%$ for cornstarch, $\phi_f\approx 53\%$ for CaCO$_3$, and $\phi_f\approx 59\%$ for glass beads. The results shown in Fig.\ \ref{particles_in_water} indicate that the packing efficiency of different materials suspended in a given liquid can be significantly different. From Figs.\ \ref{cornstarch_in_fluids} and \ref{particles_in_water}, we can conclude that the packing efficiency of particles in liquids can potentially affect the modulus of the material and the visibility of the cracks, among other properties. It should be noted that the packing fraction of porous media is not well-defined, and here we we calculate the volume fraction by the ratio of the granule particle volume to the total volume of liquid and granule particle in a given suspension for the sake of simplicity.

The ultimate packing fraction obtained after desiccation can potentially affect the maximum strain attainable upon drying, the modulus and tensile strength of the sample, the adhesion to the underlying substrate, and the visibility of the cracks. For example, after centrifuging prepared suspensions, we found that cornstarch particles pack significantly more densely ($\phi_f\approx 0.51$) in IPA than either water ($\phi_f\approx 0.45$) or silicone oil ($\phi_f\approx 0.43$), as shown in Fig.\ \ref{cornstarch_in_fluids}. In addition, both cornstarch and CaCO$_3$ pack more loosely than spherical glass beads in water (Fig.\ \ref{particles_in_water}). 

\begin{figure}
\begin{center}
\includegraphics[width=4.5 in]{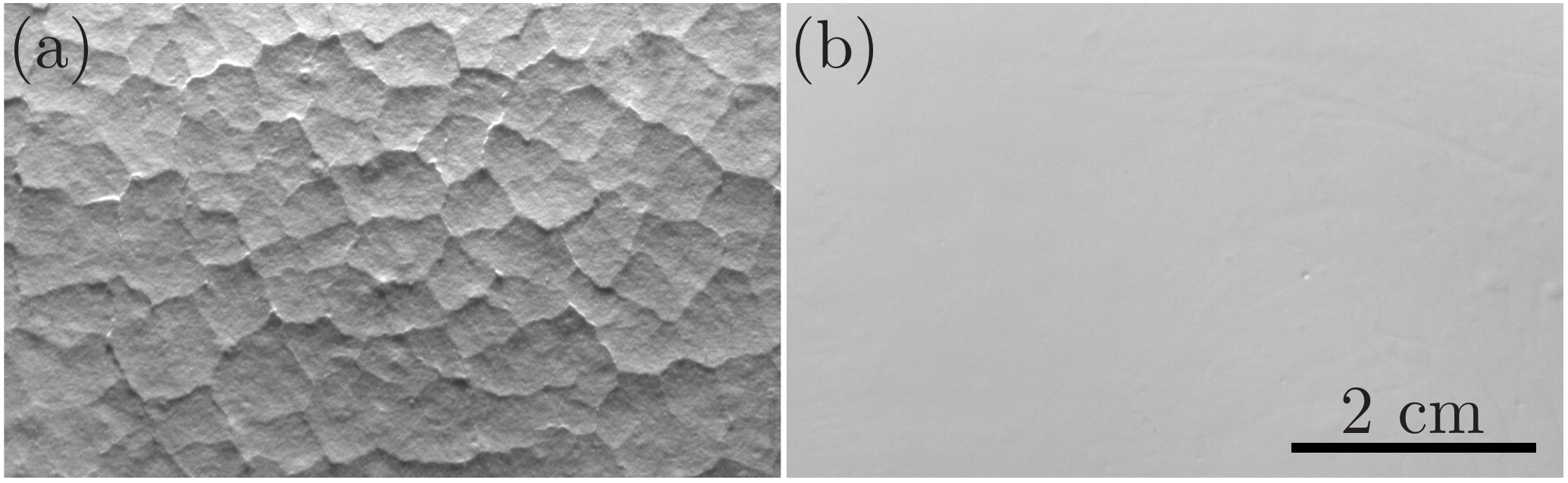}
\caption[Effect of suspending fluids on drying behavior of particulate suspensions]{Fully desiccated suspensions of cornstarch-IPA (a) and cornstarch-silicone oil (b). The two suspensions have the same initial volume (124 ml), and the same initial volume fraction ($\phi_i=40\%$). The film thicknesses of the samples in (a) and (b) are $\approx 6$ mm. The images are enhanced for better visualizations, and the scale bar applies to both images.
\label{fluids_effect}
}
\end{center} 
\end{figure}

We also observed a stark contrast in the crack patterns for cornstarch in both IPA and silicone oil as shown in Fig.\ \ref{fluids_effect}. For IPA suspensions, we only observed very fine surface cracks in thick samples ($h\gtrsim$ 1 cm), which did not penetrate more than $\approx$ 1 mm into the desiccated material (Fig.\ \ref{fluids_effect}a). We suspect that this is due to high packing density of cornstarch in IPA (Fig.\ \ref{particles_in_water}), so that only the surface layer could obtain sufficient strain to crack upon desiccation. Suspensions of cornstarch in silicone oil did not display any visible cracks for most thicknesses used in our experiments (Fig.\ \ref{fluids_effect}b), and only small cracks for very thick samples ($h\approx$ 2 cm). Even after full desiccation, the surface of these films looked like smooth paste  (Fig.\ \ref{fluids_effect}b), suggesting that the particles retained some sort of sticky interactions, possibly due to residual silicone oil adhered to the surface. Even weak, attractive interactions are expected to have a significant effect on granular packings and their mechanical properties for large system sizes \cite{koeze2018sticky}. This hypothesis is consistent with the low packing density of cornstarch in silicone oil (Fig.\ \ref{particles_in_water}), suggesting that the particles may have a strong affinity for the silicone oil. 

\begin{figure}
\begin{center}
\includegraphics[width=4.5 in]{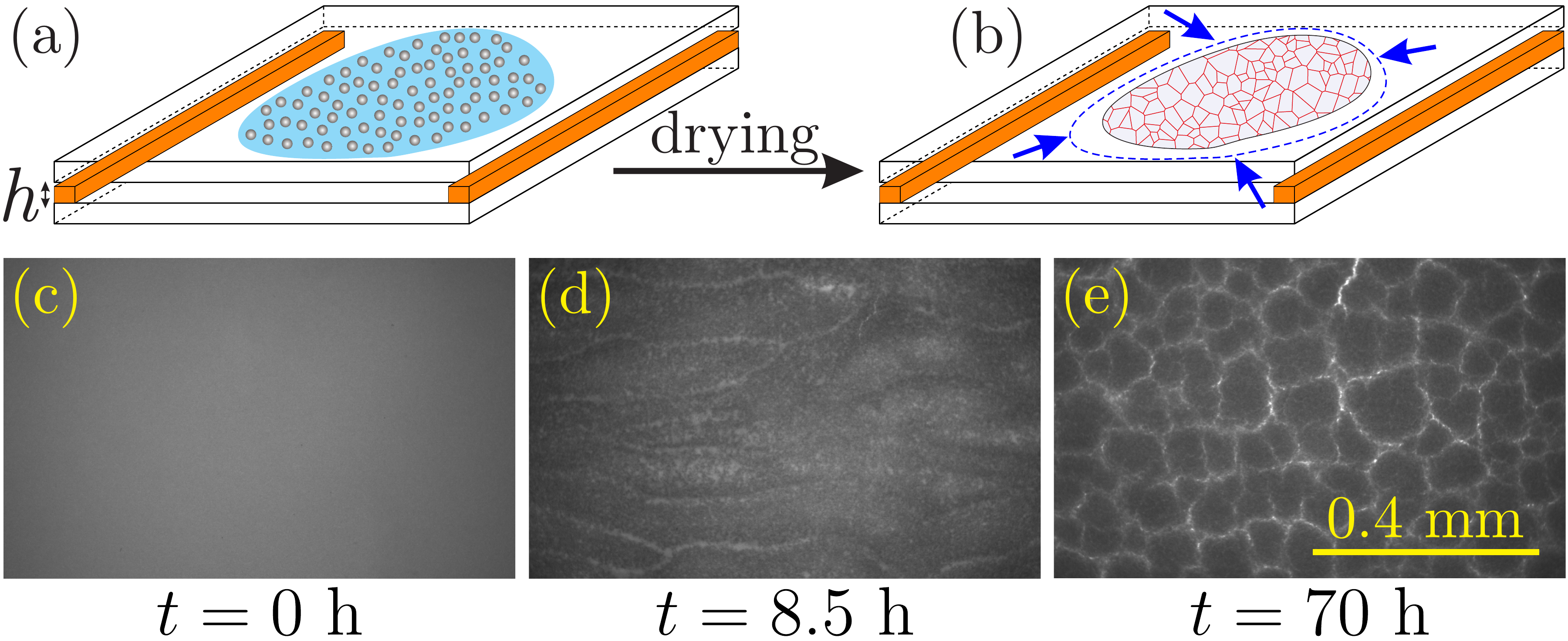}
\caption[Schematic for drying cornstarch-water suspensions in quasi-two-dimensional chambers]{(a) Experimental setup for drying cornstarch-water suspensions in quasi-two-dimensional chambers with thickness $h$. (b) The polygonal cracks in the final dried film are indicated by the red polygons. The dashed blue line represents the profile of the initial suspension, and polygonal cracks (red) appear after the shrinkage of the initial suspension (blue arrows).
Images in (c), (d) and (e) show the drying stages: (c) the initial suspension in the thin chamber, (d) the percolation of dried regions, which appear darker in color since they scatter more light, and (e) the final polygonal crack patterns in the dried film. The scale bar applies to all of the three images. 
\label{sketch_thin_chamber}
} 
\end{center} 
\end{figure}

\subsection{Cornstarch-water suspensions in thin chambers}
\label{sec:secondary_cracks_thin_chamber}
The small-scale secondary cracks observed in Fig.\ \ref{hierarchical_cracks}i are a unique feature of cornstarch-water suspensions, and did not show evidence of a critical thickness ($h_c$), in contrast to primary cracks. In order to explore the thickness dependence of the secondary cracks, we prepared cornstarch-water suspensions with different initial volume fractions, $\phi_i$, and then deposited the suspensions into the thin, quasi-two-dimensional chambers, as shown in Figs.\ \ref{sketch_thin_chamber}a and \ref{sketch_thin_chamber}b. We observed two drying stages: initially a compaction front invades throughout the film; then a second drying stage ``percolates'' throughout the film with a characteristic branching pattern, leading to the formation of liquid, capillary bridges between particles. Finally, the liquid bridges dried up followed by the formation of polygonal cracks after several days (see Figs.\ \ref{sketch_thin_chamber}c-\ref{sketch_thin_chamber}e). 

\begin{figure}
\begin{center}
\includegraphics[width=1\textwidth]{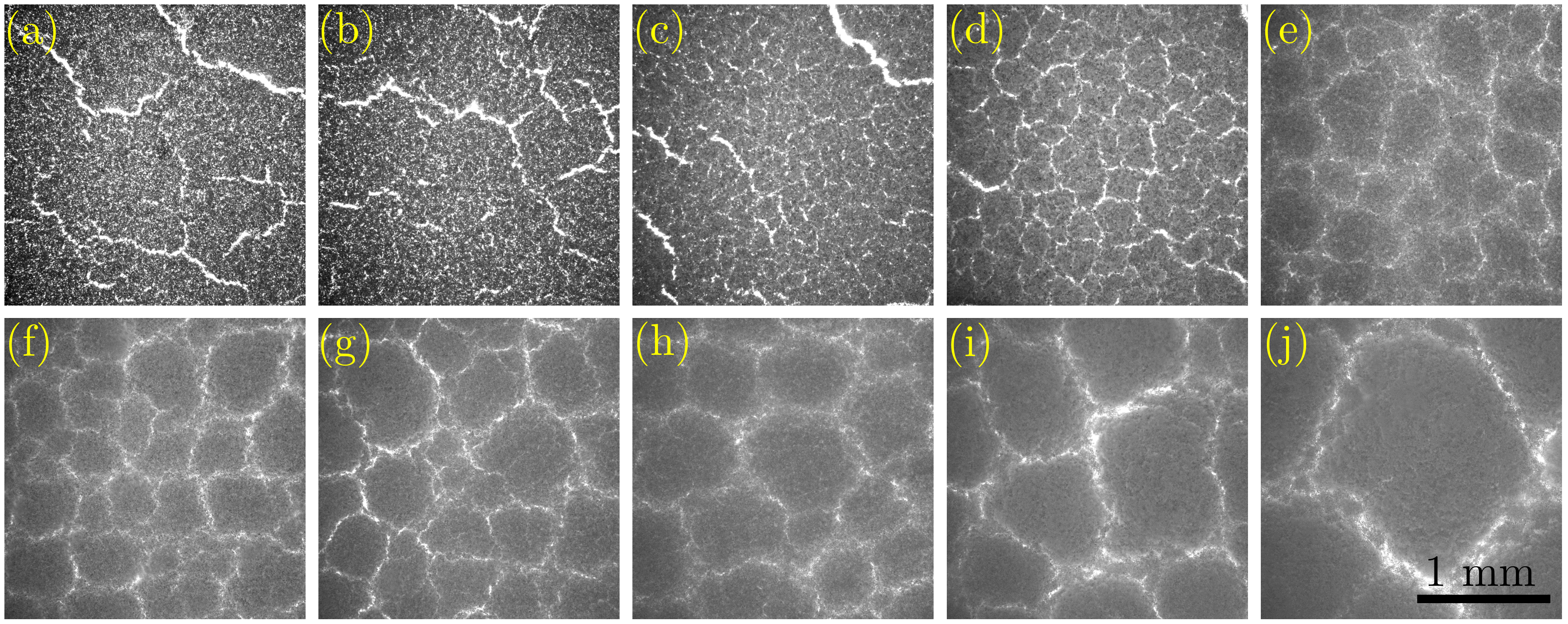}
\caption[Polygonal crack patterns of fully-desiccated cornstarch-water suspensions in thin chambers]{Polygonal crack patterns of fully-desiccated cornstarch-water suspensions in thin chambers with initial particle volume fraction $\phi_i=26\%$. Panels (a) to (h) represent the film thicknesses of 10, 25, 50, 100, 250, 400, 500, 600, 750, and 1000 $\mu$m, respectively. The scale bar applies to all images.
\label{dryingpattern}
} 
\end{center} 
\end{figure}

Figure\ \ref{dryingpattern} shows images of the polygonal cracks observed in the dried films of a cornstarch-water suspension ($\phi_i= 26\%$) in chambers with increasing $h$. Note here the thickness of the chamber is safely taken as the thickness of the dried film since the final dried film was attached to the top and bottom surfaces of the chambers, as shrinkage mostly occurred in the plane (Fig.\ \ref{sketch_thin_chamber}b). In very thin chambers ($h \simeq R$), ``dendritic'' fracture patterns are observed, as shown in Figs.\ \ref{dryingpattern}a and \ref{dryingpattern}b. Since the thickness of the chamber is comparable to the particle size, these patterns were sensitive to the flow of the suspending liquid as evaporation occurred. As $h$ exceeds a critical value, the sensitivity to flow ceased, and regular, polygonal cracks appeared, as shown in Figs.\ \ref{dryingpattern}c-\ref{dryingpattern}j. This trend holds true for all values of $\phi_i$ used in the experiments.

It should be noted here that we also prepared suspensions of glass beads and CaCO$_3$ particles in thin chambers with water, however, no cracks appear during drying of the suspensions in these chambers. This supports the hypothesis that the secondary cracks are due to a distinct drying mechanism involving deswelling of the hygroscopic cornstarch particles (Section\ \ref{sec:particle_deswelling}).

\subsection{Universal scaling of multiscale polygonal cracks}
\label{sec:universal_scaling}

\begin{figure}[!]
\begin{center}
\includegraphics[width=0.7\textwidth]{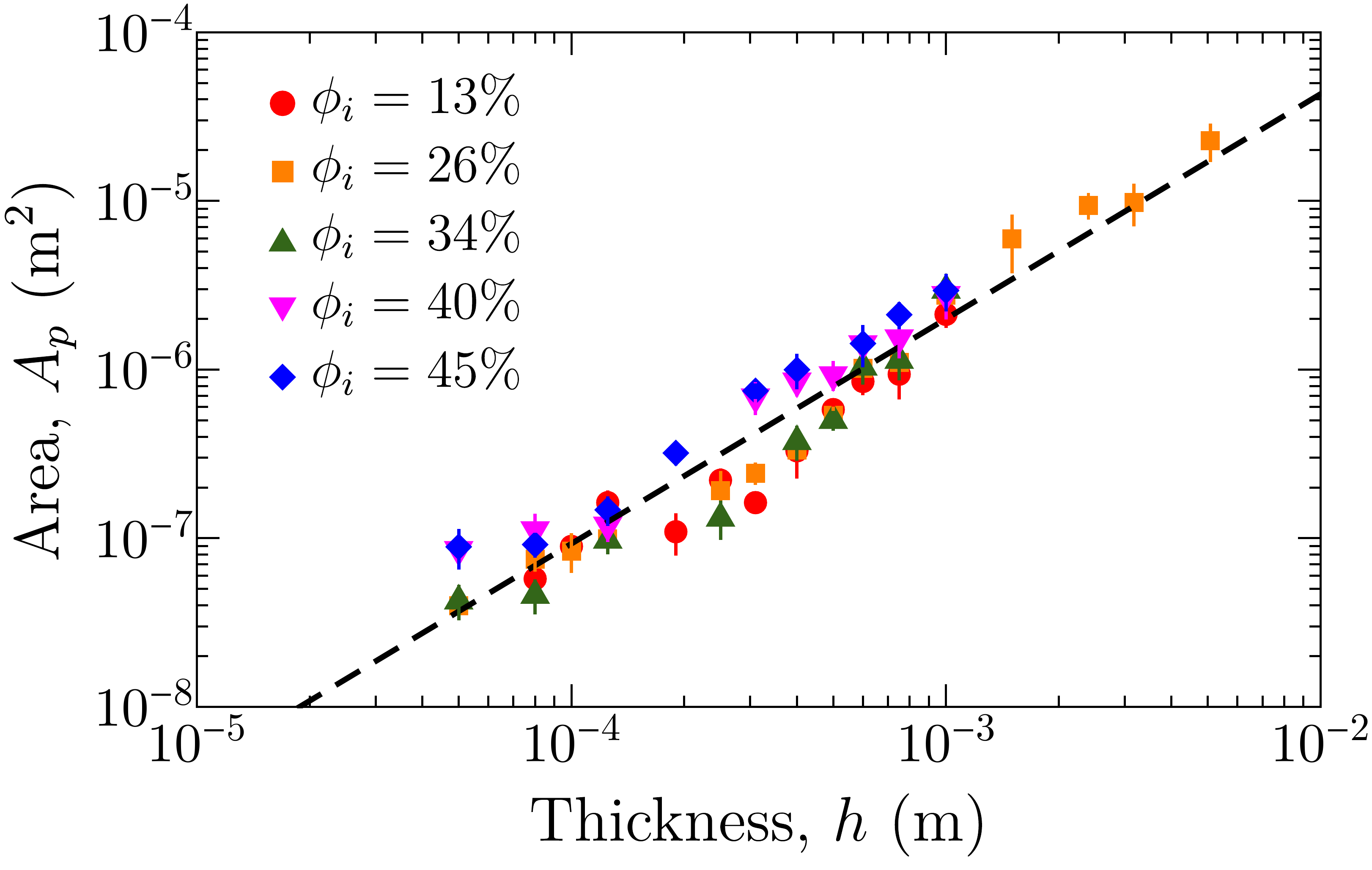}
\caption[Scaling behavior of polygon crack area $A_p$ vs. film thickness $h$ for cornstarch-water suspensions with various initial volume fractions $\phi_i$ after drying in thin quasi-2D chambers]{Scaling behavior of polygon crack area $A_p$ vs. film thickness $h$ for cornstarch-water suspensions with various initial volume fractions $\phi_i$ after drying in thin quasi-2D chambers. The dashed black line represents the best fit for the data, $A_p$ = 0.02 m$^{2/3}$ $h^{4/3}$. The error bars are from the standard deviation of multiple measurements.
\label{cornstarch2D}
} 
\end{center} 
\end{figure}
\begin{figure}[!]
\begin{center}
\includegraphics[width=1\textwidth]{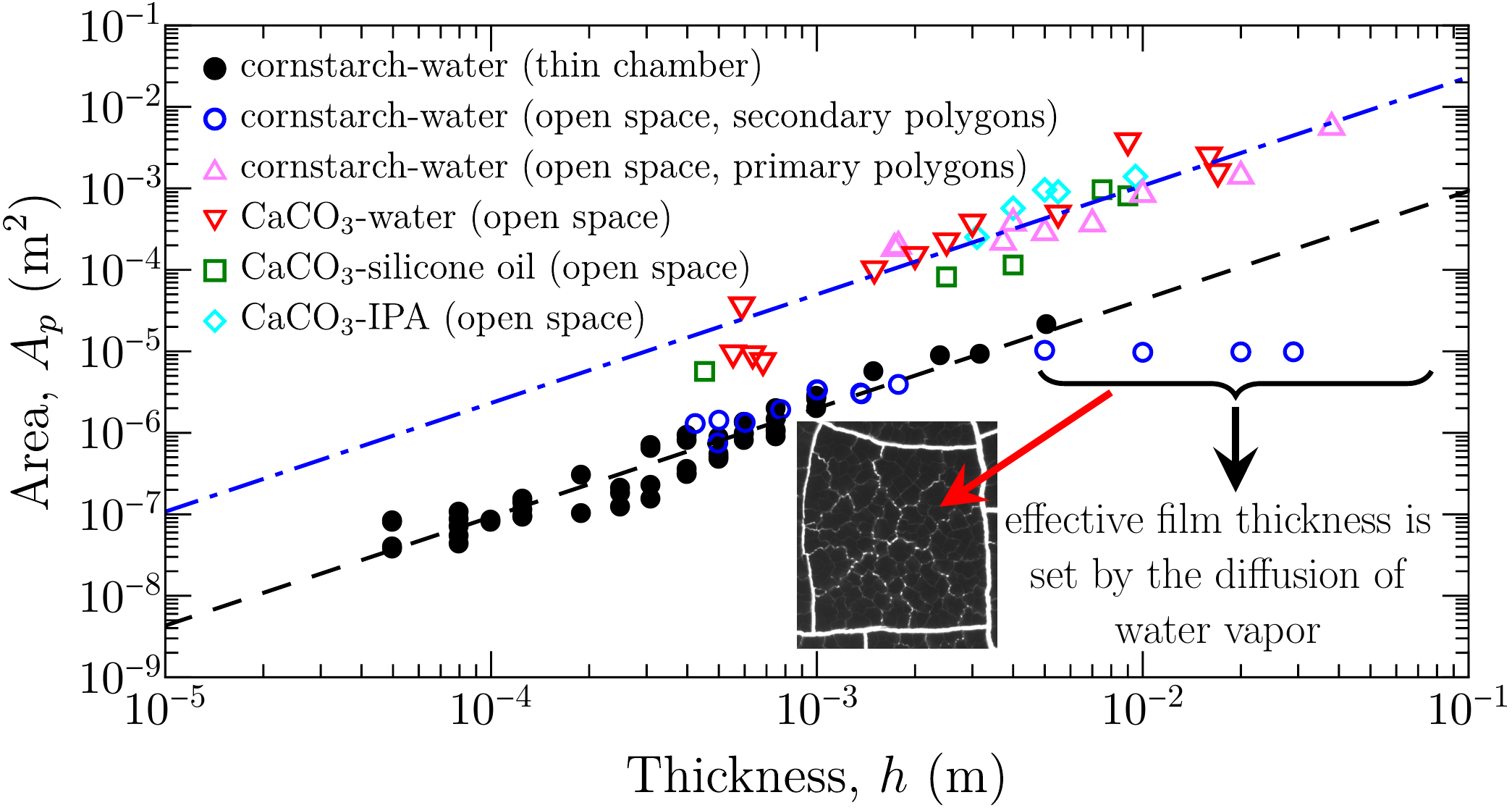}
\caption[Characteristic area ($A_p$) vs. the film thickness ($h$) of the multiscale polygonal cracks observed in both thin chambers and petri dishes]{Characteristic area ($A_p$) vs. the film thickness ($h$) of the multiscale polygonal cracks observed in both thin chambers and petri dishes. The open symbols represent the data obtained in petri dishes, whereas the solid symbol represent cornstarch-water suspensions dried in thin chambers with various $\phi_i$, from Fig.\ \ref{cornstarch2D}. The open blue circles represent small-scale, secondary cracks observed in cornstarch-water suspensions dried in petri dishes. The dot-dashed blue and dashed black lines represent $A_p=\alpha h^{4/3}$, where $\alpha$ = 0.5 m$^{2/3}$ and 0.02 m$^{2/3}$, respectively.
\label{scaling_all_cracks}
} 
\end{center} 
\end{figure}

Figures\ \ref{allcracks} and \ref{dryingpattern} show that the characteristic size of all observed polygonal cracks, in both petri dishes and thin chambers, increases with $h$. Although there are many ways to characterize the polygonal patterns, such as the average number of edges, or the aspect ratio, we simply measured the average area $A_p$ of the polygonal cracks. The results are shown in Fig.\ \ref{scaling_all_cracks}. The solid symbols represent cornstarch-water suspensions desiccated in thin chambers. The data were nearly independent of the initial volume fraction deposited in the chamber (Fig.\ \ref{cornstarch2D}). The open symbols represent polygons observed in petri dishes. More specifically, the open blue circles represent the secondary cracks in dried cornstarch-water suspensions in petri dishes, which overlap with the data from the thin chambers. For very thick suspensions of cornstarch and water, the area of the small-scale, secondary polygons saturated, and did not increase further. As we will show in Section\ \ref{effective}, the deswelling of the particles proceeds as a drying front that penetrates diffusively into the material. Thus, the effective thickness associated with the crack formation depends on the diffusion of water vapor from the film.

As mentioned previously, numerous authors have investigated the thickness dependence of the characteristic crack spacing or polygon area in desiccated suspensions \cite{groisman1994experimental,allain1995regular,komatsu1997pattern,shorlin2000alumina,lee2004drying,ma2012possible}. However, for polygonal crack networks, most of these studies cover a very limited range in thicknesses, so that a comprehensive picture of the thickness dependence of polygonal cracks is lacking. Although there is significant variation in the data from any single set of experiments, taken together, our results in Fig.\ \ref{scaling_all_cracks} strongly suggest:
\begin{align}
&A_p=\alpha h^{4/3}
\end{align}
over a wide range of thickness, for different types of polygonal cracks, in different experimental geometries, and for different liquid-particle combinations. This scaling law is indicated by the dot-dashed blue and dashed black lines in Fig.\ \ref{scaling_all_cracks}. Although the prefactor, $\alpha$, is distinct for primary and secondary cracks in different materials, the data suggest that the exponent is universal. Recently, Flores \cite{flores2017mean} derived this simple scaling law using continuum elastic theory, and a balance of surface energy and elastic energy for the initiation of cracking. We will repeat this argument here since we will make small alterations to the expression for $\alpha$.

\begin{figure}[!]
\begin{center}
\includegraphics[width=3.5 in]{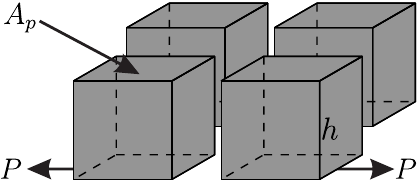}
\caption[Schematic for cracking in a material driven by tensile stress at the substrate]{Cracking in a material driven by tensile stress at the substrate. Here, $h$ is the film thickness, $A_p$ is the characteristic area of the polygonal cracks, $P$ is the tensile stress generated due to the adhesion of the film to the substrate. Adapted from Ref.\ \cite{flores2017mean}.
\label{crack_sketch}} 
\end{center} 
\end{figure}

Figure\ \ref{crack_sketch} shows a section of a thin film after the formation of cracks \cite{flores2017mean}. The thickness of the film is $h$, and the characteristic area of the polygons is $A_p$. The main tensile stress, $P$, acts on the bottom surface, where the film is adhered to the substrate. We assume that cracks will form when the energy cost of creating new surfaces at the sides of a polygon is equal to the elastic energy released during cracking:
\begin{align}
&\kappa\gamma \sqrt{A_p}h\sim\dfrac{V}{2E}\left<\sigma\right>^2,
\label{grif}
\end{align}
where $\gamma$ is the surface tension of the newly-created interfaces which have a typical area $\sqrt{A_p} h$, and $\kappa$ is the ratio of the perimeter to the area of the polygons. Here, $\left<\sigma\right>$ is the volume-averaged tensile stress in the film, $E$ is Young's modulus, and $V=A_p h$ is the volume of one polygon.

We can relate $\left<\sigma\right>$ to $P$ using the fact that the average stress in a volume element can be related to an integral over the forces acting on its boundaries \cite{landau1986theory}:
\begin{align}
&\left<\sigma_{ij}\right>=\dfrac{1}{2 V}\oint_S \left( P_i x_j+P_j x_i \right)\mathrm{d}S.
\label{intstress}
\end{align}
Since the main stress is applied at the substrate, Eq.\ \ref{intstress} provides the approximate scaling:
\begin{align}
&\left<\sigma\right>\approx P\dfrac{\sqrt{A_p}}{h}.
\label{stressscale}
\end{align}
Although we have not rigorously evaluated the integral for a thin, adhered film, one obtains the same result, $\left<\sigma\right>\propto 1/h$, by considering a similar problem, a thin, spherical elastic shell under uniform pressure. In this case, which can be solved exactly, the tangential, tensile stress scales in the same way as Eq.\ \ref{stressscale} \cite{lautrup2011}.

Combining Eqs.\ \ref{grif} and \ref{stressscale}, we arrive at the predicted scaling relation:
\begin{align}
&A_p=\left(\dfrac{2\kappa\gamma E}{P^2}\right)^{2/3}h^{4/3}.
\label{polygon_vs_h}
\end{align}

\begin{figure}[!tbph]
\begin{center}
\includegraphics[width=4 in]{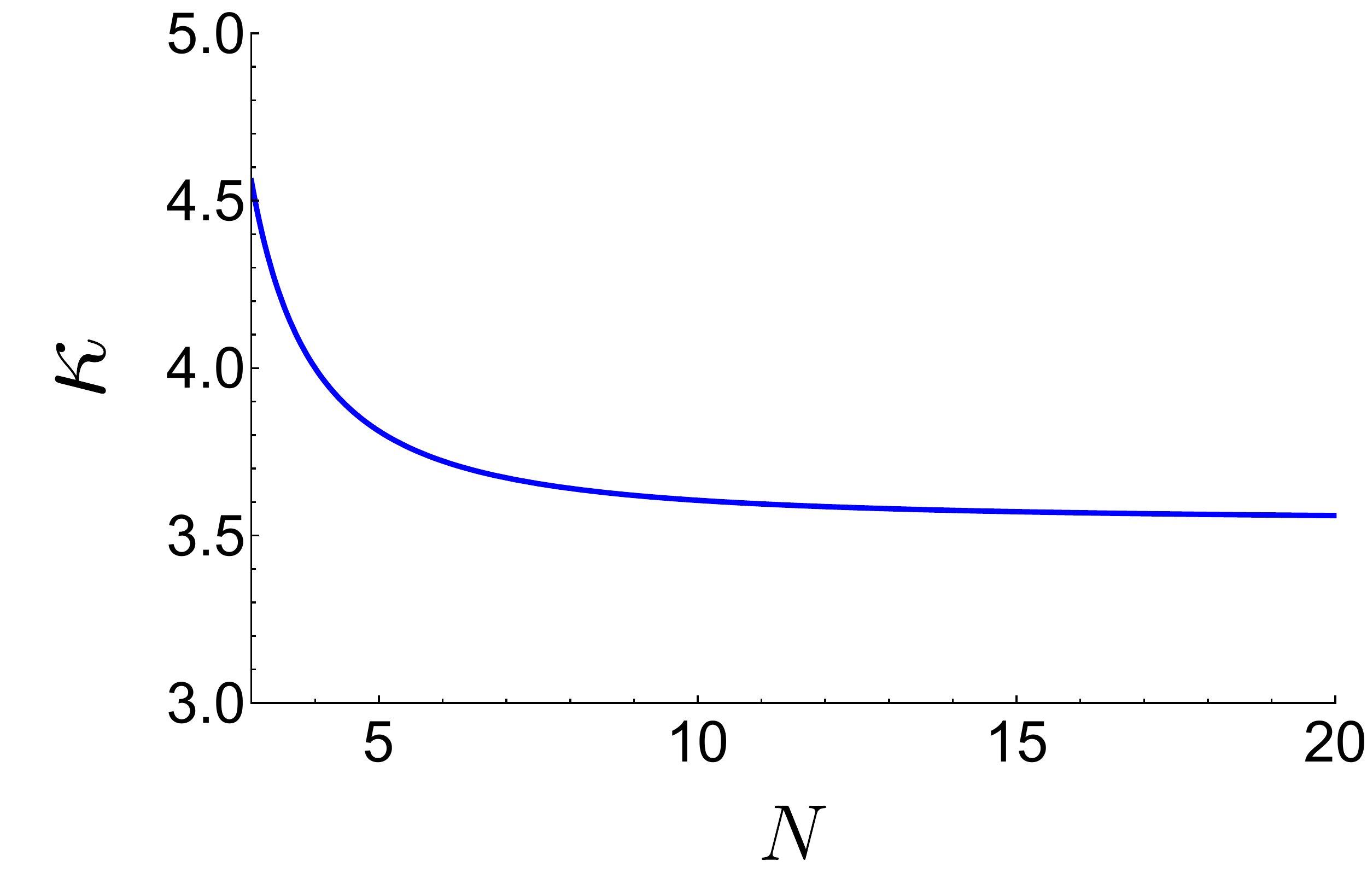}
\caption[The variation of $\kappa$ with respect to $N$]{The variations of $\kappa$ with respect to the number of polygon side $N$, i.e., Eq.\ \ref{geometry_factor}.
} 
\label{sacling_prefactor}
\end{center} 
\end{figure}

Given the surface tension of the new, ``wet'' interfaces, $\gamma$, the prefactor, $\alpha=(2\kappa\gamma E/P^2)^{2/3}$, is determined by the modulus of the material when cracks form, and stress at the substrate generated by shrinkage, $P$. When cracks form, $P$ will essentially be the yield stress, and will be smaller than $E$ \cite{mahaut2008rheology}. Most of the polygons we observe are convex. In this case, we can estimate $\kappa$ by assuming they are regular polygons, where $\kappa$ has an analytic expression:
\begin{align}
&\kappa=2\sqrt{\dfrac{N}{\cot(\pi/N)}}.
\label{geometry_factor}
\end{align}
Here, $N$ is the number of sides of the polygon. As shown in Fig.\ \ref{sacling_prefactor}, for $N$ = 3, $\kappa\approx$ 4.56. As $N\rightarrow\infty$, $\kappa\rightarrow$ 3.54. Thus, in our further discussion, we will assume $\kappa\approx$ 4 for simplicity in estimating the prefactor $\alpha$.

In order to provide some quantitative measurement of the modulus, we prepared different mixtures at different drying stages, and used a rheometer to measure the applied normal force upon indenting the material with a stainless steel ball of radius $R_b\approx 9.5$ mm using the indentation load-displacement method \cite{pharr1992generality,oliver2004measurement}. More specifically, in this method, a complete cycle of loading and unloading of the indenter into the material is performed to measure the elastic and plastic contributions to the deformation. Generally, the loading phase contains both elastic and plastic effect, whereas the initial unloading is assumed to be approximately pure elastic, such that the Young's modulus can be calculated using the linearly-elastic Hertzian contact theory \cite{doerner1986method,phadikar2012establishing}.

The stiffness (slope) for the initial unloading is \cite{oliver2004measurement,pharr1992generality,broitman2017indentation}: 
\begin{align}
\label{unloading_stiffness}
\frac{\textrm{d}F_N}{\textrm{d}y}\Bigm|_{y=y_m}&=\frac{2}{\sqrt{\pi}}E^*\sqrt{A_{\textrm{proj}}},\\
\label{modulusrelation}
E^*&=\frac{E}{1-\nu^2},
\end{align}
where $F_N$ is the normal force, $y$ is the indentation depth, $y_m$ is the maximum displacement when the indenter is fully loaded, $A_{\textrm{proj}}=\pi y_m(2R_b-y_m)$ is the projected circular area of the contact for a sphere indented into a half-space by a distance $y=y_m$, $E$ is Young's modulus, and $\nu$ is the Poisson's ratio of the film ($\approx 0.5$).

\begin{figure}[!tbph]
\begin{center}
\includegraphics[width=4 in]{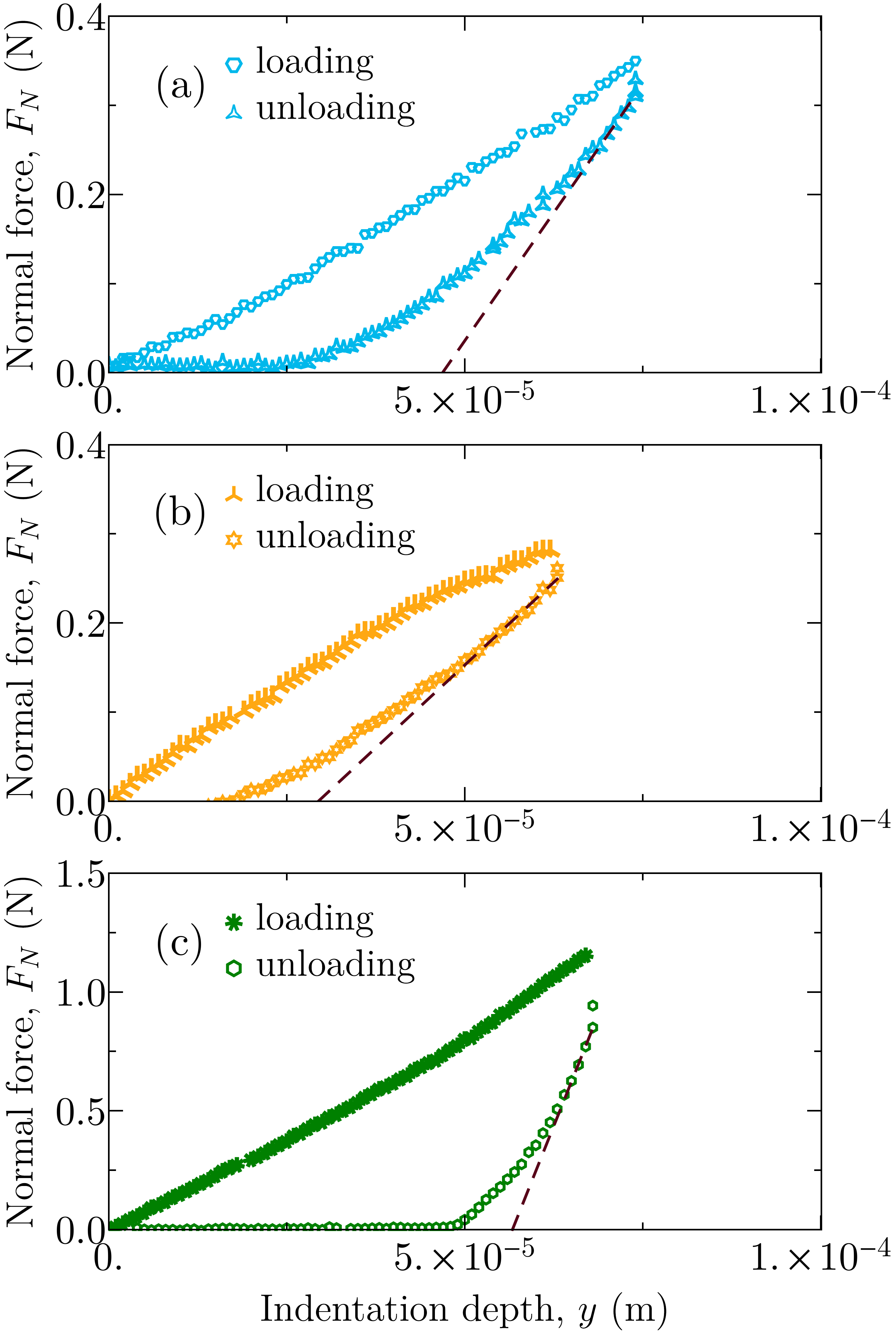}
\caption[Modulus measurement of particulate suspensions at different drying stages]{Indentation load-displacement measurement of particulate suspensions at different drying stages. (a), (b) and (c) show the data of a cornstarch-water film after primary cracks appear, a cornstarch-water film when secondary cracks appear, and a CaCO$_3$-water film when cracks appear, respectively. The dashed lines represent the eye guide for the best fits of slope of the initial unloading at $y=y_m$ using Eq.\ \ref{unloading_stiffness}, and the corresponding modulus values are $E$ = $3.0\times 10^6$ Pa (a), $2.6\times 10^6$ Pa (b), and $2.0\times 10^7$ Pa (c).
} 
\label{modulus}
\end{center} 
\end{figure}
\begin{figure}
\centering
\includegraphics[width=4 in]{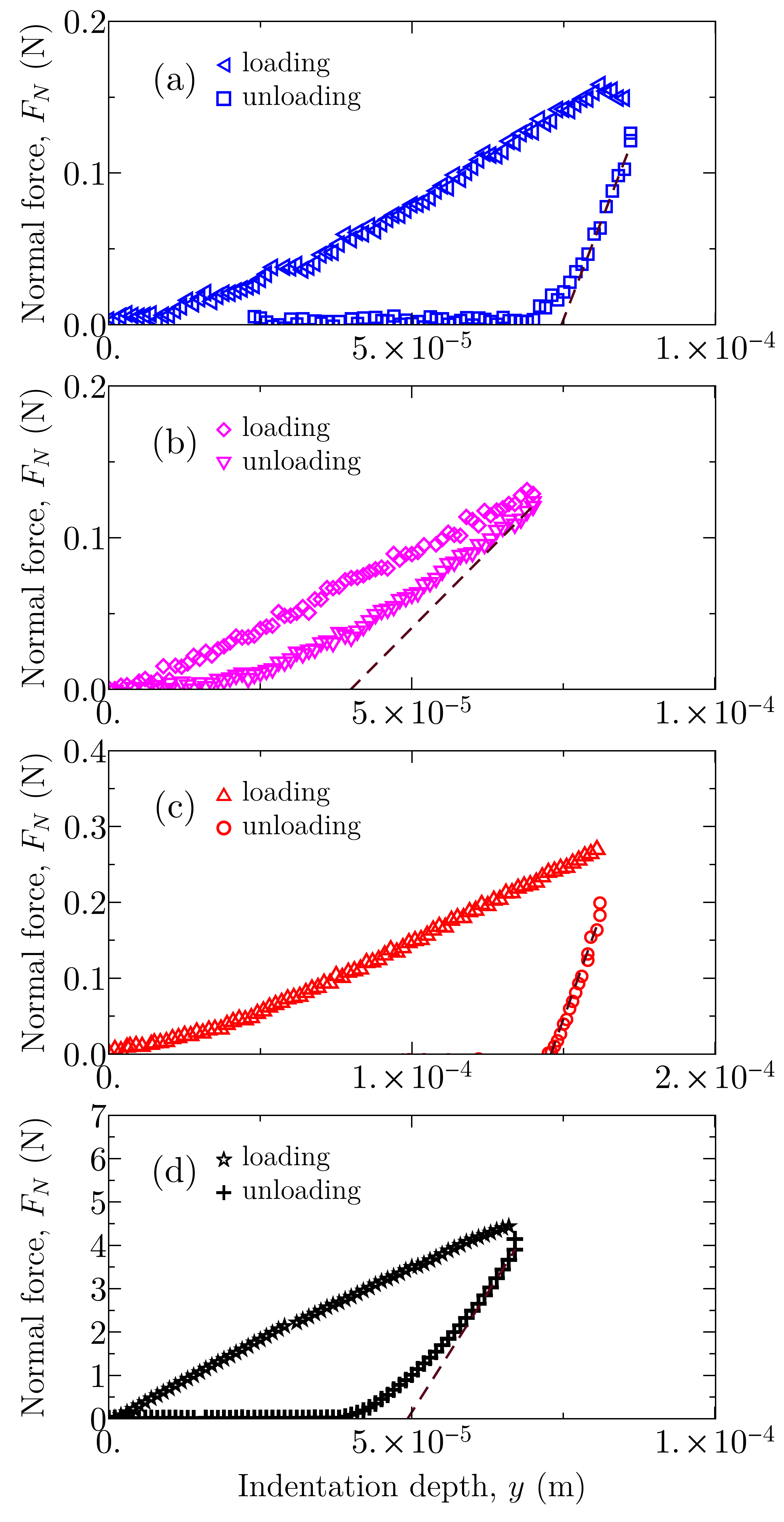}
\caption[Modulus measurement of fully desiccated particulate suspensions]{Indentation-loading displacement tests of different films at different drying stages. (a), (b), (c) and (d) show the data on fully-desiccated cornstarch-IPA film, cornstarch-water film, cornstarch-silicone oil film, and CaCO$_3$-water film, respectively. The dashed lines represent the best fits of the unloading data using Eq.\ \ref{unloading_stiffness}.}
\label{plasticity}
\end{figure}

Figure\ \ref{modulus} shows the loading and unloading versus displacement for different particulate films when primary and secondary cracks form. Note for each tested film, the entire film thickness is at least 100 times greater than the indentation depth $y_m$, and the maximum load was held for about 10 s before unloading started. By fitting the slope of the initial unloading data at $y=y_m$ using Eq.\ \ref{unloading_stiffness}, the Young's modulus of the films fall in the range from 2.6 to 20 MPa. These values are consistent with similar modulii measured in non-Brownian dense suspensions and soils \cite{moller2007shear,mahaut2008rheology,kezdi1974,obrzud2018}. In addition, we also measured the Young's modulus of fully-desiccated films, and the results are shown in Fig.\ \ref{plasticity}, in which the modulus of films is calculated to be as follows: $3.3 \times 10^6$ Pa (cornstarch-IPA film, see Fig.\ \ref{plasticity}a), $1.0 \times 10^6$ Pa (cornstarch-water film, see Fig.\ \ref{plasticity}b), $3.3 \times 10^6$ Pa (cornstarch-silicone oil film, see Fig.\ \ref{plasticity}c), $6.9 \times 10^7$ Pa (CaCO$_3$-water, see Fig.\ \ref{plasticity}d). The errors of the nonlinear fitting parameters using Eq.\ \ref{unloading_stiffness} range from 0.3\% to 2.0\%. The results of Figs.\ \ref{modulus}a and \ref{modulus}b are consistent with experimental observations that small-scale secondary cracks have a smaller prefactor, as shown in Fig.\ \ref{scaling_all_cracks}, although this alone does not explain the difference (0.02 m$^{2/3}$ versus 0.5 m$^{2/3}$, Fig.\ \ref{scaling_all_cracks}).

The value of the substrate stress, $P$, is more difficult to measure. The maximum adhesion to the substrate clearly affects the crack pattern, as shown in Fig.\ \ref{boundary_stress_effect}, so that a stronger maximum stress at the substrate decreases the polygon area, as indicated by Eq.\ \ref{polygon_vs_h}. Even for strong adhesion, shrinkage in the film will continue until $P$ is approximately the yield stress of the particle network.

For a random, close-packed particle network, the yield stress, $Y$, can be estimated as \cite{goehring2013plasticity}:
\begin{equation}
Y=\dfrac{\phi M F_\mathrm{max}}{4\pi R^2},
\label{yield_stress}
\end{equation}
where $\phi$ is the volume fraction of particles, $M$ is the coordination number, and $R$ is the particle radius (see Table\ \ref{Drying_tab:1}). $F_{\mathrm{max}}$ is the maximum force between particles. For the initiation of primary cracks, the suspension is still saturated with liquid, so we can assume that $F_{\mathrm{max}}/R^2$ can be simply estimated as the capillary pressure, $\gamma/R$. Assuming typical values of the parameters from Table\ \ref{Drying_tab:1}, Eq.\ \ref{yield_stress} gives:
\begin{equation}
Y\sim 0.27 \dfrac{\gamma}{R}.
\label{simplify_yield_stress}
\end{equation}

\begin{figure}[!]
\begin{center}
\includegraphics[width=1\textwidth]{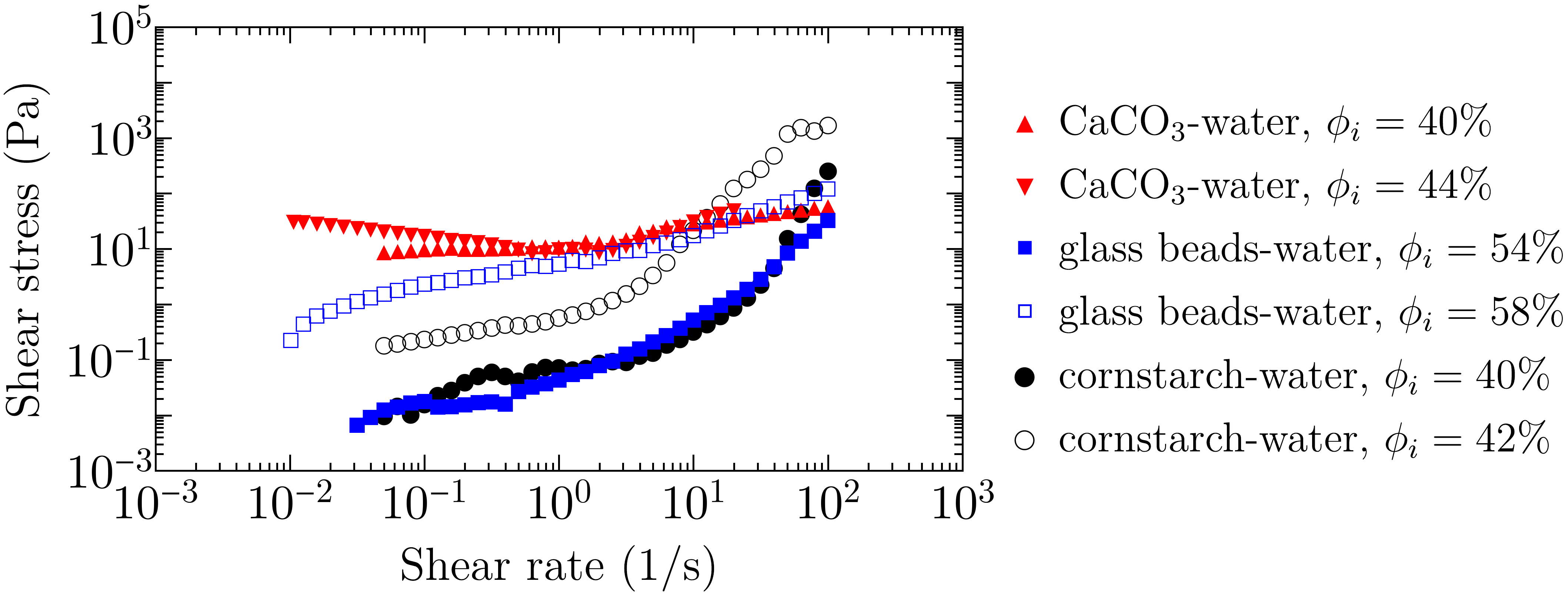}
\caption[Rheology measurements of particulate suspensions with different initial volume fractions $\phi_i$]{Rheology measurements of particulate suspensions with different initial volume fractions $\phi_i$.
\label{rheology_measurement}
} 
\end{center} 
\end{figure}

This result is not surprising for a wet sample if the inter-particle adhesion in the bulk liquid is small, i.e. the stress required to form a crack by pulling particles apart is of order the capillary forces holding them together. To confirm this, we have performed rheological measurements on cornstarch-water, CaCO$_3$-water, and glass beads-water suspensions with different volume fractions, and the results are shown in Fig.\ \ref{rheology_measurement}. Although shear thickening is observed for larger volume fractions, the maximum shear stress is smaller than $\gamma/R$ \cite{brown2012thickening}, showing that capillary forces are larger than any inter-particle force in the bulk liquid.

When a crack forms, we can assume that the boundary stress will be of the same order as the yield stress, so that $P\sim Y$. Plugging the values of $E$, $\gamma$, and $R$ for the large-scale cracks in cornstarch-water samples, and assuming $\kappa\approx 4$, Eq.\ \ref{polygon_vs_h} yields $\alpha\approx 0.4$ m$^{2/3}$, which agrees well with dot-dashed blue line in Fig.\ \ref{scaling_all_cracks}. Given the variability in the modulus $E$ between different suspensions, we do not currently have a way of collapsing all the data in Fig.\ \ref{scaling_all_cracks} for the primary cracks. For wet samples, the modulus will likely depend on the surface tension, particle size, the modulus of the particles, the particle shape, and the inter-particle friction. Nevertheless, given this large parameter space, the good agreement with Eq.\ \ref{polygon_vs_h} suggests that the polygonal crack pattern can be quantitatively understood for a range of different particles and liquids, provided some knowledge of the modulus and yield stress of the suspension.

The small-scale, secondary polygonal cracks of cornstarch-water suspensions can be observed in both open petri dishes (Fig.\ \ref{hierarchical_cracks}) and thin chambers (Fig.\ \ref{dryingpattern}). This suggests that the formation of small-scale, secondary cracks does not sensitively depend on the drying geometry, and that capillary interactions are not a dominant force, in contrast to the primary cracks. Thus, local particle adhesion likely determines the yield stress of the material. As shown in Section\ \ref{sec:particle_deswelling}, the cornstarch particles are swollen with water, so we do not currently have a way to estimate this adhesion. In addition, the factor of $\gamma$ in the stress balance (Eq.\ \ref{grif}) would be related to the surface energies of the particle-particle adhesion \cite{goehring2013plasticity}. Since $F_{\mathrm{max}}\propto\gamma$ for adhesive forces, then $\alpha\propto (\gamma/F_{\mathrm{max}}^2)^{2/3}\propto 1/\gamma^{2/3}$. We can then conclude that this adhesion of swollen particles must be stronger than capillary interactions since the prefactor is smaller for small-scale cracks. 

\subsection{Effective film thickness for cracks in thick cornstarch-water suspensions}
\label{effective}
\begin{figure}
\begin{center}
\includegraphics[width=4.2 in]{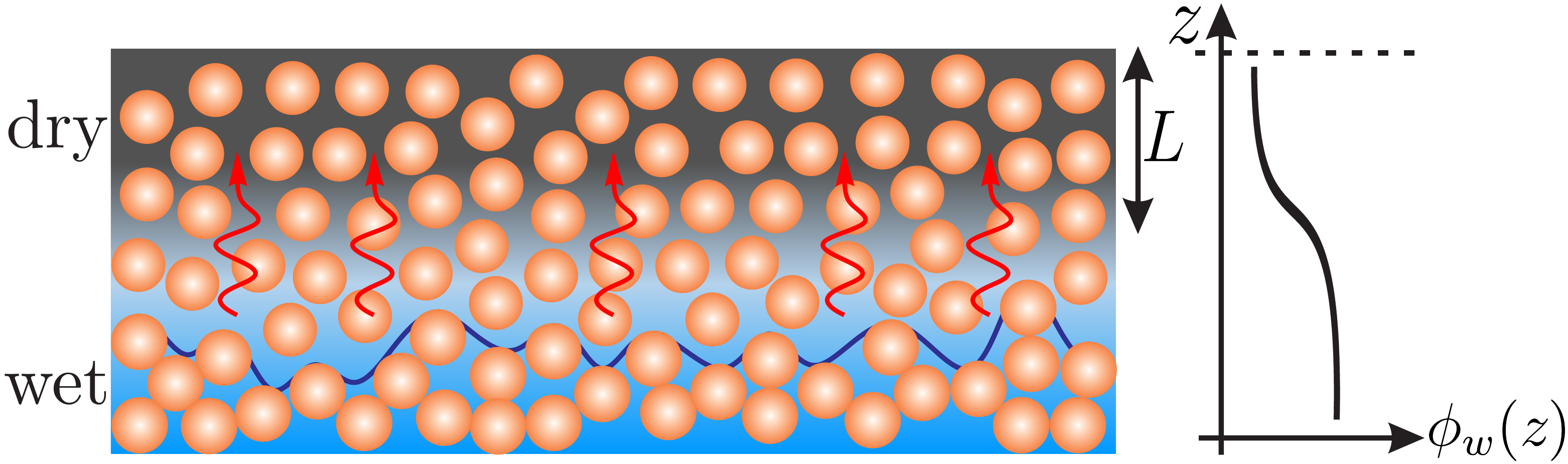}
\caption[Schematic for the effective film thickness during drying thick samples of cornstarch-water suspensions]{The boundary between dry and wet layers during drying cornstarch-water suspensions is set by the diffusion of water vapor, as indicated by the red arrows in thick films. $L$ represents the characteristic length scale of the dry layer, i.e., the effective film thickness, and $\phi_w(z)$ is the water volume fraction along the vertical direction $z$. 
\label{boundary}
} 
\end{center} 
\end{figure}

In Fig.\ \ref{scaling_all_cracks}, we showed how the area of the small-scale, secondary polygons in cornstarch-water suspensions saturated for large values of $h$ (blue open circles). Here we show that this saturation of $A_p$ is set by the diffusion of the water vapor in the later drying stage. The individual particles will remain swollen until their environment is sufficiently dry. In an open particle network where evaporation is occurring from above, water transport is initially limited by viscosity as the liquid is pulled through the porous network according to Darcy's law. As evaporation proceeds, eventually the diffusion of water vapor through the top of the sample limits the transport. This diffusion-limited transport is likely to set a boundary between wet and dry layers, leading to an effective film thickness, $L$, as illustrated in Fig.\ \ref{boundary}. 

It has been suggested that the water vapor content can be described by a nonlinear, one-dimensional effective diffusion equation \cite{goehring2009drying,goehring2009nonequilibrium}:
\begin{equation}
\frac{\partial \phi_w}{\partial t}=\frac{\partial }{\partial z} \left[D(\phi_w) \frac{\partial \phi_w}{\partial z} \right],
\label{diffusion}
\end{equation}
where $\phi_w=\phi_w(z,t)$ is the spatio-temporal variation of the local volume fraction of water content, and $D(\phi_w)$ is local diffusivity of the water vapor, and can be expressed as \cite{pel2002analytic,goehring2009nonequilibrium}:
\begin{equation}
D(\phi_w)=\left( \int_{h}^{z} \dfrac{\partial \phi_w }{\partial t} \mathrm{d}z' \right)\bigg/\left( \dfrac{\partial \phi_w }{\partial z}  \right),
\label{diffusivity}
\end{equation}
where $0\leq z\leq h$, assuming a no-flux boundary condition at the lower boundary $z=h$. With Eq.\ \ref{diffusivity}, Goehring \textit{et al.} \cite{goehring2009nonequilibrium} measured an average $D(\phi_w) \sim 10^{-9}$ m$^2$/s with $\phi_w$ ranging from 0.1 -- 0.3 g/cm$^3$. A similar value has also been reported by M{\"u}ller \cite{muller1998starch}. In our experiment the characteristic time scale for the formation of secondary cracks in petri dishes is $T\sim$ 24 h, thus the characteristic length scale of the dry layer $L$ can be estimated as $L\sim \sqrt{TD}\sim$ 1 cm, which shows excellent agreement with the saturation thickness for small-scale cracks shown in Fig.\ \ref{scaling_all_cracks}. It should be noted that in our experiment, the cornstarch-water suspensions were dried under room temperature without introducing extra heat, so that $D(\phi_w)$ should be a bit smaller than 10$^{-9}$ m$^2$/s.

\section{Summary and outlook}

In summary, we experimentally investigated polygonal crack patterns in desiccated, particulate suspensions composed of various liquid and particle combinations. The thicknesses of the films ranged from $h$ = 10 $\mu$m to 4 cm. There are two major results of this work. First, the appearance of multiple, distinct length scales associated with cracks results from distinct shrinkage mechanisms during the drying process. Whereas larger, capillary-induced crack patterns occurred in many of the liquid-particle combinations, such multiscale crack patterns only appeared in dried suspensions of cornstarch and water due to deswelling of the hygroscopic starch particles. As Fig.\ \ref{multiscale_polygonal_cracks} shows, similar multiscale crack patterns can be observed in meter-scale, planetary terrain. This finding alone may help interpret geomorphological history from surface images, even though knowledge of the relevant material properties may not be known.

Second, the characteristic area of the polygons, for all observed cracks, is consistent with a power law scaling: $A_p=\alpha h^{4/3}$, where the prefactor is determined by a balance of surface energy ($\gamma$), film modulus ($E$), and boundary stress ($P$): $\alpha\sim(2\kappa\gamma E/P^2)^{2/3}$ \cite{flores2017mean}. The values of these parameters depend on the dominant particle-particle interaction forces at play during the initiation of cracking. By quantifying the modulus and equating $P$ with the yield stress, we are able to quantitatively predict $\alpha$ for primary cracks. We note that although this scaling law is consistent with some previous predictions \cite{komatsu1997pattern}, other authors have reported a quadratic relationship between $A_p$ and $h$ \cite{groisman1994experimental,shorlin2000alumina,leung2000pattern,leung2010criticality}. However, nearly all experimental studies report less than one order of magnitude in thickness variation. In our analysis, we have assumed that for small strains, our films can be considered homogeneous, elastic materials, and it is possible that effects such as Brownian motion, particle density fluctuations, or sticky particle interactions may explain differences observed in the literature for crack spacing and polygon area. We leave this hypothesis to future studies of other materials with different particle interactions.

Although our experiments are limited to laboratory scales, the scaling law, $A_p\propto h^{4/3}$ reproduces reasonable values for polygon areas on larger scales. For example, if we assume that polygonal cracks commonly observed in wet mud with $R\approx$ 60 $\mu$m are mainly due to capillary pressure during drying, and a typical modulus of 5 MPa \cite{kezdi1974,obrzud2018}, then Eqs.\ \ref{polygon_vs_h}, \ref{geometry_factor}, \ref{yield_stress}, and \ref{simplify_yield_stress} give $\sqrt{A_p}\approx$ 1 m for a crack depth of $h\approx$ 20 cm. For polygonal crack patterns on much larger scales, such as those show in Fig.\ \ref{multiscale_polygonal_cracks}b, we note that the polygon area may saturate due to heterogeneity in the material properties with depth. In this case it is likely that the modulus of the material is much larger, or that the stress induced during shrinkage is much smaller, in order to produce very large polygon areas. In addition, in the scaling, the boundary stress applied by the substrate requires more detailed characterizations for substrates with diverse conditions for various particle-liquid combinations, and we leave these interesting points for our future studies.

It is expected that the scaling law we found in the laboratory experiments can enhance our general understanding of how nonequilibrium dynamics give rise to crack patterns that we encounter in our daily lives, and it could also provide a novel route for predicting the history and material properties of particular lands (e.g., Fig.\ \ref{multiscale_polygonal_cracks}b) by simply analyzing the images.

\chapter{Sedimentation of non-Brownian particles}
\label{sedimentation_pattern}

\section{Introduction}
In Chapters \ref{Leidenfrost_stars} and \ref{drying_cracks}, I discussed pattern formation driven by hydrodynamic and mechanical instabilities in fluid and granular systems, respectively. In addition, when solid particles are immersed in liquid, they are prone to gravitational settling in the liquid as long as the density of the particle is greater than that of the liquid. This sedimentation process gives rise to a larger number of intriguing natural structures, and is still a main research focus in fundamental fluid mechanics, mechanical, chemical, and biological engineering. In this chapter, I will report some experimental results on the dynamics of sedimentation of non-Brownian particles and the resulting pattern formation driven by such dynamics. 

Sedimentation is a ubiquitous and crucial process both in nature and industry, for instance the accumulation of geological deposits \cite{pettijohn1957sedimentary} (e.g., the structure shown in Fig.\ \ref{sedimentation_in_nature}), the separation of particles with different sizes or densities \cite{selim1983sedimentation}, and water treatment \cite{guo2009removal}. Driven by its importance, sedimentation has been extensively investigated and considerable progress has been made, however it is still far from being well understood \cite{batchelor1972sedimentation,caflisch1985variance,koch1989instability,herzhaft1999experimental,brenner1999screening,davis1985sedimentation,
ramaswamy2001issues,guazzelli2011fluctuations,piazza2014settled}. The difficulty in understanding sedimentation stems from the long-range, many-body hydrodynamic nature of particle interactions as mentioned in Section\ \ref{patterns_sedimentation}, which usually leads to problems that are difficult to solve mathematically.

Despite the complexity of sedimentation, some well-known results have been successfully proposed and experimentally verified. For instance, the settling velocity of a cloud of particles will always be faster than that of a single particle, which is due to cooperative effect of hydrodynamic interactions between particles \cite{herzhaft1996experimental,guazzelli2011physical}. One of the vital conflicting problems in sedimentation is particle density and velocity fluctuations during sedimentation, which are likely related to the particle volume fraction, system size, side wall effect and inertia \cite{caflisch1985variance,xue1992diffusion,nicolai1995effect,segre1997long}. In addition, the density nonuniformity also plays a key role in the upward swimming for the microorganisms with a heavy bottom \cite{pedley2010collective,pedley1992hydrodynamic,maddock1994mechanics,pedley1990new,guasto2012fluid,pennington1990consequences,roberts1970motion,lauga2016bacterial,elgeti2015physics}. The basic principle behind this particular locomotion is that the net effect of the competing gravitational and hydrodynamic torques exerted on the microorganisms generates an upward motion while swimming \cite{guasto2012fluid}.

\begin{figure}
\begin{center}
\includegraphics[width=1\textwidth]{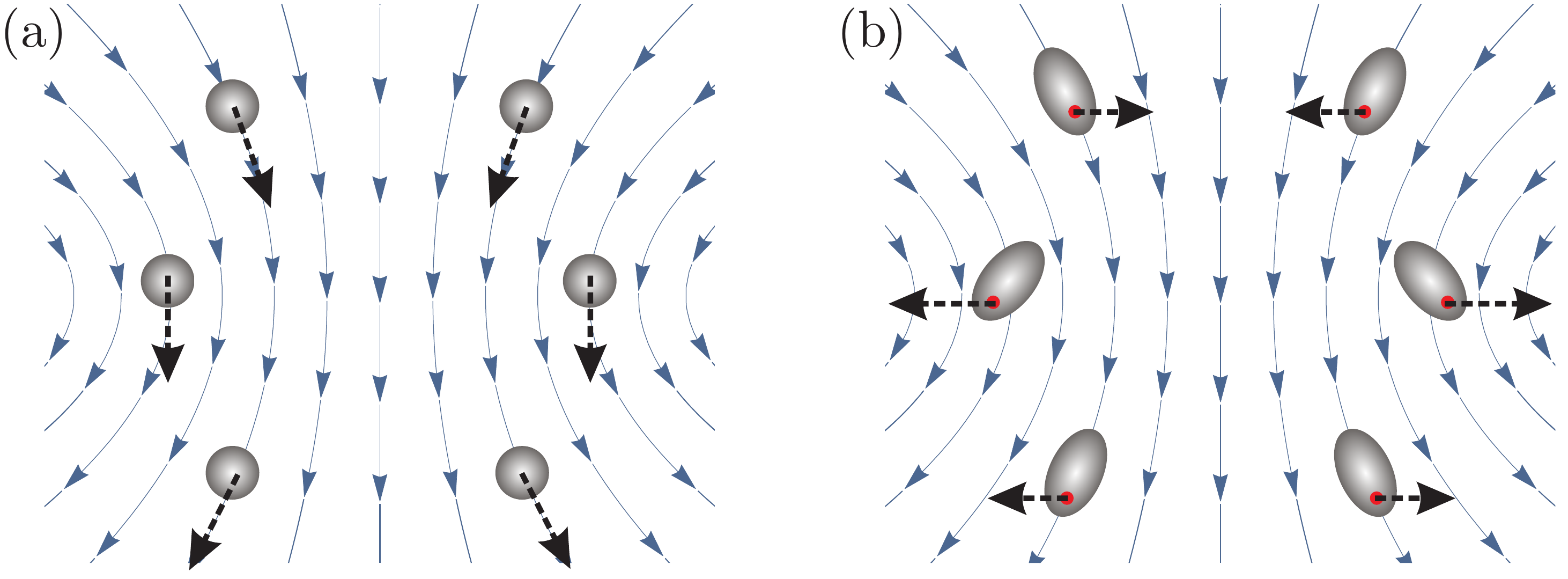}
\caption[Response to fluid disturbance for particles with different density distributions]{Response to fluid disturbance for particles with different mass distributions. The spherical particles with uniform density distributions (i.e., particle center of mass coincides with the geometry center) respond isotropically to the flow disturbance generated by the density fluctuations, leading to the cancellation of influx and outflux, and the velocity of spherical particles aligns with the fluid flow as indicated by the black arrows in (a). (b) Shows responses of prolate particles whose center of mass (red spot) is different from the geometry center. Due to the imbalance of hydrodynamic and gravitational torques, the particles respond anitropically to the flow disturbance generated by multiple particles, leading to a net lateral motion as indicated by the black arrows in (b). This figure is reproduced with permission from Goldfriend \textit{et al.,} \cite{goldfriend2017screening}, copyright (2017) by the American Physical Society.} 
\label{response_to_flow}
\end{center} 
\end{figure}

The dynamics of sedimentation are also strongly dependent on the geometry of the sedimenting objects \cite{manghi2006hydrodynamic}. For example, a pair of disk-like or semi-spherical particles exhibit periodic orbits \cite{jung2006periodic,chajwa2018kepler}, and usually non-spherical particles are subject to complicated settling dynamics due the coupling between the translational  and orientational degrees
of freedom even for a pair of particles \cite{kim1985sedimentation,goldfriend2015hydrodynamic}, and the sedimentation dynamics for many-particle systems are even more complicated as shown in Refs. \cite{butler2002dynamic,herzhaft1996experimental}. In what follows, I will show that symmetric rod-like particles with uniform density align with the fluid flow during sedimentation. Generally, for a rigid object moving in a fluid, the velocity $\bm{U}$ at position $\bm{r}$ can be written as: 
\begin{equation}
\bm{U=U^{s}+\omega^s\times r},
\label{sedi_velocity}
\end{equation}
where $\bm{U^s}$ is the translation velocity and $\bm{\omega^s}$ is angular velocity. 

For a particle sedimenting in a fluid, the fluid stress field will generate a hydrodynamic force $\bm{F}$ and a hydrodynamic torque $\bm{T}$:
\begin{gather}
\label{hydrodynamic_force}
\bm{F}=\int_{Sp} \bm{\sigma \cdot n} dS,\\
\label{hydrodynamic_torque}
\bm{T}=\int_{Sp} \bm{r \times \sigma \cdot n} dS,
\end{gather}  
where $\bm{\sigma}$ is the fluid stress tensor, and can be expressed as:
\begin{equation}
\sigma_{ij}=-p\delta_{ij}+\tau_{viscous}=-p\delta_{ij}+2\eta\varepsilon_{ij},
\label{stress_tensor}
\end{equation}
where $\eta$ is the fluid dynamic viscosity, $\tau_{viscous}$ is the viscous stress tensor, $p$ is the absolute pressure which is defined as $p=(-1/3)\sigma_{ii}$, $\delta_{ij}$ is the Kronecker delta function such that $\delta_{ij}=1$ if $i=j$, otherwise $\delta_{ij}=0$, $\varepsilon_{ij}$ is the strain rate tensor and can be expressed in a matrix form in a coordinate system with the axes of 1, 2 and 3 as:
\begin{equation}
\varepsilon=\begin{pmatrix}
 \varepsilon_{11} & \varepsilon_{12} & \varepsilon_{13}\\ 
 \varepsilon_{21} & \varepsilon_{22} & \varepsilon_{23}\\ 
 \varepsilon_{31} & \varepsilon_{32} & \varepsilon_{33}
 \end{pmatrix}.
 \end{equation}
Since $\varepsilon_{ij}$ is generally symmetric, there is $\varepsilon_{ij}=\varepsilon_{ji}$. For any material undergoing displacement, the strain rate tensor is related to velocity gradient: 
\begin{equation}
\varepsilon_{ij}=\varepsilon_{ji}=\frac{1}{2}\left( \frac{\partial u_i}{\partial x_j}+\frac{\partial u_j}{\partial x_i} \right),
\label{strain_tensor}
\end{equation}
where $u$ is the velocity and $x$ is the direction.

In Eqs.\ \ref{hydrodynamic_force} and \ref{hydrodynamic_torque}, $\bm{n}$ is the outward unit vector normal to the particle surface, and $S_p$ denotes particle surface. In particular, for a spherical particle with radius $a$ in a uniform flow stream $\bm{U^\infty}$, there is $F=6\pi\eta a U^\infty$ (see Eq.\ \ref{force_balance}), which is Stokes drag force \cite{stokes1851effect}, and for the same particle in a rotational flow with angular velocity $\bm{\omega^\infty \times r}$ there is $T=8\pi \eta a^3\omega^\infty$.

\begin{figure}
\begin{center}
\includegraphics[width=2.5in]{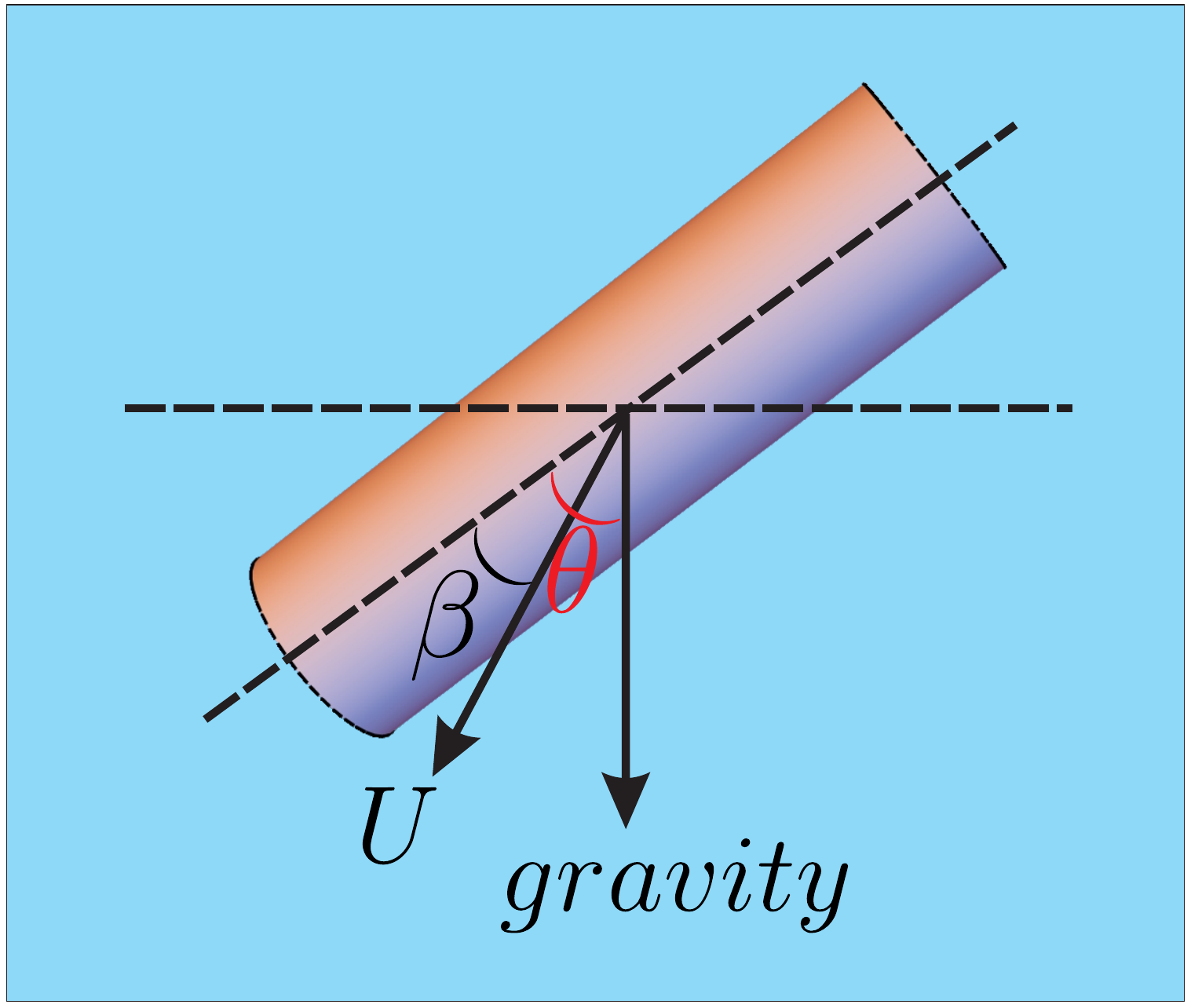}
\caption[Sketch of the sedimentation of a rod-like object]{Sketch of the sedimentation of a rod-like object.} 
\label{rod_sedimentation}
\end{center} 
\end{figure}

Assuming a no-slip boundary condition, the relationship between the object velocity, hydrodynamic force $\bm{F}$ and torque $\bm{T}$ exerted on the object can be expressed in a matrix form \cite{manghi2006hydrodynamic,lauga2009hydrodynamics}: 

\begin{equation}
\boldmath
\begin{pmatrix}
 U^s  \\ 
\omega ^s
 \end{pmatrix}
=
\begin{pmatrix}
\mu_{t} & \mu_{tr} \\ 
\mu_{rt} &\mu_{r}
 \end{pmatrix}
\begin{pmatrix}
 F  \\ 
T
 \end{pmatrix},
 \end{equation}
where the tensors $\bm{\mu_t}$ and $\bm{\mu_r}$ are the translation and rotation mobility tensors, respectively, and $\bm{\mu_{tr}}$ and $\bm{\mu_{rt}}$ are the coupling tensors between the translation and rotation motions. By dimensional analysis we can easily arrive $[\bm{\mu_t}]\sim (\eta L)^{-1}$, $[\bm{\mu_r}]\sim (\eta L^3)^{-1}$, and $[\bm{\mu_{tr}}]\sim (\eta L^2)^{-1}$, where $L$ is the characteristic length scale and $\eta$ is fluid dynamic viscosity. More specifically, for symmetric objects the mobility tensors can be diagonalized, i.e., $\bm{\mu_{tr}=\mu_{rt}=0}$, thus for symmetric rod-like objects we have \cite{guazzelli2011physical}:
\begin{equation}
\bm{\mu_t}=\left( \eta L\right)^{-1}
\begin{pmatrix}
 \mu_{\perp} & 0 & 0\\ 
 0 & \mu_{\perp} & 0\\ 
 0 & 0 & \mu_{\parallel}
 \end{pmatrix},
 \end{equation}
where the dimensionless quantities $ \mu_{\perp}$ and $\mu_{\parallel}$ represent perpendicular and parallel mobility constants, respectively, and the values of  $ \mu_{\perp}$ and $\mu_{\parallel}$ depend on the geometry of the particle, and for long rod-like object they obey $ \mu_{\parallel} \simeq 2\mu_{\perp}$, meaning that an rod-like object with a horizontal orientation moves about twice as fast as an identical object with a vertical orientation. Therefore, the angle $\beta$ between the symmetric axis of the rod-like object and the object velocity can be expressed as \cite{guyon2001physical}:
\begin{equation}
\tan \beta=\frac{U_{\perp}}{U_{\parallel}}=\frac{\mu_{\perp}mg\cos\theta}{\mu_{\parallel}mg\sin\theta}\simeq\frac{\tan\theta}{2},
\label{orientation_angle}
 \end{equation}
where $\theta$ is the angle between the symmetric axis of the object and the vertical direction as schematically shown in Fig.\ \ref{rod_sedimentation}. This means for a rod-like object sedimenting in a viscous fluid with the velocity of $U_{stokes}$ will always remain initial orientation, i.e., $\theta$ is constant, which is consistent with the predictions in Refs. \cite{oberbeck1876uber,jeffery1922motion}

Recently a theoretical study by Goldfriend \textit{et al.} \cite{goldfriend2017screening} showed that particle density distributions can strongly influence the velocity and density fluctuations, and in some certain regimes particles with nonuniform density distribution are expected to suppress velocity and density fluctuations. Figure\ \ref{response_to_flow} sketches that the coupling between the density distribution and the translational and rotational motion of the settling particles can lead to distinct sedimentation dynamics. Due to spherical symmetry, the spherical particles with uniform density isotropically respond to the flow disturbance generated by multiple particles, leading to the mutual cancellation of influx and outflux of the particles, and the velocity of spherical particles align with the fluid flow as indicated by the black arrows in Fig.\ \ref{response_to_flow}a. In contrast, the prolate particles with the center of mass (red spots in Fig.\ \ref{response_to_flow}b) different from the geometry center will respond anistropically to the flow disturbance due to the competition between the hydrodynamic and gravitational torques. Consequently, this will create a net outflux, i.e., lateral motion, of particles as illustrated in Fig.\ \ref{response_to_flow}b.

This theoretical analysis is based on linear stability analysis, in which the authors assumed a velocity fluctuations along the horizontal direction, and this results in an imbalance between hydrodynamic and gravitational torques. For spherical particles as shown in Fig.\ \ref{response_to_flow}a, this torque imbalance simply contributes to the rotational motion of the particles. However, surprisingly the prolate particles with nonuniform density distributions as shown in Fig.\ \ref{response_to_flow}b are predicted to undergo a lateral motion driven by the anistropic response to the flow disturbance, which strongly depends on the orientation and geometry of the particles. The detailed derivation for differentiated responses to the flow disturbance of the particles with different density distributions and geometries is shown in Ref. \cite{goldfriend2017screening}. It should noticed that the linear stability analysis in Ref. \cite{goldfriend2017screening} is only restricted to fluctuations in the horizontal direction, while in what follows we show experimentally that the mechanisms shown in Fig.\ \ref{response_to_flow} also apply when the particles are moving in three dimensions.

As stated above, it is of vital significance to understand and control sedimentation both for the sake of fundamental curiosity (e.g., many-body physics) and practical applications (e.g., manipulation of particulate films with uniform density and homogeneities, understanding large-scale geologic sedimentation pattern). The theoretical ideas by varying the particle density distribution as shown in Fig.\ \ref{response_to_flow} \cite{goldfriend2017screening} is a one way to realize that. For this purpose, we experimentally investigated the settling dynamics of non-Brownian particles with different density distributions at low-Reynolds numbers ($10^{-3}$) in quasi-two-dimensional (quasi-2D) and three-dimensional (3D) space. The particles are composed of equal-sized aluminum and steel balls glued together in various configurations. In the experiment, we observed that particles with uniform and nonuniform density distributions respond differently during sedimentation. A single doublet made of equal-sized aluminum and steel balls aligns with gravity, in contrast to a doublet made of two aluminum balls, which aligns with the fluid flow. For a pair of identical particles made of equal-sized aluminum and steel balls, we observed an effective repulsion between particles during sedimentation in both two-body and many-body systems, however, the particles made of identical balls show variable settling dynamics. In addition, for three or more pairs of particles, more complex dynamics were observed that are likely due to the many-body hydrodynamic nature, yet repulsion between the particles with nonuniform densities still played a key role. It is expected that density-distribution-dependent sedimentation can provide a novel tunable route towards the control of uniformity of particulate films after sedimentation, although more work is needed to fully characterize the particle dynamics and interactions.

\section{Experimental setup}
\label{experimental_setting_sedimentation}
\begin{figure}
\begin{center}
\includegraphics[width=3in]{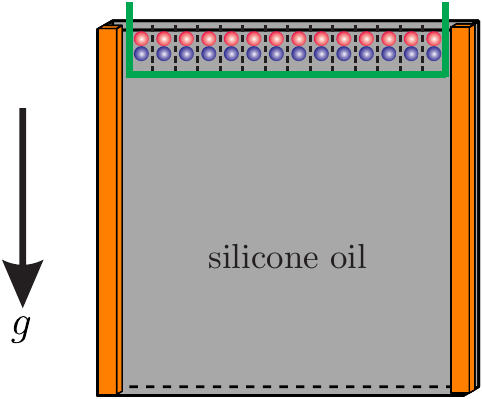}
\caption[Sketch of quasi-2D chamber for sedimentation]{Sketch of the quasi-2D chamber for sedimentation. The prepared particles are placed in the holes located on the top of the chamber and the holes have the same diameter which is slightly larger than the diameter of Al or St balls as indicated by the dashed black lines. The gate (green lines) is then manually opened to let the particles sediment. The dimension of the 2D chamber is approximately 15 cm (length) $\times$ 19 cm (height) $\times$ 0.3 cm (thickness).} 
\label{2D_chamber_sedimentation}
\end{center} 
\end{figure}

\begin{figure}
\begin{center}
\includegraphics[width=4 in]{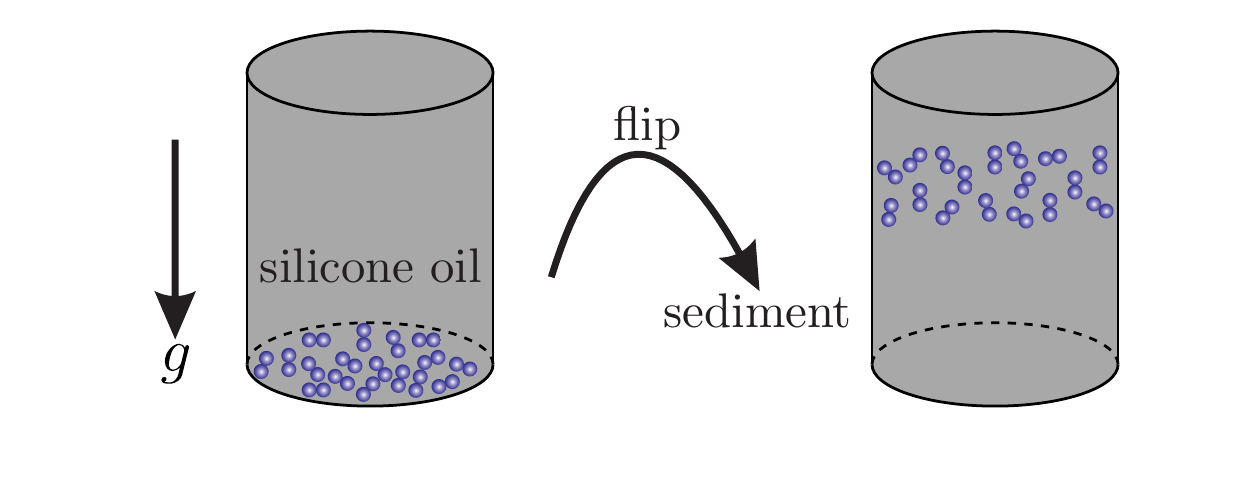}
\caption[Sketch of 3D chamber for sedimentation]{Sketch of the 3D chamber for sedimentation. The particles were initially placed inside the 3D chamber, and silicone oil was poured into the chamber which was sealed later to make sure no bubbles exist in the chamber. Then the chamber was flipped to let particles start settling under gravity. The area fraction of particle put inside the chamber is $\approx0.3$. The dimension of the 3D chamber is approximately 13 cm (diameter) $\times$ 23 cm (height).} 
\label{3D_container}
\end{center} 
\end{figure}

In the experiment, we used equal-sized (diameter $d\approx$ 2 mm) aluminum (Al) and steel (St) balls with densities of $8\times10^3$ kg/m$^3$ and $2.8\times10^3$ kg/m$^3$, respectively. In order to prepare particles with various density distributions, we glued three different types of particle doublets, namely Al-St, Al-Al, and St-St. We used silicone oils with viscosity of 1000 cSt and 10000 cSt, respectively as fluid for sedimentation. Transparent acrylic sheets and blocks were used to construct quasi-two-dimensional (quasi-2D) and three dimensional (3D) chambers as shown in Figs.\ \ref{2D_chamber_sedimentation} and \ref{3D_container}, respectively. For sedimentation in quasi-2D chambers, a USB 3.0 digital video camera (Point Grey) connected to a macro lens was used to image the sedimentation behavior from the front, whereas for the sedimentation in 3D chambers, the camera was held from the top to image the sedimentation pattern of particles. Some recorded images were enhanced using NIH ImageJ software for better visualization. The software Trackpy (an open source Python tracking software) was used to track down the trajectories of particles for the experiment in 2D chambers, and the particle diameter ($\approx 2$ mm) corresponds to 24 pixels in the images.

Given the size of the particles and the viscosity of silicone oil, the Reynolds number as defined in Chapter\ \ref{Leidenfrost_stars} is calculated to be $10^{-3}<Re<10^{-2}$, suggesting that inertial effects can be ignored. The P\'{e}clet number is defined as $Pe=\frac{a U}{D}$, which characterizes the relative importance of advective effects to Brownian motion where $U$ is the characteristic velocity (i.e., $U_{stokes}$), $a$ is the particle radius, and $D$ is the diffusion coefficient given by $D=\frac{k_B T}{6\pi\eta a}$, where $k_B$ is the Boltzmann's constant, and $\eta$ as fluid dynamic viscosity. Since the particles used in the experiment are milimeter-scale, we can estimate $Pe\sim 10^{12}$, suggesting that the particle motion was purely determined by hydrodynamics. 

\section{Results and discussion}

\subsection{Effect of density distribution on sedimentation}

We first test the settling dynamics of single particles with uniform and nonuniform density distributions in a 3D chamber. We used single particles of Al-Al and Al-St in 3D chambers for sedimentation, and then tracked the settling dynamics of the particles. 
\begin{figure}
\begin{center}
\includegraphics[width=4.5 in]{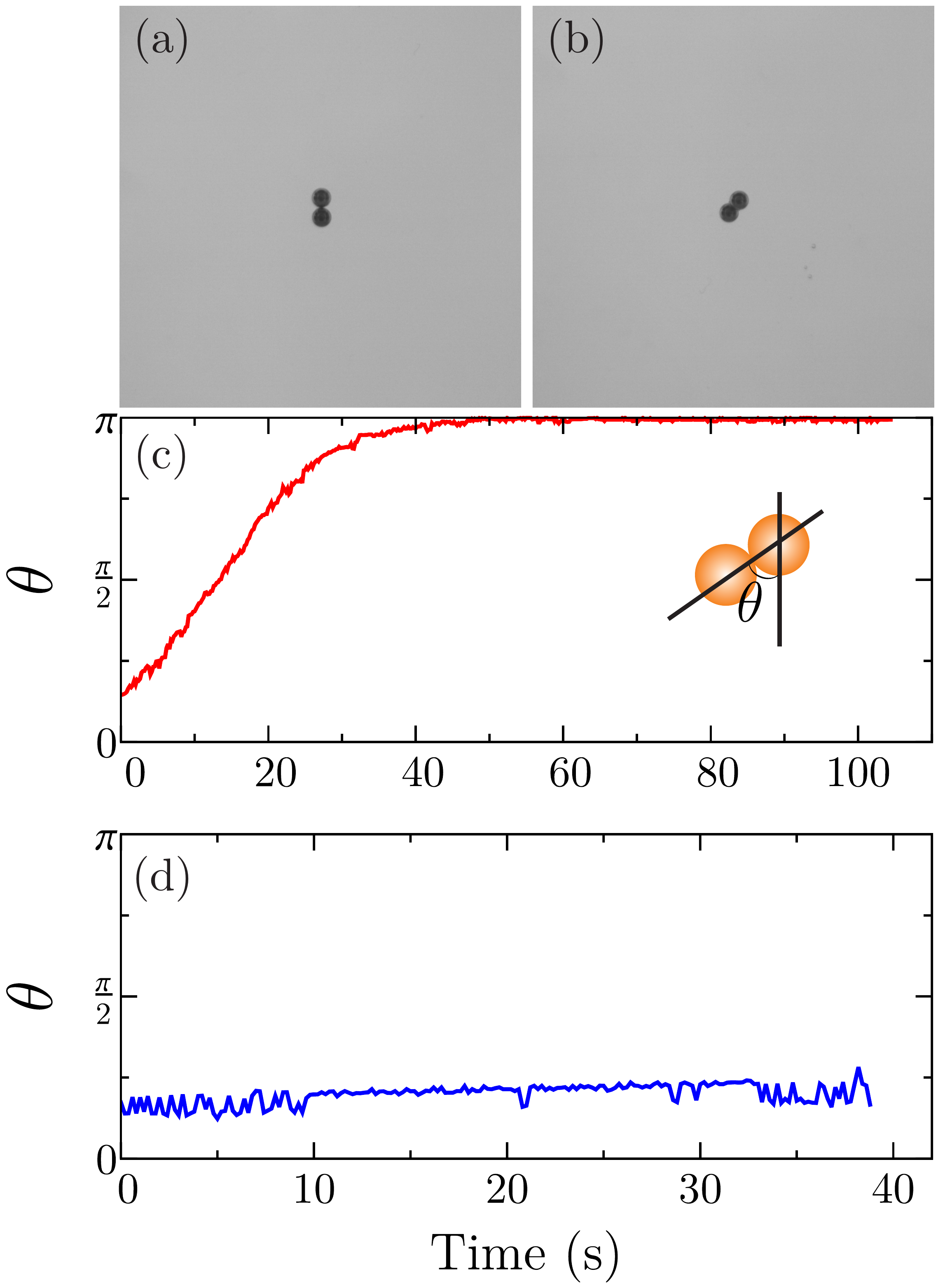}
\caption[Sedimentation of particles with different density distributions]{Sedimentation of particles with different density distributions. (a) and (b) show the images of a single Al-St and Al-Al particles during sedimentation in silicone oil with viscosity of 10000 cSt and 1000 cSt, respectively. (c) and (d) show the variation of angle ($\theta$) between the geometry center axis and the vertical direction with respect to time during the sedimentation of the particles in (a) and (b), respectively. The inset of (c) shows the definition of $\theta$.} 
\label{sedimentation_different_mass_distribution}
\end{center} 
\end{figure}

Figures\ \ref{sedimentation_different_mass_distribution}a and \ref{sedimentation_different_mass_distribution}b show the images of a single Al-St and Al-Al particle at a certain moment during sedimentation, respectively. In the sedimentation of a single Al-St particle, the heavier part was initially pointed upwards and lighter part downwards. After releasing the particle, the heavier part will induce a torque in order to lower the center of mass, leading the particle to flip. Afterwards, the Al-St particle will align with the external force, i.e., gravity. This process is confirmed by characterizing the variations of angle ($\theta$) between the vertical direction and the symmetric axis with respect to time as shown in Fig.\ \ref{sedimentation_different_mass_distribution}c. In contrast to the settling dynamics of Al-St particle, during sedimentation the single Al-Al particle does not align with gravity but rather aligns with the fluid flow. This means that the angle $\theta$ will remain the same as the initial angle during sedimentation as indicated by Eq.\ \ref{orientation_angle}. The variation of $\theta$ with time is plotted in Fig.\ \ref{sedimentation_different_mass_distribution}d, in which fluctuations in $\theta$ is likely due to noise and anomalous disturbances in the fluid, however, the average value of $\theta$ almost remains constant ($\approx\pi$/8), and generally this angle depends on the initial conditions of the sedimentation. Note that the trajectories shown in Figs.\ \ref{sedimentation_different_mass_distribution}c and \ref{sedimentation_different_mass_distribution}d represent the variation of the angle between the geometry center axis of the doublet and the vertical direction as indicated in the inset of Fig.\ \ref{sedimentation_different_mass_distribution}c.

\subsection{Sedimentation of multiple particles in quasi two dimensions}
\begin{figure}
\begin{center}
\includegraphics[width=4.5 in]{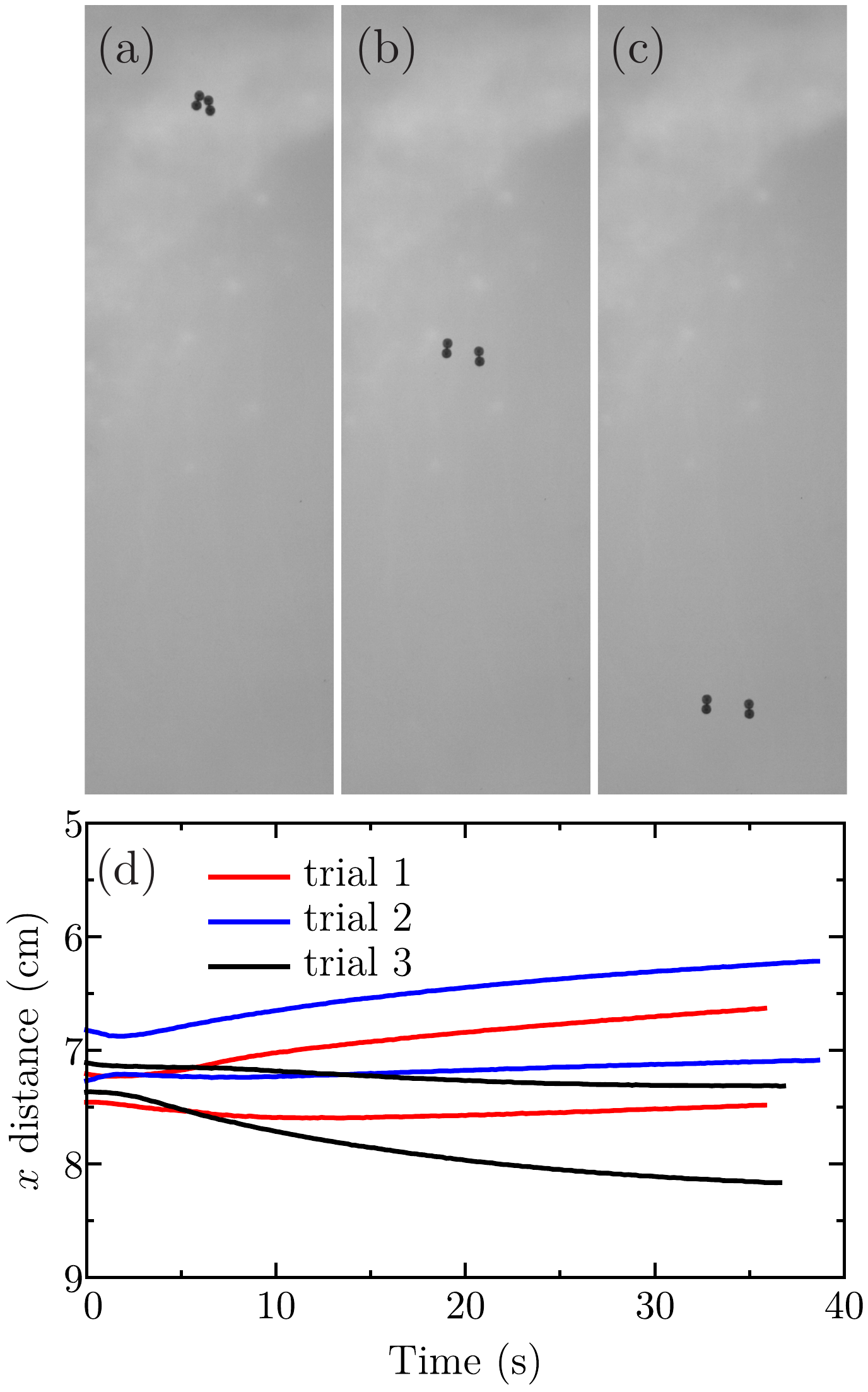}
\caption[Sedimentation of two Al-St particles in a quasi-2D chamber]{Sedimentation of two Al-St particles in a quasi-2D chamber. (a), (b) and (c) show images of two Al-St particles during sedimentation at different times. (d) shows the trajectories of the particles during sedimentation for three different trials.} 
\label{repulsion_nonuniform}
\end{center} 
\end{figure}
\begin{figure}
\begin{center}
\includegraphics[width=4.5 in]{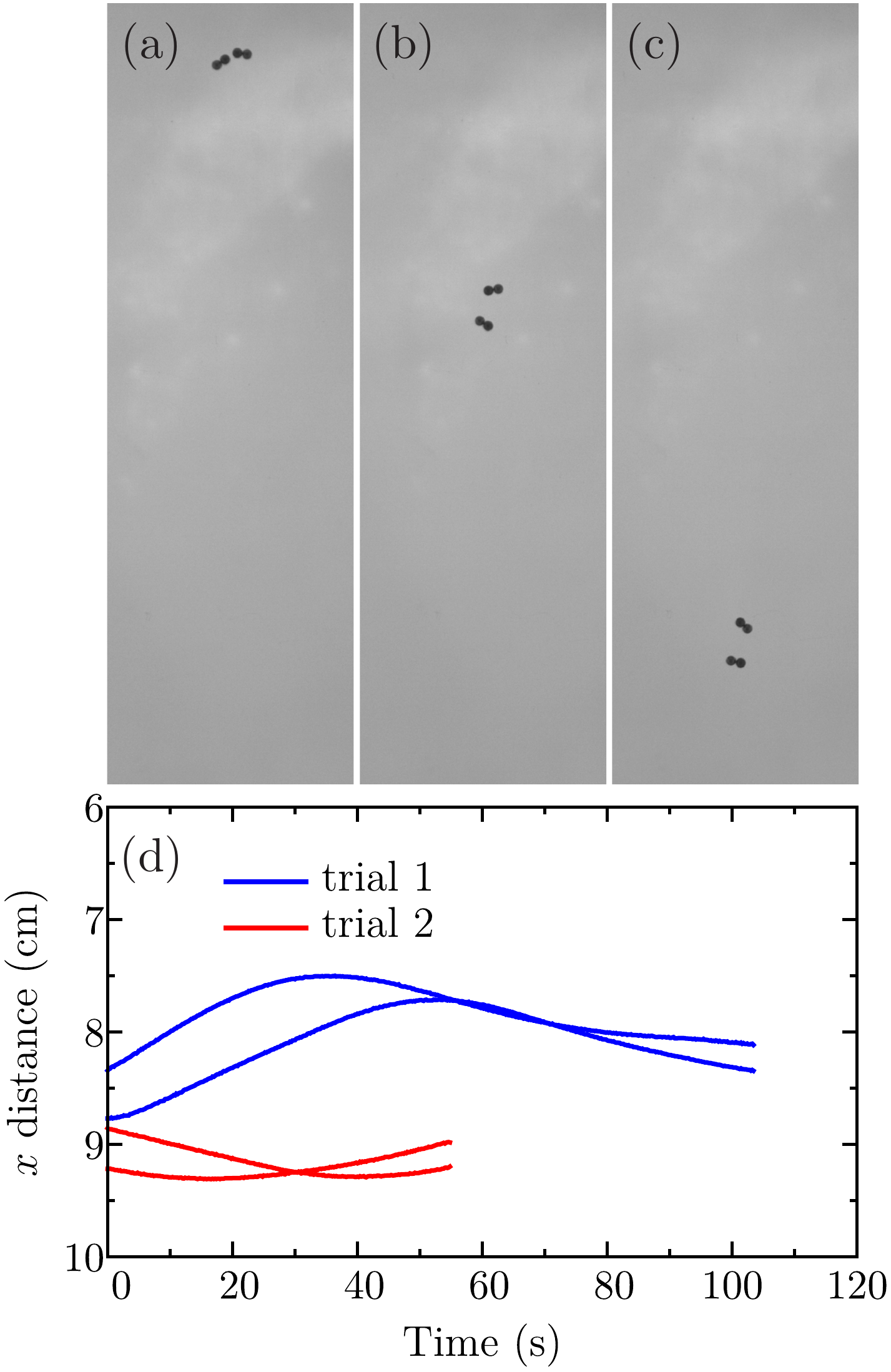}
\caption[Sedimentation of two Al-Al particles in a quasi-2D chamber]{Sedimentation of two Al-Al particles in a quasi-2D chamber. (a), (b) and (c) show images of two Al-Al particles during sedimentation at different times. (d) shows the trajectories of the particles during sedimentation for two different trials.} 
\label{sedimentation_Al-Al}
\end{center} 
\end{figure}
\begin{figure}[!]
\begin{center}
\includegraphics[width=1\textwidth]{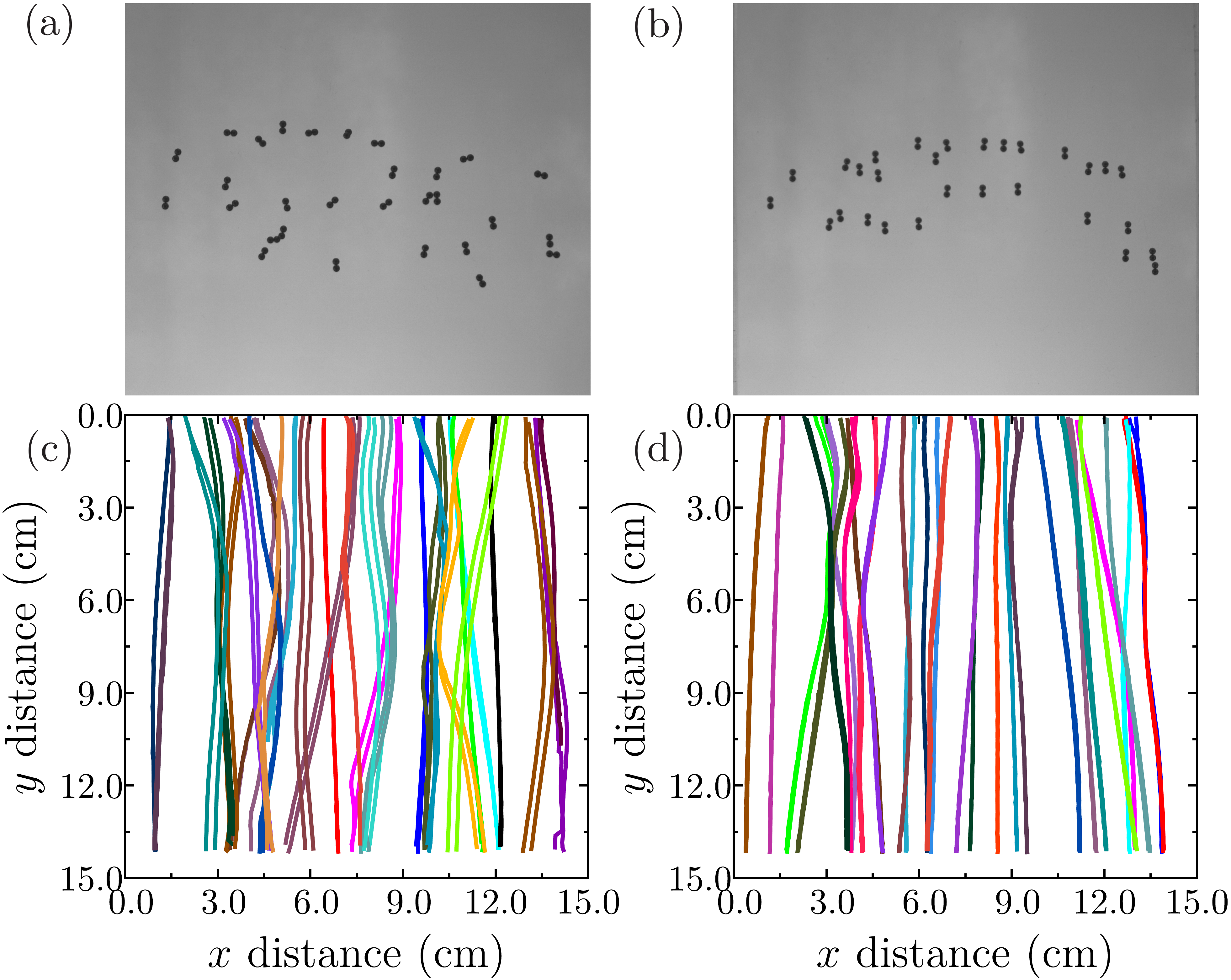}
\caption[Sedimentation of many particles in a quasi-2D chamber]{Sedimentation of many particles in a quasi-2D chamber. (a) and (b) show the images of many St-St particles and Al-St particles at a certain moment during sedimentation, respectively. (c) and (d) show the trajectories of the particles in (a) and (b) during sedimentation, respectively. In (c) and (d) each particle has two trajectories with the same color, corresponding to the two balls that compose the doublet, the trajectories of different particles are plotted in different colors.} 
\label{sedimentation_many_pairs}
\end{center} 
\end{figure}

We also systematically investigated the sedimentation of multiple particles with different density distributions in quasi-2D chambers. Figures\ \ref{repulsion_nonuniform}a to \ref{repulsion_nonuniform}c show the images of two identical Al-St particles during sedimentation at different times, which clearly indicate that the separation between the particles increases as sedimentation proceeds. This trend can be confirmed by tracking the trajectories of the particles during sedimentation, and it holds true for different trials as indicated in Fig.\ \ref{repulsion_nonuniform}d. Note that the trajectories with the same color in Fig.\ \ref{repulsion_nonuniform}c two doublet particles in the same trial. In addition, the increase of separation between particles suggests that there is an effective repulsion between particles, which agrees with the mechanism shown in Fig.\ \ref{response_to_flow}b.

For comparison, we also investigated the sedimentation of two identical Al-Al particles. Figures\ \ref{sedimentation_Al-Al}a to \ref{sedimentation_Al-Al}c show the images of two identical Al-Al particles during sedimentation at different times, in which it is clearly seen that the orientations of each particle and separation between the two particles are more irregular and complicated than that of Al-St particles shown in Fig.\ \ref{repulsion_nonuniform}. Figure\ \ref{sedimentation_Al-Al}c shows the trajectories of the particles for two different trials, we can see that the particles' trajectories often overlap though they pass the overlapping point at different times. In addition, we also occasionally observed flipping of Al-St particles during sedimentation of two such particles. Although there are only two bodies, the complicated dynamics of the sedimentation of two Al-Al particles are likely due to the non-spherical geometry. 

\begin{figure}
\begin{center}
\includegraphics[width=1\textwidth]{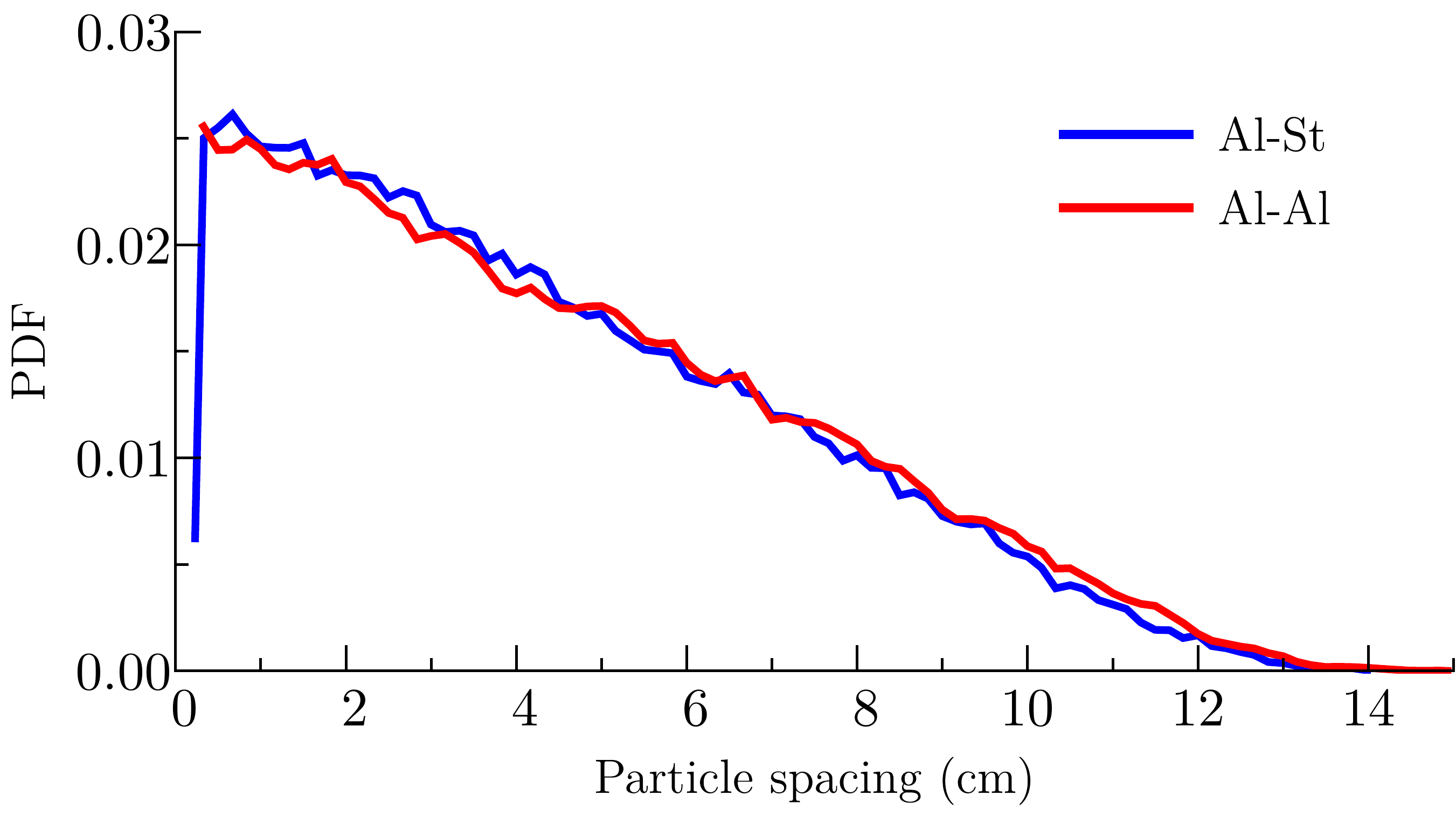}
\caption[Probability density function (PDF) of particle spacing for sedimentation of many particles in a quasi-2D chamber.]{Probability density function (PDF) of particle spacing for sedimentation of many Al-Al particles and Al-St particles in a quasi-2D chamber.} 
\label{statistics_sedimentation_many_pairs}
\end{center} 
\end{figure}

Next we investigated the settling dynamics of many particles with different density distributions in quasi-2D chambers. Figures\ \ref{sedimentation_many_pairs}a and \ref{sedimentation_many_pairs}b show the images of 29 St-St and Al-St at given times during sedimentation in a quasi-2D chamber, respectively. It is apparent that both the spacing and orientation of the St-St particles are more irregular than that of Al-St particles. This property can also be confirmed by Figs.\ \ref{sedimentation_many_pairs}c and \ref{sedimentation_many_pairs}d, which show the sedimentation trajectories of particles in Figs.\ \ref{sedimentation_many_pairs}a and \ref{sedimentation_many_pairs}b, respectively.

In order to quantity the interactions between particles during sedimentation as shown in Figs.\ \ref{sedimentation_many_pairs}a and \ref{sedimentation_many_pairs}b, we investigated the statistics of the particle spacing. Figure\ \ref{statistics_sedimentation_many_pairs} shows the probability density function (PDF) of particle spacing for the sedimentation shown in Figs.\ \ref{sedimentation_many_pairs}a and \ref{sedimentation_many_pairs}b. The distributions shown in Fig.\ \ref{statistics_sedimentation_many_pairs} are essentially the pair correlation function in one dimension. It is also clear that the distribution decays to zero at the particle spacing of $\approx$ 14 cm, which is simply due to the finite size of the experimental system. In other words, if the system size was infinite, the distribution will eventually saturate to a certain non-zero value, typically normalized to unity.

As shown in Fig.\ \ref{statistics_sedimentation_many_pairs}, the main difference between Al-Al particles and Al-St particles is that at a given small particle spacing ($\approx$ 0.4 cm), the probability of finding another St-St particle (blue curve) is higher than that of Al-St particle (nearly zero, red curve). The distinct behaviors essentially suggest that there is an effective repulsion between Al-St particles during sedimentation, which is absent in St-St particles. Therefore, the results shown in Fig.\ \ref{statistics_sedimentation_many_pairs} validate the mechanisms shown in Fig.\ \ref{response_to_flow} despite the many-body nature of the system. It should also be noted that the statistics shown in Fig.\ \ref{statistics_sedimentation_many_pairs} are based on the average of five trials for both particle configurations.

\subsection{Sedimentation of multiple particles in three dimensions}
\begin{figure}
\begin{center}
\includegraphics[width=1\textwidth]{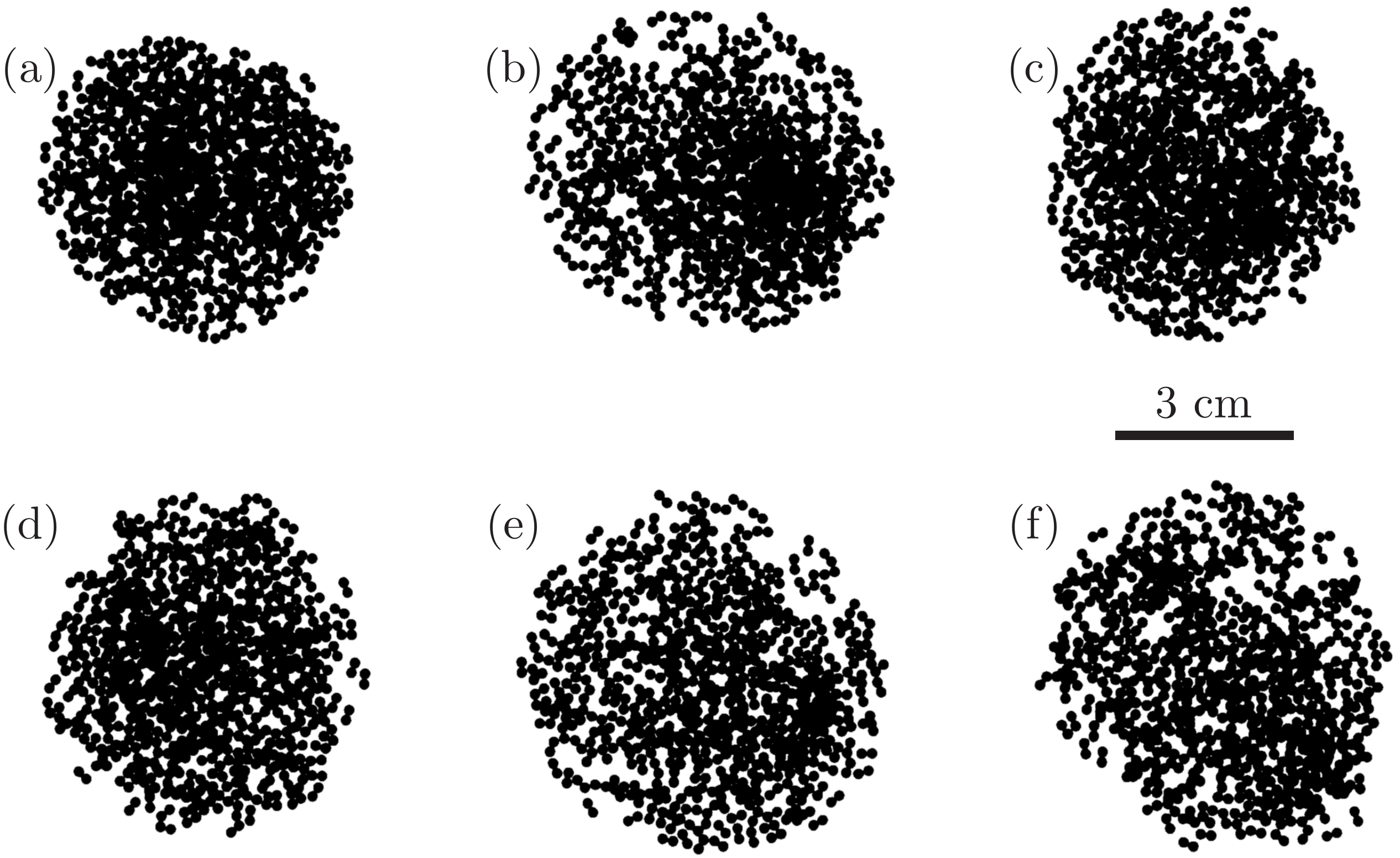}
\caption[Images of the sedimentation pattern of Al-Al particles in 3D]{Images of the sedimentation pattern of Al-Al particles in 3D. The images are enhanced for better visualization, and the scale bar applies to all images. In the images the particle diameter ($\approx 2$ mm) corresponds to 20 pixels.} 
\label{sedimentation_Al_Al_3D}
\end{center} 
\end{figure}
\begin{figure}
\begin{center}
\includegraphics[width=1\textwidth]{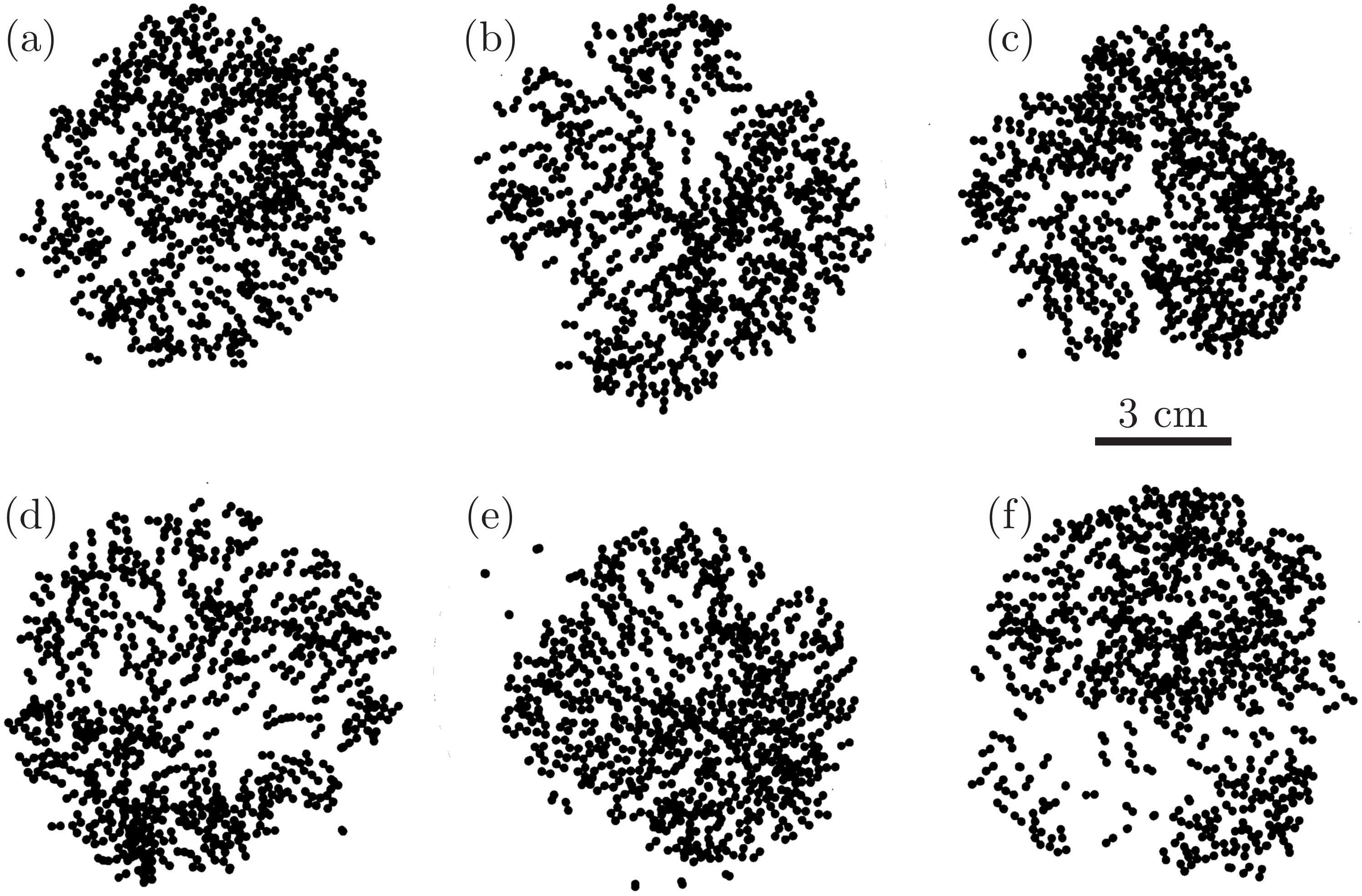}
\caption[Images of the sedimentation pattern of Al-St particles in 3D]{Images of the sedimentation pattern of Al-St particles in 3D. The images are enhanced for better visualization, and the scale bar applies to all images. In the images the particle diameter ($\approx 2$ mm) corresponds to 20 pixels. Note that the number of particles in each image here is the same as that shown in Fig.\ \ref{sedimentation_Al_Al_3D}.}
\label{sedimentation_Al_steel_3D}
\end{center} 
\end{figure}

Given the distinct sedimentation statistics for Al-Al and Al-St particles in quasi-2D chambers as shown in Fig.\ \ref{statistics_sedimentation_many_pairs}, we extended our experiment to three dimensions as illustrated in Fig.\ \ref{3D_container}, which is more similar to real-world situations. In the experiment, we prepared the same number of pairs of Al-Al and Al-St particles, and deposited the particles into two separate 3D chambers (see Section\ \ref{experimental_setting_sedimentation} for experimental details), then we investigated particle spatial distributions on the bottom of the chamber after sedimentation.. 

\begin{figure}
\begin{center}
\includegraphics[width=1\textwidth]{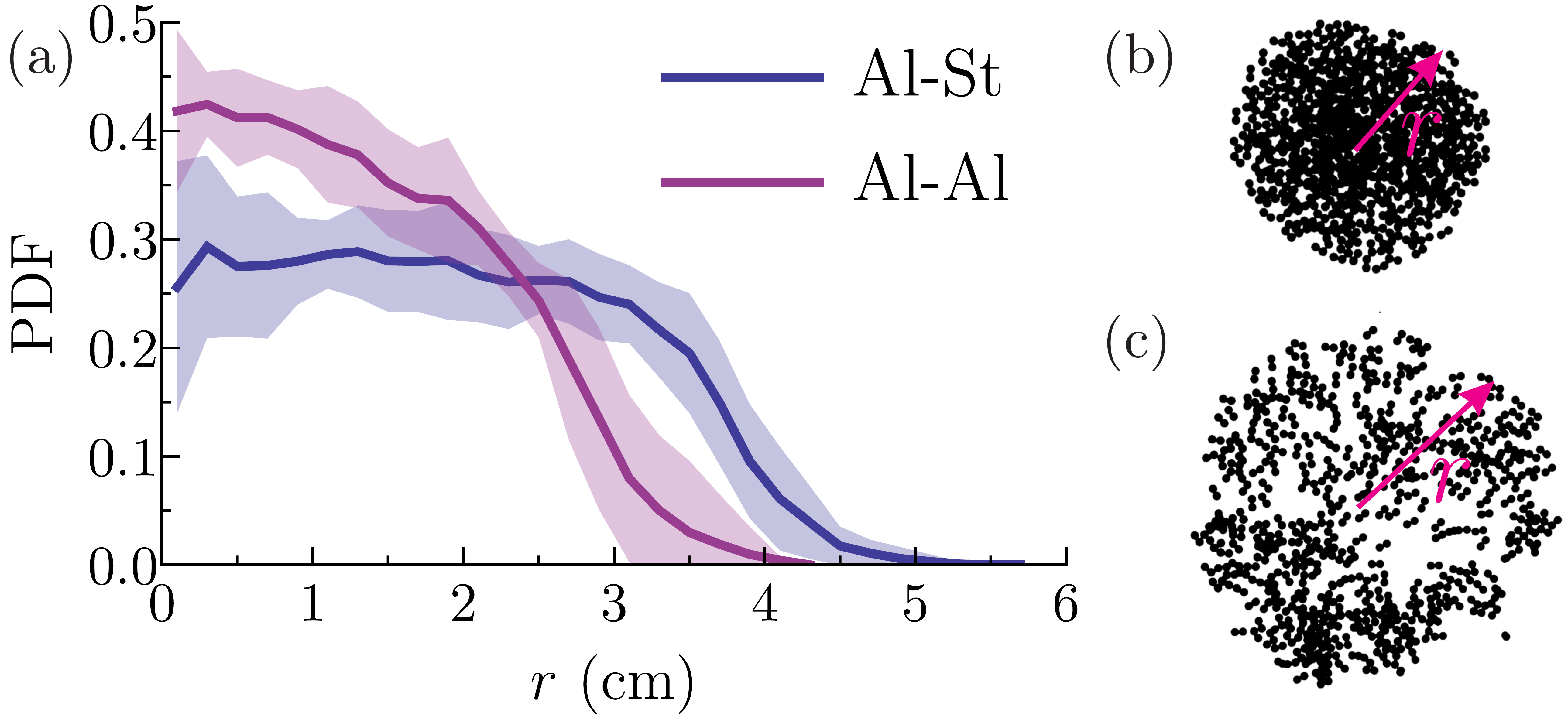}
\caption[Probability density function (PDF) of particle clusters after sedimentation in 3D]{(a) Probability density function (PDF) of particle cluster size in the radial direction for the sedimentation pattern of particles in 3D, the shaded part for each curve represents the standard deviations for multiple images. (b) and (c) show the typical images of particle cluster of Al-Al and Al-St particles in 3D, respectively.} 
\label{histogram_sedimentation_3D}
\end{center} 
\end{figure}

For the sedimentation of multiple particles in 3D, clusters of particle obstruct the view and make tracking all particle impossible. Therefore, we took images of the sedimentation pattern on the bottom of the chamber. We performed more than 100 trials for the sedimentation of Al-Al and At-St particles, respectively, and  some typical images are shown in Figures\ \ref{sedimentation_Al_Al_3D} and \ref{sedimentation_Al_steel_3D}. By simple comparison, we can conclude that the average density of Al-St particles is smaller than that of Al-Al particles.

In order to quantify the average density of the particle layer after sedimentation, we plot the probability density function (PDF) of the particle position in the radial direction, as shown in Fig.\ \ref{histogram_sedimentation_3D}a, in which integral of the distribution is normalized to unity. It is apparent in Fig.\ \ref{histogram_sedimentation_3D}a that the distribution curve for Al-St particles is more stretched than that of Al-Al particles, meaning that the cluster size of Al-St particles is larger than that of Al-Al particle clusters, which can also be confirmed by Figs.\ \ref{histogram_sedimentation_3D}b and \ref{histogram_sedimentation_3D}c. It should be noted that each distribution curve shown in Fig.\ \ref{histogram_sedimentation_3D}a is based on 10 separate images, and the shaded area for each curve represents the standard deviation for 10 images, and the distinction between the distribution curves of Al-St and Al-Al particle clusters is pretty obvious. The larger cluster size of Al-St particles is due to the hydrodynamic repulsion interaction between the particles, which agrees with the theoretical prediction shown in Fig.\ \ref{response_to_flow}, though the theory is formulated in 2D \cite{goldfriend2017screening}. The tendency of Al-Al particles to form smaller clusters suggests that the particles are unstable to density fluctuations, and consequently the cluster of Al-St particles is more uniform than that of Al-Al particles (see Figs.\ \ref{histogram_sedimentation_3D}b and \ref{histogram_sedimentation_3D}c), which agrees with experimental, theoretical and numerical studies on the sedimentation of non-spherical particles in dilute suspensions \cite{koch1989instability,herzhaft1996experimental,helzel2017kinetic,guazzelli2011fluctuations,wang2009numerical,gustavsson2009gravity,tornberg2006numerical}.

\begin{figure} [!tbph]
\begin{center}
\includegraphics[width=0.7\textwidth]{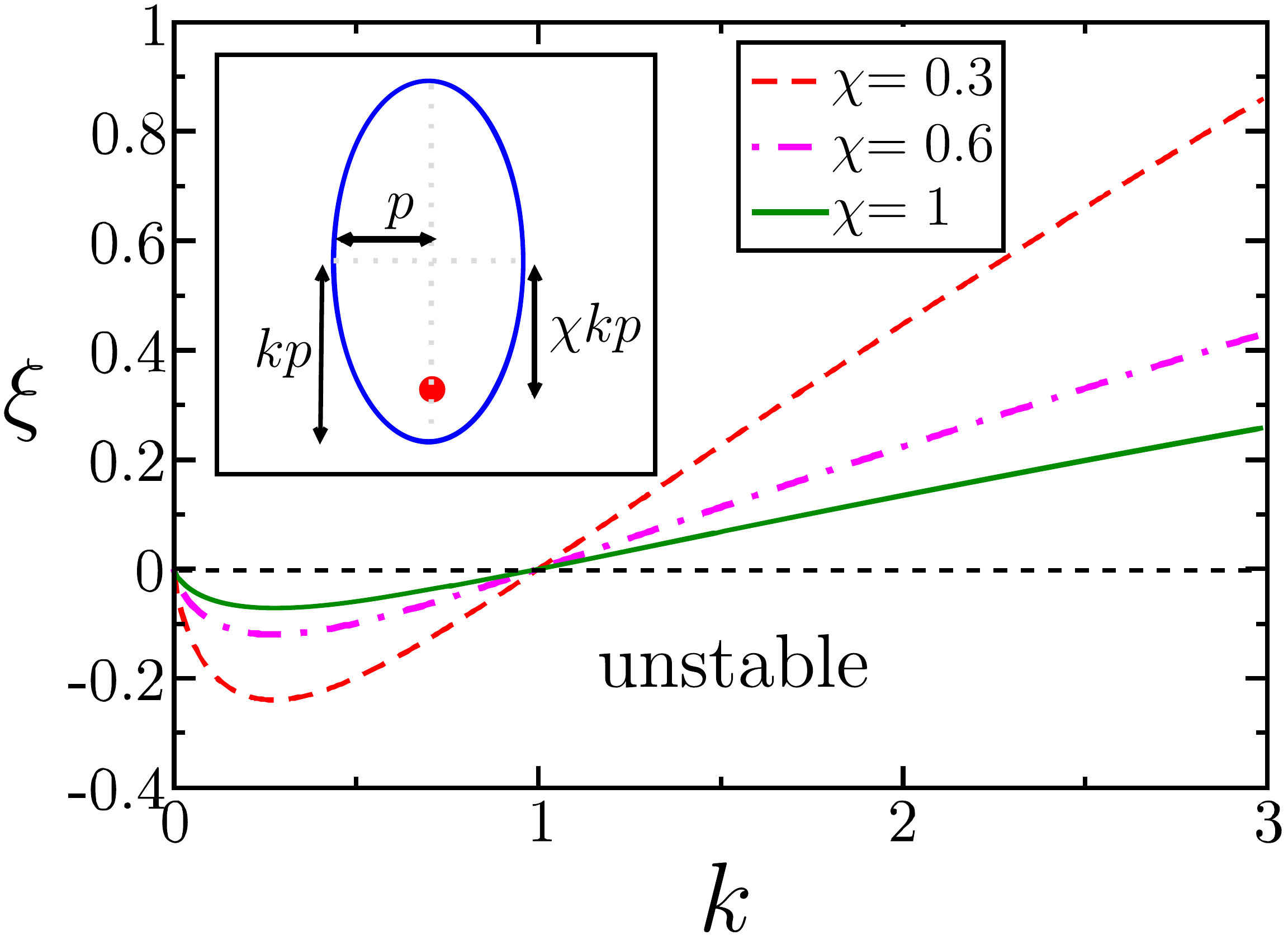}
\caption[Theoretical prediction of the effect of geometry and density distribution of particles on the sedimentation dynamics]{Theoretical prediction of the effect of geometry and density distribution of particles on the sedimentation dynamics based on linear stability analysis. The vertical axis $\xi$ represents the response parameter of the particle to the fluid flow, and the horizontal axis $k$ represents the aspect ratio of the particle. The inset sketches the geometry and the density distribution of the particle, in which the red spot denotes the center of mass of the particle, the parameter $\chi$ characterizes how the center of mass is off-centered from the geometry center. The three curves show the predictions for three different values of $\chi$, in which $k>1$ corresponds to prolate particles, and repulsion can always be observed, $k=1$ corresponds to spherical particles and there will be no repulsion as shown in Fig.\ \ref{response_to_flow}a, $k<1$ corresponds to oblate particles and the response of the particles to the flow disturbance is unstable, thus no repulsion between particles can be observed. This figure is reproduced with permission from Ref. \cite{goldfriend2017screening}, copyright (2017) by the American Physical Society. } 
\label{diagram_particle_geometry}
\end{center} 
\end{figure}

\section{Summary and outlook}

In summary, we experimentally investigated sedimentation of particles with different density distribution in quasi-2D and 3D. For the sedimentation of a single particle with uniform density distribution, the particle aligns with the fluid flow, whereas a single particle with nonuniform density distribution aligns with gravity. In quasi-2D chambers, we observed an effective repulsion during sedimentation of a pair of doublets with nonuniform density distribution. In contrast, the sedimentation dynamics of a pair of doublets with uniform density distribution are more complex, we did not observe an repulsion between such particles and we occasional observed flipping during sedimentation. In addition, we found that the repulsion still plays a key role in the sedimentation of many particles with nonuniform density distributions, which is confirmed by the probability density function of particle spacing shown in Figure\ \ref{statistics_sedimentation_many_pairs}.

For the experiment in 3D, we quantified the uniformity of the final sedimented layer of particles, and found that the particles with nonuniform mass density form layers of lower number density when compared to particles with uniform mass density as indicated in Figs.\ \ref{sedimentation_Al_Al_3D}, \ref{sedimentation_Al_steel_3D} and \ref{histogram_sedimentation_3D}. This suggests that the repulsion between particles with nonuniform density distribution still plays a role in 3D scenario. Our experimental results on the sedimentation of particles with different density distributions in 2D system validate the main predictions by Goldfriend \textit{et al.} \cite{goldfriend2017screening}. Moreover, the theoretical analysis in Ref.\ \cite{goldfriend2017screening} is restricted in 2D, whereas our results in 3D chambers also show that those mechanisms also apply to 3D. We expect our experimental results on the density-distribution-dependent sedimentation dynamics could provide a novel tunable route towards self-assembly and controlling the uniformity of particulate films after sedimentation.

The future direction in this project is to vary the geometry and density distribution of the particles. Figure\ \ref{diagram_particle_geometry} shows the phase diagram for the effective response parameter $\xi$ with respect to the aspect ratio of the particle $\kappa$ based on linear stability analysis \cite{goldfriend2017screening}. Theoretically, when $\xi>0$, the density fluctuation of particles will be suppressed, which is driven by the effective repulsion between particles with nonuniform density distributions or the self-alignability of particles. Whereas for the scenario of $\xi<0$, there is no suppression of density fluctuation and the uniform suspension becomes unstable to density fluctuations, which in turn means that no effective repulsion should be observed between particles with $\kappa<1$ during sedimentation. For spherical particles, $\kappa=1$, and the theory becomes invalid to predict the stability of the suspension. In addition, it is also interesting to investigate the effect of system size on the dynamics of sedimentation, which is still controversial hitherto \cite{caflisch1985variance,nicolai1995effect,ladd1996hydrodynamic,segre1997long,ladd1997sedimentation}, and we leave these open questions for future research.

\chapter{Summary}
\label{summary}

During my Ph.D. research activities, I experimentally investigated the nonequilibrium dynamics of pattern formation by studying a few laboratory-accessible examples in fluid and granular systems. I primarily focused on the formation of diverse spatial patterns in different experiments, and the mechanisms we uncover represent distinct pattern formation phenomena with different physics. This is perhaps not surprising since they are all nonequilibrium in nature. Although the nonequilibrium dynamics that give rise to patterns in real world can be rather complicated given the complex environment they are located in, we expect our findings to enhance our understanding of how nonequilibrium dynamics determine tremendous groups of patterns in nature.

In this thesis, I particularly reported three stories about spatial pattern formation under nonequilibrium conditions, and the main conclusions in each story are summarized as follows: 

\newpage
\noindent$\bullet$\textbf{Oscillations of Leidenfrost drops}

In fluid dynamics, most interesting and intriguing patterns are driven by multiphase-interaction-induced instabilities, such as the examples shown in Fig.\ \ref{hydrodynamic_instability_pattern}. Leidenfrost drops are known to form star-shaped oscillations, however, the complicated interplay between the solid, liquid and vapor phases as well as the highly nonequilibrium mass and heat transfer in Leidenfrost phenomenon has made the underlying mechanism for the star-shaped oscillatory pattern remain unclear. In the experiment, by focusing on the star-shaped oscillation frequency of Leidenfrost drops and pressure variations in the sustaining vapor layer, we found that the star-shaped oscillations are parametrically driven by the pressure in the vapor layer. We furthermore assumed that the pressure variations in the vapor are induced by the capillary waves underneath the drop, which was confirmed by imaging the bottom surface of the drop. The capillary waves with characteristic wavelength are generated by a large shear stress at the liquid-vapor interface due to the rapid flow of evaporated vapor, and travel form the drop center to the edge. This leads to the local variations of vapor film thickness, which generates the pressure variations in the vapor. Therefore, we conclude that hydrodynamic coupling between drop interface and vapor flow leads to the star-shaped oscillations of Leidenforst drops. We also explored potential effects of thermal convection in the liquid, and found it plays a minor role in Leidenfrost drop oscillations. In addition to the study of the star-shaped oscillations found in large Leidenfrost drops, we also gave an analytical explanation for the small-amplitude axisymmetric oscillations (\textit{breathing mode}) observed in small Leidenforst drops by a simple balance of gravitational and surface tension forces. We expect the nonequilibrium dynamics we uncover behind the patterned oscillations in both large and small Leidenfrost drops are of help to understand the coupling dynamics in systems with a supporting gas layer contacting a liquid interface and a solid surface.\\

\noindent$\bullet$\textbf{Polygonal desiccation crack patterns}

Polygonal desiccation crack patterns are commonly observed in nature, and they span multiple orders of magnitude in length scale. The formation and selection of crack patterns strongly depend on nonequilibrium multiphase interactions between fluid and solid particles as well as the drying kinetics (see Fig.\ \ref{colloidal_cracks}). To date, there is still an outstanding question that whether crack patterns on different length scales share similar dynamics. To answer this question, in the experiment we dried particulate suspensions composed of various liquid and particle combinations, and found that large-scale polygonal cracks occurred in many liquid-particle combinations are driven by the shrinkage of the particulate films (capillary-induced), whereas in cornstarch-water suspensions we found that deswelling of the hygroscopic starch particles leads to hierarchical polygonal crack patterns. In addition, we found a universal power law relating polygonal crack area, film thickness, film modulus and boundary stress for multiscale polygonal crack patterns, which can help interpret geomorphological history from surface images without knowing the material property and enable to manipulate crack patterns for specific purposes. Therefore, the nonequilibrium dynamics for the formation of multiscale polygonal crack patterns we uncover in the laboratory experiment can serve as a framework for understanding polygonal crack patterns from microscopic to geologic scales. \\

\noindent$\bullet$\textbf{Sedimentation of non-Brownian particles}

Sedimentation is a crucial process both in nature and industry, which is however not well understood due to complex coupling between the particle orientation and many-body, long-range hydrodynamic interactions. In the laboratory experiment, I investigated the sedimentation dynamics of non-Brownian particles with different density distributions. By focusing on the statistical properties of the spacial patterns of particles during sedimentation (2D) and after sedimentation (3D), we observed an effective repulsion between particles with nonuniform density distribution, in contrast to particles with uniform density distribution. These results indeed highlight the important role of particle density distribution in determining the highly nonequilibrium dynamic process of sedimentation, and can shed light into a novel pathway for large-scale self-assembly of uniform particulate films by sedimentation, and furthermore can enable to shape geomorphology manually as desired.

The underlying mechanisms we uncovered for the spatial pattern formation in the above three laboratory experimental systems represent the typical examples about how nonequilibrium dynamics give rise to pattern formation in fluid and granular systems. Although realistically the situations for pattern formation are supposed to be more complicated, our results in the limited experiment can help to model studies on the nonequilibrium dynamics that more resemble the real situations of pattern formation in nature.

\clearpage 
\phantomsection
\addcontentsline{toc}{chapter}{Bibliography}
\bibliography{nonequilibrium_patterns}

\begin{thebibliography}{100}

\bibitem{libbrecht2005physics}
K.~G. Libbrecht, The physics of snow crystals, \emph{Rep. Prog. Phys.},
  \textbf{68}, 855 (2005).

\bibitem{charru2013sand}
F.~Charru, B.~Andreotti, and P.~Claudin, Sand ripples and dunes, \emph{Annu.
  Rev. Fluid Mech.}, \textbf{45}, 469 (2013).

\bibitem{cross2009pattern}
M.~Cross and H.~Greenside, \emph{Pattern Formation and Dynamics in
  Nonequilibrium Systems} (Cambridge University Press) (2009).

\bibitem{goehring2014cracking}
L.~Goehring and S.~W. Morris, Cracking mud, freezing dirt, and breaking rocks,
  \emph{Phys. Today}, \textbf{67}, 39 (2014).

\bibitem{hofmann2015hexagonal}
M.~Hofmann, R.~Anderssohn, H.-A. Bahr, H.-J. Wei{\ss}, and J.~Nellesen, Why
  hexagonal basalt columns?, \emph{Phys. Rev. Lett.}, \textbf{115}, 154301
  (2015).

\bibitem{libbrecht2008snowflakes}
K.~Libbrecht, \emph{Snowflakes} (Voyageur Press) (2008).

\bibitem{goehring2008scaling}
L.~Goehring and S.~W. Morris, Scaling of columnar joints in basalt, \emph{J.
  Geophys. Res.: Solid Earth}, \textbf{113}, B10203 (2008).

\bibitem{koch1994biological}
A.~J. Koch and H.~Meinhardt, Biological pattern formation: from basic
  mechanisms to complex structures, \emph{Rev. Mod. Phys.}, \textbf{66}, 1481
  (1994).

\bibitem{ouyang1991transition}
Q.~Ouyang and H.~L. Swinney, Transition from a uniform state to hexagonal and
  striped turing patterns, \emph{Nature}, \textbf{352}, 610 (1991).

\bibitem{sachs2005pattern}
T.~Sachs, \emph{Pattern Formation in Plant Tissues} (Cambridge University
  Press) (2005).

\bibitem{nakamasu2009interactions}
A.~Nakamasu, G.~Takahashi, A.~Kanbe, and S.~Kondo, Interactions between
  zebrafish pigment cells responsible for the generation of turing patterns,
  \emph{Proc. Natl. Acad. Sci. U.S.A}, \textbf{106}, 8429 (2009).

\bibitem{boettiger2009neural}
A.~Boettiger, B.~Ermentrout, and G.~Oster, The neural origins of shell
  structure and pattern in aquatic mollusks, \emph{Proc. Natl. Acad. Sci.
  U.S.A}, \textbf{106}, 6837 (2009).

\bibitem{fujita2006pattern}
H.~Fujita and A.~Mochizuki, Pattern formation of leaf veins by the positive
  feedback regulation between auxin flow and auxin efflux carrier, \emph{J.
  Theor. Biol.}, \textbf{241}, 541 (2006).

\bibitem{milinkovitch2013crocodile}
M.~C. Milinkovitch, L.~Manukyan, A.~Debry, N.~Di-Po{\"\i}, S.~Martin, D.~Singh,
  D.~Lambert, and M.~Zwicker, Crocodile head scales are not developmental units
  but emerge from physical cracking, \emph{Science}, \textbf{339}, 78 (2013).

\bibitem{witten1981diffusion}
T.~A. Witten and L.~M. Sander, Diffusion-limited aggregation, a kinetic
  critical phenomenon, \emph{Phys. Rev. Lett.}, \textbf{47}, 1400 (1981).

\bibitem{garcia2012growth}
J.-M. Garcia-Ruiz, E.~Louis, P.~Meakin, and L.~M. Sander, \emph{Growth Patterns
  in Physical Sciences and Biology} (Springer Science \& Business Media)
  (2012).

\bibitem{cahn1961spinodal}
J.~W. Cahn, On spinodal decomposition, \emph{Acta Mater.}, \textbf{9}, 795
  (1961).

\bibitem{cahn1965phase}
J.~W. Cahn, Phase separation by spinodal decomposition in isotropic systems,
  \emph{J. Chem. Phys.}, \textbf{42}, 93 (1965).

\bibitem{langer1980instabilities}
J.~S. Langer, Instabilities and pattern formation in crystal growth, \emph{Rev.
  Mod. Phys.}, \textbf{52}, 1 (1980).

\bibitem{cross1993pattern}
M.~C. Cross and P.~C. Hohenberg, {Pattern formation outside of equilibrium},
  \emph{Rev. Mod. Phys.}, \textbf{65}, 851 (1993).

\bibitem{aranson2006patterns}
I.~S. Aranson and L.~S. Tsimring, Patterns and collective behavior in granular
  media: Theoretical concepts, \emph{Rev. Mod. Phys.}, \textbf{78}, 641 (2006).

\bibitem{hoyle2006pattern}
R.~Hoyle and R.~B. Hoyle, \emph{Pattern Formation: An Introduction to Methods}
  (Cambridge University Press) (2006).

\bibitem{lamb1993hydrodynamics}
H.~Lamb, \emph{Hydrodynamics} (Cambridge University Press) (1975).

\bibitem{drazin1982hydrodynamic}
P.~G. Drazin and W.~H. Reid, \emph{Hydrodynamic Stability} (Cambridge
  University Press) (2004).

\bibitem{charru2011hydrodynamic}
F.~Charru, \emph{Hydrodynamic Instabilities} (Cambridge University Press)
  (2011).

\bibitem{guyon2001physical}
E.~Guyon, J.-P. Hulin, L.~Petit, and C.~D. Mitescu, \emph{Physical
  Hydrodynamics} (Oxford University Press) (2015).

\bibitem{bischofberger2014fingering}
I.~Bischofberger, R.~Ramachandran, and S.~R. Nagel, Fingering versus stability
  in the limit of zero interfacial tension, \emph{Nat. Commun.}, \textbf{5},
  5265 (2014).

\bibitem{ma2015many}
X.~Ma, J.-J. Li\'{e}tor-Santos, and J.~C. Burton, {The many faces of a
  Leidenfrost drop}, \emph{Phys. Fluids}, \textbf{27}, 091109 (2015).

\bibitem{benard1901tourbillons}
H.~B{\'e}nard, {Les tourbillons cellulaires dans une nappe
  liquide.-M{\'e}thodes optiques d'observation et d'enregistrement},
  \emph{Journal de Physique Th{\'e}orique et Appliqu{\'e}e}, \textbf{10}, 254
  (1901).

\bibitem{rayleigh1916lix}
L.~Rayleigh, {LIX. On convection currents in a horizontal layer of fluid, when
  the higher temperature is on the under side}, \emph{Philos. Mag.},
  \textbf{32}, 529 (1916).

\bibitem{marangoni1871ausbreitung}
C.~Marangoni, {{\"U}ber die Ausbreitung der Tropfen einer Fl{\"u}ssigkeit auf
  der Oberfl{\"a}che einer anderen}, \emph{Ann. Phys.}, \textbf{219}, 337
  (1871).

\bibitem{schatz2001experiments}
M.~F. Schatz and G.~P. Neitzel, {Experiments on thermocapillary instabilities},
  \emph{Annu. Rev. Fluid Mech.}, \textbf{33}, 93 (2001).

\bibitem{maroto2007introductory}
J.~A. Maroto, V.~P{\'e}rez-Munuzuri, and M.~S. Romero-Cano, {Introductory
  analysis of B{\'e}nard--Marangoni convection}, \emph{Eur. J. Phys.},
  \textbf{28}, 311 (2007).

\bibitem{saffman1958penetration}
P.~G. Saffman and G.~I. Taylor, The penetration of a fluid into a porous medium
  or hele-shaw cell containing a more viscous liquid, \emph{Proc. R. Soc. Lond.
  A}, \textbf{245}, 312 (1958).

\bibitem{homsy1987viscous}
G.~M. Homsy, Viscous fingering in porous media, \emph{Annu. Rev. Fluid Mech.},
  \textbf{19}, 271 (1987).

\bibitem{marcus1988numerical}
P.~S. Marcus, Numerical simulation of jupiter's great red spot, \emph{Nature},
  \textbf{331}, 693 (1988).

\bibitem{foullon2011magnetic}
C.~Foullon, E.~Verwichte, V.~M. Nakariakov, K.~Nykyri, and C.~J. Farrugia,
  Magnetic kelvin-helmholtz instability at the sun, \emph{Astrophys. J. Lett.},
  \textbf{729}, L8 (2011).

\bibitem{smyth2012ocean}
W.~D. Smyth and J.~N. Moum, Ocean mixing by kelvin-helmholtz instability,
  \emph{Oceanography}, \textbf{25}, 140 (2012).

\bibitem{biance2003leidenfrost}
A.-L. Biance, C.~Clanet, and D.~Qu{\'e}r{\'e}, {Leidenfrost drops}, \emph{Phys.
  Fluids}, \textbf{15}, 1632 (2003).

\bibitem{quere2013leidenfrost}
D.~Qu{\'e}r{\'e}, {Leidenfrost dynamics}, \emph{Annu. Rev. Fluid Mech.},
  \textbf{45}, 197 (2013).

\bibitem{ma2017star}
X.~Ma, J.-J. Li{\'e}tor-Santos, and J.~C. Burton, {Star-shaped oscillations of
  Leidenfrost drops}, \emph{Phys. Rev. Fluids}, \textbf{2}, 031602 (2017).

\bibitem{ma2018self}
X.~Ma and J.~C. Burton, {Self-organized oscillations of Leidenfrost drops},
  \emph{J. Fluid Mech.}, \textbf{846}, 263 (2018).

\bibitem{jing2012formation}
G.~Jing and J.~Ma, Formation of circular crack pattern in deposition
  self-assembled by drying nanoparticle suspension, \emph{J. Phys. Chem. B},
  \textbf{116}, 6225 (2012).

\bibitem{lazarus2011craquelures}
V.~Lazarus and L.~Pauchard, From craquelures to spiral crack patterns:
  influence of layer thickness on the crack patterns induced by desiccation,
  \emph{Soft Matter}, \textbf{7}, 2552 (2011).

\bibitem{dufresne2003flow}
E.~R. Dufresne, E.~I. Corwin, N.~A. Greenblatt, J.~Ashmore, D.~Y. Wang, A.~D.
  Dinsmore, J.~X. Cheng, X.~S. Xie, J.~W. Hutchinson, and D.~A. Weitz, Flow and
  fracture in drying nanoparticle suspensions, \emph{Phys. Rev. Lett.},
  \textbf{91}, 224501 (2003).

\bibitem{routh2013drying}
A.~F. Routh, Drying of thin colloidal films, \emph{Rep. Prog. Phys.},
  \textbf{76}, 046603 (2013).

\bibitem{brutin2011pattern}
D.~Brutin, B.~Sobac, B.~Loquet, and J.~Sampol, Pattern formation in drying
  drops of blood, \emph{J. Fluid Mech.}, \textbf{667}, 85 (2011).

\bibitem{giorgiutti2015striped}
F.~Giorgiutti-Dauphin{\'e} and L.~Pauchard, Striped patterns induced by
  delamination of drying colloidal films, \emph{Soft Matter}, \textbf{11}, 1397
  (2015).

\bibitem{goehring2017drying}
L.~Goehring, J.~Li, and P.-C. Kiatkirakajorn, Drying paint: from micro-scale
  dynamics to mechanical instabilities, \emph{Phil. Trans. R. Soc. A},
  \textbf{375}, 20160161 (2017).

\bibitem{goehring2010evolution}
L.~Goehring, R.~Conroy, A.~Akhter, W.~J. Clegg, and A.~F. Routh, Evolution of
  mud-crack patterns during repeated drying cycles, \emph{Soft Matter},
  \textbf{6}, 3562 (2010).

\bibitem{Maarry2010mars}
M.~R. {El Maarry}, W.~J. Markiewicz, M.~T. Mellon, W.~Goetz, J.~M. Dohm, and
  A.~Pack, Crater floor polygons: Desiccation patterns of ancient lakes on
  {M}ars?, \emph{J. Geophys. Res.}, \textbf{115}, E10006 (2010).

\bibitem{el2014potential}
M.~R. El-Maarry, W.~Watters, N.~K. McKeown, J.~Carter, E.~N. Dobrea, J.~L.
  Bishop, A.~Pommerol, and N.~Thomas, Potential desiccation cracks on mars: A
  synthesis from modeling, analogue-field studies, and global observations,
  \emph{Icarus}, \textbf{241}, 248 (2014).

\bibitem{prosser2012avoiding}
J.~H. Prosser, T.~Brugarolas, S.~Lee, A.~J. Nolte, and D.~Lee, Avoiding cracks
  in nanoparticle films, \emph{Nano Lett.}, \textbf{12}, 5287 (2012).

\bibitem{hatton2010assembly}
B.~Hatton, L.~Mishchenko, S.~Davis, K.~H. Sandhage, and J.~Aizenberg, Assembly
  of large-area, highly ordered, crack-free inverse opal films, \emph{Proc.
  Natl. Acad. Sci. U.S.A}, \textbf{107}, 10354 (2010).

\bibitem{liu2016surface}
T.~Liu, H.~Luo, J.~Ma, W.~Xie, Y.~Wang, and G.~Jing, Surface roughness induced
  cracks of the deposition film from drying colloidal suspension, \emph{Eur.
  Phys. J. E}, \textbf{39}, 24 (2016).

\bibitem{nam2012patterning}
K.~H. Nam, I.~H. Park, and S.~H. Ko, Patterning by controlled cracking,
  \emph{Nature}, \textbf{485}, 221 (2012).

\bibitem{zeid2013influence}
W.~B. Zeid and D.~Brutin, Influence of relative humidity on spreading, pattern
  formation and adhesion of a drying drop of whole blood, \emph{Colloids Surf.
  A}, \textbf{430}, 1 (2013).

\bibitem{russel1991colloidal}
W.~B. Russel, D.~A. Saville, and W.~R. Schowalter, \emph{Colloidal Dispersions}
  (Cambridge University Press) (1991).

\bibitem{de1999granular}
P.-G. de~Gennes, Granular matter: a tentative view, \emph{Rev. Mod. Phys.},
  \textbf{71}, S374 (1999).

\bibitem{andreotti2013granular}
B.~Andreotti, Y.~Forterre, and O.~Pouliquen, \emph{Granular Media: Between
  Fluid and Solid} (Cambridge University Press) (2013).

\bibitem{herminghaus2013wet}
S.~Herminghaus, \emph{Wet Granular Matter: A Truly Complex Fluid} (World
  Scientific) (2013).

\bibitem{franklin2016handbook}
S.~V. Franklin and M.~D. Shattuck, \emph{Handbook of Granular Materials} (CRC
  Press) (2016).

\bibitem{goehring2015desiccation}
L.~Goehring, A.~Nakahara, T.~Dutta, S.~Tarafdar, and S.~Kitsunezaki,
  \emph{Desiccation Cracks and Their Patterns: Formation and Modelling in
  Science and Nature} (John Wiley \& Sons) (2015).

\bibitem{allain1995regular}
C.~Allain and L.~Limat, Regular patterns of cracks formed by directional drying
  of a collodial suspension, \emph{Phys. Rev. Lett.}, \textbf{74}, 2981 (1995).

\bibitem{inasawa2012self}
S.~Inasawa and Y.~Yamaguchi, Self-organized pattern formation of cracks
  perpendicular to the drying direction of a colloidal suspension, \emph{Soft
  Matter}, \textbf{8}, 2416 (2012).

\bibitem{goehring2011wavy}
L.~Goehring, W.~J. Clegg, and A.~F. Routh, Wavy cracks in drying colloidal
  films, \emph{Soft Matter}, \textbf{7}, 7984 (2011).

\bibitem{nandakishore2016crack}
P.~Nandakishore and L.~Goehring, Crack patterns over uneven substrates,
  \emph{Soft Matter}, \textbf{12}, 2253 (2016).

\bibitem{kiatkirakajorn2015formation}
P.-C. Kiatkirakajorn and L.~Goehring, Formation of shear bands in drying
  colloidal dispersions, \emph{Phys. Rev. Lett.}, \textbf{115}, 088302 (2015).

\bibitem{neda2002spiral}
Z.~N\'{e}da, K.-t. Leung, L.~J\'{o}zsa, and M.~Ravasz, Spiral cracks in drying
  precipitates, \emph{Phys. Rev. Lett.}, \textbf{88}, 095502 (2002).

\bibitem{vermorel2010radial}
R.~Vermorel, N.~Vandenberghe, and E.~Villermaux, Radial cracks in perforated
  thin sheets, \emph{Phys. Rev. Lett.}, \textbf{104}, 175502 (2010).

\bibitem{darwich2012highly}
S.~Darwich, K.~Mougin, and H.~Haidara, From highly ramified, large scale
  dendrite patterns of drying “alginate/au nps” solutions to capillary
  fabrication of lab-scale composite hydrogel microfibers, \emph{Soft Matter},
  \textbf{8}, 1155 (2012).

\bibitem{xu2010imaging}
Y.~Xu, W.~C. Engl, E.~R. Jerison, K.~J. Wallenstein, C.~Hyland, L.~A. Wilen,
  and E.~R. Dufresne, Imaging in-plane and normal stresses near an interface
  crack using traction force microscopy, \emph{Proc. Natl. Acad. Sci. U.S.A},
  \textbf{107}, 14964 (2010).

\bibitem{giorgiutti2016painting}
F.~Giorgiutti-Dauphin{\'e} and L.~Pauchard, Painting cracks: A way to
  investigate the pictorial matter, \emph{J. Appl. Phys.}, \textbf{120}, 065107
  (2016).

\bibitem{bohn2005hierarchicalI}
S.~Bohn, L.~Pauchard, and Y.~Couder, Hierarchical crack pattern as formed by
  successive domain divisions, \emph{Phys. Rev. E}, \textbf{71}, 046214 (2005).

\bibitem{bohn2005hierarchicalII}
S.~Bohn, J.~Platkiewicz, B.~Andreotti, M.~Adda-Bedia, and Y.~Couder,
  Hierarchical crack pattern as formed by successive domain divisions. ii. from
  disordered to deterministic behavior, \emph{Phys. Rev. E}, \textbf{71},
  046215 (2005).

\bibitem{pauchard2003morphologies}
L.~Pauchard, M.~Adda-Bedia, C.~Allain, and Y.~Couder, Morphologies resulting
  from the directional propagation of fractures, \emph{Phys. Rev. E},
  \textbf{67}, 027103 (2003).

\bibitem{boulogne2013annular}
F.~Boulogne, L.~Pauchard, and F.~Giorgiutti-Dauphin{\'e}, Annular cracks in
  thin films of nanoparticle suspensions drying on a fiber, \emph{Europhys.
  Lett.}, \textbf{102}, 39002 (2013).

\bibitem{allen1987desiccation}
J.~R.~L. Allen, Desiccation of mud in the temperate intertidal zone: studies
  from the severn estuary and eastern england, \emph{Philos. Trans. Royal Soc.
  B}, \textbf{315}, 127 (1987).

\bibitem{gauthier2010shrinkage}
G.~Gauthier, V.~Lazarus, and L.~Pauchard, Shrinkage star-shaped cracks:
  Explaining the transition from 90 degrees to 120 degrees, \emph{Europhys.
  Lett.}, \textbf{89}, 26002 (2010).

\bibitem{nawaz2008effects}
Q.~Nawaz and Y.~Rharbi, Effects of the nanomechanical properties of polymer
  nanoparticles on crack patterns during drying of colloidal suspensions,
  \emph{Macromolecules}, \textbf{41}, 59283 (2008).

\bibitem{pauchard1999influence}
L.~Pauchard, F.~Parisse, and C.~Allain, Influence of salt content on crack
  patterns formed through colloidal suspension desiccation, \emph{Phys. Rev.
  E}, \textbf{59}, 3737 (1999).

\bibitem{liu2014tuning}
T.~Liu, H.~Luo, J.~Ma, P.~Wang, L.~Wang, and G.~Jing, Tuning crack pattern by
  phase separation in the drying of binary colloid--polymer suspension,
  \emph{Phys. Lett. A}, \textbf{378}, 1191 (2014).

\bibitem{nakahara2006transition}
A.~Nakahara and Y.~Matsuo, Transition in the pattern of cracks resulting from
  memory effects in paste, \emph{Phys. Rev. E}, \textbf{74}, 045102 (2006).

\bibitem{nakahara2006imprinting}
A.~Nakahara and Y.~Matsuo, Imprinting memory into paste to control crack
  formation in drying process, \emph{J. Stat. Mech. Theory Exp.},
  \textbf{2006}, P07016 (2006).

\bibitem{boulogne2012effect}
F.~Boulogne, L.~Pauchard, and F.~Giorgiutti-Dauphin{\'e}, Effect of a
  non-volatile cosolvent on crack patterns induced by desiccation of a
  colloidal gel, \emph{Soft Matter}, \textbf{8}, 8505 (2012).

\bibitem{giorgiutti2014elapsed}
F.~Giorgiutti-Dauphin{\'e} and L.~Pauchard, Elapsed time for crack formation
  during drying, \emph{Eur. Phys. J. E}, \textbf{37}, 39 (2014).

\bibitem{khatun2012electric}
T.~Khatun, M.~D. Choudhury, T.~Dutta, and S.~Tarafdar, Electric-field-induced
  crack patterns: Experiments and simulation, \emph{Phys. Rev. E}, \textbf{86},
  016114 (2012).

\bibitem{pauchard2008crack}
L.~Pauchard, F.~Elias, P.~Boltenhagen, A.~Cebers, and J.~C. Bacri, When a crack
  is oriented by a magnetic field, \emph{Phys. Rev. E}, \textbf{77}, 021402
  (2008).

\bibitem{goehring2013plasticity}
L.~Goehring, W.~J. Clegg, and A.~F. Routh, Plasticity and fracture in drying
  colloidal films, \emph{Phys. Rev. Lett.}, \textbf{110}, 024301 (2013).

\bibitem{ma2018universal}
X.~Ma, J.~Lowensohn, and J.~C. Burton, Universal scaling of polygonal
  desiccation crack patterns, \emph{Phys. Rev. E}, \textbf{99}, 012802 (2019).

\bibitem{allen1970physical}
J.~R.~L. Allen, \emph{Physical Processes of Sedimentation} (American Elsevier
  Pub. Co.) (1970).

\bibitem{vanoni2006sedimentation}
V.~A. Vanoni, Sedimentation engineering (American Society of Civil Engineers)
  (2006).

\bibitem{batchelor1972sedimentation}
G.~K. Batchelor, Sedimentation in a dilute dispersion of spheres, \emph{J.
  Fluid Mech.}, \textbf{52}, 245 (1972).

\bibitem{davis1985sedimentation}
R.~H. Davis and A.~Acrivos, Sedimentation of noncolloidal particles at low
  reynolds numbers, \emph{Annn. Rev. Fluid Mech.}, \textbf{17}, 91 (1985).

\bibitem{guazzelli2011fluctuations}
E.~Guazzelli and J.~Hinch, Fluctuations and instability in sedimentation,
  \emph{Annn. Rev. Fluid Mech.}, \textbf{43}, 97 (2011).

\bibitem{piazza2014settled}
R.~Piazza, Settled and unsettled issues in particle settling, \emph{Rep. Prog.
  Phys.}, \textbf{77}, 056602 (2014).

\bibitem{mucha2004model}
P.~J. Mucha, S.-Y. Tee, D.~A. Weitz, B.~I. Shraiman, and M.~P. Brenner, A model
  for velocity fluctuations in sedimentation, \emph{J. Fluid Mech.},
  \textbf{501}, 71 (2004).

\bibitem{goldfriend2017screening}
T.~Goldfriend, H.~Diamant, and T.~A. Witten, Screening, hyperuniformity, and
  instability in the sedimentation of irregular objects, \emph{Phys. Rev.
  Lett.}, \textbf{118}, 158005 (2017).

\bibitem{lautrup2011}
B.~Lautrup, \emph{Physics of Continuous Matter} (CRC Press, Boca Raton, FL)
  (2011).

\bibitem{guazzelli2011physical}
E.~Guazzelli and J.~F. Morris, \emph{A Physical Introduction to Suspension
  Dynamics} (Cambridge University Press) (2011).

\bibitem{jeffery1922motion}
G.~B. Jeffery, The motion of ellipsoidal particles immersed in a viscous fluid,
  \emph{Proc. R. Soc. Lond. A}, \textbf{102}, 161 (1922).

\bibitem{xue1992diffusion}
J.-Z. Xue, E.~Herbolzheimer, M.~A. Rutgers, W.~B. Russel, and P.~M. Chaikin,
  Diffusion, dispersion, and settling of hard spheres, \emph{Phys. Rev. Lett.},
  \textbf{69}, 1715 (1992).

\bibitem{segre1997long}
P.~N. Segre, E.~Herbolzheimer, and P.~M. Chaikin, Long-range correlations in
  sedimentation, \emph{Phys. Rev. Lett.}, \textbf{79}, 2574 (1997).

\bibitem{ramaswamy2001issues}
S.~Ramaswamy, Issues in the statistical mechanics of steady sedimentation,
  \emph{Adv. Phys.}, \textbf{50}, 297 (2001).

\bibitem{padding2004hydrodynamic}
J.~T. Padding and A.~A. Louis, Hydrodynamic and brownian fluctuations in
  sedimenting suspensions, \emph{Phys. Rev. Lett.}, \textbf{93}, 220601 (2004).

\bibitem{jung2006periodic}
S.~Jung, S.~E. Spagnolie, K.~Parikh, M.~Shelley, and A.-K. Tornberg, Periodic
  sedimentation in a stokesian fluid, \emph{Phys. Rev. E}, \textbf{74}, 035302
  (2006).

\bibitem{hinch1977averaged}
E.~J. Hinch, An averaged-equation approach to particle interactions in a fluid
  suspension, \emph{J. Fluid Mech.}, \textbf{83}, 695 (1977).

\bibitem{koch1991screening}
D.~L. Koch and E.~S.~G. Shaqfeh, Screening in sedimenting suspensions, \emph{J.
  Fluid Mech.}, \textbf{224}, 275 (1991).

\bibitem{caflisch1985variance}
R.~E. Caflisch and J.~H.~C. Luke, Variance in the sedimentation speed of a
  suspension, \emph{Phys. Fluids}, \textbf{28}, 759 (1985).

\bibitem{brenner1999screening}
M.~P. Brenner, Screening mechanisms in sedimentation, \emph{Phys. Fluids},
  \textbf{11}, 754 (1999).

\bibitem{hunter2012physics}
G.~L. Hunter and E.~R. Weeks, The physics of the colloidal glass transition,
  \emph{Rep. Prog. Phys.}, \textbf{75}, 066501 (2012).

\bibitem{chajwa2018kepler}
R.~Chajwa, N.~Menon, and S.~Ramaswamy, Kepler orbits of settling discs,
  \emph{arXiv:1803.10269} (2018).

\bibitem{leidenfrost1756aquae}
J.~G. Leidenfrost, \emph{{De aquae communis nonnullis qualitatibus tractatus}}
  (Ovenius) (1756).

\bibitem{burton2012geometry}
J.~C. Burton, A.~L. Sharpe, R.~C.~A. van~der Veen, A.~Franco, and S.~R. Nagel,
  {Geometry of the vapor layer under a Leidenfrost drop}, \emph{Phys. Rev.
  Lett.}, \textbf{109}, 074301 (2012).

\bibitem{myers2009mathematical}
T.~G. Myers and J.~P.~F. Charpin, {A mathematical model of the Leidenfrost
  effect on an axisymmetric droplet}, \emph{Phys. Fluids}, \textbf{21}, 063101
  (2009).

\bibitem{pomeau2012leidenfrost}
Y.~Pomeau, M.~Le~Berre, F.~Celestini, and T.~Frisch, {The Leidenfrost effect:
  From quasi-spherical droplets to puddles}, \emph{C. R. Mecanique},
  \textbf{340}, 867 (2012).

\bibitem{xu2013hydrodynamics}
X.~Xu and T.~Qian, {Hydrodynamics of Leidenfrost droplets in one-component
  fluids}, \emph{Phys. Rev. E}, \textbf{87}, 043013 (2013).

\bibitem{sobac2014leidenfrost}
B.~Sobac, A.~Rednikov, S.~Dorbolo, and P.~Colinet, {Leidenfrost effect:
  Accurate drop shape modeling and refined scaling laws}, \emph{Phys. Rev. E},
  \textbf{90}, 053011 (2014).

\bibitem{hidalgo2016leidenfrost}
S.~Hidalgo-Caballero, Y.~Escobar-Ortega, and F.~Pacheco-V{\'a}zquez,
  {Leidenfrost phenomenon on conical surfaces}, \emph{Phys. Rev. Fluids},
  \textbf{1}, 051902 (2016).

\bibitem{maquet2016leidenfrost}
L.~Maquet, B.~Sobac, B.~Darbois-Texier, A.~Duchesne, M.~Brandenbourger,
  A.~Rednikov, P.~Colinet, and S.~Dorbolo, {Leidenfrost drops on a heated
  liquid pool}, \emph{Phys. Rev. Fluids}, \textbf{1}, 053902 (2016).

\bibitem{wong2017non}
C.~Y.~H. Wong, M.~Adda-Bedia, and D.~Vella, {Non-wetting drops at liquid
  interfaces: from liquid marbles to Leidenfrost drops}, \emph{Soft Matter},
  \textbf{13}, 5250 (2017).

\bibitem{duchemin2005static}
L.~Duchemin, J.~R. Lister, and U.~Lange, {Static shapes of levitated viscous
  drops}, \emph{J. Fluid Mech.}, \textbf{533}, 161 (2005).

\bibitem{lister2008shape}
J.~R. Lister, A.~B. Thompson, A.~Perriot, and L.~Duchemin, {Shape and stability
  of axisymmetric levitated viscous drops}, \emph{J. Fluid Mech.},
  \textbf{617}, 167 (2008).

\bibitem{snoeijer2009maximum}
J.~H. Snoeijer, P.~Brunet, and J.~Eggers, {Maximum size of drops levitated by
  an air cushion}, \emph{Phys. Rev. E}, \textbf{79}, 036307 (2009).

\bibitem{bouwhuis2013oscillating}
W.~Bouwhuis, K.~G. Winkels, I.~R. Peters, P.~Brunet, D.~van~der Meer, and J.~H.
  Snoeijer, {Oscillating and star-shaped drops levitated by an airflow},
  \emph{Phys. Rev. E}, \textbf{88}, 023017 (2013).

\bibitem{trinh2014curvature}
P.~H. Trinh, H.~Kim, N.~Hammoud, P.~D. Howell, S.~J. Chapman, and H.~A. Stone,
  {Curvature suppresses the Rayleigh-Taylor instability}, \emph{Phys. Fluids},
  \textbf{26}, 051704 (2014).

\bibitem{raux2015successive}
P.~S. Raux, G.~Dupeux, C.~Clanet, and D.~Qu{\'e}r{\'e}, {Successive
  instabilities of confined Leidenfrost puddles}, \emph{Europhys. Lett.},
  \textbf{112}, 26002 (2015).

\bibitem{maquet2015leidenfrost}
L.~Maquet, M.~Brandenbourger, B.~Sobac, A.-L. Biance, P.~Colinet, and
  S.~Dorbolo, {Leidenfrost drops: Effect of gravity}, \emph{Europhys. Lett.},
  \textbf{110}, 24001 (2015).

\bibitem{van2019asymptotic}
M.~A.~J. van Limbeek, B.~Sobac, A.~Rednikov, P.~Colinet, and J.~H. Snoeijer,
  {Asymptotic theory for a Leidenfrost drop on a liquid pool}, \emph{J. Fluid
  Mech.}, \textbf{863}, 1157 (2019).

\bibitem{vakarelski2011drag}
I.~U. Vakarelski, J.~O. Marston, D.~Y.~C. Chan, and S.~T. Thoroddsen, {Drag
  reduction by Leidenfrost vapor layers}, \emph{Phys. Rev. Lett.},
  \textbf{106}, 214501 (2011).

\bibitem{vakarelski2012stabilization}
I.~U. Vakarelski, N.~A. Patankar, J.~O. Marston, D.~Y.~C. Chan, and S.~T.
  Thoroddsen, {Stabilization of Leidenfrost vapour layer by textured
  superhydrophobic surfaces}, \emph{Nature}, \textbf{489}, 274 (2012).

\bibitem{vakarelski2014leidenfrost}
I.~U. Vakarelski, D.~Y.~C. Chan, and S.~T. Thoroddsen, {Leidenfrost vapour
  layer moderation of the drag crisis and trajectories of superhydrophobic and
  hydrophilic spheres falling in water}, \emph{Soft Matter}, \textbf{10}, 5662
  (2014).

\bibitem{linke2006self}
H.~Linke, B.~J. Alem{\'a}n, L.~D. Melling, M.~J. Taormina, M.~J. Francis, C.~C.
  Dow-Hygelund, V.~Narayanan, R.~P. Taylor, and A.~Stout, {Self-propelled
  Leidenfrost droplets}, \emph{Phys. Rev. Lett.}, \textbf{96}, 154502 (2006).

\bibitem{dupeux2011viscous}
G.~Dupeux, M.~Le~Merrer, G.~Lagubeau, C.~Clanet, S.~Hardt, and
  D.~Qu{\'e}r{\'e}, {Viscous mechanism for Leidenfrost propulsion on a
  ratchet}, \emph{Europhys. Lett.}, \textbf{96}, 58001 (2011).

\bibitem{dupeux2011trapping}
G.~Dupeux, M.~Le~Merrer, C.~Clanet, and D.~Qu{\'e}r{\'e}, {Trapping Leidenfrost
  drops with crenelations}, \emph{Phys. Rev. Lett.}, \textbf{107}, 114503
  (2011).

\bibitem{lagubeau2011leidenfrost}
G.~Lagubeau, M.~Le~Merrer, C.~Clanet, and D.~Qu{\'e}r{\'e}, {Leidenfrost on a
  ratchet}, \emph{Nat. Phys.}, \textbf{7}, 395 (2011).

\bibitem{cousins2012ratchet}
T.~R. Cousins, R.~E. Goldstein, J.~W. Jaworski, and A.~I. Pesci, {A ratchet
  trap for Leidenfrost drops}, \emph{J. Fluid Mech.}, \textbf{696}, 215 (2012).

\bibitem{li2016directional}
J.~Li, Y.~Hou, Y.~Liu, C.~Hao, M.~Li, M.~K. Chaudhury, S.~Yao, and Z.~Wang,
  {Directional transport of high-temperature Janus droplets mediated by
  structural topography}, \emph{Nat. Phys.}, \textbf{12}, 606 (2016).

\bibitem{soto2016surfing}
D.~Soto, G.~Lagubeau, C.~Clanet, and D.~Qu{\'e}r{\'e}, Surfing on a
  herringbone, \emph{Phys. Rev. Fluids}, \textbf{1}, 013902 (2016).

\bibitem{sobac2017self}
B.~Sobac, A.~Rednikov, S.~Dorbolo, and P.~Colinet, {Self-propelled Leidenfrost
  drops on a thermal gradient: A theoretical study}, \emph{Phys. Fluids},
  \textbf{29}, 082101 (2017).

\bibitem{bouillant2018leidenfrost}
A.~Bouillant, T.~Mouterde, P.~Bourrianne, A.~Lagarde, C.~Clanet, and
  D.~Qu{\'e}r{\'e}, Leidenfrost wheels, \emph{Nat. Phys.}, \textbf{14}, 1188
  (2018).

\bibitem{biance2006elasticity}
A.-L. Biance, F.~Chevy, C.~Clanet, G.~Lagubeau, and D.~Qu{\'e}r{\'e}, {On the
  elasticity of an inertial liquid shock}, \emph{J. Fluid Mech.}, \textbf{554},
  47 (2006).

\bibitem{tran2012drop}
T.~Tran, H.~J.~J. Staat, A.~Prosperetti, C.~Sun, and D.~Lohse, {Drop impact on
  superheated surfaces}, \emph{Phys. Rev. Lett.}, \textbf{108}, 036101 (2012).

\bibitem{castanet2015drop}
G.~Castanet, O.~Caballina, and F.~Lemoine, {Drop spreading at the impact in the
  Leidenfrost boiling}, \emph{Phys. Fluids}, \textbf{27}, 063302 (2015).

\bibitem{shirota2016dynamic}
M.~Shirota, M.~A.~J. van Limbeek, C.~Sun, A.~Prosperetti, and D.~Lohse,
  {Dynamic Leidenfrost effect: relevant time and length scales}, \emph{Phys.
  Rev. Lett.}, \textbf{116}, 064501 (2016).

\bibitem{abdelaziz2013green}
R.~Abdelaziz, D.~Disci-Zayed, M.~K. Hedayati, J.-H. P{\"o}hls, A.~U. Zillohu,
  B.~Erkartal, V.~S.~K. Chakravadhanula, V.~Duppel, L.~Kienle, and M.~Elbahri,
  {Green chemistry and nanofabrication in a levitated Leidenfrost drop},
  \emph{Nat. Commun.}, \textbf{4}, 2400 (2013).

\bibitem{elbahri2007anti}
M.~Elbahri, D.~Paretkar, K.~Hirmas, S.~Jebril, and R.~Adelung, Anti-lotus
  effect for nanostructuring at the leidenfrost temperature, \emph{Adv. Mat.},
  \textbf{19}, 1262 (2007).

\bibitem{maquet2014organization}
L.~Maquet, P.~Colinet, and S.~Dorbolo, Organization of microbeads in
  leidenfrost drops, \emph{Soft Matter}, \textbf{10}, 4061 (2014).

\bibitem{moreau2019explosive}
F.~Moreau, P.~Colinet, and S.~Dorbolo, {Explosive Leidenfrost droplets},
  \emph{Phys. Rev. Fluids}, \textbf{4}, 013602 (2019).

\bibitem{bain2016accelerated}
R.~M. Bain, C.~J. Pulliam, F.~Thery, and R.~G. Cooks, {Accelerated chemical
  reactions and organic synthesis in Leidenfrost droplets}, \emph{Angew. Chem.
  Int. Ed.}, \textbf{55}, 10478 (2016).

\bibitem{kadota2007microexplosion}
T.~Kadota, H.~Tanaka, D.~Segawa, S.~Nakaya, and H.~Yamasaki, {Microexplosion of
  an emulsion droplet during Leidenfrost burning}, \emph{Proc. Combust. Inst.},
  \textbf{31}, 2125 (2007).

\bibitem{bernardin1999leidenfrost}
J.~D. Bernardin and I.~Mudawar, {The Leidenfrost point: experimental study and
  assessment of existing models}, \emph{J. Heat Transfer.}, \textbf{121}, 894
  (1999).

\bibitem{talari2018leidenfrost}
V.~Talari, P.~Behar, Y.~Lu, E.~Haryadi, and D.~Liu, {Leidenfrost drops on
  micro/nanostructured surfaces}, \emph{Front. Energy}, \textbf{12}, 1 (2018).

\bibitem{shahriari2014heat}
A.~Shahriari, J.~Wurz, and V.~Bahadur, {Heat transfer enhancement accompanying
  Leidenfrost state suppression at ultrahigh temperatures}, \emph{Langmuir},
  \textbf{30}, 12074 (2014).

\bibitem{agapov2014length}
R.~L. Agapov, J.~B. Boreyko, D.~P. Briggs, B.~R. Srijanto, S.~T. Retterer,
  C.~P. Collier, and N.~V. Lavrik, Length scale of leidenfrost ratchet switches
  droplet directionality, \emph{Nanoscale}, \textbf{6}, 9293 (2014).

\bibitem{waitukaitis2017coupling}
S.~R. Waitukaitis, A.~Zuiderwijk, A.~Souslov, C.~Coulais, and M.~van Hecke,
  {Coupling the Leidenfrost effect and elastic deformations to power sustained
  bouncing}, \emph{Nat. Phys.}, \textbf{13}, 1095 (2017).

\bibitem{wells2015sublimation}
G.~G. Wells, R.~Ledesma-Aguilar, G.~McHale, and K.~Sefiane, A sublimation heat
  engine, \emph{Nat. Commun.}, \textbf{6}, 6390 (2015).

\bibitem{van1992physics}
H.~Van~Dam, {Physics of nuclear reactor safety}, \emph{Rep. Prog. Phys.},
  \textbf{55}, 2025 (1992).

\bibitem{caswell2014dynamics}
T.~A. Caswell, {Dynamics of the vapor layer below a Leidenfrost drop},
  \emph{Phys. Rev. E}, \textbf{90}, 013014 (2014).

\bibitem{rayleigh1879capillary}
L.~Rayleigh, {On the capillary phenomena of jets}, \emph{Proc. R. Soc. London},
  \textbf{29}, 71 (1879).

\bibitem{holter1952vibrations}
N.~J. Holter and W.~R. Glasscock, {Vibrations of evaporating liquid drops},
  \emph{J. Acoust. Soc. Am.}, \textbf{24}, 682 (1952).

\bibitem{adachi1984vibration}
K.~Adachi and R.~Takaki, {Vibration of a flattened drop. I. Observation},
  \emph{J. Phys. Soc. Jpn.}, \textbf{53}, 4184 (1984).

\bibitem{strier2000nitrogen}
D.~E. Strier, A.~A. Duarte, H.~Ferrari, and G.~B. Mindlin, {Nitrogen stars:
  morphogenesis of a liquid drop}, \emph{Physica A}, \textbf{283}, 261 (2000).

\bibitem{snezhko2008pulsating}
A.~Snezhko, E.~B. Jacob, and I.~S. Aranson, {Pulsating--gliding transition in
  the dynamics of levitating liquid nitrogen droplets}, \emph{New J. Phys.},
  \textbf{10}, 043034 (2008).

\bibitem{Maleidenfrost2016}
X.~Ma, J.-J. Li{\'e}tor-Santos, and J.~C. Burton, {Star-shaped oscillations of
  Leidenfrost drops}, \emph{Phys. Rev. Fluids}, \textbf{2}, 031602 (2017).

\bibitem{noblin2005triplon}
X.~Noblin, A.~Buguin, and F.~Brochard-Wyart, {Triplon modes of puddles},
  \emph{Phys. Rev. Lett.}, \textbf{94}, 166102 (2005).

\bibitem{noblin2009vibrations}
X.~Noblin, A.~Buguin, and F.~Brochard-Wyart, {Vibrations of sessile drops},
  \emph{Eur. Phys. J. Spec. Top.}, \textbf{166}, 7 (2009).

\bibitem{shen2010parametric}
C.~L. Shen, W.~J. Xie, and B.~Wei, {Parametric resonance in acoustically
  levitated water drops}, \emph{Phys. Lett. A}, \textbf{374}, 2301 (2010).

\bibitem{shen2010parametrically}
C.~L. Shen, W.~J. Xie, and B.~Wei, {Parametrically excited sectorial
  oscillation of liquid drops floating in ultrasound}, \emph{Phys. Rev. E},
  \textbf{81}, 046305 (2010).

\bibitem{mampallil2013electrowetting}
D.~Mampallil, H.~B. Eral, A.~Staicu, F.~Mugele, and D.~van~den Ende,
  {Electrowetting-driven oscillating drops sandwiched between two substrates},
  \emph{Phys. Rev. E}, \textbf{88}, 053015 (2013).

\bibitem{brunet2011star}
P.~Brunet and J.~H. Snoeijer, {Star-drops formed by periodic excitation and on
  an air cushion--A short review}, \emph{Eur. Phys. J. Spec. Top.},
  \textbf{192}, 207 (2011).

\bibitem{takaki1985vibration}
R.~Takaki and K.~Adachi, {Vibration of a flattened drop. II. Normal mode
  analysis}, \emph{J. Phys. Soc. Jpn.}, \textbf{54}, 2462 (1985).

\bibitem{tokugawa1994mechanism}
N.~Tokugawa and R.~Takaki, {Mechanism of self-induced vibration of a liquid
  drop based on the surface tension fluctuation}, \emph{J. Phys. Soc. Jpn.},
  \textbf{63}, 1758 (1994).

\bibitem{lemmon2011nist}
E.~W. Lemmon, M.~O. McLinden, D.~G. Friend, P.~J. Linstrom, and W.~G. Mallard,
  {NIST chemistry WebBook, Nist standard reference database number 69},
  \emph{(National Institute of Standards and Technology, Gaithersburg, MD,
  2011), http://webbook.nist.gov} (2011).

\bibitem{burton2010experimental}
J.~C. Burton, F.~M. Huisman, P.~Alison, D.~Rogerson, and P.~Taborek,
  {Experimental and numerical investigation of the equilibrium geometry of
  liquid lenses}, \emph{Langmuir}, \textbf{26}, 15316 (2010).

\bibitem{yoshiyasu1996self}
N.~Yoshiyasu, K.~Matsuda, and R.~Takaki, {Self-induced vibration of a water
  drop placed on an oscillating plate}, \emph{J. Phys. Soc. Jpn.}, \textbf{65},
  2068 (1996).

\bibitem{becker1991experimental}
E.~Becker, W.~J. Hiller, and T.~A. Kowalewski, {Experimental and theoretical
  investigation of large-amplitude oscillations of liquid droplets}, \emph{J.
  Fluid Mech.}, \textbf{231}, 189 (1991).

\bibitem{smith2010modulation}
W.~R. Smith, {Modulation equations for strongly nonlinear oscillations of an
  incompressible viscous drop}, \emph{J. Fluid Mech.}, \textbf{654}, 141
  (2010).

\bibitem{landau1976course}
L.~D. Landau and E.~M. Lifshitz, \emph{Course of Theoretical Physics. Vol. 1:
  Mechanics} (Elsevier) (1976).

\bibitem{miles1990parametrically}
J.~Miles and D.~Henderson, {Parametrically forced surface waves}, \emph{Annu.
  Rev. Fluid Mech.}, \textbf{22}, 143 (1990).

\bibitem{kumar1994parametric}
K.~Kumar and L.~S. Tuckerman, {Parametric instability of the interface between
  two fluids}, \emph{J. Fluid Mech.}, \textbf{279}, 49 (1994).

\bibitem{terwagne2011tibetan}
D.~Terwagne and J.~W.~M. Bush, {Tibetan singing bowls}, \emph{Nonlinearity},
  \textbf{24}, R51 (2011).

\bibitem{Miles1957}
J.~W. Miles, {On the generation of surface waves by shear flows}, \emph{J.
  Fluid Mech.}, \textbf{3}, 185 (1957).

\bibitem{Zhang1995}
X.~Zhang, {Capillary-gravity and capillary waves generated in a wind wave tank:
  observations and theories}, \emph{J. Fluid Mech.}, \textbf{289}, 51 (1995).

\bibitem{Paquier2015}
A.~Paquier, F.~Moisy, and M.~Rabaud, {Surface deformations and wave generation
  by wind blowing over a viscous liquid}, \emph{Phys. Fluids}, \textbf{27},
  122103 (2015).

\bibitem{Zeisel2008}
A.~Zeisel, M.~Stiassnie, and Y.~Agnon, {Viscous effects on wave generation by
  strong winds}, \emph{J. Fluid Mech.}, \textbf{597}, 343 (2008).

\bibitem{chang2002complex}
H.-h. Chang and E.~A. Demekhin, \emph{{Complex Wave Dynamics on Thin Films}}
  (Elsevier) (2002).

\bibitem{leal2007advanced}
L.~G. Leal, \emph{{Advanced Transport Phenomena: Fluid Mechanics and Convective
  Transport Processes}} (Cambridge University Press) (2007).

\bibitem{haumesser2002high}
P.-H. Haumesser, J.~Bancillon, M.~Daniel, M.~Perez, and J.-P. Garandet,
  {High-temperature contactless viscosity measurements by the gas--film
  levitation technique: Application to oxide and metallic glasses}, \emph{Rev.
  Sci. Instr.}, \textbf{73}, 3275 (2002).

\bibitem{paradis2005surface}
P.-F. Paradis and T.~Ishikawa, {Surface tension and viscosity measurements of
  liquid and undercooled alumina by containerless techniques}, \emph{Jpn. J.
  Appl. Phys.}, \textbf{44}, 5082 (2005).

\bibitem{ishikawa2006noncontact}
T.~Ishikawa, J.~Yu, and P.-F. Paradis, {Noncontact surface tension and
  viscosity measurements of molten oxides with a pressurized hybrid
  electrostatic-aerodynamic levitator}, \emph{Rev. Sci. Instr.}, \textbf{77},
  053901 (2006).

\bibitem{langstaff2013aerodynamic}
D.~Langstaff, M.~Gunn, G.~N. Greaves, A.~Marsing, and F.~Kargl, {Aerodynamic
  levitator furnace for measuring thermophysical properties of refractory
  liquids}, \emph{Rev. Sci. Instr.}, \textbf{84}, 124901 (2013).

\bibitem{Xu2005}
L.~Xu, W.~W. Zhang, and S.~R. Nagel, {Drop splashing on a dry smooth surface},
  \emph{Phys. Rev. Lett.}, \textbf{94}, 184505 (2005).

\bibitem{Driscoll2011}
M.~M. Driscoll and S.~R. Nagel, {Ultrafast interference imaging of air in
  splashing dynamics}, \emph{Phys. Rev. Lett.}, \textbf{107}, 154502 (2011).

\bibitem{marchand2012}
A.~Marchand, T.~S. Chan, J.~H. Snoeijer, and B.~Andreotti, {Air entrainment by
  contact lines of a solid plate plunged into a viscous fluid}, \emph{Phys.
  Rev. Lett.}, \textbf{108}, 204501 (2012).

\bibitem{Kolinski2012}
J.~M. Kolinski, S.~M. Rubinstein, S.~Mandre, M.~P. Brenner, D.~A. Weitz, and
  L.~Mahadevan, {Skating on a film of air: drops impacting on a surface},
  \emph{Phys. Rev. Lett.}, \textbf{108}, 074503 (2012).

\bibitem{Liu2013}
Y.~Liu, P.~Tan, and L.~Xu, {Compressible air entrapment in high-speed drop
  impacts on solid surfaces}, \emph{J. Fluid Mech.}, \textbf{716}, R9 (2013).

\bibitem{Liu2015}
Y.~Liu, P.~Tan, and L.~Xu, {Kelvin-Helmholtz instability in an ultrathin air
  film causes drop splashing on smooth surfaces}, \emph{Proc. Natl. Acad. Sci.
  U.S.A.}, \textbf{112}, 3280 (2015).

\bibitem{deegan1997capillary}
R.~D. Deegan, O.~Bakajin, T.~F. Dupont, G.~Huber, S.~R. Nagel, and T.~A.
  Witten, Capillary flow as the cause of ring stains from dried liquid drops,
  \emph{Nature}, \textbf{389}, 827 (1997).

\bibitem{deegan2000contact}
R.~D. Deegan, O.~Bakajin, T.~F. Dupont, G.~Huber, S.~R. Nagel, and T.~A.
  Witten, Contact line deposits in an evaporating drop, \emph{Phys. Rev. E},
  \textbf{62}, 756 (2000).

\bibitem{deegan2000pattern}
R.~D. Deegan, Pattern formation in drying drops, \emph{Phys. Rev. E},
  \textbf{61}, 475 (2000).

\bibitem{ball1999self}
P.~Ball and N.~R. Borley, \emph{The Self-made Tapestry: Pattern Formation in
  Nature} (Oxford University Press Oxford) (1999).

\bibitem{xu2009drying}
P.~Xu, A.~S. Mujumdar, and B.~Yu, Drying-induced cracks in thin film fabricated
  from colloidal dispersions, \emph{Drying Technol.}, \textbf{27}, 636 (2009).

\bibitem{goehring2013pattern}
L.~Goehring, Pattern formation in the geosciences, \emph{Phil. Trans. R. Soc.
  A}, \textbf{371}, 20120352 (2013).

\bibitem{goehring2013evolving}
L.~Goehring, Evolving fracture patterns: columnar joints, mud cracks and
  polygonal terrain, \emph{Phil. Trans. R. Soc. A}, \textbf{371}, 20120353
  (2013).

\bibitem{griffith1921vi}
A.~A. Griffith, The phenomena of rupture and flow in solids, \emph{Phil. Trans.
  R. Soc. Lond. A}, \textbf{221}, 163 (1921).

\bibitem{hutchinson1991mixed}
J.~W. Hutchinson and Z.~Suo, Mixed mode cracking in layered materials,
  \emph{Adv. Appl. Mech.}, \textbf{29}, 63 (1991).

\bibitem{fineberg1999instability}
J.~Fineberg and M.~Marder, Instability in dynamic fracture, \emph{Phys. Rep.},
  \textbf{313}, 1 (1999).

\bibitem{bouchbinder2010dynamics}
E.~Bouchbinder, J.~Fineberg, and M.~Marder, Dynamics of simple cracks,
  \emph{Annu. Rev. Condens. Matter Phys.}, \textbf{1}, 371 (2010).

\bibitem{bouchbinder2014dynamics}
E.~Bouchbinder, T.~Goldman, and J.~Fineberg, The dynamics of rapid fracture:
  instabilities, nonlinearities and length scales, \emph{Rep. Prog. Phys.},
  \textbf{77}, 046501 (2014).

\bibitem{creton2016fracture}
C.~Creton and M.~Ciccotti, Fracture and adhesion of soft materials: a review,
  \emph{Rep. Prog. Phys.}, \textbf{79}, 046601 (2016).

\bibitem{kitsunezaki2009crack}
S.~Kitsunezaki, Crack propagation speed in the drying process of paste,
  \emph{J. Phys. Soc. Jpn.}, \textbf{78}, 064801 (2009).

\bibitem{dufresne2006dynamics}
E.~R. Dufresne, D.~J. Stark, N.~A. Greenblatt, J.~X. Cheng, J.~W. Hutchinson,
  L.~Mahadevan, and D.~A. Weitz, Dynamics of fracture in drying suspensions,
  \emph{Langmuir}, \textbf{22}, 7144 (2006).

\bibitem{lidon2014dynamics}
P.~Lidon and J.-B. Salmon, Dynamics of unidirectional drying of colloidal
  dispersions, \emph{Soft Matter}, \textbf{10}, 4151 (2014).

\bibitem{man2008direct}
W.~Man and W.~B. Russel, Direct measurements of critical stresses and cracking
  in thin films of colloid dispersions, \emph{Phys. Rev. Lett.}, \textbf{100},
  198302 (2008).

\bibitem{xu2013imaging}
Y.~Xu, G.~K. German, A.~F. Mertz, and E.~R. Dufresne, Imaging stress and strain
  in the fracture of drying colloidal films, \emph{Soft Matter}, \textbf{9},
  3735 (2013).

\bibitem{tirumkudulu2005cracking}
M.~S. Tirumkudulu and W.~B. Russel, Cracking in drying latex films,
  \emph{Langmuir}, \textbf{21}, 4938 (2005).

\bibitem{singh2007cracking}
K.~B. Singh and M.~S. Tirumkudulu, Cracking in drying colloidal films,
  \emph{Phys. Rev. Lett.}, \textbf{98}, 218302 (2007).

\bibitem{chiu1993dryingI}
R.~C. Chiu, T.~J. Garino, and M.~J. Cima, Drying of granular ceramic films: I,
  effect of processing variables on cracking behavior, \emph{J. Am. Ceram.
  Soc.}, \textbf{76}, 2257 (1993).

\bibitem{chiu1993dryingII}
R.~C. Chiu and M.~J. Cima, Drying of granular ceramic films: Ii, drying stress
  and saturation uniformity, \emph{J. Am. Ceram. Soc.}, \textbf{76}, 2769
  (1993).

\bibitem{lee2004drying}
W.~P. Lee and A.~F. Routh, Why do drying films crack?, \emph{Langmuir},
  \textbf{20}, 9885 (2004).

\bibitem{smith2011effects}
M.~I. Smith and J.~S. Sharp, Effects of substrate constraint on crack pattern
  formation in thin films of colloidal polystyrene particles, \emph{Langmuir},
  \textbf{27}, 8009 (2011).

\bibitem{ma2012possible}
J.~Ma and G.~Jing, Possible origin of the crack pattern in deposition films
  formed from a drying colloidal suspension, \emph{Phys. Rev. E}, \textbf{86},
  061406 (2012).

\bibitem{komatsu1997pattern}
T.~S. Komatsu and S.-i. Sasa, Pattern selection of cracks in directionally
  drying fracture, \emph{Jpn. J. Appl. Phys.}, \textbf{36}, 391 (1997).

\bibitem{groisman1994experimental}
A.~Groisman and E.~Kaplan, An experimental study of cracking induced by
  desiccation, \emph{Europhys. Lett.}, \textbf{25}, 415 (1994).

\bibitem{shorlin2000alumina}
K.~A. Shorlin, J.~R. {de Bruyn}, M.~Graham, and S.~W. Morris, Development and
  geometry of isotropic and directional shrinkage-crack patterns, \emph{Phys.
  Rev. E}, \textbf{61}, 6950 (2000).

\bibitem{leung2000pattern}
K.-t. Leung and Z.~N{\'e}da, Pattern formation and selection in quasistatic
  fracture, \emph{Phys. Rev. Lett.}, \textbf{85}, 662 (2000).

\bibitem{leung2010criticality}
K.-t. Leung and Z.~N{\'e}da, Criticality and pattern formation in fracture by
  residual stresses, \emph{Phys. Rev. E}, \textbf{82}, 046118 (2010).

\bibitem{flores2017mean}
J.~C. Flores, Mean-field crack networks on desiccated films and their
  applications: Girl with a pearl earring, \emph{Soft Matter}, \textbf{13},
  1352 (2017).

\bibitem{akiba2017morphometric}
Y.~Akiba, J.~Magome, H.~Kobayashi, and H.~Shima, Morphometric analysis of
  polygonal cracking patterns in desiccated starch slurries, \emph{Phys. Rev.
  E}, \textbf{96}, 023003 (2017).

\bibitem{colina2000experimental}
H.~Colina and S.~Roux, Experimental model of cracking induced by drying
  shrinkage, \emph{Eur. Phys. J. E}, \textbf{1}, 189 (2000).

\bibitem{muller1998starch}
G.~M{\"u}ller, Starch columns: Analog model for basalt columns, \emph{J.
  Geophys. Res.: Solid Earth}, \textbf{103}, 15239 (1998).

\bibitem{muller1998experimental}
G.~M{\"u}ller, Experimental simulation of basalt columns, \emph{J. Volcanol.
  Geotherm. Res.}, \textbf{86}, 93 (1998).

\bibitem{toramaru2004columnar}
A.~Toramaru and T.~Matsumoto, Columnar joint morphology and cooling rate: A
  starch-water mixture experiment, \emph{J. Geophys. Res. Solid Earth},
  \textbf{109}, B02205 (2004).

\bibitem{goehring2009drying}
L.~Goehring, Drying and cracking mechanisms in a starch slurry, \emph{Phys.
  Rev. E}, \textbf{80}, 036116 (2009).

\bibitem{goehring2009nonequilibrium}
L.~Goehring, L.~Mahadevan, and S.~W. Morris, Nonequilibrium scale selection
  mechanism for columnar jointing, \emph{Proc. Natl. Acad. Sci. U.S.A.},
  \textbf{106}, 387 (2009).

\bibitem{goehring2005order}
L.~Goehring and S.~W. Morris, Order and disorder in columnar joints,
  \emph{Europhys. Lett.}, \textbf{69}, 739 (2005).

\bibitem{goehring2010solidification}
L.~Goehring, W.~J. Clegg, and A.~F. Routh, Solidification and ordering during
  directional drying of a colloidal dispersion, \emph{Langmuir}, \textbf{26},
  9269 (2010).

\bibitem{mizuguchi2005directional}
T.~Mizuguchi, A.~Nishimoto, S.~Kitsunezaki, Y.~Yamazaki, and I.~Aoki,
  Directional crack propagation of granular water systems, \emph{Phys. Rev. E},
  \textbf{71}, 056122 (2005).

\bibitem{katz2013}
B.~L. Johnson, M.~R. Holland, J.~G. Miller, and J.~I. Katz, Ultrasonic
  attenuation and speed of sound of cornstarch suspensions, \emph{J. Acoust.
  Soc. Am.}, \textbf{133}, 1399 (2013).

\bibitem{silverstein1997studies}
D.~L. Silverstein and T.~Fort, Studies in air- water interfacial area for wet
  unsaturated particulate porous media systems, \emph{Langmuir}, \textbf{13},
  4758 (1997).

\bibitem{lu2004unsaturated}
N.~Lu and W.~J. Likos, \emph{Unsaturated Soil Mechanics} (Wiley) (2004).

\bibitem{akiba2018}
Y.~Akiba and H.~Shima, {Flow velocity-dependent transition of anisotropic crack
  patterns in CaCO$_3$ pastes}, \emph{J. Phys. Soc. Jpn.}, \textbf{88}, 024001
  (2019).

\bibitem{koeze2018sticky}
D.~J. Koeze and B.~P. Tighe, Sticky matters: Jamming and rigid cluster
  statistics with attractive particle interactions, \emph{Phys. Rev. Lett.},
  \textbf{121}, 188002 (2018).

\bibitem{landau1986theory}
L.~D. Landau and E.~M. Lifshitz, \emph{Theory of Elasticity} (Elsevier, New
  York, NY) (1986).

\bibitem{mahaut2008rheology}
F.~Mahaut, X.~Chateau, P.~Coussot, and G.~Ovarlez, Yield stress and elastic
  modulus of suspensions of noncolloidal particles in yield stress fluids,
  \emph{J. Rheology}, \textbf{52}, 287 (2008).

\bibitem{pharr1992generality}
G.~M. Pharr, W.~C. Oliver, and F.~R. Brotzen, On the generality of the
  relationship among contact stiffness, contact area, and elastic modulus
  during indentation, \emph{J. Mater. Res.}, \textbf{7}, 613 (1992).

\bibitem{oliver2004measurement}
W.~C. Oliver and G.~M. Pharr, Measurement of hardness and elastic modulus by
  instrumented indentation: Advances in understanding and refinements to
  methodology, \emph{J. Mater. Res.}, \textbf{19}, 3 (2004).

\bibitem{doerner1986method}
M.~F. Doerner and W.~D. Nix, A method for interpreting the data from
  depth-sensing indentation instruments, \emph{J. Mater. Res.}, \textbf{1}, 601
  (1986).

\bibitem{phadikar2012establishing}
J.~K. Phadikar, T.~A. Bogetti, and A.~M. Karlsson, On establishing
  elastic--plastic properties of a sphere by indentation testing, \emph{Int. J.
  Solids Struct.}, \textbf{49}, 1961 (2012).

\bibitem{broitman2017indentation}
E.~Broitman, Indentation hardness measurements at macro-, micro-, and
  nanoscale: a critical overview, \emph{Tribol. Lett.}, \textbf{65}, 23 (2017).

\bibitem{moller2007shear}
P.~C.~F. M{\o}ller and D.~Bonn, The shear modulus of wet granular matter,
  \emph{Europhys. Lett.}, \textbf{80}, 38002 (2007).

\bibitem{kezdi1974}
A.~K\'{e}zdi and L.~R\'{e}th\'{a}ti, \emph{Handbook of Soil Mechanics}
  (Elsevier, Amsterdam) (1974).

\bibitem{obrzud2018}
R.~F. Obrzud and A.~Truty, \emph{The Hardening Soil model -- A Practical
  Guidebook} (Zace Services Ltd., Switzerland) (2018).

\bibitem{brown2012thickening}
E.~Brown and H.~M. Jaeger, The role of dilation and confining stresses in shear
  thickening of dense suspensions, \emph{J. Rheol.}, \textbf{56}, 875 (2012).

\bibitem{pel2002analytic}
L.~Pel, K.~A. Landman, and E.~F. Kaasschieter, Analytic solution for the
  non-linear drying problem, \emph{Int. J. Heat Mass Transf.}, \textbf{45},
  3173 (2002).

\bibitem{pettijohn1957sedimentary}
F.~J. Pettijohn, \emph{Sedimentary rocks} (Harper \& Brothers New York) (1957).

\bibitem{selim1983sedimentation}
M.~S. Selim, A.~C. Kothari, and R.~M. Turian, Sedimentation of multisized
  particles in concentrated suspensions, \emph{AlChE J.}, \textbf{29}, 1029
  (1983).

\bibitem{guo2009removal}
X.~Guo, Z.~Wu, and M.~He, Removal of antimony (v) and antimony (iii) from
  drinking water by coagulation--flocculation--sedimentation (cfs), \emph{Water
  Res.}, \textbf{43}, 4327 (2009).

\bibitem{koch1989instability}
D.~L. Koch and E.~S.~G. Shaqfeh, The instability of a dispersion of sedimenting
  spheroids, \emph{J. Fluid Mech.}, \textbf{209}, 521 (1989).

\bibitem{herzhaft1999experimental}
B.~Herzhaft and {\'E}.~Guazzelli, Experimental study of the sedimentation of
  dilute and semi-dilute suspensions of fibres, \emph{J. Fluid Mech.},
  \textbf{384}, 133 (1999).

\bibitem{herzhaft1996experimental}
B.~Herzhaft, {\'E}.~Guazzelli, M.~B. Mackaplow, and E.~S.~G. Shaqfeh,
  Experimental investigation of the sedimentation of a dilute fiber suspension,
  \emph{Phys. Rev. Lett.}, \textbf{77}, 290 (1996).

\bibitem{nicolai1995effect}
H.~Nicolai and E.~Guazzelli, Effect of the vessel size on the hydrodynamic
  diffusion of sedimenting spheres, \emph{Phys. Fluids}, \textbf{7}, 3 (1995).

\bibitem{pedley2010collective}
T.~J. Pedley, Collective behaviour of swimming micro-organisms, \emph{Exp.
  Mech.}, \textbf{50}, 1293 (2010).

\bibitem{pedley1992hydrodynamic}
T.~J. Pedley and J.~O. Kessler, Hydrodynamic phenomena in suspensions of
  swimming microorganisms, \emph{Annu. Rev. Fluid Mech.}, \textbf{24}, 313
  (1992).

\bibitem{maddock1994mechanics}
L.~Maddock, Q.~Bone, and J.~M.~V. Rayner, \emph{The mechanics and physiology of
  animal swimming} (Cambridge University Press) (1994).

\bibitem{pedley1990new}
T.~J. Pedley and J.~O. Kessler, A new continuum model for suspensions of
  gyrotactic micro-organisms, \emph{J. Fluid Mech.}, \textbf{212}, 155 (1990).

\bibitem{guasto2012fluid}
J.~S. Guasto, R.~Rusconi, and R.~Stocker, Fluid mechanics of planktonic
  microorganisms, \emph{Annu. Rev. Fluid Mech.}, \textbf{44}, 373 (2012).

\bibitem{pennington1990consequences}
J.~T. Pennington and R.~R. Strathmann, Consequences of the calcite skeletons of
  planktonic echinoderm larvae for orientation, swimming, and shape,
  \emph{Biol. Bull.}, \textbf{179}, 121 (1990).

\bibitem{roberts1970motion}
A.~M. Roberts, Motion of spermatozoa in fluid streams, \emph{Nature},
  \textbf{228}, 375 (1970).

\bibitem{lauga2016bacterial}
E.~Lauga, Bacterial hydrodynamics, \emph{Annu. Rev. Fluid Mech.}, \textbf{48},
  105 (2016).

\bibitem{elgeti2015physics}
J.~Elgeti, R.~G. Winkler, and G.~Gompper, Physics of microswimmers—single
  particle motion and collective behavior: a review, \emph{Rep. Prog. Phys.},
  \textbf{78}, 056601 (2015).

\bibitem{manghi2006hydrodynamic}
M.~Manghi, X.~Schlagberger, Y.-W. Kim, and R.~R. Netz, Hydrodynamic effects in
  driven soft matter, \emph{Soft Matter}, \textbf{2}, 653 (2006).

\bibitem{kim1985sedimentation}
S.~Kim, Sedimentation of two arbitrarily oriented spheroids in a viscous fluid,
  \emph{Intl. J. Multiphase Flow}, \textbf{11}, 699 (1985).

\bibitem{goldfriend2015hydrodynamic}
T.~Goldfriend, H.~Diamant, and T.~A. Witten, Hydrodynamic interactions between
  two forced objects of arbitrary shape. i. effect on alignment, \emph{Phys.
  Fluids}, \textbf{27}, 123303 (2015).

\bibitem{butler2002dynamic}
J.~E. Butler and E.~S.~G. Shaqfeh, Dynamic simulations of the inhomogeneous
  sedimentation of rigid fibres, \emph{J. Fluid Mech.}, \textbf{468}, 205
  (2002).

\bibitem{stokes1851effect}
G.~G. Stokes, On the effect of the internal friction of fluids on the motion of
  pendulums, \emph{Trans. Camb. Phil. Soc.}, \textbf{9}, 8 (1851).

\bibitem{lauga2009hydrodynamics}
E.~Lauga and T.~R. Powers, The hydrodynamics of swimming microorganisms,
  \emph{Rep. Prog. Phys.}, \textbf{72}, 096601 (2009).

\bibitem{oberbeck1876uber}
A.~Oberbeck, Uber stationare flussigkeitsbewegungen mit berucksichtigung der
  inner reibung, \emph{J. Reine Angew. Math.}, \textbf{81}, 62 (1876).

\bibitem{helzel2017kinetic}
C.~Helzel and A.~E. Tzavaras, A kinetic model for the sedimentation of
  rod--like particles, \emph{Multiscale Model Simul.}, \textbf{15}, 500 (2017).

\bibitem{wang2009numerical}
J.~Wang and A.~Layton, Numerical simulations of fiber sedimentation in
  navier-stokes flows, \emph{Commun. Comput. Phys.}, \textbf{5}, 61 (2009).

\bibitem{gustavsson2009gravity}
K.~Gustavsson and A.-K. Tornberg, Gravity induced sedimentation of slender
  fibers, \emph{Phys. Fluids}, \textbf{21}, 123301 (2009).

\bibitem{tornberg2006numerical}
A.-K. Tornberg and K.~Gustavsson, A numerical method for simulations of rigid
  fiber suspensions, \emph{J. Comput. Phys.}, \textbf{215}, 172 (2006).

\bibitem{ladd1996hydrodynamic}
A.~J.~C. Ladd, Hydrodynamic screening in sedimenting suspensions of
  non-brownian spheres, \emph{Phys. Rev. Lett.}, \textbf{76}, 1392 (1996).

\bibitem{ladd1997sedimentation}
A.~J.~C. Ladd, Sedimentation of homogeneous suspensions of non-brownian
  spheres, \emph{Phys. Fluids}, \textbf{9}, 491 (1997).

\end{thebibliography}
\bibliographystyle{pre}

\end{document}